\documentclass[apjl]{emulateapj}
\usepackage{comment}
\usepackage{ifthen}

\newcommand{\forloop}[5][1]%
{%
\setcounter{#2}{#3}%
\ifthenelse{#4}%
	{%
	#5%
	\addtocounter{#2}{#1}%
	\forloop[#1]{#2}{\value{#2}}{#4}{#5}%
	}%
	{%
	}%
}%

\newcommand{\ctbd}[1]{}

\newcommand{\lc}{light curve}
\newcommand{\lcs}{light curves}
\newcommand{\Lc}{Light curve}

\newcommand{\band}[1]{\ensuremath{#1}~band}

\newcommand{\kms}{\ensuremath{\rm km\,s^{-1}}}
\newcommand{\ms}{\ensuremath{\rm m\,s^{-1}}}

\newcommand{\gcmc}{\ensuremath{\rm g\,cm^{-3}}}
\newcommand{\ergscmsq}{\ensuremath{\rm erg\,s^{-1}\,cm^{-2}}}

\newcommand{\vsini}{\ensuremath{v \sin{i}}}
\newcommand{\feh}{\ensuremath{\rm [Fe/H]}}

\newcommand{\vmac}{\ensuremath{v_{\rm mac}}}
\newcommand{\vmic}{\ensuremath{v_{\rm mic}}}

\newcommand{\vic}{\ensuremath{V\!-\!I_C}}

\newcommand{\rsun}{\ensuremath{R_\sun}}
\newcommand{\msun}{\ensuremath{M_\sun}}
\newcommand{\lsun}{\ensuremath{L_\sun}}

\newcommand{\rstar}{\ensuremath{R_\star}}
\newcommand{\mstar}{\ensuremath{M_\star}}
\newcommand{\lstar}{\ensuremath{L_\star}}

\newcommand{\teffstar}{\ensuremath{T_{\rm eff\star}}}

\newcommand{\loggstar}{\ensuremath{\log{g_{\star}}}}

\newcommand{\rpl}{\ensuremath{R_{p}}}
\newcommand{\mpl}{\ensuremath{M_{p}}}

\newcommand{\rhopl}{\ensuremath{\rho_{p}}}

\newcommand{\arstar}{\ensuremath{a/\rstar}}
\newcommand{\zrstar}{\ensuremath{\zeta/\rstar}}

\newcommand{\rjup}{\ensuremath{R_{\rm J}}}
\newcommand{\mjup}{\ensuremath{M_{\rm J}}}

\newcommand{\refsec}[1]{\mbox{\S\ \ref{sec:#1}}}

\newcommand{\reffigl}[1]{Figure~\ref{fig:#1}}
\newcommand{\refsecl}[1]{\mbox{Section \ref{sec:#1}}}

\newcommand{\reftabl}[1]{Table~\ref{tab:#1}}

\newcommand{\reffigls}[2]{Figures~\ref{fig:#1}-\ref{fig:#2}}
\newcommand{\reftabls}[2]{Tables~\ref{tab:#1}-\ref{tab:#2}}

\newcommand{\flwof}{\mbox{FLWO 1.2\,m}}

\newcommand{\loopand}{\ifnum\value{planetcounter}=3 and \else\fi}
\newcommand{\loopcomma}{\ifnum\value{planetcounter}<3 ,\else. \fi}
\newcommand{\loopcommanoperiod}{\ifnum\value{planetcounter}<3 ,\else \space\fi}
\newcommand{\loopcommanospace}{\ifnum\value{planetcounter}<3 ,\else \fi}

\newcommand{\hatcurhtrxxxxA}{HTR315-005}                               
\newcommand{\hatcurfieldxxxxA}{267}                                    
\newcommand{\hatcurCCraxxxxA}{\ensuremath{07^{\mathrm h}35^{\mathrm m}01.97{\mathrm s}}}                             
\newcommand{\hatcurCCdecxxxxA}{\ensuremath{+17{\arcdeg}49{\arcmin}48.3{\arcsec}}}                            
\newcommand{\hatcurCCmagxxxxA}{12.422}                                 
\newcommand{\hatcurCCtwomassxxxxA}{2MASS~07350197+1749482}             
\newcommand{\hatcurCCgscxxxxA}{GSC~1364-01424}                         
\newcommand{\hatcurCCtassmvxxxxA}{12.422}                              
\newcommand{\hatcurCCtwomassJmagxxxxA}{\ensuremath{11.424\pm0.020}}    
\newcommand{\hatcurCCtwomassHmagxxxxA}{\ensuremath{11.184\pm0.022}}    
\newcommand{\hatcurCCtwomassKmagxxxxA}{\ensuremath{11.157\pm0.020}}    
\newcommand{\hatcurCCcitJmagxxxxA}{\ensuremath{11.446\pm0.021}}        
\newcommand{\hatcurCCcitHmagxxxxA}{\ensuremath{11.181\pm0.022}}        
\newcommand{\hatcurCCcitKmagxxxxA}{\ensuremath{11.181\pm0.020}}        
\newcommand{\hatcurCCbbJmagxxxxA}{\ensuremath{11.487\pm0.021}}         
\newcommand{\hatcurCCbbHmagxxxxA}{\ensuremath{11.200\pm0.023}}         
\newcommand{\hatcurCCbbKmagxxxxA}{\ensuremath{11.201\pm0.020}}         
\newcommand{\hatcurCCesoJmagxxxxA}{\ensuremath{11.488\pm0.022}}        
\newcommand{\hatcurCCesoHmagxxxxA}{\ensuremath{11.193\pm0.025}}        
\newcommand{\hatcurCCesoKmagxxxxA}{\ensuremath{11.200\pm0.021}}        
\newcommand{\hatcurCCesoJHmagxxxxA}{\ensuremath{0.295\pm0.032}}        
\newcommand{\hatcurCCesoJKmagxxxxA}{\ensuremath{0.288\pm0.008}}        
\newcommand{\hatcurCCesoHKmagxxxxA}{\ensuremath{-0.008\pm0.033}}       
\newcommand{\hatcurLCdipxxxxA}{\ensuremath{10.9}}                      
\newcommand{\hatcurLCrprstarxxxxA}{\ensuremath{0.0993\pm0.0025}}       
\newcommand{\hatcurLCbsqxxxxA}{\ensuremath{0.122_{-0.061}^{+0.069}}}   
\newcommand{\hatcurLCimpxxxxA}{\ensuremath{0.349_{-0.120}^{+0.085}}}   
\newcommand{\hatcurLCzetaxxxxA}{\ensuremath{12.75\pm0.07}}             
\newcommand{\hatcurLCdurxxxxA}{\ensuremath{0.1745\pm0.0017}}           
\newcommand{\hatcurLCdurshortxxxxA}{\ensuremath{0.1745}}               
\newcommand{\hatcurLCdurhrxxxxA}{\ensuremath{4.189\pm0.040}}           
\newcommand{\hatcurLCdurhrshortxxxxA}{\ensuremath{4.189}}              
\newcommand{\hatcurLCqxxxxA}{\ensuremath{0.0492\pm0.0005}}             
\newcommand{\hatcurLCqshortxxxxA}{\ensuremath{0.049}}                  
\newcommand{\hatcurLCingdurxxxxA}{\ensuremath{0.0178\pm0.0017}}        
\newcommand{\hatcurLCPxxxxA}{\ensuremath{3.543870\pm0.000005}}         
\newcommand{\hatcurLCPprecxxxxA}{\ensuremath{3.5438703}}               
\newcommand{\hatcurLCPshortxxxxA}{\ensuremath{3.5439}}                 
\newcommand{\hatcurLCTxxxxA}{\ensuremath{2455208.75049\pm0.00041}}     
\newcommand{\hatcurLCTAxxxxA}{\ensuremath{2454429.09902\pm0.00116}}    
\newcommand{\hatcurLCTBxxxxA}{\ensuremath{2455605.66397\pm0.00083}}    
\newcommand{\hatcurLChatnetmAxxxxA}{\ensuremath{12.0199\pm0.0002}}     
\newcommand{\hatcurLCiblendAxxxxA}{\ensuremath{0.42\pm0.07}}           
\newcommand{\hatcurLChatnetmBxxxxA}{\ensuremath{12.0197\pm0.0001}}     
\newcommand{\hatcurLCiblendBxxxxA}{\ensuremath{0.62\pm0.04}}           
\newcommand{\hatcurSMEiteffxxxxA}{\ensuremath{6325\pm100}}             
\newcommand{\hatcurSMEizfehxxxxA}{\ensuremath{0.14\pm0.1}}             
\newcommand{\hatcurSMEizfehshortxxxxA}{\ensuremath{0.14}}              
\newcommand{\hatcurSMEiloggxxxxA}{\ensuremath{4.04\pm0.1}}             
\newcommand{\hatcurSMEivsinxxxxA}{\ensuremath{12.7\pm0.5}}             
\newcommand{\hatcurSMEivmacxxxxA}{\ensuremath{4.87}}                   
\newcommand{\hatcurSMEivmicxxxxA}{\ensuremath{0.85}}                   
\newcommand{\hatcurSMEiiteffxxxxA}{\ensuremath{6430\pm100}}            
\newcommand{\hatcurSMEiizfehxxxxA}{\ensuremath{0.19\pm0.10}}           
\newcommand{\hatcurSMEiizfehshortxxxxA}{\ensuremath{0.19}}             
\newcommand{\hatcurSMEiiloggxxxxA}{\ensuremath{4.16\pm0.1}}            
\newcommand{\hatcurSMEiivsinxxxxA}{\ensuremath{12.7\pm0.5}}            
\newcommand{\hatcurSMEiivmacxxxxA}{\ensuremath{5.04}}                  
\newcommand{\hatcurSMEiivmicxxxxA}{\ensuremath{0.85}}                  
\newcommand{\hatcurDSteffxxxxA}{\ensuremath{6250\pm100}}               
\newcommand{\hatcurDSzfehxxxxA}{\ensuremath{NULL\pmNULL}}              
\newcommand{\hatcurDSloggxxxxA}{\ensuremath{4.0\pm0.25}}               
\newcommand{\hatcurDSvsinixxxxA}{\ensuremath{14.0\pm1.0}}              
\newcommand{\hatcurDSgammaxxxxA}{\ensuremath{28.42\pm0.28}}            
\newcommand{\hatcurDSnumspecxxxxA}{\ensuremath{4}}                     
\newcommand{\hatcurDSspanxxxxA}{\ensuremath{123}}                      
\newcommand{\hatcurDSrvrmsxxxxA}{\ensuremath{0.55}}                    
\newcommand{\hatcurTRESteffxxxxA}{\ensuremath{6250\pm100}}             
\newcommand{\hatcurTRESzfehxxxxA}{\ensuremath{NULL\pmNULL}}            
\newcommand{\hatcurTRESloggxxxxA}{\ensuremath{3.5\pm0.5}}              
\newcommand{\hatcurTRESvsinixxxxA}{\ensuremath{16\pm0.5}}              
\newcommand{\hatcurTRESgammaxxxxA}{\ensuremath{28.70\pmNULL}}          
\newcommand{\hatcurTRESnumspecxxxxA}{\ensuremath{1}}                   
\newcommand{\hatcurTRESspanxxxxA}{\ensuremath{1}}                      
\newcommand{\hatcurTRESrvrmsxxxxA}{\ensuremath{0.00}}                  
\newcommand{\hatcurFIESteffxxxxA}{\ensuremath{NULL\pmNULL}}            
\newcommand{\hatcurFIESzfehxxxxA}{\ensuremath{NULL\pmNULL}}            
\newcommand{\hatcurFIESloggxxxxA}{\ensuremath{NULL\pmNULL}}            
\newcommand{\hatcurFIESvsinixxxxA}{\ensuremath{NULL\pmNULL}}           
\newcommand{\hatcurFIESgammaxxxxA}{\ensuremath{NULL\pmNULL}}           
\newcommand{\hatcurFIESnumspecxxxxA}{\ensuremath{0}}                   
\newcommand{\hatcurFIESspanxxxxA}{\ensuremath{0}}                      
\newcommand{\hatcurFIESrvrmsxxxxA}{\ensuremath{0.00}}                  
\newcommand{\hatcurLBizxxxxA}{\ensuremath{0.1274}}                     
\newcommand{\hatcurLBiizxxxxA}{\ensuremath{0.3714}}                    
\newcommand{\hatcurLBiixxxxA}{\ensuremath{0.1787}}                     
\newcommand{\hatcurLBiiixxxxA}{\ensuremath{0.3812}}                    
\newcommand{\hatcurLBiIxxxxA}{\ensuremath{0.1598}}                     
\newcommand{\hatcurLBiiIxxxxA}{\ensuremath{0.3793}}                    
\newcommand{\hatcurLBigxxxxA}{\ensuremath{0.4245}}                     
\newcommand{\hatcurLBiigxxxxA}{\ensuremath{0.3293}}                    
\newcommand{\hatcurLBirxxxxA}{\ensuremath{0.2554}}                     
\newcommand{\hatcurLBiirxxxxA}{\ensuremath{0.3870}}                    
\newcommand{\hatcurLBiRxxxxA}{\ensuremath{0.2335}}                     
\newcommand{\hatcurLBiiRxxxxA}{\ensuremath{0.3871}}                    
\newcommand{\hatcurLBikepxxxxA}{\ensuremath{}}                 
\newcommand{\hatcurLBiikepxxxxA}{\ensuremath{}}                
\newcommand{\hatcurISOmxxxxA}{\ensuremath{1.40\pm0.05}}                
\newcommand{\hatcurISOmshortxxxxA}{\ensuremath{1.40}}                  
\newcommand{\hatcurISOmlongxxxxA}{\ensuremath{1.404\pm0.051}}          
\newcommand{\hatcurISOrxxxxA}{\ensuremath{1.62_{-0.06}^{+0.08}}}       
\newcommand{\hatcurISOrshortxxxxA}{\ensuremath{1.62}}                  
\newcommand{\hatcurISOrlongxxxxA}{\ensuremath{1.625_{-0.062}^{+0.081}}} 
\newcommand{\hatcurISOrhoxxxxA}{\ensuremath{0.46\pm0.05}}              
\newcommand{\hatcurISOloggxxxxA}{\ensuremath{4.16\pm0.03}}             
\newcommand{\hatcurISOlumxxxxA}{\ensuremath{4.05\pm0.48}}              
\newcommand{\hatcurISOlumshortxxxxA}{\ensuremath{4.05}}                
\newcommand{\hatcurISOmvxxxxA}{\ensuremath{3.21\pm0.14}}               
\newcommand{\hatcurISOvixxxxA}{\ensuremath{0.516\pm0.025}}             
\newcommand{\hatcurISOagexxxxA}{\ensuremath{2.0\pm0.4}}                
\newcommand{\hatcurISOsigmaxxxxA}{\ensuremath{0.00030\pm0.00006}}      
\newcommand{\hatcurISOMJxxxxA}{\ensuremath{2.38\pm0.11}}               
\newcommand{\hatcurISOMHxxxxA}{\ensuremath{2.16\pm0.10}}               
\newcommand{\hatcurISOMKxxxxA}{\ensuremath{2.12\pm0.10}}               
\newcommand{\hatcurISOJKxxxxA}{\ensuremath{0.27\pm0.02}}               
\newcommand{\hatcurISOspecxxxxA}{F8}                                   
\newcommand{\hatcurRVKxxxxA}{\ensuremath{63.6\pm10.4}}                 
\newcommand{\hatcurRVkxxxxA}{\ensuremath{0.000\pm0.000}}               
\newcommand{\hatcurRVhxxxxA}{\ensuremath{0.000\pm0.000}}               
\newcommand{\hatcurRVtronexxxxA}{\ensuremath{0.0000\pm0.0000}}         
\newcommand{\hatcurRVtrtwoxxxxA}{\ensuremath{0.0000\pm0.0000}}         
\newcommand{\hatcurRVgammaxxxxA}{\ensuremath{-7.8\pm8.3}}              
\newcommand{\hatcurRVjitterxxxxA}{\ensuremath{43.0}}                   
\newcommand{\hatcurRVfitrmsxxxxA}{\ensuremath{43.8}}                   
\newcommand{\hatcurRVeccenxxxxA}{\ensuremath{0.000\pm0.000}}           
\newcommand{\hatcurRVomegaxxxxA}{\ensuremath{0\pm0}}                   
\newcommand{\hatcurPPixxxxA}{\ensuremath{87.0\pm1.0}}                  
\newcommand{\hatcurPPgxxxxA}{\ensuremath{5.9\pm1.2}}                   
\newcommand{\hatcurPPloggxxxxA}{\ensuremath{2.77\pm0.09}}              
\newcommand{\hatcurPParxxxxA}{\ensuremath{6.74\pm0.25}}                
\newcommand{\hatcurPParelxxxxA}{\ensuremath{0.0509\pm0.0006}}          
\newcommand{\hatcurPPrhoxxxxA}{\ensuremath{0.19\pm0.04}}               
\newcommand{\hatcurPPmxxxxA}{\ensuremath{0.60\pm0.10}}                 
\newcommand{\hatcurPPmshortxxxxA}{\ensuremath{0.60}}                   
\newcommand{\hatcurPPmlongxxxxA}{\ensuremath{0.599\pm0.099}}           
\newcommand{\hatcurPPmexxxxA}{\ensuremath{190.3\pm31.4}}               
\newcommand{\hatcurPPmeshortxxxxA}{\ensuremath{190.3}}                 
\newcommand{\hatcurPPmelongxxxxA}{\ensuremath{190.30\pm31.43}}         
\newcommand{\hatcurPPrxxxxA}{\ensuremath{1.57_{-0.08}^{+0.11}}}        
\newcommand{\hatcurPPrshortxxxxA}{\ensuremath{1.57}}                   
\newcommand{\hatcurPPrlongxxxxA}{\ensuremath{1.571_{-0.081}^{+0.108}}} 
\newcommand{\hatcurPPrexxxxA}{\ensuremath{17.6_{-0.9}^{+1.2}}}         
\newcommand{\hatcurPPreshortxxxxA}{\ensuremath{17.6}}                  
\newcommand{\hatcurPPrelongxxxxA}{\ensuremath{17.61_{-0.91}^{+1.21}}}  
\newcommand{\hatcurPPmrcorrxxxxA}{\ensuremath{0.09}}                   
\newcommand{\hatcurPPteffxxxxA}{\ensuremath{1752\pm43}}                
\newcommand{\hatcurPPthetaxxxxA}{\ensuremath{0.027\pm0.005}}           
\newcommand{\hatcurPPfluxperixxxxA}{\ensuremath{2.13\pm0.21}}          
\newcommand{\hatcurPPfluxperidimxxxxA}{\ensuremath{9}}                 
\newcommand{\hatcurPPfluxapxxxxA}{\ensuremath{2.13\pm0.21}}            
\newcommand{\hatcurPPfluxapdimxxxxA}{\ensuremath{9}}                   
\newcommand{\hatcurPPfluxavgxxxxA}{\ensuremath{2.13\pm0.21}}           
\newcommand{\hatcurPPfluxavgdimxxxxA}{\ensuremath{9}}                  
\newcommand{\hatcurXsecphasexxxxA}{\ensuremath{0.5000\pm0.0000}}       
\newcommand{\hatcurXsecondaryxxxxA}{\ensuremath{2455210.522\pm0.000}}  
\newcommand{\hatcurXsecdurxxxxA}{\ensuremath{0.1745\pm0.0017}}         
\newcommand{\hatcurXsecingdurxxxxA}{\ensuremath{0.0178\pm0.0017}}      
\newcommand{\hatcurPPphiconjxxxxA}{\ensuremath{0.2500\pm0.0000}}       
\newcommand{\hatcurPPperixxxxA}{\ensuremath{2455207.86\pm0.00}}        
\newcommand{\hatcurPPaequivxxxxA}{\ensuremath{0.0253\pm0.0012}}        
\newcommand{\hatcurPPtcircxxxxA}{\ensuremath{56.4_{-15.2}^{+20.8}}}    
\newcommand{\hatcurPPtinfallxxxxA}{\ensuremath{1099.9_{-226.6}^{+343.9}}} 
\newcommand{\hatcurXdistxxxxA}{\ensuremath{654\pm30}}                  
\newcommand{\hatcurXAvxxxxA}{\ensuremath{0.171\pm0.135}}               
\newcommand{\hatcurXdistredxxxxA}{\ensuremath{642\pm29}}               
\newcommand{\hatcurCCpmraxxxxA}{\ensuremath{5.0\pm1.7}}                
\newcommand{\hatcurCCpmdecxxxxA}{\ensuremath{-8.3\pm0.8}}              
\newcommand{\hatcurCCpmxxxxA}{\ensuremath{9.68969\pm1.87883}}          

\newcommand{\hatcurhtrxxxxB}{HTR159-019}                               
\newcommand{\hatcurfieldxxxxB}{159}                                    
\newcommand{\hatcurCCraxxxxB}{\ensuremath{22^{\mathrm h}22^{\mathrm m}03.00{\mathrm s}}}                             
\newcommand{\hatcurCCdecxxxxB}{\ensuremath{+45{\arcdeg}27{\arcmin}26.6{\arcsec}}}                            
\newcommand{\hatcurCCmagxxxxB}{11.699}                                 
\newcommand{\hatcurCCtwomassxxxxB}{2MASS~22220308+4527265}             
\newcommand{\hatcurCCgscxxxxB}{GSC~3607-01028}                         
\newcommand{\hatcurCCtassmvxxxxB}{11.699}                              
\newcommand{\hatcurCCtwomassJmagxxxxB}{\ensuremath{10.367\pm0.023}}    
\newcommand{\hatcurCCtwomassHmagxxxxB}{\ensuremath{10.085\pm0.018}}    
\newcommand{\hatcurCCtwomassKmagxxxxB}{\ensuremath{10.009\pm0.018}}    
\newcommand{\hatcurCCcitJmagxxxxB}{\ensuremath{10.384\pm0.023}}        
\newcommand{\hatcurCCcitHmagxxxxB}{\ensuremath{10.080\pm0.019}}        
\newcommand{\hatcurCCcitKmagxxxxB}{\ensuremath{10.033\pm0.018}}        
\newcommand{\hatcurCCbbJmagxxxxB}{\ensuremath{10.433\pm0.025}}         
\newcommand{\hatcurCCbbHmagxxxxB}{\ensuremath{10.101\pm0.020}}         
\newcommand{\hatcurCCbbKmagxxxxB}{\ensuremath{10.053\pm0.018}}         
\newcommand{\hatcurCCesoJmagxxxxB}{\ensuremath{10.434\pm0.026}}        
\newcommand{\hatcurCCesoHmagxxxxB}{\ensuremath{10.097\pm0.021}}        
\newcommand{\hatcurCCesoKmagxxxxB}{\ensuremath{10.052\pm0.019}}        
\newcommand{\hatcurCCesoJHmagxxxxB}{\ensuremath{0.338\pm0.032}}        
\newcommand{\hatcurCCesoJKmagxxxxB}{\ensuremath{0.383\pm0.032}}        
\newcommand{\hatcurCCesoHKmagxxxxB}{\ensuremath{0.044\pm0.009}}        
\newcommand{\hatcurLCdipxxxxB}{\ensuremath{4.4}}                       
\newcommand{\hatcurLCrprstarxxxxB}{\ensuremath{0.0807\pm0.0014}}       
\newcommand{\hatcurLCbsqxxxxB}{\ensuremath{0.030_{-0.019}^{+0.045}}}   
\newcommand{\hatcurLCimpxxxxB}{\ensuremath{0.174_{-0.077}^{+0.091}}}   
\newcommand{\hatcurLCzetaxxxxB}{\ensuremath{8.48\pm0.04}}              
\newcommand{\hatcurLCdurxxxxB}{\ensuremath{0.2557\pm0.0014}}           
\newcommand{\hatcurLCdurshortxxxxB}{\ensuremath{0.2557}}               
\newcommand{\hatcurLCdurhrxxxxB}{\ensuremath{6.138\pm0.034}}           
\newcommand{\hatcurLCdurhrshortxxxxB}{\ensuremath{6.138}}              
\newcommand{\hatcurLCqxxxxB}{\ensuremath{0.0574\pm0.0003}}             
\newcommand{\hatcurLCqshortxxxxB}{\ensuremath{0.057}}                  
\newcommand{\hatcurLCingdurxxxxB}{\ensuremath{0.0196\pm0.0009}}        
\newcommand{\hatcurLCPxxxxB}{\ensuremath{4.457243\pm0.000010}}         
\newcommand{\hatcurLCPprecxxxxB}{\ensuremath{4.4572427}}               
\newcommand{\hatcurLCPshortxxxxB}{\ensuremath{4.4572}}                 
\newcommand{\hatcurLCTxxxxB}{\ensuremath{2455813.17584\pm0.00054}}     
\newcommand{\hatcurLCTAxxxxB}{\ensuremath{2453321.57723\pm0.00553}}    
\newcommand{\hatcurLCTBxxxxB}{\ensuremath{2455839.91930\pm0.00056}}    
\newcommand{\hatcurSMEiteffxxxxB}{\ensuremath{6140\pm100}}             
\newcommand{\hatcurSMEizfehxxxxB}{\ensuremath{0.25\pm0.1}}             
\newcommand{\hatcurSMEizfehshortxxxxB}{\ensuremath{0.25}}              
\newcommand{\hatcurSMEiloggxxxxB}{\ensuremath{4.04\pm0.1}}             
\newcommand{\hatcurSMEivsinxxxxB}{\ensuremath{6.7\pm0.5}}              
\newcommand{\hatcurSMEivmacxxxxB}{\ensuremath{4.59}}                   
\newcommand{\hatcurSMEivmicxxxxB}{\ensuremath{0.85}}                   
\newcommand{\hatcurSMEiiteffxxxxB}{\ensuremath{6080\pm100}}            
\newcommand{\hatcurSMEiizfehxxxxB}{\ensuremath{0.22\pm0.10}}           
\newcommand{\hatcurSMEiizfehshortxxxxB}{\ensuremath{0.22}}             
\newcommand{\hatcurSMEiiloggxxxxB}{\ensuremath{3.92\pm0.1}}            
\newcommand{\hatcurSMEiivsinxxxxB}{\ensuremath{6.9\pm0.5}}             
\newcommand{\hatcurSMEiivmacxxxxB}{\ensuremath{4.50}}                  
\newcommand{\hatcurSMEiivmicxxxxB}{\ensuremath{0.85}}                  
\newcommand{\hatcurDSteffxxxxB}{\ensuremath{NULL\pmNULL}}              
\newcommand{\hatcurDSzfehxxxxB}{\ensuremath{NULL\pmNULL}}              
\newcommand{\hatcurDSloggxxxxB}{\ensuremath{NULL\pmNULL}}              
\newcommand{\hatcurDSvsinixxxxB}{\ensuremath{NULL\pmNULL}}             
\newcommand{\hatcurDSgammaxxxxB}{\ensuremath{NULL\pmNULL}}             
\newcommand{\hatcurDSnumspecxxxxB}{\ensuremath{0}}                     
\newcommand{\hatcurDSspanxxxxB}{\ensuremath{0}}                        
\newcommand{\hatcurDSrvrmsxxxxB}{\ensuremath{0.00}}                    
\newcommand{\hatcurTRESteffxxxxB}{\ensuremath{6000\pm100}}             
\newcommand{\hatcurTRESzfehxxxxB}{\ensuremath{0.3\pm0.1}}              
\newcommand{\hatcurTRESloggxxxxB}{\ensuremath{4.0\pm0.2}}              
\newcommand{\hatcurTRESvsinixxxxB}{\ensuremath{9.2\pm0.7}}             
\newcommand{\hatcurTRESgammaxxxxB}{\ensuremath{-25.0\pm0.1}}           
\newcommand{\hatcurTRESnumspecxxxxB}{\ensuremath{5}}                   
\newcommand{\hatcurTRESspanxxxxB}{\ensuremath{84}}                     
\newcommand{\hatcurTRESrvrmsxxxxB}{\ensuremath{0.17}}                  
\newcommand{\hatcurFIESteffxxxxB}{\ensuremath{NULL\pmNULL}}            
\newcommand{\hatcurFIESzfehxxxxB}{\ensuremath{NULL\pmNULL}}            
\newcommand{\hatcurFIESloggxxxxB}{\ensuremath{NULL\pmNULL}}            
\newcommand{\hatcurFIESvsinixxxxB}{\ensuremath{NULL\pmNULL}}           
\newcommand{\hatcurFIESgammaxxxxB}{\ensuremath{NULL\pmNULL}}           
\newcommand{\hatcurFIESnumspecxxxxB}{\ensuremath{0}}                   
\newcommand{\hatcurFIESspanxxxxB}{\ensuremath{0}}                      
\newcommand{\hatcurFIESrvrmsxxxxB}{\ensuremath{0.00}}                  
\newcommand{\hatcurLBizxxxxB}{\ensuremath{0.1634}}                     
\newcommand{\hatcurLBiizxxxxB}{\ensuremath{0.3584}}                    
\newcommand{\hatcurLBiixxxxB}{\ensuremath{0.2190}}                     
\newcommand{\hatcurLBiiixxxxB}{\ensuremath{0.3650}}                    
\newcommand{\hatcurLBiIxxxxB}{\ensuremath{0.1987}}                     
\newcommand{\hatcurLBiiIxxxxB}{\ensuremath{0.3646}}                    
\newcommand{\hatcurLBigxxxxB}{\ensuremath{0.4963}}                     
\newcommand{\hatcurLBiigxxxxB}{\ensuremath{0.2799}}                    
\newcommand{\hatcurLBirxxxxB}{\ensuremath{0.3051}}                     
\newcommand{\hatcurLBiirxxxxB}{\ensuremath{0.3623}}                    
\newcommand{\hatcurLBiRxxxxB}{\ensuremath{0.2809}}                     
\newcommand{\hatcurLBiiRxxxxB}{\ensuremath{0.3644}}                    
\newcommand{\hatcurLBikepxxxxB}{\ensuremath{}}                 
\newcommand{\hatcurLBiikepxxxxB}{\ensuremath{}}                
\newcommand{\hatcurISOmxxxxB}{\ensuremath{1.51_{-0.11}^{+0.05}}}       
\newcommand{\hatcurISOmshortxxxxB}{\ensuremath{1.51}}                  
\newcommand{\hatcurISOmlongxxxxB}{\ensuremath{1.512_{-0.109}^{+0.045}}} 
\newcommand{\hatcurISOrxxxxB}{\ensuremath{2.21\pm0.06}}                
\newcommand{\hatcurISOrshortxxxxB}{\ensuremath{2.21}}                  
\newcommand{\hatcurISOrlongxxxxB}{\ensuremath{2.206\pm0.061}}          
\newcommand{\hatcurISOrhoxxxxB}{\ensuremath{0.20_{-0.01}^{+0.01}}}     
\newcommand{\hatcurISOloggxxxxB}{\ensuremath{3.93\pm0.02}}             
\newcommand{\hatcurISOlumxxxxB}{\ensuremath{6.00\pm0.61}}              
\newcommand{\hatcurISOlumshortxxxxB}{\ensuremath{6.00}}                
\newcommand{\hatcurISOmvxxxxB}{\ensuremath{2.83\pm0.13}}               
\newcommand{\hatcurISOvixxxxB}{\ensuremath{0.609\pm0.029}}             
\newcommand{\hatcurISOagexxxxB}{\ensuremath{2.7_{-0.3}^{+0.9}}}        
\newcommand{\hatcurISOsigmaxxxxB}{\ensuremath{0.00020\pm0.00002}}      
\newcommand{\hatcurISOMJxxxxB}{\ensuremath{1.83\pm0.09}}               
\newcommand{\hatcurISOMHxxxxB}{\ensuremath{1.54\pm0.07}}               
\newcommand{\hatcurISOMKxxxxB}{\ensuremath{1.49\pm0.07}}               
\newcommand{\hatcurISOJKxxxxB}{\ensuremath{0.33\pm0.02}}               
\newcommand{\hatcurISOspecxxxxB}{F8}                                   
\newcommand{\hatcurRVKxxxxB}{\ensuremath{58.1\pm2.9}}                  
\newcommand{\hatcurRVkxxxxB}{\ensuremath{0.000\pm0.000}}               
\newcommand{\hatcurRVhxxxxB}{\ensuremath{0.000\pm0.000}}               
\newcommand{\hatcurRVtronexxxxB}{\ensuremath{0.0000\pm0.0000}}         
\newcommand{\hatcurRVtrtwoxxxxB}{\ensuremath{0.0000\pm0.0000}}         
\newcommand{\hatcurRVgammaxxxxB}{\ensuremath{-16.6\pm2.1}}             
\newcommand{\hatcurRVjitterxxxxB}{\ensuremath{6.2}}                    
\newcommand{\hatcurRVfitrmsxxxxB}{\ensuremath{7.2}}                    
\newcommand{\hatcurRVeccenxxxxB}{\ensuremath{0.000\pm0.000}}           
\newcommand{\hatcurRVomegaxxxxB}{\ensuremath{0\pm0}}                   
\newcommand{\hatcurPPixxxxB}{\ensuremath{88.3\pm0.9}}                  
\newcommand{\hatcurPPgxxxxB}{\ensuremath{5.1\pm0.4}}                   
\newcommand{\hatcurPPloggxxxxB}{\ensuremath{2.71\pm0.03}}              
\newcommand{\hatcurPParxxxxB}{\ensuremath{5.92_{-0.14}^{+0.06}}}       
\newcommand{\hatcurPParelxxxxB}{\ensuremath{0.0608_{-0.0015}^{+0.0006}}} 
\newcommand{\hatcurPPrhoxxxxB}{\ensuremath{0.15\pm0.01}}               
\newcommand{\hatcurPPmxxxxB}{\ensuremath{0.62\pm0.04}}                 
\newcommand{\hatcurPPmshortxxxxB}{\ensuremath{0.62}}                   
\newcommand{\hatcurPPmlongxxxxB}{\ensuremath{0.615\pm0.038}}           
\newcommand{\hatcurPPmexxxxB}{\ensuremath{195.5\pm12.2}}               
\newcommand{\hatcurPPmeshortxxxxB}{\ensuremath{195.5}}                 
\newcommand{\hatcurPPmelongxxxxB}{\ensuremath{195.54\pm12.16}}         
\newcommand{\hatcurPPrxxxxB}{\ensuremath{1.73\pm0.06}}                 
\newcommand{\hatcurPPrshortxxxxB}{\ensuremath{1.73}}                   
\newcommand{\hatcurPPrlongxxxxB}{\ensuremath{1.730\pm0.062}}           
\newcommand{\hatcurPPrexxxxB}{\ensuremath{19.4\pm0.7}}                 
\newcommand{\hatcurPPreshortxxxxB}{\ensuremath{19.4}}                  
\newcommand{\hatcurPPrelongxxxxB}{\ensuremath{19.39\pm0.70}}           
\newcommand{\hatcurPPmrcorrxxxxB}{\ensuremath{0.36}}                   
\newcommand{\hatcurPPteffxxxxB}{\ensuremath{1770\pm33}}                
\newcommand{\hatcurPPthetaxxxxB}{\ensuremath{0.029\pm0.002}}           
\newcommand{\hatcurPPfluxperixxxxB}{\ensuremath{2.22\pm0.17}}          
\newcommand{\hatcurPPfluxperidimxxxxB}{\ensuremath{9}}                 
\newcommand{\hatcurPPfluxapxxxxB}{\ensuremath{2.22\pm0.17}}            
\newcommand{\hatcurPPfluxapdimxxxxB}{\ensuremath{9}}                   
\newcommand{\hatcurPPfluxavgxxxxB}{\ensuremath{2.22\pm0.17}}           
\newcommand{\hatcurPPfluxavgdimxxxxB}{\ensuremath{9}}                  
\newcommand{\hatcurXsecphasexxxxB}{\ensuremath{0.5000\pm0.0000}}       
\newcommand{\hatcurXsecondaryxxxxB}{\ensuremath{2455815.404\pm0.001}}  
\newcommand{\hatcurXsecdurxxxxB}{\ensuremath{0.2557\pm0.0014}}         
\newcommand{\hatcurXsecingdurxxxxB}{\ensuremath{0.0196\pm0.0009}}      
\newcommand{\hatcurPPphiconjxxxxB}{\ensuremath{0.2500\pm0.0000}}       
\newcommand{\hatcurPPperixxxxB}{\ensuremath{2455812.06\pm0.00}}        
\newcommand{\hatcurPPaequivxxxxB}{\ensuremath{0.0248\pm0.0009}}        
\newcommand{\hatcurPPtcircxxxxB}{\ensuremath{102.4\pm15.2}}            
\newcommand{\hatcurPPtinfallxxxxB}{\ensuremath{741.2\pm72.4}}          
\newcommand{\hatcurXdistxxxxB}{\ensuremath{514\pm17}}                  
\newcommand{\hatcurXAvxxxxB}{\ensuremath{0.353\pm0.127}}               
\newcommand{\hatcurXdistredxxxxB}{\ensuremath{501\pm16}}               
\newcommand{\hatcurCCpmraxxxxB}{\ensuremath{9.2\pm0.9}}                
\newcommand{\hatcurCCpmdecxxxxB}{\ensuremath{-4.4\pm1.0}}              
\newcommand{\hatcurCCpmxxxxB}{\ensuremath{10.198\pm1.34536}}           

\newcommand{\hatcurhtrxxxxC}{HTR388-004}                               
\newcommand{\hatcurfieldxxxxC}{388}                                    
\newcommand{\hatcurCCraxxxxC}{\ensuremath{19^{\mathrm h}49^{\mathrm m}17.40{\mathrm s}}}                             
\newcommand{\hatcurCCdecxxxxC}{\ensuremath{+04{\arcdeg}40{\arcmin}20.7{\arcsec}}}                            
\newcommand{\hatcurCCmagxxxxC}{11.087}                                 
\newcommand{\hatcurCCtwomassxxxxC}{2MASS~19491743+0440207}             
\newcommand{\hatcurCCgscxxxxC}{GSC~0488-02442}                         
\newcommand{\hatcurCCtassmvxxxxC}{11.087}                              
\newcommand{\hatcurCCtwomassJmagxxxxC}{\ensuremath{10.006\pm0.027}}    
\newcommand{\hatcurCCtwomassHmagxxxxC}{\ensuremath{9.777\pm0.032}}     
\newcommand{\hatcurCCtwomassKmagxxxxC}{\ensuremath{9.728\pm0.029}}     
\newcommand{\hatcurCCcitJmagxxxxC}{\ensuremath{10.028\pm0.027}}        
\newcommand{\hatcurCCcitHmagxxxxC}{\ensuremath{9.772\pm0.032}}         
\newcommand{\hatcurCCcitKmagxxxxC}{\ensuremath{9.752\pm0.029}}         
\newcommand{\hatcurCCbbJmagxxxxC}{\ensuremath{10.070\pm0.029}}         
\newcommand{\hatcurCCbbHmagxxxxC}{\ensuremath{9.793\pm0.033}}          
\newcommand{\hatcurCCbbKmagxxxxC}{\ensuremath{9.772\pm0.029}}          
\newcommand{\hatcurCCesoJmagxxxxC}{\ensuremath{10.070\pm0.030}}        
\newcommand{\hatcurCCesoHmagxxxxC}{\ensuremath{9.788\pm0.036}}         
\newcommand{\hatcurCCesoKmagxxxxC}{\ensuremath{9.771\pm0.029}}         
\newcommand{\hatcurCCesoJHmagxxxxC}{\ensuremath{0.282\pm0.045}}        
\newcommand{\hatcurCCesoJKmagxxxxC}{\ensuremath{0.300\pm0.042}}        
\newcommand{\hatcurCCesoHKmagxxxxC}{\ensuremath{0.017\pm0.047}}        
\newcommand{\hatcurLCdipxxxxC}{\ensuremath{8.4}}                       
\newcommand{\hatcurLCrprstarxxxxC}{\ensuremath{0.1028\pm0.0016}}       
\newcommand{\hatcurLCbsqxxxxC}{\ensuremath{0.049_{-0.029}^{+0.051}}}   
\newcommand{\hatcurLCimpxxxxC}{\ensuremath{0.222_{-0.093}^{+0.088}}}   
\newcommand{\hatcurLCzetaxxxxC}{\ensuremath{13.01\pm0.06}}             
\newcommand{\hatcurLCdurxxxxC}{\ensuremath{0.1704\pm0.0012}}           
\newcommand{\hatcurLCdurshortxxxxC}{\ensuremath{0.1704}}               
\newcommand{\hatcurLCdurhrxxxxC}{\ensuremath{4.090\pm0.030}}           
\newcommand{\hatcurLCdurhrshortxxxxC}{\ensuremath{4.090}}              
\newcommand{\hatcurLCqxxxxC}{\ensuremath{0.0633\pm0.0005}}             
\newcommand{\hatcurLCqshortxxxxC}{\ensuremath{0.063}}                  
\newcommand{\hatcurLCingdurxxxxC}{\ensuremath{0.0166\pm0.0009}}        
\newcommand{\hatcurLCPxxxxC}{\ensuremath{2.694047\pm0.000004}}         
\newcommand{\hatcurLCPprecxxxxC}{\ensuremath{2.6940468}}               
\newcommand{\hatcurLCPshortxxxxC}{\ensuremath{2.6940}}                 
\newcommand{\hatcurLCTxxxxC}{\ensuremath{2454983.86167\pm0.00107}}     
\newcommand{\hatcurLCTAxxxxC}{\ensuremath{2454965.00334\pm0.00110}}    
\newcommand{\hatcurLCTBxxxxC}{\ensuremath{2455713.94836\pm0.00041}}    
\newcommand{\hatcurSMEiteffxxxxC}{\ensuremath{6007\pm100}}             
\newcommand{\hatcurSMEizfehxxxxC}{\ensuremath{0.06\pm0.1}}             
\newcommand{\hatcurSMEizfehshortxxxxC}{\ensuremath{0.06}}              
\newcommand{\hatcurSMEiloggxxxxC}{\ensuremath{3.68\pm0.06}}            
\newcommand{\hatcurSMEivsinxxxxC}{\ensuremath{20.6\pm0.5}}             
\newcommand{\hatcurSMEivmacxxxxC}{\ensuremath{4.38}}                   
\newcommand{\hatcurSMEivmicxxxxC}{\ensuremath{0.85}}                   
\newcommand{\hatcurSMEiiteffxxxxC}{\ensuremath{6390\pm100}}            
\newcommand{\hatcurSMEiizfehxxxxC}{\ensuremath{0.21\pm0.10}}           
\newcommand{\hatcurSMEiizfehshortxxxxC}{\ensuremath{0.21}}             
\newcommand{\hatcurSMEiiloggxxxxC}{\ensuremath{4.13\pm0.1}}            
\newcommand{\hatcurSMEiivsinxxxxC}{\ensuremath{19.6\pm0.5}}            
\newcommand{\hatcurSMEiivmacxxxxC}{\ensuremath{4.97}}                  
\newcommand{\hatcurSMEiivmicxxxxC}{\ensuremath{0.85}}                  
\newcommand{\hatcurDSteffxxxxC}{\ensuremath{NULL\pmNULL}}              
\newcommand{\hatcurDSzfehxxxxC}{\ensuremath{NULL\pmNULL}}              
\newcommand{\hatcurDSloggxxxxC}{\ensuremath{NULL\pmNULL}}              
\newcommand{\hatcurDSvsinixxxxC}{\ensuremath{NULL\pmNULL}}             
\newcommand{\hatcurDSgammaxxxxC}{\ensuremath{NULL\pmNULL}}             
\newcommand{\hatcurDSnumspecxxxxC}{\ensuremath{0}}                     
\newcommand{\hatcurDSspanxxxxC}{\ensuremath{0}}                        
\newcommand{\hatcurDSrvrmsxxxxC}{\ensuremath{0.00}}                    
\newcommand{\hatcurTRESteffxxxxC}{\ensuremath{6333\pm100}}             
\newcommand{\hatcurTRESzfehxxxxC}{\ensuremath{0.0\pm0.0}}              
\newcommand{\hatcurTRESloggxxxxC}{\ensuremath{3.83\pm0.25}}            
\newcommand{\hatcurTRESvsinixxxxC}{\ensuremath{25.0\pm0.5}}            
\newcommand{\hatcurTRESgammaxxxxC}{\ensuremath{31.68\pm0.61}}          
\newcommand{\hatcurTRESnumspecxxxxC}{\ensuremath{3}}                   
\newcommand{\hatcurTRESspanxxxxC}{\ensuremath{54}}                     
\newcommand{\hatcurTRESrvrmsxxxxC}{\ensuremath{1.05}}                  
\newcommand{\hatcurFIESteffxxxxC}{\ensuremath{NULL\pmNULL}}            
\newcommand{\hatcurFIESzfehxxxxC}{\ensuremath{NULL\pmNULL}}            
\newcommand{\hatcurFIESloggxxxxC}{\ensuremath{NULL\pmNULL}}            
\newcommand{\hatcurFIESvsinixxxxC}{\ensuremath{NULL\pmNULL}}           
\newcommand{\hatcurFIESgammaxxxxC}{\ensuremath{NULL\pmNULL}}           
\newcommand{\hatcurFIESnumspecxxxxC}{\ensuremath{0}}                   
\newcommand{\hatcurFIESspanxxxxC}{\ensuremath{0}}                      
\newcommand{\hatcurFIESrvrmsxxxxC}{\ensuremath{0.00}}                  
\newcommand{\hatcurLBizxxxxC}{\ensuremath{0.1380}}                     
\newcommand{\hatcurLBiizxxxxC}{\ensuremath{0.3655}}                    
\newcommand{\hatcurLBiixxxxC}{\ensuremath{0.1908}}                     
\newcommand{\hatcurLBiiixxxxC}{\ensuremath{0.3746}}                    
\newcommand{\hatcurLBiIxxxxC}{\ensuremath{0.1721}}                     
\newcommand{\hatcurLBiiIxxxxC}{\ensuremath{0.3725}}                    
\newcommand{\hatcurLBigxxxxC}{\ensuremath{0.4329}}                     
\newcommand{\hatcurLBiigxxxxC}{\ensuremath{0.3245}}                    
\newcommand{\hatcurLBirxxxxC}{\ensuremath{0.2658}}                     
\newcommand{\hatcurLBiirxxxxC}{\ensuremath{0.3814}}                    
\newcommand{\hatcurLBiRxxxxC}{\ensuremath{0.2445}}                     
\newcommand{\hatcurLBiiRxxxxC}{\ensuremath{0.3811}}                    
\newcommand{\hatcurLBikepxxxxC}{\ensuremath{}}                 
\newcommand{\hatcurLBiikepxxxxC}{\ensuremath{}}                
\newcommand{\hatcurISOmxxxxC}{\ensuremath{1.42\pm0.05}}                
\newcommand{\hatcurISOmshortxxxxC}{\ensuremath{1.42}}                  
\newcommand{\hatcurISOmlongxxxxC}{\ensuremath{1.418\pm0.047}}          
\newcommand{\hatcurISOrxxxxC}{\ensuremath{1.68_{-0.04}^{+0.06}}}       
\newcommand{\hatcurISOrshortxxxxC}{\ensuremath{1.68}}                  
\newcommand{\hatcurISOrlongxxxxC}{\ensuremath{1.683_{-0.036}^{+0.058}}} 
\newcommand{\hatcurISOrhoxxxxC}{\ensuremath{0.42_{-0.03}^{+0.02}}}     
\newcommand{\hatcurISOloggxxxxC}{\ensuremath{4.14\pm0.02}}             
\newcommand{\hatcurISOlumxxxxC}{\ensuremath{4.25\pm0.41}}              
\newcommand{\hatcurISOlumshortxxxxC}{\ensuremath{4.25}}                
\newcommand{\hatcurISOmvxxxxC}{\ensuremath{3.16\pm0.12}}               
\newcommand{\hatcurISOvixxxxC}{\ensuremath{0.526\pm0.026}}             
\newcommand{\hatcurISOagexxxxC}{\ensuremath{2.2\pm0.4}}                
\newcommand{\hatcurISOsigmaxxxxC}{\ensuremath{0.00040\pm0.00006}}      
\newcommand{\hatcurISOMJxxxxC}{\ensuremath{2.32\pm0.08}}               
\newcommand{\hatcurISOMHxxxxC}{\ensuremath{2.09\pm0.07}}               
\newcommand{\hatcurISOMKxxxxC}{\ensuremath{2.04\pm0.06}}               
\newcommand{\hatcurISOJKxxxxC}{\ensuremath{0.27\pm0.02}}               
\newcommand{\hatcurISOspecxxxxC}{F}                                    
\newcommand{\hatcurRVKxxxxC}{\ensuremath{92.5\pm11.6}}                 
\newcommand{\hatcurRVkxxxxC}{\ensuremath{0.000\pm0.000}}               
\newcommand{\hatcurRVhxxxxC}{\ensuremath{0.000\pm0.000}}               
\newcommand{\hatcurRVtronexxxxC}{\ensuremath{0.0000\pm0.0000}}         
\newcommand{\hatcurRVtrtwoxxxxC}{\ensuremath{0.0000\pm0.0000}}         
\newcommand{\hatcurRVjitterAxxxxC}{\ensuremath{33.4}}                  
\newcommand{\hatcurRVeccenxxxxC}{\ensuremath{0.000\pm0.000}}           
\newcommand{\hatcurRVomegaxxxxC}{\ensuremath{0\pm0}}                   
\newcommand{\hatcurPPixxxxC}{\ensuremath{87.7\pm1.0}}                  
\newcommand{\hatcurPPgxxxxC}{\ensuremath{6.9\pm1.0}}                   
\newcommand{\hatcurPPloggxxxxC}{\ensuremath{2.84\pm0.06}}              
\newcommand{\hatcurPParxxxxC}{\ensuremath{5.44_{-0.15}^{+0.09}}}       
\newcommand{\hatcurPParelxxxxC}{\ensuremath{0.0426\pm0.0005}}          
\newcommand{\hatcurPPrhoxxxxC}{\ensuremath{0.20\pm0.03}}               
\newcommand{\hatcurPPmxxxxC}{\ensuremath{0.80\pm0.10}}                 
\newcommand{\hatcurPPmshortxxxxC}{\ensuremath{0.80}}                   
\newcommand{\hatcurPPmlongxxxxC}{\ensuremath{0.800\pm0.102}}           
\newcommand{\hatcurPPmexxxxC}{\ensuremath{254.4\pm32.3}}               
\newcommand{\hatcurPPmeshortxxxxC}{\ensuremath{254.4}}                 
\newcommand{\hatcurPPmelongxxxxC}{\ensuremath{254.37\pm32.31}}         
\newcommand{\hatcurPPrxxxxC}{\ensuremath{1.68_{-0.05}^{+0.08}}}        
\newcommand{\hatcurPPrshortxxxxC}{\ensuremath{1.68}}                   
\newcommand{\hatcurPPrlongxxxxC}{\ensuremath{1.685_{-0.051}^{+0.076}}} 
\newcommand{\hatcurPPrexxxxC}{\ensuremath{18.9_{-0.6}^{+0.9}}}         
\newcommand{\hatcurPPreshortxxxxC}{\ensuremath{18.9}}                  
\newcommand{\hatcurPPrelongxxxxC}{\ensuremath{18.88_{-0.57}^{+0.85}}}  
\newcommand{\hatcurPPmrcorrxxxxC}{\ensuremath{0.10}}                   
\newcommand{\hatcurPPteffxxxxC}{\ensuremath{1941\pm38}}                
\newcommand{\hatcurPPthetaxxxxC}{\ensuremath{0.028\pm0.004}}           
\newcommand{\hatcurPPfluxperixxxxC}{\ensuremath{3.20\pm0.25}}          
\newcommand{\hatcurPPfluxperidimxxxxC}{\ensuremath{9}}                 
\newcommand{\hatcurPPfluxapxxxxC}{\ensuremath{3.20\pm0.25}}            
\newcommand{\hatcurPPfluxapdimxxxxC}{\ensuremath{9}}                   
\newcommand{\hatcurPPfluxavgxxxxC}{\ensuremath{3.20\pm0.25}}           
\newcommand{\hatcurPPfluxavgdimxxxxC}{\ensuremath{9}}                  
\newcommand{\hatcurXsecphasexxxxC}{\ensuremath{0.5000\pm0.0000}}       
\newcommand{\hatcurXsecondaryxxxxC}{\ensuremath{2454985.209\pm0.001}}  
\newcommand{\hatcurXsecdurxxxxC}{\ensuremath{0.1704\pm0.0012}}         
\newcommand{\hatcurXsecingdurxxxxC}{\ensuremath{0.0166\pm0.0009}}      
\newcommand{\hatcurPPphiconjxxxxC}{\ensuremath{0.2500\pm0.0000}}       
\newcommand{\hatcurPPperixxxxC}{\ensuremath{2454983.19\pm0.00}}        
\newcommand{\hatcurPPaequivxxxxC}{\ensuremath{0.0206\pm0.0008}}        
\newcommand{\hatcurPPtcircxxxxC}{\ensuremath{16.4\pm3.4}}              
\newcommand{\hatcurPPtinfallxxxxC}{\ensuremath{212.8\pm36.6}}          
\newcommand{\hatcurXdistxxxxC}{\ensuremath{351_{-9}^{+13}}}            
\newcommand{\hatcurXAvxxxxC}{\ensuremath{0.248\pm0.134}}               
\newcommand{\hatcurXdistredxxxxC}{\ensuremath{344_{-8}^{+12}}}         
\newcommand{\hatcurCCpmraxxxxC}{\ensuremath{-0.4\pm0.7}}               
\newcommand{\hatcurCCpmdecxxxxC}{\ensuremath{-2.9\pm4.5}}              
\newcommand{\hatcurCCpmxxxxC}{\ensuremath{2.92746\pm4.55412}}          
\newcommand{\hatcurCCbbHmag}[1]{\ifnum#1=39 %
\hatcurCCbbHmagxxxxA
\else
\ifnum#1=40 %
\hatcurCCbbHmagxxxxB
\else
\ifnum#1=41 %
\hatcurCCbbHmagxxxxC
\else
??????\fi
\fi
\fi
}
\newcommand{\hatcurCCbbJmag}[1]{\ifnum#1=39 %
\hatcurCCbbJmagxxxxA
\else
\ifnum#1=40 %
\hatcurCCbbJmagxxxxB
\else
\ifnum#1=41 %
\hatcurCCbbJmagxxxxC
\else
??????\fi
\fi
\fi
}
\newcommand{\hatcurCCbbKmag}[1]{\ifnum#1=39 %
\hatcurCCbbKmagxxxxA
\else
\ifnum#1=40 %
\hatcurCCbbKmagxxxxB
\else
\ifnum#1=41 %
\hatcurCCbbKmagxxxxC
\else
??????\fi
\fi
\fi
}
\newcommand{\hatcurCCcitHmag}[1]{\ifnum#1=39 %
\hatcurCCcitHmagxxxxA
\else
\ifnum#1=40 %
\hatcurCCcitHmagxxxxB
\else
\ifnum#1=41 %
\hatcurCCcitHmagxxxxC
\else
??????\fi
\fi
\fi
}
\newcommand{\hatcurCCcitJmag}[1]{\ifnum#1=39 %
\hatcurCCcitJmagxxxxA
\else
\ifnum#1=40 %
\hatcurCCcitJmagxxxxB
\else
\ifnum#1=41 %
\hatcurCCcitJmagxxxxC
\else
??????\fi
\fi
\fi
}
\newcommand{\hatcurCCcitKmag}[1]{\ifnum#1=39 %
\hatcurCCcitKmagxxxxA
\else
\ifnum#1=40 %
\hatcurCCcitKmagxxxxB
\else
\ifnum#1=41 %
\hatcurCCcitKmagxxxxC
\else
??????\fi
\fi
\fi
}
\newcommand{\hatcurCCdec}[1]{\ifnum#1=39 %
\hatcurCCdecxxxxA
\else
\ifnum#1=40 %
\hatcurCCdecxxxxB
\else
\ifnum#1=41 %
\hatcurCCdecxxxxC
\else
??????\fi
\fi
\fi
}
\newcommand{\hatcurCCesoHKmag}[1]{\ifnum#1=39 %
\hatcurCCesoHKmagxxxxA
\else
\ifnum#1=40 %
\hatcurCCesoHKmagxxxxB
\else
\ifnum#1=41 %
\hatcurCCesoHKmagxxxxC
\else
??????\fi
\fi
\fi
}
\newcommand{\hatcurCCesoHmag}[1]{\ifnum#1=39 %
\hatcurCCesoHmagxxxxA
\else
\ifnum#1=40 %
\hatcurCCesoHmagxxxxB
\else
\ifnum#1=41 %
\hatcurCCesoHmagxxxxC
\else
??????\fi
\fi
\fi
}
\newcommand{\hatcurCCesoJHmag}[1]{\ifnum#1=39 %
\hatcurCCesoJHmagxxxxA
\else
\ifnum#1=40 %
\hatcurCCesoJHmagxxxxB
\else
\ifnum#1=41 %
\hatcurCCesoJHmagxxxxC
\else
??????\fi
\fi
\fi
}
\newcommand{\hatcurCCesoJKmag}[1]{\ifnum#1=39 %
\hatcurCCesoJKmagxxxxA
\else
\ifnum#1=40 %
\hatcurCCesoJKmagxxxxB
\else
\ifnum#1=41 %
\hatcurCCesoJKmagxxxxC
\else
??????\fi
\fi
\fi
}
\newcommand{\hatcurCCesoJmag}[1]{\ifnum#1=39 %
\hatcurCCesoJmagxxxxA
\else
\ifnum#1=40 %
\hatcurCCesoJmagxxxxB
\else
\ifnum#1=41 %
\hatcurCCesoJmagxxxxC
\else
??????\fi
\fi
\fi
}
\newcommand{\hatcurCCesoKmag}[1]{\ifnum#1=39 %
\hatcurCCesoKmagxxxxA
\else
\ifnum#1=40 %
\hatcurCCesoKmagxxxxB
\else
\ifnum#1=41 %
\hatcurCCesoKmagxxxxC
\else
??????\fi
\fi
\fi
}
\newcommand{\hatcurCCgsc}[1]{\ifnum#1=39 %
\hatcurCCgscxxxxA
\else
\ifnum#1=40 %
\hatcurCCgscxxxxB
\else
\ifnum#1=41 %
\hatcurCCgscxxxxC
\else
??????\fi
\fi
\fi
}
\newcommand{\hatcurCCmag}[1]{\ifnum#1=39 %
\hatcurCCmagxxxxA
\else
\ifnum#1=40 %
\hatcurCCmagxxxxB
\else
\ifnum#1=41 %
\hatcurCCmagxxxxC
\else
??????\fi
\fi
\fi
}
\newcommand{\hatcurCCpm}[1]{\ifnum#1=39 %
\hatcurCCpmxxxxA
\else
\ifnum#1=40 %
\hatcurCCpmxxxxB
\else
\ifnum#1=41 %
\hatcurCCpmxxxxC
\else
??????\fi
\fi
\fi
}
\newcommand{\hatcurCCpmdec}[1]{\ifnum#1=39 %
\hatcurCCpmdecxxxxA
\else
\ifnum#1=40 %
\hatcurCCpmdecxxxxB
\else
\ifnum#1=41 %
\hatcurCCpmdecxxxxC
\else
??????\fi
\fi
\fi
}
\newcommand{\hatcurCCpmra}[1]{\ifnum#1=39 %
\hatcurCCpmraxxxxA
\else
\ifnum#1=40 %
\hatcurCCpmraxxxxB
\else
\ifnum#1=41 %
\hatcurCCpmraxxxxC
\else
??????\fi
\fi
\fi
}
\newcommand{\hatcurCCra}[1]{\ifnum#1=39 %
\hatcurCCraxxxxA
\else
\ifnum#1=40 %
\hatcurCCraxxxxB
\else
\ifnum#1=41 %
\hatcurCCraxxxxC
\else
??????\fi
\fi
\fi
}
\newcommand{\hatcurCCtassmv}[1]{\ifnum#1=39 %
\hatcurCCtassmvxxxxA
\else
\ifnum#1=40 %
\hatcurCCtassmvxxxxB
\else
\ifnum#1=41 %
\hatcurCCtassmvxxxxC
\else
??????\fi
\fi
\fi
}
\newcommand{\hatcurCCtwomass}[1]{\ifnum#1=39 %
\hatcurCCtwomassxxxxA
\else
\ifnum#1=40 %
\hatcurCCtwomassxxxxB
\else
\ifnum#1=41 %
\hatcurCCtwomassxxxxC
\else
??????\fi
\fi
\fi
}
\newcommand{\hatcurCCtwomassHmag}[1]{\ifnum#1=39 %
\hatcurCCtwomassHmagxxxxA
\else
\ifnum#1=40 %
\hatcurCCtwomassHmagxxxxB
\else
\ifnum#1=41 %
\hatcurCCtwomassHmagxxxxC
\else
??????\fi
\fi
\fi
}
\newcommand{\hatcurCCtwomassJmag}[1]{\ifnum#1=39 %
\hatcurCCtwomassJmagxxxxA
\else
\ifnum#1=40 %
\hatcurCCtwomassJmagxxxxB
\else
\ifnum#1=41 %
\hatcurCCtwomassJmagxxxxC
\else
??????\fi
\fi
\fi
}
\newcommand{\hatcurCCtwomassKmag}[1]{\ifnum#1=39 %
\hatcurCCtwomassKmagxxxxA
\else
\ifnum#1=40 %
\hatcurCCtwomassKmagxxxxB
\else
\ifnum#1=41 %
\hatcurCCtwomassKmagxxxxC
\else
??????\fi
\fi
\fi
}
\newcommand{\hatcurDSgamma}[1]{\ifnum#1=39 %
\hatcurDSgammaxxxxA
\else
\ifnum#1=40 %
\hatcurDSgammaxxxxB
\else
\ifnum#1=41 %
\hatcurDSgammaxxxxC
\else
??????\fi
\fi
\fi
}
\newcommand{\hatcurDSlogg}[1]{\ifnum#1=39 %
\hatcurDSloggxxxxA
\else
\ifnum#1=40 %
\hatcurDSloggxxxxB
\else
\ifnum#1=41 %
\hatcurDSloggxxxxC
\else
??????\fi
\fi
\fi
}
\newcommand{\hatcurDSnumspec}[1]{\ifnum#1=39 %
\hatcurDSnumspecxxxxA
\else
\ifnum#1=40 %
\hatcurDSnumspecxxxxB
\else
\ifnum#1=41 %
\hatcurDSnumspecxxxxC
\else
??????\fi
\fi
\fi
}
\newcommand{\hatcurDSrvrms}[1]{\ifnum#1=39 %
\hatcurDSrvrmsxxxxA
\else
\ifnum#1=40 %
\hatcurDSrvrmsxxxxB
\else
\ifnum#1=41 %
\hatcurDSrvrmsxxxxC
\else
??????\fi
\fi
\fi
}
\newcommand{\hatcurDSspan}[1]{\ifnum#1=39 %
\hatcurDSspanxxxxA
\else
\ifnum#1=40 %
\hatcurDSspanxxxxB
\else
\ifnum#1=41 %
\hatcurDSspanxxxxC
\else
??????\fi
\fi
\fi
}
\newcommand{\hatcurDSteff}[1]{\ifnum#1=39 %
\hatcurDSteffxxxxA
\else
\ifnum#1=40 %
\hatcurDSteffxxxxB
\else
\ifnum#1=41 %
\hatcurDSteffxxxxC
\else
??????\fi
\fi
\fi
}
\newcommand{\hatcurDSvsini}[1]{\ifnum#1=39 %
\hatcurDSvsinixxxxA
\else
\ifnum#1=40 %
\hatcurDSvsinixxxxB
\else
\ifnum#1=41 %
\hatcurDSvsinixxxxC
\else
??????\fi
\fi
\fi
}
\newcommand{\hatcurDSzfeh}[1]{\ifnum#1=39 %
\hatcurDSzfehxxxxA
\else
\ifnum#1=40 %
\hatcurDSzfehxxxxB
\else
\ifnum#1=41 %
\hatcurDSzfehxxxxC
\else
??????\fi
\fi
\fi
}
\newcommand{\hatcurfield}[1]{\ifnum#1=39 %
\hatcurfieldxxxxA
\else
\ifnum#1=40 %
\hatcurfieldxxxxB
\else
\ifnum#1=41 %
\hatcurfieldxxxxC
\else
??????\fi
\fi
\fi
}
\newcommand{\hatcurFIESgamma}[1]{\ifnum#1=39 %
\hatcurFIESgammaxxxxA
\else
\ifnum#1=40 %
\hatcurFIESgammaxxxxB
\else
\ifnum#1=41 %
\hatcurFIESgammaxxxxC
\else
??????\fi
\fi
\fi
}
\newcommand{\hatcurFIESlogg}[1]{\ifnum#1=39 %
\hatcurFIESloggxxxxA
\else
\ifnum#1=40 %
\hatcurFIESloggxxxxB
\else
\ifnum#1=41 %
\hatcurFIESloggxxxxC
\else
??????\fi
\fi
\fi
}
\newcommand{\hatcurFIESnumspec}[1]{\ifnum#1=39 %
\hatcurFIESnumspecxxxxA
\else
\ifnum#1=40 %
\hatcurFIESnumspecxxxxB
\else
\ifnum#1=41 %
\hatcurFIESnumspecxxxxC
\else
??????\fi
\fi
\fi
}
\newcommand{\hatcurFIESrvrms}[1]{\ifnum#1=39 %
\hatcurFIESrvrmsxxxxA
\else
\ifnum#1=40 %
\hatcurFIESrvrmsxxxxB
\else
\ifnum#1=41 %
\hatcurFIESrvrmsxxxxC
\else
??????\fi
\fi
\fi
}
\newcommand{\hatcurFIESspan}[1]{\ifnum#1=39 %
\hatcurFIESspanxxxxA
\else
\ifnum#1=40 %
\hatcurFIESspanxxxxB
\else
\ifnum#1=41 %
\hatcurFIESspanxxxxC
\else
??????\fi
\fi
\fi
}
\newcommand{\hatcurFIESteff}[1]{\ifnum#1=39 %
\hatcurFIESteffxxxxA
\else
\ifnum#1=40 %
\hatcurFIESteffxxxxB
\else
\ifnum#1=41 %
\hatcurFIESteffxxxxC
\else
??????\fi
\fi
\fi
}
\newcommand{\hatcurFIESvsini}[1]{\ifnum#1=39 %
\hatcurFIESvsinixxxxA
\else
\ifnum#1=40 %
\hatcurFIESvsinixxxxB
\else
\ifnum#1=41 %
\hatcurFIESvsinixxxxC
\else
??????\fi
\fi
\fi
}
\newcommand{\hatcurFIESzfeh}[1]{\ifnum#1=39 %
\hatcurFIESzfehxxxxA
\else
\ifnum#1=40 %
\hatcurFIESzfehxxxxB
\else
\ifnum#1=41 %
\hatcurFIESzfehxxxxC
\else
??????\fi
\fi
\fi
}
\newcommand{\hatcurhtr}[1]{\ifnum#1=39 %
\hatcurhtrxxxxA
\else
\ifnum#1=40 %
\hatcurhtrxxxxB
\else
\ifnum#1=41 %
\hatcurhtrxxxxC
\else
??????\fi
\fi
\fi
}
\newcommand{\hatcurISOage}[1]{\ifnum#1=39 %
\hatcurISOagexxxxA
\else
\ifnum#1=40 %
\hatcurISOagexxxxB
\else
\ifnum#1=41 %
\hatcurISOagexxxxC
\else
??????\fi
\fi
\fi
}
\newcommand{\hatcurISOJK}[1]{\ifnum#1=39 %
\hatcurISOJKxxxxA
\else
\ifnum#1=40 %
\hatcurISOJKxxxxB
\else
\ifnum#1=41 %
\hatcurISOJKxxxxC
\else
??????\fi
\fi
\fi
}
\newcommand{\hatcurISOlogg}[1]{\ifnum#1=39 %
\hatcurISOloggxxxxA
\else
\ifnum#1=40 %
\hatcurISOloggxxxxB
\else
\ifnum#1=41 %
\hatcurISOloggxxxxC
\else
??????\fi
\fi
\fi
}
\newcommand{\hatcurISOlum}[1]{\ifnum#1=39 %
\hatcurISOlumxxxxA
\else
\ifnum#1=40 %
\hatcurISOlumxxxxB
\else
\ifnum#1=41 %
\hatcurISOlumxxxxC
\else
??????\fi
\fi
\fi
}
\newcommand{\hatcurISOlumshort}[1]{\ifnum#1=39 %
\hatcurISOlumshortxxxxA
\else
\ifnum#1=40 %
\hatcurISOlumshortxxxxB
\else
\ifnum#1=41 %
\hatcurISOlumshortxxxxC
\else
??????\fi
\fi
\fi
}
\newcommand{\hatcurISOm}[1]{\ifnum#1=39 %
\hatcurISOmxxxxA
\else
\ifnum#1=40 %
\hatcurISOmxxxxB
\else
\ifnum#1=41 %
\hatcurISOmxxxxC
\else
??????\fi
\fi
\fi
}
\newcommand{\hatcurISOMH}[1]{\ifnum#1=39 %
\hatcurISOMHxxxxA
\else
\ifnum#1=40 %
\hatcurISOMHxxxxB
\else
\ifnum#1=41 %
\hatcurISOMHxxxxC
\else
??????\fi
\fi
\fi
}
\newcommand{\hatcurISOMJ}[1]{\ifnum#1=39 %
\hatcurISOMJxxxxA
\else
\ifnum#1=40 %
\hatcurISOMJxxxxB
\else
\ifnum#1=41 %
\hatcurISOMJxxxxC
\else
??????\fi
\fi
\fi
}
\newcommand{\hatcurISOMK}[1]{\ifnum#1=39 %
\hatcurISOMKxxxxA
\else
\ifnum#1=40 %
\hatcurISOMKxxxxB
\else
\ifnum#1=41 %
\hatcurISOMKxxxxC
\else
??????\fi
\fi
\fi
}
\newcommand{\hatcurISOmlong}[1]{\ifnum#1=39 %
\hatcurISOmlongxxxxA
\else
\ifnum#1=40 %
\hatcurISOmlongxxxxB
\else
\ifnum#1=41 %
\hatcurISOmlongxxxxC
\else
??????\fi
\fi
\fi
}
\newcommand{\hatcurISOmshort}[1]{\ifnum#1=39 %
\hatcurISOmshortxxxxA
\else
\ifnum#1=40 %
\hatcurISOmshortxxxxB
\else
\ifnum#1=41 %
\hatcurISOmshortxxxxC
\else
??????\fi
\fi
\fi
}
\newcommand{\hatcurISOmv}[1]{\ifnum#1=39 %
\hatcurISOmvxxxxA
\else
\ifnum#1=40 %
\hatcurISOmvxxxxB
\else
\ifnum#1=41 %
\hatcurISOmvxxxxC
\else
??????\fi
\fi
\fi
}
\newcommand{\hatcurISOr}[1]{\ifnum#1=39 %
\hatcurISOrxxxxA
\else
\ifnum#1=40 %
\hatcurISOrxxxxB
\else
\ifnum#1=41 %
\hatcurISOrxxxxC
\else
??????\fi
\fi
\fi
}
\newcommand{\hatcurISOrho}[1]{\ifnum#1=39 %
\hatcurISOrhoxxxxA
\else
\ifnum#1=40 %
\hatcurISOrhoxxxxB
\else
\ifnum#1=41 %
\hatcurISOrhoxxxxC
\else
??????\fi
\fi
\fi
}
\newcommand{\hatcurISOrlong}[1]{\ifnum#1=39 %
\hatcurISOrlongxxxxA
\else
\ifnum#1=40 %
\hatcurISOrlongxxxxB
\else
\ifnum#1=41 %
\hatcurISOrlongxxxxC
\else
??????\fi
\fi
\fi
}
\newcommand{\hatcurISOrshort}[1]{\ifnum#1=39 %
\hatcurISOrshortxxxxA
\else
\ifnum#1=40 %
\hatcurISOrshortxxxxB
\else
\ifnum#1=41 %
\hatcurISOrshortxxxxC
\else
??????\fi
\fi
\fi
}
\newcommand{\hatcurISOsigma}[1]{\ifnum#1=39 %
\hatcurISOsigmaxxxxA
\else
\ifnum#1=40 %
\hatcurISOsigmaxxxxB
\else
\ifnum#1=41 %
\hatcurISOsigmaxxxxC
\else
??????\fi
\fi
\fi
}
\newcommand{\hatcurISOspec}[1]{\ifnum#1=39 %
\hatcurISOspecxxxxA
\else
\ifnum#1=40 %
\hatcurISOspecxxxxB
\else
\ifnum#1=41 %
\hatcurISOspecxxxxC
\else
??????\fi
\fi
\fi
}
\newcommand{\hatcurISOvi}[1]{\ifnum#1=39 %
\hatcurISOvixxxxA
\else
\ifnum#1=40 %
\hatcurISOvixxxxB
\else
\ifnum#1=41 %
\hatcurISOvixxxxC
\else
??????\fi
\fi
\fi
}
\newcommand{\hatcurLBig}[1]{\ifnum#1=39 %
\hatcurLBigxxxxA
\else
\ifnum#1=40 %
\hatcurLBigxxxxB
\else
\ifnum#1=41 %
\hatcurLBigxxxxC
\else
??????\fi
\fi
\fi
}
\newcommand{\hatcurLBii}[1]{\ifnum#1=39 %
\hatcurLBiixxxxA
\else
\ifnum#1=40 %
\hatcurLBiixxxxB
\else
\ifnum#1=41 %
\hatcurLBiixxxxC
\else
??????\fi
\fi
\fi
}
\newcommand{\hatcurLBiI}[1]{\ifnum#1=39 %
\hatcurLBiIxxxxA
\else
\ifnum#1=40 %
\hatcurLBiIxxxxB
\else
\ifnum#1=41 %
\hatcurLBiIxxxxC
\else
??????\fi
\fi
\fi
}
\newcommand{\hatcurLBir}[1]{\ifnum#1=39 %
\hatcurLBirxxxxA
\else
\ifnum#1=40 %
\hatcurLBirxxxxB
\else
\ifnum#1=41 %
\hatcurLBirxxxxC
\else
??????\fi
\fi
\fi
}
\newcommand{\hatcurLBiR}[1]{\ifnum#1=39 %
\hatcurLBiRxxxxA
\else
\ifnum#1=40 %
\hatcurLBiRxxxxB
\else
\ifnum#1=41 %
\hatcurLBiRxxxxC
\else
??????\fi
\fi
\fi
}
\newcommand{\hatcurLBiig}[1]{\ifnum#1=39 %
\hatcurLBiigxxxxA
\else
\ifnum#1=40 %
\hatcurLBiigxxxxB
\else
\ifnum#1=41 %
\hatcurLBiigxxxxC
\else
??????\fi
\fi
\fi
}
\newcommand{\hatcurLBiii}[1]{\ifnum#1=39 %
\hatcurLBiiixxxxA
\else
\ifnum#1=40 %
\hatcurLBiiixxxxB
\else
\ifnum#1=41 %
\hatcurLBiiixxxxC
\else
??????\fi
\fi
\fi
}
\newcommand{\hatcurLBiiI}[1]{\ifnum#1=39 %
\hatcurLBiiIxxxxA
\else
\ifnum#1=40 %
\hatcurLBiiIxxxxB
\else
\ifnum#1=41 %
\hatcurLBiiIxxxxC
\else
??????\fi
\fi
\fi
}
\newcommand{\hatcurLBiikep}[1]{\ifnum#1=39 %
\hatcurLBiikepxxxxA
\else
\ifnum#1=40 %
\hatcurLBiikepxxxxB
\else
\ifnum#1=41 %
\hatcurLBiikepxxxxC
\else
??????\fi
\fi
\fi
}
\newcommand{\hatcurLBiiz}[1]{\ifnum#1=39 %
\hatcurLBiizxxxxA
\else
\ifnum#1=40 %
\hatcurLBiizxxxxB
\else
\ifnum#1=41 %
\hatcurLBiizxxxxC
\else
??????\fi
\fi
\fi
}
\newcommand{\hatcurLBiir}[1]{\ifnum#1=39 %
\hatcurLBiirxxxxA
\else
\ifnum#1=40 %
\hatcurLBiirxxxxB
\else
\ifnum#1=41 %
\hatcurLBiirxxxxC
\else
??????\fi
\fi
\fi
}
\newcommand{\hatcurLBiiR}[1]{\ifnum#1=39 %
\hatcurLBiiRxxxxA
\else
\ifnum#1=40 %
\hatcurLBiiRxxxxB
\else
\ifnum#1=41 %
\hatcurLBiiRxxxxC
\else
??????\fi
\fi
\fi
}
\newcommand{\hatcurLBikep}[1]{\ifnum#1=39 %
\hatcurLBikepxxxxA
\else
\ifnum#1=40 %
\hatcurLBikepxxxxB
\else
\ifnum#1=41 %
\hatcurLBikepxxxxC
\else
??????\fi
\fi
\fi
}
\newcommand{\hatcurLBiz}[1]{\ifnum#1=39 %
\hatcurLBizxxxxA
\else
\ifnum#1=40 %
\hatcurLBizxxxxB
\else
\ifnum#1=41 %
\hatcurLBizxxxxC
\else
??????\fi
\fi
\fi
}
\newcommand{\hatcurLCbsq}[1]{\ifnum#1=39 %
\hatcurLCbsqxxxxA
\else
\ifnum#1=40 %
\hatcurLCbsqxxxxB
\else
\ifnum#1=41 %
\hatcurLCbsqxxxxC
\else
??????\fi
\fi
\fi
}
\newcommand{\hatcurLCdip}[1]{\ifnum#1=39 %
\hatcurLCdipxxxxA
\else
\ifnum#1=40 %
\hatcurLCdipxxxxB
\else
\ifnum#1=41 %
\hatcurLCdipxxxxC
\else
??????\fi
\fi
\fi
}
\newcommand{\hatcurLCdur}[1]{\ifnum#1=39 %
\hatcurLCdurxxxxA
\else
\ifnum#1=40 %
\hatcurLCdurxxxxB
\else
\ifnum#1=41 %
\hatcurLCdurxxxxC
\else
??????\fi
\fi
\fi
}
\newcommand{\hatcurLCdurhr}[1]{\ifnum#1=39 %
\hatcurLCdurhrxxxxA
\else
\ifnum#1=40 %
\hatcurLCdurhrxxxxB
\else
\ifnum#1=41 %
\hatcurLCdurhrxxxxC
\else
??????\fi
\fi
\fi
}
\newcommand{\hatcurLCdurhrshort}[1]{\ifnum#1=39 %
\hatcurLCdurhrshortxxxxA
\else
\ifnum#1=40 %
\hatcurLCdurhrshortxxxxB
\else
\ifnum#1=41 %
\hatcurLCdurhrshortxxxxC
\else
??????\fi
\fi
\fi
}
\newcommand{\hatcurLCdurshort}[1]{\ifnum#1=39 %
\hatcurLCdurshortxxxxA
\else
\ifnum#1=40 %
\hatcurLCdurshortxxxxB
\else
\ifnum#1=41 %
\hatcurLCdurshortxxxxC
\else
??????\fi
\fi
\fi
}
\newcommand{\hatcurLChatnetmA}[1]{\ifnum#1=39 %
\hatcurLChatnetmAxxxxA
\else
\ifnum#1=40 %
\hatcurLChatnetmAxxxxB
\else
\ifnum#1=41 %
\hatcurLChatnetmAxxxxC
\else
??????\fi
\fi
\fi
}
\newcommand{\hatcurLChatnetmB}[1]{\ifnum#1=39 %
\hatcurLChatnetmBxxxxA
\else
\ifnum#1=40 %
\hatcurLChatnetmBxxxxB
\else
\ifnum#1=41 %
\hatcurLChatnetmBxxxxC
\else
??????\fi
\fi
\fi
}
\newcommand{\hatcurLCiblendA}[1]{\ifnum#1=39 %
\hatcurLCiblendAxxxxA
\else
\ifnum#1=40 %
\hatcurLCiblendAxxxxB
\else
\ifnum#1=41 %
\hatcurLCiblendAxxxxC
\else
??????\fi
\fi
\fi
}
\newcommand{\hatcurLCiblendB}[1]{\ifnum#1=39 %
\hatcurLCiblendBxxxxA
\else
\ifnum#1=40 %
\hatcurLCiblendBxxxxB
\else
\ifnum#1=41 %
\hatcurLCiblendBxxxxC
\else
??????\fi
\fi
\fi
}
\newcommand{\hatcurLCimp}[1]{\ifnum#1=39 %
\hatcurLCimpxxxxA
\else
\ifnum#1=40 %
\hatcurLCimpxxxxB
\else
\ifnum#1=41 %
\hatcurLCimpxxxxC
\else
??????\fi
\fi
\fi
}
\newcommand{\hatcurLCingdur}[1]{\ifnum#1=39 %
\hatcurLCingdurxxxxA
\else
\ifnum#1=40 %
\hatcurLCingdurxxxxB
\else
\ifnum#1=41 %
\hatcurLCingdurxxxxC
\else
??????\fi
\fi
\fi
}
\newcommand{\hatcurLCP}[1]{\ifnum#1=39 %
\hatcurLCPxxxxA
\else
\ifnum#1=40 %
\hatcurLCPxxxxB
\else
\ifnum#1=41 %
\hatcurLCPxxxxC
\else
??????\fi
\fi
\fi
}
\newcommand{\hatcurLCPprec}[1]{\ifnum#1=39 %
\hatcurLCPprecxxxxA
\else
\ifnum#1=40 %
\hatcurLCPprecxxxxB
\else
\ifnum#1=41 %
\hatcurLCPprecxxxxC
\else
??????\fi
\fi
\fi
}
\newcommand{\hatcurLCPshort}[1]{\ifnum#1=39 %
\hatcurLCPshortxxxxA
\else
\ifnum#1=40 %
\hatcurLCPshortxxxxB
\else
\ifnum#1=41 %
\hatcurLCPshortxxxxC
\else
??????\fi
\fi
\fi
}
\newcommand{\hatcurLCq}[1]{\ifnum#1=39 %
\hatcurLCqxxxxA
\else
\ifnum#1=40 %
\hatcurLCqxxxxB
\else
\ifnum#1=41 %
\hatcurLCqxxxxC
\else
??????\fi
\fi
\fi
}
\newcommand{\hatcurLCqshort}[1]{\ifnum#1=39 %
\hatcurLCqshortxxxxA
\else
\ifnum#1=40 %
\hatcurLCqshortxxxxB
\else
\ifnum#1=41 %
\hatcurLCqshortxxxxC
\else
??????\fi
\fi
\fi
}
\newcommand{\hatcurLCrprstar}[1]{\ifnum#1=39 %
\hatcurLCrprstarxxxxA
\else
\ifnum#1=40 %
\hatcurLCrprstarxxxxB
\else
\ifnum#1=41 %
\hatcurLCrprstarxxxxC
\else
??????\fi
\fi
\fi
}
\newcommand{\hatcurLCT}[1]{\ifnum#1=39 %
\hatcurLCTxxxxA
\else
\ifnum#1=40 %
\hatcurLCTxxxxB
\else
\ifnum#1=41 %
\hatcurLCTxxxxC
\else
??????\fi
\fi
\fi
}
\newcommand{\hatcurLCTA}[1]{\ifnum#1=39 %
\hatcurLCTAxxxxA
\else
\ifnum#1=40 %
\hatcurLCTAxxxxB
\else
\ifnum#1=41 %
\hatcurLCTAxxxxC
\else
??????\fi
\fi
\fi
}
\newcommand{\hatcurLCTB}[1]{\ifnum#1=39 %
\hatcurLCTBxxxxA
\else
\ifnum#1=40 %
\hatcurLCTBxxxxB
\else
\ifnum#1=41 %
\hatcurLCTBxxxxC
\else
??????\fi
\fi
\fi
}
\newcommand{\hatcurLCzeta}[1]{\ifnum#1=39 %
\hatcurLCzetaxxxxA
\else
\ifnum#1=40 %
\hatcurLCzetaxxxxB
\else
\ifnum#1=41 %
\hatcurLCzetaxxxxC
\else
??????\fi
\fi
\fi
}
\newcommand{\hatcurPPaequiv}[1]{\ifnum#1=39 %
\hatcurPPaequivxxxxA
\else
\ifnum#1=40 %
\hatcurPPaequivxxxxB
\else
\ifnum#1=41 %
\hatcurPPaequivxxxxC
\else
??????\fi
\fi
\fi
}
\newcommand{\hatcurPPar}[1]{\ifnum#1=39 %
\hatcurPParxxxxA
\else
\ifnum#1=40 %
\hatcurPParxxxxB
\else
\ifnum#1=41 %
\hatcurPParxxxxC
\else
??????\fi
\fi
\fi
}
\newcommand{\hatcurPParel}[1]{\ifnum#1=39 %
\hatcurPParelxxxxA
\else
\ifnum#1=40 %
\hatcurPParelxxxxB
\else
\ifnum#1=41 %
\hatcurPParelxxxxC
\else
??????\fi
\fi
\fi
}
\newcommand{\hatcurPPfluxap}[1]{\ifnum#1=39 %
\hatcurPPfluxapxxxxA
\else
\ifnum#1=40 %
\hatcurPPfluxapxxxxB
\else
\ifnum#1=41 %
\hatcurPPfluxapxxxxC
\else
??????\fi
\fi
\fi
}
\newcommand{\hatcurPPfluxapdim}[1]{\ifnum#1=39 %
\hatcurPPfluxapdimxxxxA
\else
\ifnum#1=40 %
\hatcurPPfluxapdimxxxxB
\else
\ifnum#1=41 %
\hatcurPPfluxapdimxxxxC
\else
??????\fi
\fi
\fi
}
\newcommand{\hatcurPPfluxavg}[1]{\ifnum#1=39 %
\hatcurPPfluxavgxxxxA
\else
\ifnum#1=40 %
\hatcurPPfluxavgxxxxB
\else
\ifnum#1=41 %
\hatcurPPfluxavgxxxxC
\else
??????\fi
\fi
\fi
}
\newcommand{\hatcurPPfluxavgdim}[1]{\ifnum#1=39 %
\hatcurPPfluxavgdimxxxxA
\else
\ifnum#1=40 %
\hatcurPPfluxavgdimxxxxB
\else
\ifnum#1=41 %
\hatcurPPfluxavgdimxxxxC
\else
??????\fi
\fi
\fi
}
\newcommand{\hatcurPPfluxperi}[1]{\ifnum#1=39 %
\hatcurPPfluxperixxxxA
\else
\ifnum#1=40 %
\hatcurPPfluxperixxxxB
\else
\ifnum#1=41 %
\hatcurPPfluxperixxxxC
\else
??????\fi
\fi
\fi
}
\newcommand{\hatcurPPfluxperidim}[1]{\ifnum#1=39 %
\hatcurPPfluxperidimxxxxA
\else
\ifnum#1=40 %
\hatcurPPfluxperidimxxxxB
\else
\ifnum#1=41 %
\hatcurPPfluxperidimxxxxC
\else
??????\fi
\fi
\fi
}
\newcommand{\hatcurPPg}[1]{\ifnum#1=39 %
\hatcurPPgxxxxA
\else
\ifnum#1=40 %
\hatcurPPgxxxxB
\else
\ifnum#1=41 %
\hatcurPPgxxxxC
\else
??????\fi
\fi
\fi
}
\newcommand{\hatcurPPi}[1]{\ifnum#1=39 %
\hatcurPPixxxxA
\else
\ifnum#1=40 %
\hatcurPPixxxxB
\else
\ifnum#1=41 %
\hatcurPPixxxxC
\else
??????\fi
\fi
\fi
}
\newcommand{\hatcurPPlogg}[1]{\ifnum#1=39 %
\hatcurPPloggxxxxA
\else
\ifnum#1=40 %
\hatcurPPloggxxxxB
\else
\ifnum#1=41 %
\hatcurPPloggxxxxC
\else
??????\fi
\fi
\fi
}
\newcommand{\hatcurPPm}[1]{\ifnum#1=39 %
\hatcurPPmxxxxA
\else
\ifnum#1=40 %
\hatcurPPmxxxxB
\else
\ifnum#1=41 %
\hatcurPPmxxxxC
\else
??????\fi
\fi
\fi
}
\newcommand{\hatcurPPme}[1]{\ifnum#1=39 %
\hatcurPPmexxxxA
\else
\ifnum#1=40 %
\hatcurPPmexxxxB
\else
\ifnum#1=41 %
\hatcurPPmexxxxC
\else
??????\fi
\fi
\fi
}
\newcommand{\hatcurPPmelong}[1]{\ifnum#1=39 %
\hatcurPPmelongxxxxA
\else
\ifnum#1=40 %
\hatcurPPmelongxxxxB
\else
\ifnum#1=41 %
\hatcurPPmelongxxxxC
\else
??????\fi
\fi
\fi
}
\newcommand{\hatcurPPmeshort}[1]{\ifnum#1=39 %
\hatcurPPmeshortxxxxA
\else
\ifnum#1=40 %
\hatcurPPmeshortxxxxB
\else
\ifnum#1=41 %
\hatcurPPmeshortxxxxC
\else
??????\fi
\fi
\fi
}
\newcommand{\hatcurPPmlong}[1]{\ifnum#1=39 %
\hatcurPPmlongxxxxA
\else
\ifnum#1=40 %
\hatcurPPmlongxxxxB
\else
\ifnum#1=41 %
\hatcurPPmlongxxxxC
\else
??????\fi
\fi
\fi
}
\newcommand{\hatcurPPmrcorr}[1]{\ifnum#1=39 %
\hatcurPPmrcorrxxxxA
\else
\ifnum#1=40 %
\hatcurPPmrcorrxxxxB
\else
\ifnum#1=41 %
\hatcurPPmrcorrxxxxC
\else
??????\fi
\fi
\fi
}
\newcommand{\hatcurPPmshort}[1]{\ifnum#1=39 %
\hatcurPPmshortxxxxA
\else
\ifnum#1=40 %
\hatcurPPmshortxxxxB
\else
\ifnum#1=41 %
\hatcurPPmshortxxxxC
\else
??????\fi
\fi
\fi
}
\newcommand{\hatcurPPperi}[1]{\ifnum#1=39 %
\hatcurPPperixxxxA
\else
\ifnum#1=40 %
\hatcurPPperixxxxB
\else
\ifnum#1=41 %
\hatcurPPperixxxxC
\else
??????\fi
\fi
\fi
}
\newcommand{\hatcurPPphiconj}[1]{\ifnum#1=39 %
\hatcurPPphiconjxxxxA
\else
\ifnum#1=40 %
\hatcurPPphiconjxxxxB
\else
\ifnum#1=41 %
\hatcurPPphiconjxxxxC
\else
??????\fi
\fi
\fi
}
\newcommand{\hatcurPPr}[1]{\ifnum#1=39 %
\hatcurPPrxxxxA
\else
\ifnum#1=40 %
\hatcurPPrxxxxB
\else
\ifnum#1=41 %
\hatcurPPrxxxxC
\else
??????\fi
\fi
\fi
}
\newcommand{\hatcurPPre}[1]{\ifnum#1=39 %
\hatcurPPrexxxxA
\else
\ifnum#1=40 %
\hatcurPPrexxxxB
\else
\ifnum#1=41 %
\hatcurPPrexxxxC
\else
??????\fi
\fi
\fi
}
\newcommand{\hatcurPPrelong}[1]{\ifnum#1=39 %
\hatcurPPrelongxxxxA
\else
\ifnum#1=40 %
\hatcurPPrelongxxxxB
\else
\ifnum#1=41 %
\hatcurPPrelongxxxxC
\else
??????\fi
\fi
\fi
}
\newcommand{\hatcurPPreshort}[1]{\ifnum#1=39 %
\hatcurPPreshortxxxxA
\else
\ifnum#1=40 %
\hatcurPPreshortxxxxB
\else
\ifnum#1=41 %
\hatcurPPreshortxxxxC
\else
??????\fi
\fi
\fi
}
\newcommand{\hatcurPPrho}[1]{\ifnum#1=39 %
\hatcurPPrhoxxxxA
\else
\ifnum#1=40 %
\hatcurPPrhoxxxxB
\else
\ifnum#1=41 %
\hatcurPPrhoxxxxC
\else
??????\fi
\fi
\fi
}
\newcommand{\hatcurPPrlong}[1]{\ifnum#1=39 %
\hatcurPPrlongxxxxA
\else
\ifnum#1=40 %
\hatcurPPrlongxxxxB
\else
\ifnum#1=41 %
\hatcurPPrlongxxxxC
\else
??????\fi
\fi
\fi
}
\newcommand{\hatcurPPrshort}[1]{\ifnum#1=39 %
\hatcurPPrshortxxxxA
\else
\ifnum#1=40 %
\hatcurPPrshortxxxxB
\else
\ifnum#1=41 %
\hatcurPPrshortxxxxC
\else
??????\fi
\fi
\fi
}
\newcommand{\hatcurPPtcirc}[1]{\ifnum#1=39 %
\hatcurPPtcircxxxxA
\else
\ifnum#1=40 %
\hatcurPPtcircxxxxB
\else
\ifnum#1=41 %
\hatcurPPtcircxxxxC
\else
??????\fi
\fi
\fi
}
\newcommand{\hatcurPPteff}[1]{\ifnum#1=39 %
\hatcurPPteffxxxxA
\else
\ifnum#1=40 %
\hatcurPPteffxxxxB
\else
\ifnum#1=41 %
\hatcurPPteffxxxxC
\else
??????\fi
\fi
\fi
}
\newcommand{\hatcurPPtheta}[1]{\ifnum#1=39 %
\hatcurPPthetaxxxxA
\else
\ifnum#1=40 %
\hatcurPPthetaxxxxB
\else
\ifnum#1=41 %
\hatcurPPthetaxxxxC
\else
??????\fi
\fi
\fi
}
\newcommand{\hatcurPPtinfall}[1]{\ifnum#1=39 %
\hatcurPPtinfallxxxxA
\else
\ifnum#1=40 %
\hatcurPPtinfallxxxxB
\else
\ifnum#1=41 %
\hatcurPPtinfallxxxxC
\else
??????\fi
\fi
\fi
}
\newcommand{\hatcurRVeccen}[1]{\ifnum#1=39 %
\hatcurRVeccenxxxxA
\else
\ifnum#1=40 %
\hatcurRVeccenxxxxB
\else
\ifnum#1=41 %
\hatcurRVeccenxxxxC
\else
??????\fi
\fi
\fi
}
\newcommand{\hatcurRVfitrms}[1]{\ifnum#1=39 %
\hatcurRVfitrmsxxxxA
\else
\ifnum#1=40 %
\hatcurRVfitrmsxxxxB
\else
\ifnum#1=41 %
\hatcurRVfitrmsxxxxC
\else
??????\fi
\fi
\fi
}
\newcommand{\hatcurRVgamma}[1]{\ifnum#1=39 %
\hatcurRVgammaxxxxA
\else
\ifnum#1=40 %
\hatcurRVgammaxxxxB
\else
\ifnum#1=41 %
\hatcurRVgammaxxxxC
\else
??????\fi
\fi
\fi
}
\newcommand{\hatcurRVh}[1]{\ifnum#1=39 %
\hatcurRVhxxxxA
\else
\ifnum#1=40 %
\hatcurRVhxxxxB
\else
\ifnum#1=41 %
\hatcurRVhxxxxC
\else
??????\fi
\fi
\fi
}
\newcommand{\hatcurRVjitter}[1]{\ifnum#1=39 %
\hatcurRVjitterxxxxA
\else
\ifnum#1=40 %
\hatcurRVjitterxxxxB
\else
\ifnum#1=41 %
\hatcurRVjitterAxxxxC
\else
??????\fi
\fi
\fi
}
\newcommand{\hatcurRVk}[1]{\ifnum#1=39 %
\hatcurRVkxxxxA
\else
\ifnum#1=40 %
\hatcurRVkxxxxB
\else
\ifnum#1=41 %
\hatcurRVkxxxxC
\else
??????\fi
\fi
\fi
}
\newcommand{\hatcurRVK}[1]{\ifnum#1=39 %
\hatcurRVKxxxxA
\else
\ifnum#1=40 %
\hatcurRVKxxxxB
\else
\ifnum#1=41 %
\hatcurRVKxxxxC
\else
??????\fi
\fi
\fi
}
\newcommand{\hatcurRVomega}[1]{\ifnum#1=39 %
\hatcurRVomegaxxxxA
\else
\ifnum#1=40 %
\hatcurRVomegaxxxxB
\else
\ifnum#1=41 %
\hatcurRVomegaxxxxC
\else
??????\fi
\fi
\fi
}
\newcommand{\hatcurRVtrone}[1]{\ifnum#1=39 %
\hatcurRVtronexxxxA
\else
\ifnum#1=40 %
\hatcurRVtronexxxxB
\else
\ifnum#1=41 %
\hatcurRVtronexxxxC
\else
??????\fi
\fi
\fi
}
\newcommand{\hatcurRVtrtwo}[1]{\ifnum#1=39 %
\hatcurRVtrtwoxxxxA
\else
\ifnum#1=40 %
\hatcurRVtrtwoxxxxB
\else
\ifnum#1=41 %
\hatcurRVtrtwoxxxxC
\else
??????\fi
\fi
\fi
}
\newcommand{\hatcurSMEiilogg}[1]{\ifnum#1=39 %
\hatcurSMEiiloggxxxxA
\else
\ifnum#1=40 %
\hatcurSMEiiloggxxxxB
\else
\ifnum#1=41 %
\hatcurSMEiiloggxxxxC
\else
??????\fi
\fi
\fi
}
\newcommand{\hatcurSMEiiteff}[1]{\ifnum#1=39 %
\hatcurSMEiiteffxxxxA
\else
\ifnum#1=40 %
\hatcurSMEiiteffxxxxB
\else
\ifnum#1=41 %
\hatcurSMEiiteffxxxxC
\else
??????\fi
\fi
\fi
}
\newcommand{\hatcurSMEiivmac}[1]{\ifnum#1=39 %
\hatcurSMEiivmacxxxxA
\else
\ifnum#1=40 %
\hatcurSMEiivmacxxxxB
\else
\ifnum#1=41 %
\hatcurSMEiivmacxxxxC
\else
??????\fi
\fi
\fi
}
\newcommand{\hatcurSMEiivmic}[1]{\ifnum#1=39 %
\hatcurSMEiivmicxxxxA
\else
\ifnum#1=40 %
\hatcurSMEiivmicxxxxB
\else
\ifnum#1=41 %
\hatcurSMEiivmicxxxxC
\else
??????\fi
\fi
\fi
}
\newcommand{\hatcurSMEiivsin}[1]{\ifnum#1=39 %
\hatcurSMEiivsinxxxxA
\else
\ifnum#1=40 %
\hatcurSMEiivsinxxxxB
\else
\ifnum#1=41 %
\hatcurSMEiivsinxxxxC
\else
??????\fi
\fi
\fi
}
\newcommand{\hatcurSMEiizfeh}[1]{\ifnum#1=39 %
\hatcurSMEiizfehxxxxA
\else
\ifnum#1=40 %
\hatcurSMEiizfehxxxxB
\else
\ifnum#1=41 %
\hatcurSMEiizfehxxxxC
\else
??????\fi
\fi
\fi
}
\newcommand{\hatcurSMEiizfehshort}[1]{\ifnum#1=39 %
\hatcurSMEiizfehshortxxxxA
\else
\ifnum#1=40 %
\hatcurSMEiizfehshortxxxxB
\else
\ifnum#1=41 %
\hatcurSMEiizfehshortxxxxC
\else
??????\fi
\fi
\fi
}
\newcommand{\hatcurSMEilogg}[1]{\ifnum#1=39 %
\hatcurSMEiloggxxxxA
\else
\ifnum#1=40 %
\hatcurSMEiloggxxxxB
\else
\ifnum#1=41 %
\hatcurSMEiloggxxxxC
\else
??????\fi
\fi
\fi
}
\newcommand{\hatcurSMEiteff}[1]{\ifnum#1=39 %
\hatcurSMEiteffxxxxA
\else
\ifnum#1=40 %
\hatcurSMEiteffxxxxB
\else
\ifnum#1=41 %
\hatcurSMEiteffxxxxC
\else
??????\fi
\fi
\fi
}
\newcommand{\hatcurSMEivmac}[1]{\ifnum#1=39 %
\hatcurSMEivmacxxxxA
\else
\ifnum#1=40 %
\hatcurSMEivmacxxxxB
\else
\ifnum#1=41 %
\hatcurSMEivmacxxxxC
\else
??????\fi
\fi
\fi
}
\newcommand{\hatcurSMEivmic}[1]{\ifnum#1=39 %
\hatcurSMEivmicxxxxA
\else
\ifnum#1=40 %
\hatcurSMEivmicxxxxB
\else
\ifnum#1=41 %
\hatcurSMEivmicxxxxC
\else
??????\fi
\fi
\fi
}
\newcommand{\hatcurSMEivsin}[1]{\ifnum#1=39 %
\hatcurSMEivsinxxxxA
\else
\ifnum#1=40 %
\hatcurSMEivsinxxxxB
\else
\ifnum#1=41 %
\hatcurSMEivsinxxxxC
\else
??????\fi
\fi
\fi
}
\newcommand{\hatcurSMEizfeh}[1]{\ifnum#1=39 %
\hatcurSMEizfehxxxxA
\else
\ifnum#1=40 %
\hatcurSMEizfehxxxxB
\else
\ifnum#1=41 %
\hatcurSMEizfehxxxxC
\else
??????\fi
\fi
\fi
}
\newcommand{\hatcurSMEizfehshort}[1]{\ifnum#1=39 %
\hatcurSMEizfehshortxxxxA
\else
\ifnum#1=40 %
\hatcurSMEizfehshortxxxxB
\else
\ifnum#1=41 %
\hatcurSMEizfehshortxxxxC
\else
??????\fi
\fi
\fi
}
\newcommand{\hatcurTRESgamma}[1]{\ifnum#1=39 %
\hatcurTRESgammaxxxxA
\else
\ifnum#1=40 %
\hatcurTRESgammaxxxxB
\else
\ifnum#1=41 %
\hatcurTRESgammaxxxxC
\else
??????\fi
\fi
\fi
}
\newcommand{\hatcurTRESlogg}[1]{\ifnum#1=39 %
\hatcurTRESloggxxxxA
\else
\ifnum#1=40 %
\hatcurTRESloggxxxxB
\else
\ifnum#1=41 %
\hatcurTRESloggxxxxC
\else
??????\fi
\fi
\fi
}
\newcommand{\hatcurTRESnumspec}[1]{\ifnum#1=39 %
\hatcurTRESnumspecxxxxA
\else
\ifnum#1=40 %
\hatcurTRESnumspecxxxxB
\else
\ifnum#1=41 %
\hatcurTRESnumspecxxxxC
\else
??????\fi
\fi
\fi
}
\newcommand{\hatcurTRESrvrms}[1]{\ifnum#1=39 %
\hatcurTRESrvrmsxxxxA
\else
\ifnum#1=40 %
\hatcurTRESrvrmsxxxxB
\else
\ifnum#1=41 %
\hatcurTRESrvrmsxxxxC
\else
??????\fi
\fi
\fi
}
\newcommand{\hatcurTRESspan}[1]{\ifnum#1=39 %
\hatcurTRESspanxxxxA
\else
\ifnum#1=40 %
\hatcurTRESspanxxxxB
\else
\ifnum#1=41 %
\hatcurTRESspanxxxxC
\else
??????\fi
\fi
\fi
}
\newcommand{\hatcurTRESteff}[1]{\ifnum#1=39 %
\hatcurTRESteffxxxxA
\else
\ifnum#1=40 %
\hatcurTRESteffxxxxB
\else
\ifnum#1=41 %
\hatcurTRESteffxxxxC
\else
??????\fi
\fi
\fi
}
\newcommand{\hatcurTRESvsini}[1]{\ifnum#1=39 %
\hatcurTRESvsinixxxxA
\else
\ifnum#1=40 %
\hatcurTRESvsinixxxxB
\else
\ifnum#1=41 %
\hatcurTRESvsinixxxxC
\else
??????\fi
\fi
\fi
}
\newcommand{\hatcurTRESzfeh}[1]{\ifnum#1=39 %
\hatcurTRESzfehxxxxA
\else
\ifnum#1=40 %
\hatcurTRESzfehxxxxB
\else
\ifnum#1=41 %
\hatcurTRESzfehxxxxC
\else
??????\fi
\fi
\fi
}
\newcommand{\hatcurXdist}[1]{\ifnum#1=39 %
\hatcurXdistxxxxA
\else
\ifnum#1=40 %
\hatcurXdistxxxxB
\else
\ifnum#1=41 %
\hatcurXdistxxxxC
\else
??????\fi
\fi
\fi
}
\newcommand{\hatcurXdistred}[1]{\ifnum#1=39 %
\hatcurXdistredxxxxA
\else
\ifnum#1=40 %
\hatcurXdistredxxxxB
\else
\ifnum#1=41 %
\hatcurXdistredxxxxC
\else
??????\fi
\fi
\fi
}
\newcommand{\hatcurXAv}[1]{\ifnum#1=39 %
\hatcurXAvxxxxA
\else
\ifnum#1=40 %
\hatcurXAvxxxxB
\else
\ifnum#1=41 %
\hatcurXAvxxxxC
\else
??????\fi
\fi
\fi
}
\newcommand{\hatcurXsecdur}[1]{\ifnum#1=39 %
\hatcurXsecdurxxxxA
\else
\ifnum#1=40 %
\hatcurXsecdurxxxxB
\else
\ifnum#1=41 %
\hatcurXsecdurxxxxC
\else
??????\fi
\fi
\fi
}
\newcommand{\hatcurXsecingdur}[1]{\ifnum#1=39 %
\hatcurXsecingdurxxxxA
\else
\ifnum#1=40 %
\hatcurXsecingdurxxxxB
\else
\ifnum#1=41 %
\hatcurXsecingdurxxxxC
\else
??????\fi
\fi
\fi
}
\newcommand{\hatcurXsecondary}[1]{\ifnum#1=39 %
\hatcurXsecondaryxxxxA
\else
\ifnum#1=40 %
\hatcurXsecondaryxxxxB
\else
\ifnum#1=41 %
\hatcurXsecondaryxxxxC
\else
??????\fi
\fi
\fi
}
\newcommand{\hatcurXsecphase}[1]{\ifnum#1=39 %
\hatcurXsecphasexxxxA
\else
\ifnum#1=40 %
\hatcurXsecphasexxxxB
\else
\ifnum#1=41 %
\hatcurXsecphasexxxxC
\else
??????\fi
\fi
\fi
}

\newcommand{\hatcurhtrxxxxeccenA}{HTR315-005}                          
\newcommand{\hatcurfieldxxxxeccenA}{267}                               
\newcommand{\hatcurCCraxxxxeccenA}{\ensuremath{07^{\mathrm h}35^{\mathrm m}01.97{\mathrm s}}}                        
\newcommand{\hatcurCCdecxxxxeccenA}{\ensuremath{+17{\arcdeg}49{\arcmin}48.3{\arcsec}}}                       
\newcommand{\hatcurCCmagxxxxeccenA}{12.422}                            
\newcommand{\hatcurCCtwomassxxxxeccenA}{2MASS~07350197+1749482}        
\newcommand{\hatcurCCgscxxxxeccenA}{GSC~1364-01424}                    
\newcommand{\hatcurCCtassmvxxxxeccenA}{12.422}                         
\newcommand{\hatcurCCtwomassJmagxxxxeccenA}{\ensuremath{11.424\pm0.020}} 
\newcommand{\hatcurCCtwomassHmagxxxxeccenA}{\ensuremath{11.184\pm0.022}} 
\newcommand{\hatcurCCtwomassKmagxxxxeccenA}{\ensuremath{11.157\pm0.020}} 
\newcommand{\hatcurCCcitJmagxxxxeccenA}{\ensuremath{11.446\pm0.021}}   
\newcommand{\hatcurCCcitHmagxxxxeccenA}{\ensuremath{11.181\pm0.022}}   
\newcommand{\hatcurCCcitKmagxxxxeccenA}{\ensuremath{11.181\pm0.020}}   
\newcommand{\hatcurCCbbJmagxxxxeccenA}{\ensuremath{11.487\pm0.021}}    
\newcommand{\hatcurCCbbHmagxxxxeccenA}{\ensuremath{11.200\pm0.023}}    
\newcommand{\hatcurCCbbKmagxxxxeccenA}{\ensuremath{11.201\pm0.020}}    
\newcommand{\hatcurCCesoJmagxxxxeccenA}{\ensuremath{11.488\pm0.022}}   
\newcommand{\hatcurCCesoHmagxxxxeccenA}{\ensuremath{11.193\pm0.025}}   
\newcommand{\hatcurCCesoKmagxxxxeccenA}{\ensuremath{11.200\pm0.021}}   
\newcommand{\hatcurCCesoJHmagxxxxeccenA}{\ensuremath{0.295\pm0.032}}   
\newcommand{\hatcurCCesoJKmagxxxxeccenA}{\ensuremath{0.288\pm0.008}}   
\newcommand{\hatcurCCesoHKmagxxxxeccenA}{\ensuremath{-0.008\pm0.033}}  
\newcommand{\hatcurLCdipxxxxeccenA}{\ensuremath{10.9}}                 
\newcommand{\hatcurLCrprstarxxxxeccenA}{\ensuremath{0.0991\pm0.0023}}  
\newcommand{\hatcurLCbsqxxxxeccenA}{\ensuremath{0.120_{-0.061}^{+0.070}}} 
\newcommand{\hatcurLCimpxxxxeccenA}{\ensuremath{0.346_{-0.122}^{+0.085}}} 
\newcommand{\hatcurLCzetaxxxxeccenA}{\ensuremath{12.76\pm0.08}}        
\newcommand{\hatcurLCdurxxxxeccenA}{\ensuremath{0.1743\pm0.0018}}      
\newcommand{\hatcurLCdurshortxxxxeccenA}{\ensuremath{0.1743}}          
\newcommand{\hatcurLCdurhrxxxxeccenA}{\ensuremath{4.184\pm0.042}}      
\newcommand{\hatcurLCdurhrshortxxxxeccenA}{\ensuremath{4.184}}         
\newcommand{\hatcurLCqxxxxeccenA}{\ensuremath{0.0492\pm0.0005}}        
\newcommand{\hatcurLCqshortxxxxeccenA}{\ensuremath{0.049}}             
\newcommand{\hatcurLCingdurxxxxeccenA}{\ensuremath{0.0176\pm0.0016}}   
\newcommand{\hatcurLCPxxxxeccenA}{\ensuremath{3.543870\pm0.000005}}    
\newcommand{\hatcurLCPprecxxxxeccenA}{\ensuremath{3.5438704}}          
\newcommand{\hatcurLCPshortxxxxeccenA}{\ensuremath{3.5439}}            
\newcommand{\hatcurLCTxxxxeccenA}{\ensuremath{2455187.48732\pm0.00043}} 
\newcommand{\hatcurLCTAxxxxeccenA}{\ensuremath{2454429.09909\pm0.00116}} 
\newcommand{\hatcurLCTBxxxxeccenA}{\ensuremath{2455605.66402\pm0.00085}} 
\newcommand{\hatcurLChatnetmAxxxxeccenA}{\ensuremath{12.0199\pm0.0002}} 
\newcommand{\hatcurLCiblendAxxxxeccenA}{\ensuremath{0.42\pm0.07}}      
\newcommand{\hatcurLChatnetmBxxxxeccenA}{\ensuremath{12.0197\pm0.0001}} 
\newcommand{\hatcurLCiblendBxxxxeccenA}{\ensuremath{0.63\pm0.04}}      
\newcommand{\hatcurSMEiteffxxxxeccenA}{\ensuremath{6325\pm100}}        
\newcommand{\hatcurSMEizfehxxxxeccenA}{\ensuremath{0.14\pm0.1}}        
\newcommand{\hatcurSMEizfehshortxxxxeccenA}{\ensuremath{0.14}}         
\newcommand{\hatcurSMEiloggxxxxeccenA}{\ensuremath{4.04\pm0.1}}        
\newcommand{\hatcurSMEivsinxxxxeccenA}{\ensuremath{12.7\pm0.5}}        
\newcommand{\hatcurSMEivmacxxxxeccenA}{\ensuremath{4.87}}              
\newcommand{\hatcurSMEivmicxxxxeccenA}{\ensuremath{0.85}}              
\newcommand{\hatcurSMEiiteffxxxxeccenA}{\ensuremath{6420\pm100}}       
\newcommand{\hatcurSMEiizfehxxxxeccenA}{\ensuremath{0.18\pm0.10}}      
\newcommand{\hatcurSMEiizfehshortxxxxeccenA}{\ensuremath{0.18}}        
\newcommand{\hatcurSMEiiloggxxxxeccenA}{\ensuremath{4.15\pm0.1}}       
\newcommand{\hatcurSMEiivsinxxxxeccenA}{\ensuremath{12.7\pm0.5}}       
\newcommand{\hatcurSMEiivmacxxxxeccenA}{\ensuremath{5.03}}             
\newcommand{\hatcurSMEiivmicxxxxeccenA}{\ensuremath{0.85}}             
\newcommand{\hatcurDSteffxxxxeccenA}{\ensuremath{6250\pm100}}          
\newcommand{\hatcurDSzfehxxxxeccenA}{\ensuremath{NULL\pmNULL}}         
\newcommand{\hatcurDSloggxxxxeccenA}{\ensuremath{4.0\pm0.25}}          
\newcommand{\hatcurDSvsinixxxxeccenA}{\ensuremath{14.0\pm1.0}}         
\newcommand{\hatcurDSgammaxxxxeccenA}{\ensuremath{28.42\pm0.28}}       
\newcommand{\hatcurDSnumspecxxxxeccenA}{\ensuremath{4}}                
\newcommand{\hatcurDSspanxxxxeccenA}{\ensuremath{123}}                 
\newcommand{\hatcurDSrvrmsxxxxeccenA}{\ensuremath{0.55}}               
\newcommand{\hatcurTRESteffxxxxeccenA}{\ensuremath{6250\pm100}}        
\newcommand{\hatcurTRESzfehxxxxeccenA}{\ensuremath{NULL\pmNULL}}       
\newcommand{\hatcurTRESloggxxxxeccenA}{\ensuremath{3.5\pm0.5}}         
\newcommand{\hatcurTRESvsinixxxxeccenA}{\ensuremath{16\pm0.5}}         
\newcommand{\hatcurTRESgammaxxxxeccenA}{\ensuremath{28.70\pmNULL}}     
\newcommand{\hatcurTRESnumspecxxxxeccenA}{\ensuremath{1}}              
\newcommand{\hatcurTRESspanxxxxeccenA}{\ensuremath{1}}                 
\newcommand{\hatcurTRESrvrmsxxxxeccenA}{\ensuremath{0.00}}             
\newcommand{\hatcurFIESteffxxxxeccenA}{\ensuremath{NULL\pmNULL}}       
\newcommand{\hatcurFIESzfehxxxxeccenA}{\ensuremath{NULL\pmNULL}}       
\newcommand{\hatcurFIESloggxxxxeccenA}{\ensuremath{NULL\pmNULL}}       
\newcommand{\hatcurFIESvsinixxxxeccenA}{\ensuremath{NULL\pmNULL}}      
\newcommand{\hatcurFIESgammaxxxxeccenA}{\ensuremath{NULL\pmNULL}}      
\newcommand{\hatcurFIESnumspecxxxxeccenA}{\ensuremath{0}}              
\newcommand{\hatcurFIESspanxxxxeccenA}{\ensuremath{0}}                 
\newcommand{\hatcurFIESrvrmsxxxxeccenA}{\ensuremath{0.00}}             
\newcommand{\hatcurLBizxxxxeccenA}{\ensuremath{0.1283}}                
\newcommand{\hatcurLBiizxxxxeccenA}{\ensuremath{0.3709}}               
\newcommand{\hatcurLBiixxxxeccenA}{\ensuremath{0.1796}}                
\newcommand{\hatcurLBiiixxxxeccenA}{\ensuremath{0.3807}}               
\newcommand{\hatcurLBiIxxxxeccenA}{\ensuremath{0.1607}}                
\newcommand{\hatcurLBiiIxxxxeccenA}{\ensuremath{0.3789}}               
\newcommand{\hatcurLBigxxxxeccenA}{\ensuremath{0.4260}}                
\newcommand{\hatcurLBiigxxxxeccenA}{\ensuremath{0.3281}}               
\newcommand{\hatcurLBirxxxxeccenA}{\ensuremath{0.2564}}                
\newcommand{\hatcurLBiirxxxxeccenA}{\ensuremath{0.3864}}               
\newcommand{\hatcurLBiRxxxxeccenA}{\ensuremath{0.2345}}                
\newcommand{\hatcurLBiiRxxxxeccenA}{\ensuremath{0.3865}}               
\newcommand{\hatcurLBikepxxxxeccenA}{\ensuremath{}}            
\newcommand{\hatcurLBiikepxxxxeccenA}{\ensuremath{}}           
\newcommand{\hatcurISOmxxxxeccenA}{\ensuremath{1.40_{-0.07}^{+0.10}}}  
\newcommand{\hatcurISOmshortxxxxeccenA}{\ensuremath{1.40}}             
\newcommand{\hatcurISOmlongxxxxeccenA}{\ensuremath{1.400_{-0.069}^{+0.102}}} 
\newcommand{\hatcurISOrxxxxeccenA}{\ensuremath{1.62_{-0.17}^{+0.29}}}  
\newcommand{\hatcurISOrshortxxxxeccenA}{\ensuremath{1.62}}             
\newcommand{\hatcurISOrlongxxxxeccenA}{\ensuremath{1.622_{-0.167}^{+0.294}}} 
\newcommand{\hatcurISOrhoxxxxeccenA}{\ensuremath{0.46_{-0.13}^{+0.18}}} 
\newcommand{\hatcurISOloggxxxxeccenA}{\ensuremath{4.16\pm0.10}}        
\newcommand{\hatcurISOlumxxxxeccenA}{\ensuremath{4.01_{-0.81}^{+1.78}}} 
\newcommand{\hatcurISOlumshortxxxxeccenA}{\ensuremath{4.01}}           
\newcommand{\hatcurISOmvxxxxeccenA}{\ensuremath{3.22\pm0.31}}          
\newcommand{\hatcurISOvixxxxeccenA}{\ensuremath{0.519\pm0.025}}        
\newcommand{\hatcurISOagexxxxeccenA}{\ensuremath{2.0_{-0.6}^{+0.5}}}   
\newcommand{\hatcurISOsigmaxxxxeccenA}{\ensuremath{0.00030\pm0.00015}} 
\newcommand{\hatcurISOMJxxxxeccenA}{\ensuremath{2.39\pm0.30}}          
\newcommand{\hatcurISOMHxxxxeccenA}{\ensuremath{2.17\pm0.29}}          
\newcommand{\hatcurISOMKxxxxeccenA}{\ensuremath{2.12\pm0.29}}          
\newcommand{\hatcurISOJKxxxxeccenA}{\ensuremath{0.27\pm0.02}}          
\newcommand{\hatcurISOspecxxxxeccenA}{F8}                              
\newcommand{\hatcurRVKxxxxeccenA}{\ensuremath{63.8\pm10.6}}            
\newcommand{\hatcurRVrkxxxxeccenA}{\ensuremath{0.094\pm0.189}}         
\newcommand{\hatcurRVrhxxxxeccenA}{\ensuremath{-0.008\pm0.263}}        
\newcommand{\hatcurRVkxxxxeccenA}{\ensuremath{0.022\pm0.074}}          
\newcommand{\hatcurRVhxxxxeccenA}{\ensuremath{-0.001\pm0.114}}         
\newcommand{\hatcurRVtronexxxxeccenA}{\ensuremath{0.0000\pm0.0000}}    
\newcommand{\hatcurRVtrtwoxxxxeccenA}{\ensuremath{0.0000\pm0.0000}}    
\newcommand{\hatcurRVgammaxxxxeccenA}{\ensuremath{-8.8\pm8.4}}         
\newcommand{\hatcurRVjitterxxxxeccenA}{\ensuremath{42.7}}              
\newcommand{\hatcurRVfitrmsxxxxeccenA}{\ensuremath{43.6}}              
\newcommand{\hatcurRVeccenxxxxeccenA}{\ensuremath{0.094\pm0.086}}      
\newcommand{\hatcurRVomegaxxxxeccenA}{\ensuremath{188\pm115}}          
\newcommand{\hatcurPPixxxxeccenA}{\ensuremath{87.1_{-1.5}^{+1.1}}}     
\newcommand{\hatcurPPgxxxxeccenA}{\ensuremath{5.9_{-1.6}^{+2.1}}}      
\newcommand{\hatcurPPloggxxxxeccenA}{\ensuremath{2.77\pm0.14}}         
\newcommand{\hatcurPParxxxxeccenA}{\ensuremath{6.74\pm0.77}}           
\newcommand{\hatcurPParelxxxxeccenA}{\ensuremath{0.0509_{-0.0009}^{+0.0012}}} 
\newcommand{\hatcurPPrhoxxxxeccenA}{\ensuremath{0.19_{-0.06}^{+0.10}}} 
\newcommand{\hatcurPPmxxxxeccenA}{\ensuremath{0.60\pm0.10}}            
\newcommand{\hatcurPPmshortxxxxeccenA}{\ensuremath{0.60}}              
\newcommand{\hatcurPPmlongxxxxeccenA}{\ensuremath{0.596\pm0.099}}      
\newcommand{\hatcurPPmexxxxeccenA}{\ensuremath{189.4\pm31.5}}          
\newcommand{\hatcurPPmeshortxxxxeccenA}{\ensuremath{189.4}}            
\newcommand{\hatcurPPmelongxxxxeccenA}{\ensuremath{189.38\pm31.54}}    
\newcommand{\hatcurPPrxxxxeccenA}{\ensuremath{1.56_{-0.17}^{+0.29}}}   
\newcommand{\hatcurPPrshortxxxxeccenA}{\ensuremath{1.56}}              
\newcommand{\hatcurPPrlongxxxxeccenA}{\ensuremath{1.565_{-0.169}^{+0.292}}} 
\newcommand{\hatcurPPrexxxxeccenA}{\ensuremath{17.5_{-1.9}^{+3.3}}}    
\newcommand{\hatcurPPreshortxxxxeccenA}{\ensuremath{17.5}}             
\newcommand{\hatcurPPrelongxxxxeccenA}{\ensuremath{17.54_{-1.89}^{+3.27}}} 
\newcommand{\hatcurPPmrcorrxxxxeccenA}{\ensuremath{0.08}}              
\newcommand{\hatcurPPteffxxxxeccenA}{\ensuremath{1751_{-87}^{+137}}}   
\newcommand{\hatcurPPthetaxxxxeccenA}{\ensuremath{0.027\pm0.006}}      
\newcommand{\hatcurPPfluxperixxxxeccenA}{\ensuremath{2.46_{-0.33}^{+2.70}}} 
\newcommand{\hatcurPPfluxperidimxxxxeccenA}{\ensuremath{9}}            
\newcommand{\hatcurPPfluxapxxxxeccenA}{\ensuremath{1.87_{-0.48}^{+0.31}}} 
\newcommand{\hatcurPPfluxapdimxxxxeccenA}{\ensuremath{9}}              
\newcommand{\hatcurPPfluxavgxxxxeccenA}{\ensuremath{2.12_{-0.38}^{+0.85}}} 
\newcommand{\hatcurPPfluxavgdimxxxxeccenA}{\ensuremath{9}}             
\newcommand{\hatcurXsecphasexxxxeccenA}{\ensuremath{0.5143\pm0.0475}}  
\newcommand{\hatcurXsecondaryxxxxeccenA}{\ensuremath{2455189.310\pm0.168}} 
\newcommand{\hatcurXsecdurxxxxeccenA}{\ensuremath{0.1741\pm0.0378}}    
\newcommand{\hatcurXsecingdurxxxxeccenA}{\ensuremath{0.0176\pm0.0084}} 
\newcommand{\hatcurPPphiconjxxxxeccenA}{\ensuremath{0.0921\pm0.2704}}  
\newcommand{\hatcurPPperixxxxeccenA}{\ensuremath{2455187.16\pm0.96}}   
\newcommand{\hatcurPPaequivxxxxeccenA}{\ensuremath{0.0254\pm0.0030}}   
\newcommand{\hatcurPPtcircxxxxeccenA}{\ensuremath{54.3_{-26.0}^{+41.9}}} 
\newcommand{\hatcurPPtinfallxxxxeccenA}{\ensuremath{1106.3_{-440.7}^{+907.0}}} 
\newcommand{\hatcurXdistxxxxeccenA}{\ensuremath{653_{-67}^{+118}}}     
\newcommand{\hatcurXAvxxxxeccenA}{\ensuremath{0.156_{-0.111}^{+0.152}}} 
\newcommand{\hatcurXdistredxxxxeccenA}{\ensuremath{641_{-66}^{+115}}}  
\newcommand{\hatcurCCpmraxxxxeccenA}{\ensuremath{5.0\pm1.7}}           
\newcommand{\hatcurCCpmdecxxxxeccenA}{\ensuremath{-8.3\pm0.8}}         
\newcommand{\hatcurCCpmxxxxeccenA}{\ensuremath{9.68969\pm1.87883}}     

\newcommand{\hatcurhtrxxxxeccenB}{HTR159-019}                          
\newcommand{\hatcurfieldxxxxeccenB}{159}                               
\newcommand{\hatcurCCraxxxxeccenB}{\ensuremath{22^{\mathrm h}22^{\mathrm m}03.00{\mathrm s}}}                        
\newcommand{\hatcurCCdecxxxxeccenB}{\ensuremath{+45{\arcdeg}27{\arcmin}26.6{\arcsec}}}                       
\newcommand{\hatcurCCmagxxxxeccenB}{11.699}                            
\newcommand{\hatcurCCtwomassxxxxeccenB}{2MASS~22220308+4527265}        
\newcommand{\hatcurCCgscxxxxeccenB}{GSC~3607-01028}                    
\newcommand{\hatcurCCtassmvxxxxeccenB}{11.699}                         
\newcommand{\hatcurCCtwomassJmagxxxxeccenB}{\ensuremath{10.367\pm0.023}} 
\newcommand{\hatcurCCtwomassHmagxxxxeccenB}{\ensuremath{10.085\pm0.018}} 
\newcommand{\hatcurCCtwomassKmagxxxxeccenB}{\ensuremath{10.009\pm0.018}} 
\newcommand{\hatcurCCcitJmagxxxxeccenB}{\ensuremath{10.384\pm0.023}}   
\newcommand{\hatcurCCcitHmagxxxxeccenB}{\ensuremath{10.080\pm0.019}}   
\newcommand{\hatcurCCcitKmagxxxxeccenB}{\ensuremath{10.033\pm0.018}}   
\newcommand{\hatcurCCbbJmagxxxxeccenB}{\ensuremath{10.433\pm0.025}}    
\newcommand{\hatcurCCbbHmagxxxxeccenB}{\ensuremath{10.101\pm0.020}}    
\newcommand{\hatcurCCbbKmagxxxxeccenB}{\ensuremath{10.053\pm0.018}}    
\newcommand{\hatcurCCesoJmagxxxxeccenB}{\ensuremath{10.434\pm0.026}}   
\newcommand{\hatcurCCesoHmagxxxxeccenB}{\ensuremath{10.097\pm0.021}}   
\newcommand{\hatcurCCesoKmagxxxxeccenB}{\ensuremath{10.052\pm0.019}}   
\newcommand{\hatcurCCesoJHmagxxxxeccenB}{\ensuremath{0.338\pm0.032}}   
\newcommand{\hatcurCCesoJKmagxxxxeccenB}{\ensuremath{0.383\pm0.032}}   
\newcommand{\hatcurCCesoHKmagxxxxeccenB}{\ensuremath{0.044\pm0.009}}   
\newcommand{\hatcurLCdipxxxxeccenB}{\ensuremath{4.4}}                  
\newcommand{\hatcurLCrprstarxxxxeccenB}{\ensuremath{0.0807\pm0.0014}}  
\newcommand{\hatcurLCbsqxxxxeccenB}{\ensuremath{0.028_{-0.017}^{+0.043}}} 
\newcommand{\hatcurLCimpxxxxeccenB}{\ensuremath{0.168_{-0.073}^{+0.090}}} 
\newcommand{\hatcurLCzetaxxxxeccenB}{\ensuremath{8.49\pm0.04}}         
\newcommand{\hatcurLCdurxxxxeccenB}{\ensuremath{0.2553\pm0.0014}}      
\newcommand{\hatcurLCdurshortxxxxeccenB}{\ensuremath{0.2553}}          
\newcommand{\hatcurLCdurhrxxxxeccenB}{\ensuremath{6.126\pm0.033}}      
\newcommand{\hatcurLCdurhrshortxxxxeccenB}{\ensuremath{6.126}}         
\newcommand{\hatcurLCqxxxxeccenB}{\ensuremath{0.0573\pm0.0003}}        
\newcommand{\hatcurLCqshortxxxxeccenB}{\ensuremath{0.057}}             
\newcommand{\hatcurLCingdurxxxxeccenB}{\ensuremath{0.0196\pm0.0008}}   
\newcommand{\hatcurLCPxxxxeccenB}{\ensuremath{4.457243\pm0.000010}}    
\newcommand{\hatcurLCPprecxxxxeccenB}{\ensuremath{4.4572431}}          
\newcommand{\hatcurLCPshortxxxxeccenB}{\ensuremath{4.4572}}            
\newcommand{\hatcurLCTxxxxeccenB}{\ensuremath{2455817.63309\pm0.00054}} 
\newcommand{\hatcurLCTAxxxxeccenB}{\ensuremath{2453321.57691\pm0.00566}} 
\newcommand{\hatcurLCTBxxxxeccenB}{\ensuremath{2455839.91931\pm0.00056}} 
\newcommand{\hatcurSMEiteffxxxxeccenB}{\ensuremath{6140\pm100}}        
\newcommand{\hatcurSMEizfehxxxxeccenB}{\ensuremath{0.25\pm0.1}}        
\newcommand{\hatcurSMEizfehshortxxxxeccenB}{\ensuremath{0.25}}         
\newcommand{\hatcurSMEiloggxxxxeccenB}{\ensuremath{4.04\pm0.1}}        
\newcommand{\hatcurSMEivsinxxxxeccenB}{\ensuremath{6.7\pm0.5}}         
\newcommand{\hatcurSMEivmacxxxxeccenB}{\ensuremath{4.59}}              
\newcommand{\hatcurSMEivmicxxxxeccenB}{\ensuremath{0.85}}              
\newcommand{\hatcurSMEiiteffxxxxeccenB}{\ensuremath{6040\pm100}}       
\newcommand{\hatcurSMEiizfehxxxxeccenB}{\ensuremath{0.19\pm0.10}}      
\newcommand{\hatcurSMEiizfehshortxxxxeccenB}{\ensuremath{0.19}}        
\newcommand{\hatcurSMEiiloggxxxxeccenB}{\ensuremath{3.83\pm0.1}}       
\newcommand{\hatcurSMEiivsinxxxxeccenB}{\ensuremath{7.0\pm0.5}}        
\newcommand{\hatcurSMEiivmacxxxxeccenB}{\ensuremath{4.43}}             
\newcommand{\hatcurSMEiivmicxxxxeccenB}{\ensuremath{0.85}}             
\newcommand{\hatcurDSteffxxxxeccenB}{\ensuremath{NULL\pmNULL}}         
\newcommand{\hatcurDSzfehxxxxeccenB}{\ensuremath{NULL\pmNULL}}         
\newcommand{\hatcurDSloggxxxxeccenB}{\ensuremath{NULL\pmNULL}}         
\newcommand{\hatcurDSvsinixxxxeccenB}{\ensuremath{NULL\pmNULL}}        
\newcommand{\hatcurDSgammaxxxxeccenB}{\ensuremath{NULL\pmNULL}}        
\newcommand{\hatcurDSnumspecxxxxeccenB}{\ensuremath{0}}                
\newcommand{\hatcurDSspanxxxxeccenB}{\ensuremath{0}}                   
\newcommand{\hatcurDSrvrmsxxxxeccenB}{\ensuremath{0.00}}               
\newcommand{\hatcurTRESteffxxxxeccenB}{\ensuremath{6000\pm100}}        
\newcommand{\hatcurTRESzfehxxxxeccenB}{\ensuremath{0.3\pm0.1}}         
\newcommand{\hatcurTRESloggxxxxeccenB}{\ensuremath{4.0\pm0.2}}         
\newcommand{\hatcurTRESvsinixxxxeccenB}{\ensuremath{9.2\pm0.7}}        
\newcommand{\hatcurTRESgammaxxxxeccenB}{\ensuremath{-25.0\pm0.1}}      
\newcommand{\hatcurTRESnumspecxxxxeccenB}{\ensuremath{5}}              
\newcommand{\hatcurTRESspanxxxxeccenB}{\ensuremath{84}}                
\newcommand{\hatcurTRESrvrmsxxxxeccenB}{\ensuremath{0.17}}             
\newcommand{\hatcurFIESteffxxxxeccenB}{\ensuremath{NULL\pmNULL}}       
\newcommand{\hatcurFIESzfehxxxxeccenB}{\ensuremath{NULL\pmNULL}}       
\newcommand{\hatcurFIESloggxxxxeccenB}{\ensuremath{NULL\pmNULL}}       
\newcommand{\hatcurFIESvsinixxxxeccenB}{\ensuremath{NULL\pmNULL}}      
\newcommand{\hatcurFIESgammaxxxxeccenB}{\ensuremath{NULL\pmNULL}}      
\newcommand{\hatcurFIESnumspecxxxxeccenB}{\ensuremath{0}}              
\newcommand{\hatcurFIESspanxxxxeccenB}{\ensuremath{0}}                 
\newcommand{\hatcurFIESrvrmsxxxxeccenB}{\ensuremath{0.00}}             
\newcommand{\hatcurLBizxxxxeccenB}{\ensuremath{0.1659}}                
\newcommand{\hatcurLBiizxxxxeccenB}{\ensuremath{0.3570}}               
\newcommand{\hatcurLBiixxxxeccenB}{\ensuremath{0.2213}}                
\newcommand{\hatcurLBiiixxxxeccenB}{\ensuremath{0.3636}}               
\newcommand{\hatcurLBiIxxxxeccenB}{\ensuremath{0.2010}}                
\newcommand{\hatcurLBiiIxxxxeccenB}{\ensuremath{0.3633}}               
\newcommand{\hatcurLBigxxxxeccenB}{\ensuremath{0.5021}}                
\newcommand{\hatcurLBiigxxxxeccenB}{\ensuremath{0.2752}}               
\newcommand{\hatcurLBirxxxxeccenB}{\ensuremath{0.3084}}                
\newcommand{\hatcurLBiirxxxxeccenB}{\ensuremath{0.3601}}               
\newcommand{\hatcurLBiRxxxxeccenB}{\ensuremath{0.2839}}                
\newcommand{\hatcurLBiiRxxxxeccenB}{\ensuremath{0.3624}}               
\newcommand{\hatcurLBikepxxxxeccenB}{\ensuremath{}}            
\newcommand{\hatcurLBiikepxxxxeccenB}{\ensuremath{}}           
\newcommand{\hatcurISOmxxxxeccenB}{\ensuremath{1.50\pm0.10}}           
\newcommand{\hatcurISOmshortxxxxeccenB}{\ensuremath{1.50}}             
\newcommand{\hatcurISOmlongxxxxeccenB}{\ensuremath{1.504\pm0.103}}     
\newcommand{\hatcurISOrxxxxeccenB}{\ensuremath{2.42\pm0.15}}           
\newcommand{\hatcurISOrshortxxxxeccenB}{\ensuremath{2.42}}             
\newcommand{\hatcurISOrlongxxxxeccenB}{\ensuremath{2.422\pm0.154}}     
\newcommand{\hatcurISOrhoxxxxeccenB}{\ensuremath{0.15\pm0.02}}         
\newcommand{\hatcurISOloggxxxxeccenB}{\ensuremath{3.85\pm0.04}}        
\newcommand{\hatcurISOlumxxxxeccenB}{\ensuremath{6.98_{-0.97}^{+1.27}}} 
\newcommand{\hatcurISOlumshortxxxxeccenB}{\ensuremath{6.98}}           
\newcommand{\hatcurISOmvxxxxeccenB}{\ensuremath{2.67\pm0.18}}          
\newcommand{\hatcurISOvixxxxeccenB}{\ensuremath{0.620\pm0.029}}        
\newcommand{\hatcurISOagexxxxeccenB}{\ensuremath{2.9_{-0.6}^{+0.8}}}   
\newcommand{\hatcurISOsigmaxxxxeccenB}{\ensuremath{0.00010\pm0.00005}} 
\newcommand{\hatcurISOMJxxxxeccenB}{\ensuremath{1.65\pm0.15}}          
\newcommand{\hatcurISOMHxxxxeccenB}{\ensuremath{1.36\pm0.15}}          
\newcommand{\hatcurISOMKxxxxeccenB}{\ensuremath{1.31\pm0.14}}          
\newcommand{\hatcurISOJKxxxxeccenB}{\ensuremath{0.34\pm0.02}}          
\newcommand{\hatcurISOspecxxxxeccenB}{F8}                              
\newcommand{\hatcurRVKxxxxeccenB}{\ensuremath{55.4\pm2.5}}             
\newcommand{\hatcurRVrkxxxxeccenB}{\ensuremath{0.041\pm0.064}}         
\newcommand{\hatcurRVrhxxxxeccenB}{\ensuremath{0.300_{-0.137}^{+0.074}}} 
\newcommand{\hatcurRVkxxxxeccenB}{\ensuremath{0.012\pm0.018}}          
\newcommand{\hatcurRVhxxxxeccenB}{\ensuremath{0.092\pm0.051}}          
\newcommand{\hatcurRVtronexxxxeccenB}{\ensuremath{0.0000\pm0.0000}}    
\newcommand{\hatcurRVtrtwoxxxxeccenB}{\ensuremath{0.0000\pm0.0000}}    
\newcommand{\hatcurRVgammaxxxxeccenB}{\ensuremath{-15.9\pm1.8}}        
\newcommand{\hatcurRVjitterxxxxeccenB}{\ensuremath{4.1}}               
\newcommand{\hatcurRVfitrmsxxxxeccenB}{\ensuremath{5.3}}               
\newcommand{\hatcurRVeccenxxxxeccenB}{\ensuremath{0.095\pm0.048}}      
\newcommand{\hatcurRVomegaxxxxeccenB}{\ensuremath{83\pm42}}            
\newcommand{\hatcurPPixxxxeccenB}{\ensuremath{88.0_{-1.2}^{+0.9}}}     
\newcommand{\hatcurPPgxxxxeccenB}{\ensuremath{4.0\pm0.6}}              
\newcommand{\hatcurPPloggxxxxeccenB}{\ensuremath{2.60\pm0.06}}         
\newcommand{\hatcurPParxxxxeccenB}{\ensuremath{5.39\pm0.28}}           
\newcommand{\hatcurPParelxxxxeccenB}{\ensuremath{0.0607\pm0.0014}}     
\newcommand{\hatcurPPrhoxxxxeccenB}{\ensuremath{0.11\pm0.02}}          
\newcommand{\hatcurPPmxxxxeccenB}{\ensuremath{0.58\pm0.04}}            
\newcommand{\hatcurPPmshortxxxxeccenB}{\ensuremath{0.58}}              
\newcommand{\hatcurPPmlongxxxxeccenB}{\ensuremath{0.584\pm0.035}}      
\newcommand{\hatcurPPmexxxxeccenB}{\ensuremath{185.6\pm11.2}}          
\newcommand{\hatcurPPmeshortxxxxeccenB}{\ensuremath{185.6}}            
\newcommand{\hatcurPPmelongxxxxeccenB}{\ensuremath{185.60\pm11.21}}    
\newcommand{\hatcurPPrxxxxeccenB}{\ensuremath{1.90\pm0.13}}            
\newcommand{\hatcurPPrshortxxxxeccenB}{\ensuremath{1.90}}              
\newcommand{\hatcurPPrlongxxxxeccenB}{\ensuremath{1.900\pm0.127}}      
\newcommand{\hatcurPPrexxxxeccenB}{\ensuremath{21.3\pm1.4}}            
\newcommand{\hatcurPPreshortxxxxeccenB}{\ensuremath{21.3}}             
\newcommand{\hatcurPPrelongxxxxeccenB}{\ensuremath{21.30\pm1.43}}      
\newcommand{\hatcurPPmrcorrxxxxeccenB}{\ensuremath{0.03}}              
\newcommand{\hatcurPPteffxxxxeccenB}{\ensuremath{1843\pm60}}           
\newcommand{\hatcurPPthetaxxxxeccenB}{\ensuremath{0.025\pm0.002}}      
\newcommand{\hatcurPPfluxperixxxxeccenB}{\ensuremath{3.17_{-0.57}^{+0.90}}} 
\newcommand{\hatcurPPfluxperidimxxxxeccenB}{\ensuremath{9}}            
\newcommand{\hatcurPPfluxapxxxxeccenB}{\ensuremath{2.16\pm0.17}}       
\newcommand{\hatcurPPfluxapdimxxxxeccenB}{\ensuremath{9}}              
\newcommand{\hatcurPPfluxavgxxxxeccenB}{\ensuremath{2.60\pm0.34}}      
\newcommand{\hatcurPPfluxavgdimxxxxeccenB}{\ensuremath{9}}             
\newcommand{\hatcurXsecphasexxxxeccenB}{\ensuremath{0.5075\pm0.0117}}  
\newcommand{\hatcurXsecondaryxxxxeccenB}{\ensuremath{2455819.895\pm0.052}} 
\newcommand{\hatcurXsecdurxxxxeccenB}{\ensuremath{0.3048\pm0.0300}}    
\newcommand{\hatcurXsecingdurxxxxeccenB}{\ensuremath{0.0239\pm0.0029}} 
\newcommand{\hatcurPPphiconjxxxxeccenB}{\ensuremath{0.0165_{-0.0516}^{+0.0896}}} 
\newcommand{\hatcurPPperixxxxeccenB}{\ensuremath{2455817.56\pm0.33}}   
\newcommand{\hatcurPPaequivxxxxeccenB}{\ensuremath{0.0230\pm0.0014}}   
\newcommand{\hatcurPPtcircxxxxeccenB}{\ensuremath{57.3_{-17.4}^{+25.9}}} 
\newcommand{\hatcurPPtinfallxxxxeccenB}{\ensuremath{492.0_{-98.2}^{+134.2}}} 
\newcommand{\hatcurXdistxxxxeccenB}{\ensuremath{561\pm37}}             
\newcommand{\hatcurXAvxxxxeccenB}{\ensuremath{0.317\pm0.126}}          
\newcommand{\hatcurXdistredxxxxeccenB}{\ensuremath{548\pm36}}          
\newcommand{\hatcurCCpmraxxxxeccenB}{\ensuremath{9.2\pm0.9}}           
\newcommand{\hatcurCCpmdecxxxxeccenB}{\ensuremath{-4.4\pm1.0}}         
\newcommand{\hatcurCCpmxxxxeccenB}{\ensuremath{10.198\pm1.34536}}      

\newcommand{\hatcurhtrxxxxeccenC}{HTR388-004}                          
\newcommand{\hatcurfieldxxxxeccenC}{388}                               
\newcommand{\hatcurCCraxxxxeccenC}{\ensuremath{19^{\mathrm h}49^{\mathrm m}17.40{\mathrm s}}}                        
\newcommand{\hatcurCCdecxxxxeccenC}{\ensuremath{+04{\arcdeg}40{\arcmin}20.7{\arcsec}}}                       
\newcommand{\hatcurCCmagxxxxeccenC}{11.087}                            
\newcommand{\hatcurCCtwomassxxxxeccenC}{2MASS~19491743+0440207}        
\newcommand{\hatcurCCgscxxxxeccenC}{GSC~0488-02442}                    
\newcommand{\hatcurCCtassmvxxxxeccenC}{11.087}                         
\newcommand{\hatcurCCtwomassJmagxxxxeccenC}{\ensuremath{10.006\pm0.027}} 
\newcommand{\hatcurCCtwomassHmagxxxxeccenC}{\ensuremath{9.777\pm0.032}} 
\newcommand{\hatcurCCtwomassKmagxxxxeccenC}{\ensuremath{9.728\pm0.029}} 
\newcommand{\hatcurCCcitJmagxxxxeccenC}{\ensuremath{10.028\pm0.027}}   
\newcommand{\hatcurCCcitHmagxxxxeccenC}{\ensuremath{9.772\pm0.032}}    
\newcommand{\hatcurCCcitKmagxxxxeccenC}{\ensuremath{9.752\pm0.029}}    
\newcommand{\hatcurCCbbJmagxxxxeccenC}{\ensuremath{10.070\pm0.029}}    
\newcommand{\hatcurCCbbHmagxxxxeccenC}{\ensuremath{9.793\pm0.033}}     
\newcommand{\hatcurCCbbKmagxxxxeccenC}{\ensuremath{9.772\pm0.029}}     
\newcommand{\hatcurCCesoJmagxxxxeccenC}{\ensuremath{10.070\pm0.030}}   
\newcommand{\hatcurCCesoHmagxxxxeccenC}{\ensuremath{9.788\pm0.036}}    
\newcommand{\hatcurCCesoKmagxxxxeccenC}{\ensuremath{9.771\pm0.029}}    
\newcommand{\hatcurCCesoJHmagxxxxeccenC}{\ensuremath{0.282\pm0.045}}   
\newcommand{\hatcurCCesoJKmagxxxxeccenC}{\ensuremath{0.300\pm0.042}}   
\newcommand{\hatcurCCesoHKmagxxxxeccenC}{\ensuremath{0.017\pm0.047}}   
\newcommand{\hatcurLCdipxxxxeccenC}{\ensuremath{8.4}}                  
\newcommand{\hatcurLCrprstarxxxxeccenC}{\ensuremath{0.1030\pm0.0016}}  
\newcommand{\hatcurLCbsqxxxxeccenC}{\ensuremath{0.056_{-0.033}^{+0.053}}} 
\newcommand{\hatcurLCimpxxxxeccenC}{\ensuremath{0.237_{-0.097}^{+0.088}}} 
\newcommand{\hatcurLCzetaxxxxeccenC}{\ensuremath{12.99\pm0.06}}        
\newcommand{\hatcurLCdurxxxxeccenC}{\ensuremath{0.1708\pm0.0013}}      
\newcommand{\hatcurLCdurshortxxxxeccenC}{\ensuremath{0.1708}}          
\newcommand{\hatcurLCdurhrxxxxeccenC}{\ensuremath{4.098\pm0.031}}      
\newcommand{\hatcurLCdurhrshortxxxxeccenC}{\ensuremath{4.098}}         
\newcommand{\hatcurLCqxxxxeccenC}{\ensuremath{0.0634\pm0.0005}}        
\newcommand{\hatcurLCqshortxxxxeccenC}{\ensuremath{0.063}}             
\newcommand{\hatcurLCingdurxxxxeccenC}{\ensuremath{0.0168\pm0.0010}}   
\newcommand{\hatcurLCPxxxxeccenC}{\ensuremath{2.694047\pm0.000005}}    
\newcommand{\hatcurLCPprecxxxxeccenC}{\ensuremath{2.6940466}}          
\newcommand{\hatcurLCPshortxxxxeccenC}{\ensuremath{2.6940}}            
\newcommand{\hatcurLCTxxxxeccenC}{\ensuremath{2454983.86167\pm0.00107}} 
\newcommand{\hatcurLCTAxxxxeccenC}{\ensuremath{2454965.00335\pm0.00110}} 
\newcommand{\hatcurLCTBxxxxeccenC}{\ensuremath{2455713.94831\pm0.00042}} 
\newcommand{\hatcurLChatnetmAxxxxeccenC}{\ensuremath{10.9743\pm0.0001}} 
\newcommand{\hatcurLCiblendAxxxxeccenC}{\ensuremath{0.75\pm0.04}}      
\newcommand{\hatcurLChatnetmBxxxxeccenC}{\ensuremath{10.0105\pm0.0008}} 
\newcommand{\hatcurLCiblendBxxxxeccenC}{\ensuremath{-0.04\pm0.00}}     
\newcommand{\hatcurSMEiteffxxxxeccenC}{\ensuremath{6007\pm100}}        
\newcommand{\hatcurSMEizfehxxxxeccenC}{\ensuremath{0.06\pm0.1}}        
\newcommand{\hatcurSMEizfehshortxxxxeccenC}{\ensuremath{0.06}}         
\newcommand{\hatcurSMEiloggxxxxeccenC}{\ensuremath{3.68\pm0.06}}       
\newcommand{\hatcurSMEivsinxxxxeccenC}{\ensuremath{20.6\pm0.5}}        
\newcommand{\hatcurSMEivmacxxxxeccenC}{\ensuremath{4.38}}              
\newcommand{\hatcurSMEivmicxxxxeccenC}{\ensuremath{0.85}}              
\newcommand{\hatcurSMEiiteffxxxxeccenC}{\ensuremath{6480\pm100}}       
\newcommand{\hatcurSMEiizfehxxxxeccenC}{\ensuremath{0.24\pm0.10}}      
\newcommand{\hatcurSMEiizfehshortxxxxeccenC}{\ensuremath{0.24}}        
\newcommand{\hatcurSMEiiloggxxxxeccenC}{\ensuremath{4.24\pm0.15}}      
\newcommand{\hatcurSMEiivsinxxxxeccenC}{\ensuremath{19.3\pm0.5}}       
\newcommand{\hatcurSMEiivmacxxxxeccenC}{\ensuremath{5.11}}             
\newcommand{\hatcurSMEiivmicxxxxeccenC}{\ensuremath{0.85}}             
\newcommand{\hatcurDSteffxxxxeccenC}{\ensuremath{NULL\pmNULL}}         
\newcommand{\hatcurDSzfehxxxxeccenC}{\ensuremath{NULL\pmNULL}}         
\newcommand{\hatcurDSloggxxxxeccenC}{\ensuremath{NULL\pmNULL}}         
\newcommand{\hatcurDSvsinixxxxeccenC}{\ensuremath{NULL\pmNULL}}        
\newcommand{\hatcurDSgammaxxxxeccenC}{\ensuremath{NULL\pmNULL}}        
\newcommand{\hatcurDSnumspecxxxxeccenC}{\ensuremath{0}}                
\newcommand{\hatcurDSspanxxxxeccenC}{\ensuremath{0}}                   
\newcommand{\hatcurDSrvrmsxxxxeccenC}{\ensuremath{0.00}}               
\newcommand{\hatcurTRESteffxxxxeccenC}{\ensuremath{6333\pm100}}        
\newcommand{\hatcurTRESzfehxxxxeccenC}{\ensuremath{0.0\pm0.0}}         
\newcommand{\hatcurTRESloggxxxxeccenC}{\ensuremath{3.83\pm0.25}}       
\newcommand{\hatcurTRESvsinixxxxeccenC}{\ensuremath{25.0\pm0.5}}       
\newcommand{\hatcurTRESgammaxxxxeccenC}{\ensuremath{31.68\pm0.61}}     
\newcommand{\hatcurTRESnumspecxxxxeccenC}{\ensuremath{3}}              
\newcommand{\hatcurTRESspanxxxxeccenC}{\ensuremath{54}}                
\newcommand{\hatcurTRESrvrmsxxxxeccenC}{\ensuremath{1.05}}             
\newcommand{\hatcurFIESteffxxxxeccenC}{\ensuremath{NULL\pmNULL}}       
\newcommand{\hatcurFIESzfehxxxxeccenC}{\ensuremath{NULL\pmNULL}}       
\newcommand{\hatcurFIESloggxxxxeccenC}{\ensuremath{NULL\pmNULL}}       
\newcommand{\hatcurFIESvsinixxxxeccenC}{\ensuremath{NULL\pmNULL}}      
\newcommand{\hatcurFIESgammaxxxxeccenC}{\ensuremath{NULL\pmNULL}}      
\newcommand{\hatcurFIESnumspecxxxxeccenC}{\ensuremath{0}}              
\newcommand{\hatcurFIESspanxxxxeccenC}{\ensuremath{0}}                 
\newcommand{\hatcurFIESrvrmsxxxxeccenC}{\ensuremath{0.00}}             
\newcommand{\hatcurLBizxxxxeccenC}{\ensuremath{0.1282}}                
\newcommand{\hatcurLBiizxxxxeccenC}{\ensuremath{0.3709}}               
\newcommand{\hatcurLBiixxxxeccenC}{\ensuremath{0.1803}}                
\newcommand{\hatcurLBiiixxxxeccenC}{\ensuremath{0.3807}}               
\newcommand{\hatcurLBiIxxxxeccenC}{\ensuremath{0.1618}}                
\newcommand{\hatcurLBiiIxxxxeccenC}{\ensuremath{0.3785}}               
\newcommand{\hatcurLBigxxxxeccenC}{\ensuremath{0.4182}}                
\newcommand{\hatcurLBiigxxxxeccenC}{\ensuremath{0.3352}}               
\newcommand{\hatcurLBirxxxxeccenC}{\ensuremath{0.2544}}                
\newcommand{\hatcurLBiirxxxxeccenC}{\ensuremath{0.3882}}               
\newcommand{\hatcurLBiRxxxxeccenC}{\ensuremath{0.2333}}                
\newcommand{\hatcurLBiiRxxxxeccenC}{\ensuremath{0.3878}}               
\newcommand{\hatcurLBikepxxxxeccenC}{\ensuremath{}}            
\newcommand{\hatcurLBiikepxxxxeccenC}{\ensuremath{}}           
\newcommand{\hatcurISOmxxxxeccenC}{\ensuremath{1.40\pm0.07}}           
\newcommand{\hatcurISOmshortxxxxeccenC}{\ensuremath{1.40}}             
\newcommand{\hatcurISOmlongxxxxeccenC}{\ensuremath{1.405\pm0.066}}     
\newcommand{\hatcurISOrxxxxeccenC}{\ensuremath{1.53_{-0.13}^{+0.18}}}  
\newcommand{\hatcurISOrshortxxxxeccenC}{\ensuremath{1.53}}             
\newcommand{\hatcurISOrlongxxxxeccenC}{\ensuremath{1.525_{-0.133}^{+0.177}}} 
\newcommand{\hatcurISOrhoxxxxeccenC}{\ensuremath{0.56_{-0.13}^{+0.17}}} 
\newcommand{\hatcurISOloggxxxxeccenC}{\ensuremath{4.22\pm0.07}}        
\newcommand{\hatcurISOlumxxxxeccenC}{\ensuremath{3.68_{-0.68}^{+1.00}}} 
\newcommand{\hatcurISOlumshortxxxxeccenC}{\ensuremath{3.68}}           
\newcommand{\hatcurISOmvxxxxeccenC}{\ensuremath{3.31\pm0.25}}          
\newcommand{\hatcurISOvixxxxeccenC}{\ensuremath{0.506\pm0.025}}        
\newcommand{\hatcurISOagexxxxeccenC}{\ensuremath{1.5\pm0.6}}           
\newcommand{\hatcurISOsigmaxxxxeccenC}{\ensuremath{0.00050\pm0.00015}} 
\newcommand{\hatcurISOMJxxxxeccenC}{\ensuremath{2.50\pm0.22}}          
\newcommand{\hatcurISOMHxxxxeccenC}{\ensuremath{2.29\pm0.22}}          
\newcommand{\hatcurISOMKxxxxeccenC}{\ensuremath{2.25\pm0.22}}          
\newcommand{\hatcurISOJKxxxxeccenC}{\ensuremath{0.26\pm0.02}}          
\newcommand{\hatcurISOspecxxxxeccenC}{F}                               
\newcommand{\hatcurRVKxxxxeccenC}{\ensuremath{95.6\pm10.4}}            
\newcommand{\hatcurRVrkxxxxeccenC}{\ensuremath{-0.189\pm0.100}}        
\newcommand{\hatcurRVrhxxxxeccenC}{\ensuremath{-0.293_{-0.120}^{+0.244}}} 
\newcommand{\hatcurRVkxxxxeccenC}{\ensuremath{-0.068\pm0.039}}         
\newcommand{\hatcurRVhxxxxeccenC}{\ensuremath{-0.105\pm0.086}}         
\newcommand{\hatcurRVtronexxxxeccenC}{\ensuremath{0.0000\pm0.0000}}    
\newcommand{\hatcurRVtrtwoxxxxeccenC}{\ensuremath{0.0000\pm0.0000}}    
\newcommand{\hatcurRVjitterAxxxxeccenC}{\ensuremath{27.5}}             
\newcommand{\hatcurRVeccenxxxxeccenC}{\ensuremath{0.139\pm0.063}}      
\newcommand{\hatcurRVomegaxxxxeccenC}{\ensuremath{237\pm38}}           
\newcommand{\hatcurPPixxxxeccenC}{\ensuremath{87.9\pm0.9}}             
\newcommand{\hatcurPPgxxxxeccenC}{\ensuremath{8.6\pm1.7}}              
\newcommand{\hatcurPPloggxxxxeccenC}{\ensuremath{2.93\pm0.09}}         
\newcommand{\hatcurPParxxxxeccenC}{\ensuremath{5.97\pm0.52}}           
\newcommand{\hatcurPParelxxxxeccenC}{\ensuremath{0.0424\pm0.0007}}     
\newcommand{\hatcurPPrhoxxxxeccenC}{\ensuremath{0.28_{-0.07}^{+0.09}}} 
\newcommand{\hatcurPPmxxxxeccenC}{\ensuremath{0.81\pm0.09}}            
\newcommand{\hatcurPPmshortxxxxeccenC}{\ensuremath{0.81}}              
\newcommand{\hatcurPPmlongxxxxeccenC}{\ensuremath{0.812\pm0.094}}      
\newcommand{\hatcurPPmexxxxeccenC}{\ensuremath{258.0\pm30.0}}          
\newcommand{\hatcurPPmeshortxxxxeccenC}{\ensuremath{258.0}}            
\newcommand{\hatcurPPmelongxxxxeccenC}{\ensuremath{257.96\pm29.96}}    
\newcommand{\hatcurPPrxxxxeccenC}{\ensuremath{1.53_{-0.14}^{+0.19}}}   
\newcommand{\hatcurPPrshortxxxxeccenC}{\ensuremath{1.53}}              
\newcommand{\hatcurPPrlongxxxxeccenC}{\ensuremath{1.529_{-0.136}^{+0.185}}} 
\newcommand{\hatcurPPrexxxxeccenC}{\ensuremath{17.1_{-1.5}^{+2.1}}}    
\newcommand{\hatcurPPreshortxxxxeccenC}{\ensuremath{17.1}}             
\newcommand{\hatcurPPrelongxxxxeccenC}{\ensuremath{17.14_{-1.52}^{+2.08}}} 
\newcommand{\hatcurPPmrcorrxxxxeccenC}{\ensuremath{0.32}}              
\newcommand{\hatcurPPteffxxxxeccenC}{\ensuremath{1879\pm88}}           
\newcommand{\hatcurPPthetaxxxxeccenC}{\ensuremath{0.032\pm0.004}}      
\newcommand{\hatcurPPfluxperixxxxeccenC}{\ensuremath{3.74_{-0.33}^{+0.86}}} 
\newcommand{\hatcurPPfluxperidimxxxxeccenC}{\ensuremath{9}}            
\newcommand{\hatcurPPfluxapxxxxeccenC}{\ensuremath{2.18\pm0.61}}       
\newcommand{\hatcurPPfluxapdimxxxxeccenC}{\ensuremath{9}}              
\newcommand{\hatcurPPfluxavgxxxxeccenC}{\ensuremath{2.82_{-0.44}^{+0.65}}} 
\newcommand{\hatcurPPfluxavgdimxxxxeccenC}{\ensuremath{9}}             
\newcommand{\hatcurXsecphasexxxxeccenC}{\ensuremath{0.4566\pm0.0251}}  
\newcommand{\hatcurXsecondaryxxxxeccenC}{\ensuremath{2454985.092\pm0.068}} 
\newcommand{\hatcurXsecdurxxxxeccenC}{\ensuremath{0.1397\pm0.0241}}    
\newcommand{\hatcurXsecingdurxxxxeccenC}{\ensuremath{0.0135\pm0.0027}} 
\newcommand{\hatcurPPphiconjxxxxeccenC}{\ensuremath{-0.3741_{-0.0611}^{+0.2354}}} 
\newcommand{\hatcurPPperixxxxeccenC}{\ensuremath{2454984.87\pm0.46}}   
\newcommand{\hatcurPPaequivxxxxeccenC}{\ensuremath{0.0221\pm0.0021}}   
\newcommand{\hatcurPPtcircxxxxeccenC}{\ensuremath{23.7\pm8.4}}         
\newcommand{\hatcurPPtinfallxxxxeccenC}{\ensuremath{336.1_{-113.7}^{+213.4}}} 
\newcommand{\hatcurXdistxxxxeccenC}{\ensuremath{319_{-28}^{+37}}}      
\newcommand{\hatcurXAvxxxxeccenC}{\ensuremath{0.332\pm0.134}}          
\newcommand{\hatcurXdistredxxxxeccenC}{\ensuremath{311_{-27}^{+36}}}   
\newcommand{\hatcurCCpmraxxxxeccenC}{\ensuremath{-0.4\pm0.7}}          
\newcommand{\hatcurCCpmdecxxxxeccenC}{\ensuremath{-2.9\pm4.5}}         
\newcommand{\hatcurCCpmxxxxeccenC}{\ensuremath{2.92746\pm4.55412}}     
\newcommand{\hatcurCCbbHmageccen}[1]{\ifnum#1=39 %
\hatcurCCbbHmagxxxxeccenA
\else
\ifnum#1=40 %
\hatcurCCbbHmagxxxxeccenB
\else
\ifnum#1=41 %
\hatcurCCbbHmagxxxxeccenC
\else
??????\fi
\fi
\fi
}
\newcommand{\hatcurCCbbJmageccen}[1]{\ifnum#1=39 %
\hatcurCCbbJmagxxxxeccenA
\else
\ifnum#1=40 %
\hatcurCCbbJmagxxxxeccenB
\else
\ifnum#1=41 %
\hatcurCCbbJmagxxxxeccenC
\else
??????\fi
\fi
\fi
}
\newcommand{\hatcurCCbbKmageccen}[1]{\ifnum#1=39 %
\hatcurCCbbKmagxxxxeccenA
\else
\ifnum#1=40 %
\hatcurCCbbKmagxxxxeccenB
\else
\ifnum#1=41 %
\hatcurCCbbKmagxxxxeccenC
\else
??????\fi
\fi
\fi
}
\newcommand{\hatcurCCcitHmageccen}[1]{\ifnum#1=39 %
\hatcurCCcitHmagxxxxeccenA
\else
\ifnum#1=40 %
\hatcurCCcitHmagxxxxeccenB
\else
\ifnum#1=41 %
\hatcurCCcitHmagxxxxeccenC
\else
??????\fi
\fi
\fi
}
\newcommand{\hatcurCCcitJmageccen}[1]{\ifnum#1=39 %
\hatcurCCcitJmagxxxxeccenA
\else
\ifnum#1=40 %
\hatcurCCcitJmagxxxxeccenB
\else
\ifnum#1=41 %
\hatcurCCcitJmagxxxxeccenC
\else
??????\fi
\fi
\fi
}
\newcommand{\hatcurCCcitKmageccen}[1]{\ifnum#1=39 %
\hatcurCCcitKmagxxxxeccenA
\else
\ifnum#1=40 %
\hatcurCCcitKmagxxxxeccenB
\else
\ifnum#1=41 %
\hatcurCCcitKmagxxxxeccenC
\else
??????\fi
\fi
\fi
}
\newcommand{\hatcurCCdececcen}[1]{\ifnum#1=39 %
\hatcurCCdecxxxxeccenA
\else
\ifnum#1=40 %
\hatcurCCdecxxxxeccenB
\else
\ifnum#1=41 %
\hatcurCCdecxxxxeccenC
\else
??????\fi
\fi
\fi
}
\newcommand{\hatcurCCesoHKmageccen}[1]{\ifnum#1=39 %
\hatcurCCesoHKmagxxxxeccenA
\else
\ifnum#1=40 %
\hatcurCCesoHKmagxxxxeccenB
\else
\ifnum#1=41 %
\hatcurCCesoHKmagxxxxeccenC
\else
??????\fi
\fi
\fi
}
\newcommand{\hatcurCCesoHmageccen}[1]{\ifnum#1=39 %
\hatcurCCesoHmagxxxxeccenA
\else
\ifnum#1=40 %
\hatcurCCesoHmagxxxxeccenB
\else
\ifnum#1=41 %
\hatcurCCesoHmagxxxxeccenC
\else
??????\fi
\fi
\fi
}
\newcommand{\hatcurCCesoJHmageccen}[1]{\ifnum#1=39 %
\hatcurCCesoJHmagxxxxeccenA
\else
\ifnum#1=40 %
\hatcurCCesoJHmagxxxxeccenB
\else
\ifnum#1=41 %
\hatcurCCesoJHmagxxxxeccenC
\else
??????\fi
\fi
\fi
}
\newcommand{\hatcurCCesoJKmageccen}[1]{\ifnum#1=39 %
\hatcurCCesoJKmagxxxxeccenA
\else
\ifnum#1=40 %
\hatcurCCesoJKmagxxxxeccenB
\else
\ifnum#1=41 %
\hatcurCCesoJKmagxxxxeccenC
\else
??????\fi
\fi
\fi
}
\newcommand{\hatcurCCesoJmageccen}[1]{\ifnum#1=39 %
\hatcurCCesoJmagxxxxeccenA
\else
\ifnum#1=40 %
\hatcurCCesoJmagxxxxeccenB
\else
\ifnum#1=41 %
\hatcurCCesoJmagxxxxeccenC
\else
??????\fi
\fi
\fi
}
\newcommand{\hatcurCCesoKmageccen}[1]{\ifnum#1=39 %
\hatcurCCesoKmagxxxxeccenA
\else
\ifnum#1=40 %
\hatcurCCesoKmagxxxxeccenB
\else
\ifnum#1=41 %
\hatcurCCesoKmagxxxxeccenC
\else
??????\fi
\fi
\fi
}
\newcommand{\hatcurCCgsceccen}[1]{\ifnum#1=39 %
\hatcurCCgscxxxxeccenA
\else
\ifnum#1=40 %
\hatcurCCgscxxxxeccenB
\else
\ifnum#1=41 %
\hatcurCCgscxxxxeccenC
\else
??????\fi
\fi
\fi
}
\newcommand{\hatcurCCmageccen}[1]{\ifnum#1=39 %
\hatcurCCmagxxxxeccenA
\else
\ifnum#1=40 %
\hatcurCCmagxxxxeccenB
\else
\ifnum#1=41 %
\hatcurCCmagxxxxeccenC
\else
??????\fi
\fi
\fi
}
\newcommand{\hatcurCCpmeccen}[1]{\ifnum#1=39 %
\hatcurCCpmxxxxeccenA
\else
\ifnum#1=40 %
\hatcurCCpmxxxxeccenB
\else
\ifnum#1=41 %
\hatcurCCpmxxxxeccenC
\else
??????\fi
\fi
\fi
}
\newcommand{\hatcurCCpmdececcen}[1]{\ifnum#1=39 %
\hatcurCCpmdecxxxxeccenA
\else
\ifnum#1=40 %
\hatcurCCpmdecxxxxeccenB
\else
\ifnum#1=41 %
\hatcurCCpmdecxxxxeccenC
\else
??????\fi
\fi
\fi
}
\newcommand{\hatcurCCpmraeccen}[1]{\ifnum#1=39 %
\hatcurCCpmraxxxxeccenA
\else
\ifnum#1=40 %
\hatcurCCpmraxxxxeccenB
\else
\ifnum#1=41 %
\hatcurCCpmraxxxxeccenC
\else
??????\fi
\fi
\fi
}
\newcommand{\hatcurCCraeccen}[1]{\ifnum#1=39 %
\hatcurCCraxxxxeccenA
\else
\ifnum#1=40 %
\hatcurCCraxxxxeccenB
\else
\ifnum#1=41 %
\hatcurCCraxxxxeccenC
\else
??????\fi
\fi
\fi
}
\newcommand{\hatcurCCtassmveccen}[1]{\ifnum#1=39 %
\hatcurCCtassmvxxxxeccenA
\else
\ifnum#1=40 %
\hatcurCCtassmvxxxxeccenB
\else
\ifnum#1=41 %
\hatcurCCtassmvxxxxeccenC
\else
??????\fi
\fi
\fi
}
\newcommand{\hatcurCCtwomasseccen}[1]{\ifnum#1=39 %
\hatcurCCtwomassxxxxeccenA
\else
\ifnum#1=40 %
\hatcurCCtwomassxxxxeccenB
\else
\ifnum#1=41 %
\hatcurCCtwomassxxxxeccenC
\else
??????\fi
\fi
\fi
}
\newcommand{\hatcurCCtwomassHmageccen}[1]{\ifnum#1=39 %
\hatcurCCtwomassHmagxxxxeccenA
\else
\ifnum#1=40 %
\hatcurCCtwomassHmagxxxxeccenB
\else
\ifnum#1=41 %
\hatcurCCtwomassHmagxxxxeccenC
\else
??????\fi
\fi
\fi
}
\newcommand{\hatcurCCtwomassJmageccen}[1]{\ifnum#1=39 %
\hatcurCCtwomassJmagxxxxeccenA
\else
\ifnum#1=40 %
\hatcurCCtwomassJmagxxxxeccenB
\else
\ifnum#1=41 %
\hatcurCCtwomassJmagxxxxeccenC
\else
??????\fi
\fi
\fi
}
\newcommand{\hatcurCCtwomassKmageccen}[1]{\ifnum#1=39 %
\hatcurCCtwomassKmagxxxxeccenA
\else
\ifnum#1=40 %
\hatcurCCtwomassKmagxxxxeccenB
\else
\ifnum#1=41 %
\hatcurCCtwomassKmagxxxxeccenC
\else
??????\fi
\fi
\fi
}
\newcommand{\hatcurDSgammaeccen}[1]{\ifnum#1=39 %
\hatcurDSgammaxxxxeccenA
\else
\ifnum#1=40 %
\hatcurDSgammaxxxxeccenB
\else
\ifnum#1=41 %
\hatcurDSgammaxxxxeccenC
\else
??????\fi
\fi
\fi
}
\newcommand{\hatcurDSloggeccen}[1]{\ifnum#1=39 %
\hatcurDSloggxxxxeccenA
\else
\ifnum#1=40 %
\hatcurDSloggxxxxeccenB
\else
\ifnum#1=41 %
\hatcurDSloggxxxxeccenC
\else
??????\fi
\fi
\fi
}
\newcommand{\hatcurDSnumspececcen}[1]{\ifnum#1=39 %
\hatcurDSnumspecxxxxeccenA
\else
\ifnum#1=40 %
\hatcurDSnumspecxxxxeccenB
\else
\ifnum#1=41 %
\hatcurDSnumspecxxxxeccenC
\else
??????\fi
\fi
\fi
}
\newcommand{\hatcurDSrvrmseccen}[1]{\ifnum#1=39 %
\hatcurDSrvrmsxxxxeccenA
\else
\ifnum#1=40 %
\hatcurDSrvrmsxxxxeccenB
\else
\ifnum#1=41 %
\hatcurDSrvrmsxxxxeccenC
\else
??????\fi
\fi
\fi
}
\newcommand{\hatcurDSspaneccen}[1]{\ifnum#1=39 %
\hatcurDSspanxxxxeccenA
\else
\ifnum#1=40 %
\hatcurDSspanxxxxeccenB
\else
\ifnum#1=41 %
\hatcurDSspanxxxxeccenC
\else
??????\fi
\fi
\fi
}
\newcommand{\hatcurDSteffeccen}[1]{\ifnum#1=39 %
\hatcurDSteffxxxxeccenA
\else
\ifnum#1=40 %
\hatcurDSteffxxxxeccenB
\else
\ifnum#1=41 %
\hatcurDSteffxxxxeccenC
\else
??????\fi
\fi
\fi
}
\newcommand{\hatcurDSvsinieccen}[1]{\ifnum#1=39 %
\hatcurDSvsinixxxxeccenA
\else
\ifnum#1=40 %
\hatcurDSvsinixxxxeccenB
\else
\ifnum#1=41 %
\hatcurDSvsinixxxxeccenC
\else
??????\fi
\fi
\fi
}
\newcommand{\hatcurDSzfeheccen}[1]{\ifnum#1=39 %
\hatcurDSzfehxxxxeccenA
\else
\ifnum#1=40 %
\hatcurDSzfehxxxxeccenB
\else
\ifnum#1=41 %
\hatcurDSzfehxxxxeccenC
\else
??????\fi
\fi
\fi
}
\newcommand{\hatcurfieldeccen}[1]{\ifnum#1=39 %
\hatcurfieldxxxxeccenA
\else
\ifnum#1=40 %
\hatcurfieldxxxxeccenB
\else
\ifnum#1=41 %
\hatcurfieldxxxxeccenC
\else
??????\fi
\fi
\fi
}
\newcommand{\hatcurFIESgammaeccen}[1]{\ifnum#1=39 %
\hatcurFIESgammaxxxxeccenA
\else
\ifnum#1=40 %
\hatcurFIESgammaxxxxeccenB
\else
\ifnum#1=41 %
\hatcurFIESgammaxxxxeccenC
\else
??????\fi
\fi
\fi
}
\newcommand{\hatcurFIESloggeccen}[1]{\ifnum#1=39 %
\hatcurFIESloggxxxxeccenA
\else
\ifnum#1=40 %
\hatcurFIESloggxxxxeccenB
\else
\ifnum#1=41 %
\hatcurFIESloggxxxxeccenC
\else
??????\fi
\fi
\fi
}
\newcommand{\hatcurFIESnumspececcen}[1]{\ifnum#1=39 %
\hatcurFIESnumspecxxxxeccenA
\else
\ifnum#1=40 %
\hatcurFIESnumspecxxxxeccenB
\else
\ifnum#1=41 %
\hatcurFIESnumspecxxxxeccenC
\else
??????\fi
\fi
\fi
}
\newcommand{\hatcurFIESrvrmseccen}[1]{\ifnum#1=39 %
\hatcurFIESrvrmsxxxxeccenA
\else
\ifnum#1=40 %
\hatcurFIESrvrmsxxxxeccenB
\else
\ifnum#1=41 %
\hatcurFIESrvrmsxxxxeccenC
\else
??????\fi
\fi
\fi
}
\newcommand{\hatcurFIESspaneccen}[1]{\ifnum#1=39 %
\hatcurFIESspanxxxxeccenA
\else
\ifnum#1=40 %
\hatcurFIESspanxxxxeccenB
\else
\ifnum#1=41 %
\hatcurFIESspanxxxxeccenC
\else
??????\fi
\fi
\fi
}
\newcommand{\hatcurFIESteffeccen}[1]{\ifnum#1=39 %
\hatcurFIESteffxxxxeccenA
\else
\ifnum#1=40 %
\hatcurFIESteffxxxxeccenB
\else
\ifnum#1=41 %
\hatcurFIESteffxxxxeccenC
\else
??????\fi
\fi
\fi
}
\newcommand{\hatcurFIESvsinieccen}[1]{\ifnum#1=39 %
\hatcurFIESvsinixxxxeccenA
\else
\ifnum#1=40 %
\hatcurFIESvsinixxxxeccenB
\else
\ifnum#1=41 %
\hatcurFIESvsinixxxxeccenC
\else
??????\fi
\fi
\fi
}
\newcommand{\hatcurFIESzfeheccen}[1]{\ifnum#1=39 %
\hatcurFIESzfehxxxxeccenA
\else
\ifnum#1=40 %
\hatcurFIESzfehxxxxeccenB
\else
\ifnum#1=41 %
\hatcurFIESzfehxxxxeccenC
\else
??????\fi
\fi
\fi
}
\newcommand{\hatcurhtreccen}[1]{\ifnum#1=39 %
\hatcurhtrxxxxeccenA
\else
\ifnum#1=40 %
\hatcurhtrxxxxeccenB
\else
\ifnum#1=41 %
\hatcurhtrxxxxeccenC
\else
??????\fi
\fi
\fi
}
\newcommand{\hatcurISOageeccen}[1]{\ifnum#1=39 %
\hatcurISOagexxxxeccenA
\else
\ifnum#1=40 %
\hatcurISOagexxxxeccenB
\else
\ifnum#1=41 %
\hatcurISOagexxxxeccenC
\else
??????\fi
\fi
\fi
}
\newcommand{\hatcurISOJKeccen}[1]{\ifnum#1=39 %
\hatcurISOJKxxxxeccenA
\else
\ifnum#1=40 %
\hatcurISOJKxxxxeccenB
\else
\ifnum#1=41 %
\hatcurISOJKxxxxeccenC
\else
??????\fi
\fi
\fi
}
\newcommand{\hatcurISOloggeccen}[1]{\ifnum#1=39 %
\hatcurISOloggxxxxeccenA
\else
\ifnum#1=40 %
\hatcurISOloggxxxxeccenB
\else
\ifnum#1=41 %
\hatcurISOloggxxxxeccenC
\else
??????\fi
\fi
\fi
}
\newcommand{\hatcurISOlumeccen}[1]{\ifnum#1=39 %
\hatcurISOlumxxxxeccenA
\else
\ifnum#1=40 %
\hatcurISOlumxxxxeccenB
\else
\ifnum#1=41 %
\hatcurISOlumxxxxeccenC
\else
??????\fi
\fi
\fi
}
\newcommand{\hatcurISOlumshorteccen}[1]{\ifnum#1=39 %
\hatcurISOlumshortxxxxeccenA
\else
\ifnum#1=40 %
\hatcurISOlumshortxxxxeccenB
\else
\ifnum#1=41 %
\hatcurISOlumshortxxxxeccenC
\else
??????\fi
\fi
\fi
}
\newcommand{\hatcurISOmeccen}[1]{\ifnum#1=39 %
\hatcurISOmxxxxeccenA
\else
\ifnum#1=40 %
\hatcurISOmxxxxeccenB
\else
\ifnum#1=41 %
\hatcurISOmxxxxeccenC
\else
??????\fi
\fi
\fi
}
\newcommand{\hatcurISOMHeccen}[1]{\ifnum#1=39 %
\hatcurISOMHxxxxeccenA
\else
\ifnum#1=40 %
\hatcurISOMHxxxxeccenB
\else
\ifnum#1=41 %
\hatcurISOMHxxxxeccenC
\else
??????\fi
\fi
\fi
}
\newcommand{\hatcurISOMJeccen}[1]{\ifnum#1=39 %
\hatcurISOMJxxxxeccenA
\else
\ifnum#1=40 %
\hatcurISOMJxxxxeccenB
\else
\ifnum#1=41 %
\hatcurISOMJxxxxeccenC
\else
??????\fi
\fi
\fi
}
\newcommand{\hatcurISOMKeccen}[1]{\ifnum#1=39 %
\hatcurISOMKxxxxeccenA
\else
\ifnum#1=40 %
\hatcurISOMKxxxxeccenB
\else
\ifnum#1=41 %
\hatcurISOMKxxxxeccenC
\else
??????\fi
\fi
\fi
}
\newcommand{\hatcurISOmlongeccen}[1]{\ifnum#1=39 %
\hatcurISOmlongxxxxeccenA
\else
\ifnum#1=40 %
\hatcurISOmlongxxxxeccenB
\else
\ifnum#1=41 %
\hatcurISOmlongxxxxeccenC
\else
??????\fi
\fi
\fi
}
\newcommand{\hatcurISOmshorteccen}[1]{\ifnum#1=39 %
\hatcurISOmshortxxxxeccenA
\else
\ifnum#1=40 %
\hatcurISOmshortxxxxeccenB
\else
\ifnum#1=41 %
\hatcurISOmshortxxxxeccenC
\else
??????\fi
\fi
\fi
}
\newcommand{\hatcurISOmveccen}[1]{\ifnum#1=39 %
\hatcurISOmvxxxxeccenA
\else
\ifnum#1=40 %
\hatcurISOmvxxxxeccenB
\else
\ifnum#1=41 %
\hatcurISOmvxxxxeccenC
\else
??????\fi
\fi
\fi
}
\newcommand{\hatcurISOreccen}[1]{\ifnum#1=39 %
\hatcurISOrxxxxeccenA
\else
\ifnum#1=40 %
\hatcurISOrxxxxeccenB
\else
\ifnum#1=41 %
\hatcurISOrxxxxeccenC
\else
??????\fi
\fi
\fi
}
\newcommand{\hatcurISOrhoeccen}[1]{\ifnum#1=39 %
\hatcurISOrhoxxxxeccenA
\else
\ifnum#1=40 %
\hatcurISOrhoxxxxeccenB
\else
\ifnum#1=41 %
\hatcurISOrhoxxxxeccenC
\else
??????\fi
\fi
\fi
}
\newcommand{\hatcurISOrlongeccen}[1]{\ifnum#1=39 %
\hatcurISOrlongxxxxeccenA
\else
\ifnum#1=40 %
\hatcurISOrlongxxxxeccenB
\else
\ifnum#1=41 %
\hatcurISOrlongxxxxeccenC
\else
??????\fi
\fi
\fi
}
\newcommand{\hatcurISOrshorteccen}[1]{\ifnum#1=39 %
\hatcurISOrshortxxxxeccenA
\else
\ifnum#1=40 %
\hatcurISOrshortxxxxeccenB
\else
\ifnum#1=41 %
\hatcurISOrshortxxxxeccenC
\else
??????\fi
\fi
\fi
}
\newcommand{\hatcurISOsigmaeccen}[1]{\ifnum#1=39 %
\hatcurISOsigmaxxxxeccenA
\else
\ifnum#1=40 %
\hatcurISOsigmaxxxxeccenB
\else
\ifnum#1=41 %
\hatcurISOsigmaxxxxeccenC
\else
??????\fi
\fi
\fi
}
\newcommand{\hatcurISOspececcen}[1]{\ifnum#1=39 %
\hatcurISOspecxxxxeccenA
\else
\ifnum#1=40 %
\hatcurISOspecxxxxeccenB
\else
\ifnum#1=41 %
\hatcurISOspecxxxxeccenC
\else
??????\fi
\fi
\fi
}
\newcommand{\hatcurISOvieccen}[1]{\ifnum#1=39 %
\hatcurISOvixxxxeccenA
\else
\ifnum#1=40 %
\hatcurISOvixxxxeccenB
\else
\ifnum#1=41 %
\hatcurISOvixxxxeccenC
\else
??????\fi
\fi
\fi
}
\newcommand{\hatcurLBigeccen}[1]{\ifnum#1=39 %
\hatcurLBigxxxxeccenA
\else
\ifnum#1=40 %
\hatcurLBigxxxxeccenB
\else
\ifnum#1=41 %
\hatcurLBigxxxxeccenC
\else
??????\fi
\fi
\fi
}
\newcommand{\hatcurLBiieccen}[1]{\ifnum#1=39 %
\hatcurLBiixxxxeccenA
\else
\ifnum#1=40 %
\hatcurLBiixxxxeccenB
\else
\ifnum#1=41 %
\hatcurLBiixxxxeccenC
\else
??????\fi
\fi
\fi
}
\newcommand{\hatcurLBiIeccen}[1]{\ifnum#1=39 %
\hatcurLBiIxxxxeccenA
\else
\ifnum#1=40 %
\hatcurLBiIxxxxeccenB
\else
\ifnum#1=41 %
\hatcurLBiIxxxxeccenC
\else
??????\fi
\fi
\fi
}
\newcommand{\hatcurLBireccen}[1]{\ifnum#1=39 %
\hatcurLBirxxxxeccenA
\else
\ifnum#1=40 %
\hatcurLBirxxxxeccenB
\else
\ifnum#1=41 %
\hatcurLBirxxxxeccenC
\else
??????\fi
\fi
\fi
}
\newcommand{\hatcurLBiReccen}[1]{\ifnum#1=39 %
\hatcurLBiRxxxxeccenA
\else
\ifnum#1=40 %
\hatcurLBiRxxxxeccenB
\else
\ifnum#1=41 %
\hatcurLBiRxxxxeccenC
\else
??????\fi
\fi
\fi
}
\newcommand{\hatcurLBiigeccen}[1]{\ifnum#1=39 %
\hatcurLBiigxxxxeccenA
\else
\ifnum#1=40 %
\hatcurLBiigxxxxeccenB
\else
\ifnum#1=41 %
\hatcurLBiigxxxxeccenC
\else
??????\fi
\fi
\fi
}
\newcommand{\hatcurLBiiieccen}[1]{\ifnum#1=39 %
\hatcurLBiiixxxxeccenA
\else
\ifnum#1=40 %
\hatcurLBiiixxxxeccenB
\else
\ifnum#1=41 %
\hatcurLBiiixxxxeccenC
\else
??????\fi
\fi
\fi
}
\newcommand{\hatcurLBiiIeccen}[1]{\ifnum#1=39 %
\hatcurLBiiIxxxxeccenA
\else
\ifnum#1=40 %
\hatcurLBiiIxxxxeccenB
\else
\ifnum#1=41 %
\hatcurLBiiIxxxxeccenC
\else
??????\fi
\fi
\fi
}
\newcommand{\hatcurLBiikepeccen}[1]{\ifnum#1=39 %
\hatcurLBiikepxxxxeccenA
\else
\ifnum#1=40 %
\hatcurLBiikepxxxxeccenB
\else
\ifnum#1=41 %
\hatcurLBiikepxxxxeccenC
\else
??????\fi
\fi
\fi
}
\newcommand{\hatcurLBiizeccen}[1]{\ifnum#1=39 %
\hatcurLBiizxxxxeccenA
\else
\ifnum#1=40 %
\hatcurLBiizxxxxeccenB
\else
\ifnum#1=41 %
\hatcurLBiizxxxxeccenC
\else
??????\fi
\fi
\fi
}
\newcommand{\hatcurLBiireccen}[1]{\ifnum#1=39 %
\hatcurLBiirxxxxeccenA
\else
\ifnum#1=40 %
\hatcurLBiirxxxxeccenB
\else
\ifnum#1=41 %
\hatcurLBiirxxxxeccenC
\else
??????\fi
\fi
\fi
}
\newcommand{\hatcurLBiiReccen}[1]{\ifnum#1=39 %
\hatcurLBiiRxxxxeccenA
\else
\ifnum#1=40 %
\hatcurLBiiRxxxxeccenB
\else
\ifnum#1=41 %
\hatcurLBiiRxxxxeccenC
\else
??????\fi
\fi
\fi
}
\newcommand{\hatcurLBikepeccen}[1]{\ifnum#1=39 %
\hatcurLBikepxxxxeccenA
\else
\ifnum#1=40 %
\hatcurLBikepxxxxeccenB
\else
\ifnum#1=41 %
\hatcurLBikepxxxxeccenC
\else
??????\fi
\fi
\fi
}
\newcommand{\hatcurLBizeccen}[1]{\ifnum#1=39 %
\hatcurLBizxxxxeccenA
\else
\ifnum#1=40 %
\hatcurLBizxxxxeccenB
\else
\ifnum#1=41 %
\hatcurLBizxxxxeccenC
\else
??????\fi
\fi
\fi
}
\newcommand{\hatcurLCbsqeccen}[1]{\ifnum#1=39 %
\hatcurLCbsqxxxxeccenA
\else
\ifnum#1=40 %
\hatcurLCbsqxxxxeccenB
\else
\ifnum#1=41 %
\hatcurLCbsqxxxxeccenC
\else
??????\fi
\fi
\fi
}
\newcommand{\hatcurLCdipeccen}[1]{\ifnum#1=39 %
\hatcurLCdipxxxxeccenA
\else
\ifnum#1=40 %
\hatcurLCdipxxxxeccenB
\else
\ifnum#1=41 %
\hatcurLCdipxxxxeccenC
\else
??????\fi
\fi
\fi
}
\newcommand{\hatcurLCdureccen}[1]{\ifnum#1=39 %
\hatcurLCdurxxxxeccenA
\else
\ifnum#1=40 %
\hatcurLCdurxxxxeccenB
\else
\ifnum#1=41 %
\hatcurLCdurxxxxeccenC
\else
??????\fi
\fi
\fi
}
\newcommand{\hatcurLCdurhreccen}[1]{\ifnum#1=39 %
\hatcurLCdurhrxxxxeccenA
\else
\ifnum#1=40 %
\hatcurLCdurhrxxxxeccenB
\else
\ifnum#1=41 %
\hatcurLCdurhrxxxxeccenC
\else
??????\fi
\fi
\fi
}
\newcommand{\hatcurLCdurhrshorteccen}[1]{\ifnum#1=39 %
\hatcurLCdurhrshortxxxxeccenA
\else
\ifnum#1=40 %
\hatcurLCdurhrshortxxxxeccenB
\else
\ifnum#1=41 %
\hatcurLCdurhrshortxxxxeccenC
\else
??????\fi
\fi
\fi
}
\newcommand{\hatcurLCdurshorteccen}[1]{\ifnum#1=39 %
\hatcurLCdurshortxxxxeccenA
\else
\ifnum#1=40 %
\hatcurLCdurshortxxxxeccenB
\else
\ifnum#1=41 %
\hatcurLCdurshortxxxxeccenC
\else
??????\fi
\fi
\fi
}
\newcommand{\hatcurLChatnetmAeccen}[1]{\ifnum#1=39 %
\hatcurLChatnetmAxxxxeccenA
\else
\ifnum#1=40 %
\hatcurLChatnetmAxxxxeccenB
\else
\ifnum#1=41 %
\hatcurLChatnetmAxxxxeccenC
\else
??????\fi
\fi
\fi
}
\newcommand{\hatcurLChatnetmBeccen}[1]{\ifnum#1=39 %
\hatcurLChatnetmBxxxxeccenA
\else
\ifnum#1=40 %
\hatcurLChatnetmBxxxxeccenB
\else
\ifnum#1=41 %
\hatcurLChatnetmBxxxxeccenC
\else
??????\fi
\fi
\fi
}
\newcommand{\hatcurLCiblendAeccen}[1]{\ifnum#1=39 %
\hatcurLCiblendAxxxxeccenA
\else
\ifnum#1=40 %
\hatcurLCiblendAxxxxeccenB
\else
\ifnum#1=41 %
\hatcurLCiblendAxxxxeccenC
\else
??????\fi
\fi
\fi
}
\newcommand{\hatcurLCiblendBeccen}[1]{\ifnum#1=39 %
\hatcurLCiblendBxxxxeccenA
\else
\ifnum#1=40 %
\hatcurLCiblendBxxxxeccenB
\else
\ifnum#1=41 %
\hatcurLCiblendBxxxxeccenC
\else
??????\fi
\fi
\fi
}
\newcommand{\hatcurLCimpeccen}[1]{\ifnum#1=39 %
\hatcurLCimpxxxxeccenA
\else
\ifnum#1=40 %
\hatcurLCimpxxxxeccenB
\else
\ifnum#1=41 %
\hatcurLCimpxxxxeccenC
\else
??????\fi
\fi
\fi
}
\newcommand{\hatcurLCingdureccen}[1]{\ifnum#1=39 %
\hatcurLCingdurxxxxeccenA
\else
\ifnum#1=40 %
\hatcurLCingdurxxxxeccenB
\else
\ifnum#1=41 %
\hatcurLCingdurxxxxeccenC
\else
??????\fi
\fi
\fi
}
\newcommand{\hatcurLCPeccen}[1]{\ifnum#1=39 %
\hatcurLCPxxxxeccenA
\else
\ifnum#1=40 %
\hatcurLCPxxxxeccenB
\else
\ifnum#1=41 %
\hatcurLCPxxxxeccenC
\else
??????\fi
\fi
\fi
}
\newcommand{\hatcurLCPprececcen}[1]{\ifnum#1=39 %
\hatcurLCPprecxxxxeccenA
\else
\ifnum#1=40 %
\hatcurLCPprecxxxxeccenB
\else
\ifnum#1=41 %
\hatcurLCPprecxxxxeccenC
\else
??????\fi
\fi
\fi
}
\newcommand{\hatcurLCPshorteccen}[1]{\ifnum#1=39 %
\hatcurLCPshortxxxxeccenA
\else
\ifnum#1=40 %
\hatcurLCPshortxxxxeccenB
\else
\ifnum#1=41 %
\hatcurLCPshortxxxxeccenC
\else
??????\fi
\fi
\fi
}
\newcommand{\hatcurLCqeccen}[1]{\ifnum#1=39 %
\hatcurLCqxxxxeccenA
\else
\ifnum#1=40 %
\hatcurLCqxxxxeccenB
\else
\ifnum#1=41 %
\hatcurLCqxxxxeccenC
\else
??????\fi
\fi
\fi
}
\newcommand{\hatcurLCqshorteccen}[1]{\ifnum#1=39 %
\hatcurLCqshortxxxxeccenA
\else
\ifnum#1=40 %
\hatcurLCqshortxxxxeccenB
\else
\ifnum#1=41 %
\hatcurLCqshortxxxxeccenC
\else
??????\fi
\fi
\fi
}
\newcommand{\hatcurLCrprstareccen}[1]{\ifnum#1=39 %
\hatcurLCrprstarxxxxeccenA
\else
\ifnum#1=40 %
\hatcurLCrprstarxxxxeccenB
\else
\ifnum#1=41 %
\hatcurLCrprstarxxxxeccenC
\else
??????\fi
\fi
\fi
}
\newcommand{\hatcurLCTeccen}[1]{\ifnum#1=39 %
\hatcurLCTxxxxeccenA
\else
\ifnum#1=40 %
\hatcurLCTxxxxeccenB
\else
\ifnum#1=41 %
\hatcurLCTxxxxeccenC
\else
??????\fi
\fi
\fi
}
\newcommand{\hatcurLCTAeccen}[1]{\ifnum#1=39 %
\hatcurLCTAxxxxeccenA
\else
\ifnum#1=40 %
\hatcurLCTAxxxxeccenB
\else
\ifnum#1=41 %
\hatcurLCTAxxxxeccenC
\else
??????\fi
\fi
\fi
}
\newcommand{\hatcurLCTBeccen}[1]{\ifnum#1=39 %
\hatcurLCTBxxxxeccenA
\else
\ifnum#1=40 %
\hatcurLCTBxxxxeccenB
\else
\ifnum#1=41 %
\hatcurLCTBxxxxeccenC
\else
??????\fi
\fi
\fi
}
\newcommand{\hatcurLCzetaeccen}[1]{\ifnum#1=39 %
\hatcurLCzetaxxxxeccenA
\else
\ifnum#1=40 %
\hatcurLCzetaxxxxeccenB
\else
\ifnum#1=41 %
\hatcurLCzetaxxxxeccenC
\else
??????\fi
\fi
\fi
}
\newcommand{\hatcurPPaequiveccen}[1]{\ifnum#1=39 %
\hatcurPPaequivxxxxeccenA
\else
\ifnum#1=40 %
\hatcurPPaequivxxxxeccenB
\else
\ifnum#1=41 %
\hatcurPPaequivxxxxeccenC
\else
??????\fi
\fi
\fi
}
\newcommand{\hatcurPPareccen}[1]{\ifnum#1=39 %
\hatcurPParxxxxeccenA
\else
\ifnum#1=40 %
\hatcurPParxxxxeccenB
\else
\ifnum#1=41 %
\hatcurPParxxxxeccenC
\else
??????\fi
\fi
\fi
}
\newcommand{\hatcurPPareleccen}[1]{\ifnum#1=39 %
\hatcurPParelxxxxeccenA
\else
\ifnum#1=40 %
\hatcurPParelxxxxeccenB
\else
\ifnum#1=41 %
\hatcurPParelxxxxeccenC
\else
??????\fi
\fi
\fi
}
\newcommand{\hatcurPPfluxapeccen}[1]{\ifnum#1=39 %
\hatcurPPfluxapxxxxeccenA
\else
\ifnum#1=40 %
\hatcurPPfluxapxxxxeccenB
\else
\ifnum#1=41 %
\hatcurPPfluxapxxxxeccenC
\else
??????\fi
\fi
\fi
}
\newcommand{\hatcurPPfluxapdimeccen}[1]{\ifnum#1=39 %
\hatcurPPfluxapdimxxxxeccenA
\else
\ifnum#1=40 %
\hatcurPPfluxapdimxxxxeccenB
\else
\ifnum#1=41 %
\hatcurPPfluxapdimxxxxeccenC
\else
??????\fi
\fi
\fi
}
\newcommand{\hatcurPPfluxavgeccen}[1]{\ifnum#1=39 %
\hatcurPPfluxavgxxxxeccenA
\else
\ifnum#1=40 %
\hatcurPPfluxavgxxxxeccenB
\else
\ifnum#1=41 %
\hatcurPPfluxavgxxxxeccenC
\else
??????\fi
\fi
\fi
}
\newcommand{\hatcurPPfluxavgdimeccen}[1]{\ifnum#1=39 %
\hatcurPPfluxavgdimxxxxeccenA
\else
\ifnum#1=40 %
\hatcurPPfluxavgdimxxxxeccenB
\else
\ifnum#1=41 %
\hatcurPPfluxavgdimxxxxeccenC
\else
??????\fi
\fi
\fi
}
\newcommand{\hatcurPPfluxperieccen}[1]{\ifnum#1=39 %
\hatcurPPfluxperixxxxeccenA
\else
\ifnum#1=40 %
\hatcurPPfluxperixxxxeccenB
\else
\ifnum#1=41 %
\hatcurPPfluxperixxxxeccenC
\else
??????\fi
\fi
\fi
}
\newcommand{\hatcurPPfluxperidimeccen}[1]{\ifnum#1=39 %
\hatcurPPfluxperidimxxxxeccenA
\else
\ifnum#1=40 %
\hatcurPPfluxperidimxxxxeccenB
\else
\ifnum#1=41 %
\hatcurPPfluxperidimxxxxeccenC
\else
??????\fi
\fi
\fi
}
\newcommand{\hatcurPPgeccen}[1]{\ifnum#1=39 %
\hatcurPPgxxxxeccenA
\else
\ifnum#1=40 %
\hatcurPPgxxxxeccenB
\else
\ifnum#1=41 %
\hatcurPPgxxxxeccenC
\else
??????\fi
\fi
\fi
}
\newcommand{\hatcurPPieccen}[1]{\ifnum#1=39 %
\hatcurPPixxxxeccenA
\else
\ifnum#1=40 %
\hatcurPPixxxxeccenB
\else
\ifnum#1=41 %
\hatcurPPixxxxeccenC
\else
??????\fi
\fi
\fi
}
\newcommand{\hatcurPPloggeccen}[1]{\ifnum#1=39 %
\hatcurPPloggxxxxeccenA
\else
\ifnum#1=40 %
\hatcurPPloggxxxxeccenB
\else
\ifnum#1=41 %
\hatcurPPloggxxxxeccenC
\else
??????\fi
\fi
\fi
}
\newcommand{\hatcurPPmeccen}[1]{\ifnum#1=39 %
\hatcurPPmxxxxeccenA
\else
\ifnum#1=40 %
\hatcurPPmxxxxeccenB
\else
\ifnum#1=41 %
\hatcurPPmxxxxeccenC
\else
??????\fi
\fi
\fi
}
\newcommand{\hatcurPPmeeccen}[1]{\ifnum#1=39 %
\hatcurPPmexxxxeccenA
\else
\ifnum#1=40 %
\hatcurPPmexxxxeccenB
\else
\ifnum#1=41 %
\hatcurPPmexxxxeccenC
\else
??????\fi
\fi
\fi
}
\newcommand{\hatcurPPmelongeccen}[1]{\ifnum#1=39 %
\hatcurPPmelongxxxxeccenA
\else
\ifnum#1=40 %
\hatcurPPmelongxxxxeccenB
\else
\ifnum#1=41 %
\hatcurPPmelongxxxxeccenC
\else
??????\fi
\fi
\fi
}
\newcommand{\hatcurPPmeshorteccen}[1]{\ifnum#1=39 %
\hatcurPPmeshortxxxxeccenA
\else
\ifnum#1=40 %
\hatcurPPmeshortxxxxeccenB
\else
\ifnum#1=41 %
\hatcurPPmeshortxxxxeccenC
\else
??????\fi
\fi
\fi
}
\newcommand{\hatcurPPmlongeccen}[1]{\ifnum#1=39 %
\hatcurPPmlongxxxxeccenA
\else
\ifnum#1=40 %
\hatcurPPmlongxxxxeccenB
\else
\ifnum#1=41 %
\hatcurPPmlongxxxxeccenC
\else
??????\fi
\fi
\fi
}
\newcommand{\hatcurPPmrcorreccen}[1]{\ifnum#1=39 %
\hatcurPPmrcorrxxxxeccenA
\else
\ifnum#1=40 %
\hatcurPPmrcorrxxxxeccenB
\else
\ifnum#1=41 %
\hatcurPPmrcorrxxxxeccenC
\else
??????\fi
\fi
\fi
}
\newcommand{\hatcurPPmshorteccen}[1]{\ifnum#1=39 %
\hatcurPPmshortxxxxeccenA
\else
\ifnum#1=40 %
\hatcurPPmshortxxxxeccenB
\else
\ifnum#1=41 %
\hatcurPPmshortxxxxeccenC
\else
??????\fi
\fi
\fi
}
\newcommand{\hatcurPPperieccen}[1]{\ifnum#1=39 %
\hatcurPPperixxxxeccenA
\else
\ifnum#1=40 %
\hatcurPPperixxxxeccenB
\else
\ifnum#1=41 %
\hatcurPPperixxxxeccenC
\else
??????\fi
\fi
\fi
}
\newcommand{\hatcurPPphiconjeccen}[1]{\ifnum#1=39 %
\hatcurPPphiconjxxxxeccenA
\else
\ifnum#1=40 %
\hatcurPPphiconjxxxxeccenB
\else
\ifnum#1=41 %
\hatcurPPphiconjxxxxeccenC
\else
??????\fi
\fi
\fi
}
\newcommand{\hatcurPPreccen}[1]{\ifnum#1=39 %
\hatcurPPrxxxxeccenA
\else
\ifnum#1=40 %
\hatcurPPrxxxxeccenB
\else
\ifnum#1=41 %
\hatcurPPrxxxxeccenC
\else
??????\fi
\fi
\fi
}
\newcommand{\hatcurPPreeccen}[1]{\ifnum#1=39 %
\hatcurPPrexxxxeccenA
\else
\ifnum#1=40 %
\hatcurPPrexxxxeccenB
\else
\ifnum#1=41 %
\hatcurPPrexxxxeccenC
\else
??????\fi
\fi
\fi
}
\newcommand{\hatcurPPrelongeccen}[1]{\ifnum#1=39 %
\hatcurPPrelongxxxxeccenA
\else
\ifnum#1=40 %
\hatcurPPrelongxxxxeccenB
\else
\ifnum#1=41 %
\hatcurPPrelongxxxxeccenC
\else
??????\fi
\fi
\fi
}
\newcommand{\hatcurPPreshorteccen}[1]{\ifnum#1=39 %
\hatcurPPreshortxxxxeccenA
\else
\ifnum#1=40 %
\hatcurPPreshortxxxxeccenB
\else
\ifnum#1=41 %
\hatcurPPreshortxxxxeccenC
\else
??????\fi
\fi
\fi
}
\newcommand{\hatcurPPrhoeccen}[1]{\ifnum#1=39 %
\hatcurPPrhoxxxxeccenA
\else
\ifnum#1=40 %
\hatcurPPrhoxxxxeccenB
\else
\ifnum#1=41 %
\hatcurPPrhoxxxxeccenC
\else
??????\fi
\fi
\fi
}
\newcommand{\hatcurPPrlongeccen}[1]{\ifnum#1=39 %
\hatcurPPrlongxxxxeccenA
\else
\ifnum#1=40 %
\hatcurPPrlongxxxxeccenB
\else
\ifnum#1=41 %
\hatcurPPrlongxxxxeccenC
\else
??????\fi
\fi
\fi
}
\newcommand{\hatcurPPrshorteccen}[1]{\ifnum#1=39 %
\hatcurPPrshortxxxxeccenA
\else
\ifnum#1=40 %
\hatcurPPrshortxxxxeccenB
\else
\ifnum#1=41 %
\hatcurPPrshortxxxxeccenC
\else
??????\fi
\fi
\fi
}
\newcommand{\hatcurPPtcirceccen}[1]{\ifnum#1=39 %
\hatcurPPtcircxxxxeccenA
\else
\ifnum#1=40 %
\hatcurPPtcircxxxxeccenB
\else
\ifnum#1=41 %
\hatcurPPtcircxxxxeccenC
\else
??????\fi
\fi
\fi
}
\newcommand{\hatcurPPteffeccen}[1]{\ifnum#1=39 %
\hatcurPPteffxxxxeccenA
\else
\ifnum#1=40 %
\hatcurPPteffxxxxeccenB
\else
\ifnum#1=41 %
\hatcurPPteffxxxxeccenC
\else
??????\fi
\fi
\fi
}
\newcommand{\hatcurPPthetaeccen}[1]{\ifnum#1=39 %
\hatcurPPthetaxxxxeccenA
\else
\ifnum#1=40 %
\hatcurPPthetaxxxxeccenB
\else
\ifnum#1=41 %
\hatcurPPthetaxxxxeccenC
\else
??????\fi
\fi
\fi
}
\newcommand{\hatcurPPtinfalleccen}[1]{\ifnum#1=39 %
\hatcurPPtinfallxxxxeccenA
\else
\ifnum#1=40 %
\hatcurPPtinfallxxxxeccenB
\else
\ifnum#1=41 %
\hatcurPPtinfallxxxxeccenC
\else
??????\fi
\fi
\fi
}
\newcommand{\hatcurRVecceneccen}[1]{\ifnum#1=39 %
\hatcurRVeccenxxxxeccenA
\else
\ifnum#1=40 %
\hatcurRVeccenxxxxeccenB
\else
\ifnum#1=41 %
\hatcurRVeccenxxxxeccenC
\else
??????\fi
\fi
\fi
}
\newcommand{\hatcurRVfitrmseccen}[1]{\ifnum#1=39 %
\hatcurRVfitrmsxxxxeccenA
\else
\ifnum#1=40 %
\hatcurRVfitrmsxxxxeccenB
\else
\ifnum#1=41 %
\hatcurRVfitrmsxxxxeccenC
\else
??????\fi
\fi
\fi
}
\newcommand{\hatcurRVgammaeccen}[1]{\ifnum#1=39 %
\hatcurRVgammaxxxxeccenA
\else
\ifnum#1=40 %
\hatcurRVgammaxxxxeccenB
\else
\ifnum#1=41 %
\hatcurRVgammaxxxxeccenC
\else
??????\fi
\fi
\fi
}
\newcommand{\hatcurRVrheccen}[1]{\ifnum#1=39 %
\hatcurRVrhxxxxeccenA
\else
\ifnum#1=40 %
\hatcurRVrhxxxxeccenB
\else
\ifnum#1=41 %
\hatcurRVrhxxxxeccenC
\else
??????\fi
\fi
\fi
}
\newcommand{\hatcurRVheccen}[1]{\ifnum#1=39 %
\hatcurRVhxxxxeccenA
\else
\ifnum#1=40 %
\hatcurRVhxxxxeccenB
\else
\ifnum#1=41 %
\hatcurRVhxxxxeccenC
\else
??????\fi
\fi
\fi
}
\newcommand{\hatcurRVjittereccen}[1]{\ifnum#1=39 %
\hatcurRVjitterxxxxeccenA
\else
\ifnum#1=40 %
\hatcurRVjitterxxxxeccenB
\else
\ifnum#1=41 %
\hatcurRVjitterAxxxxeccenC
\else
??????\fi
\fi
\fi
}
\newcommand{\hatcurRVrkeccen}[1]{\ifnum#1=39 %
\hatcurRVrkxxxxeccenA
\else
\ifnum#1=40 %
\hatcurRVrkxxxxeccenB
\else
\ifnum#1=41 %
\hatcurRVrkxxxxeccenC
\else
??????\fi
\fi
\fi
}
\newcommand{\hatcurRVkeccen}[1]{\ifnum#1=39 %
\hatcurRVkxxxxeccenA
\else
\ifnum#1=40 %
\hatcurRVkxxxxeccenB
\else
\ifnum#1=41 %
\hatcurRVkxxxxeccenC
\else
??????\fi
\fi
\fi
}
\newcommand{\hatcurRVKeccen}[1]{\ifnum#1=39 %
\hatcurRVKxxxxeccenA
\else
\ifnum#1=40 %
\hatcurRVKxxxxeccenB
\else
\ifnum#1=41 %
\hatcurRVKxxxxeccenC
\else
??????\fi
\fi
\fi
}
\newcommand{\hatcurRVomegaeccen}[1]{\ifnum#1=39 %
\hatcurRVomegaxxxxeccenA
\else
\ifnum#1=40 %
\hatcurRVomegaxxxxeccenB
\else
\ifnum#1=41 %
\hatcurRVomegaxxxxeccenC
\else
??????\fi
\fi
\fi
}
\newcommand{\hatcurRVtroneeccen}[1]{\ifnum#1=39 %
\hatcurRVtronexxxxeccenA
\else
\ifnum#1=40 %
\hatcurRVtronexxxxeccenB
\else
\ifnum#1=41 %
\hatcurRVtronexxxxeccenC
\else
??????\fi
\fi
\fi
}
\newcommand{\hatcurRVtrtwoeccen}[1]{\ifnum#1=39 %
\hatcurRVtrtwoxxxxeccenA
\else
\ifnum#1=40 %
\hatcurRVtrtwoxxxxeccenB
\else
\ifnum#1=41 %
\hatcurRVtrtwoxxxxeccenC
\else
??????\fi
\fi
\fi
}
\newcommand{\hatcurSMEiiloggeccen}[1]{\ifnum#1=39 %
\hatcurSMEiiloggxxxxeccenA
\else
\ifnum#1=40 %
\hatcurSMEiiloggxxxxeccenB
\else
\ifnum#1=41 %
\hatcurSMEiiloggxxxxeccenC
\else
??????\fi
\fi
\fi
}
\newcommand{\hatcurSMEiiteffeccen}[1]{\ifnum#1=39 %
\hatcurSMEiiteffxxxxeccenA
\else
\ifnum#1=40 %
\hatcurSMEiiteffxxxxeccenB
\else
\ifnum#1=41 %
\hatcurSMEiiteffxxxxeccenC
\else
??????\fi
\fi
\fi
}
\newcommand{\hatcurSMEiivmaceccen}[1]{\ifnum#1=39 %
\hatcurSMEiivmacxxxxeccenA
\else
\ifnum#1=40 %
\hatcurSMEiivmacxxxxeccenB
\else
\ifnum#1=41 %
\hatcurSMEiivmacxxxxeccenC
\else
??????\fi
\fi
\fi
}
\newcommand{\hatcurSMEiivmiceccen}[1]{\ifnum#1=39 %
\hatcurSMEiivmicxxxxeccenA
\else
\ifnum#1=40 %
\hatcurSMEiivmicxxxxeccenB
\else
\ifnum#1=41 %
\hatcurSMEiivmicxxxxeccenC
\else
??????\fi
\fi
\fi
}
\newcommand{\hatcurSMEiivsineccen}[1]{\ifnum#1=39 %
\hatcurSMEiivsinxxxxeccenA
\else
\ifnum#1=40 %
\hatcurSMEiivsinxxxxeccenB
\else
\ifnum#1=41 %
\hatcurSMEiivsinxxxxeccenC
\else
??????\fi
\fi
\fi
}
\newcommand{\hatcurSMEiizfeheccen}[1]{\ifnum#1=39 %
\hatcurSMEiizfehxxxxeccenA
\else
\ifnum#1=40 %
\hatcurSMEiizfehxxxxeccenB
\else
\ifnum#1=41 %
\hatcurSMEiizfehxxxxeccenC
\else
??????\fi
\fi
\fi
}
\newcommand{\hatcurSMEiizfehshorteccen}[1]{\ifnum#1=39 %
\hatcurSMEiizfehshortxxxxeccenA
\else
\ifnum#1=40 %
\hatcurSMEiizfehshortxxxxeccenB
\else
\ifnum#1=41 %
\hatcurSMEiizfehshortxxxxeccenC
\else
??????\fi
\fi
\fi
}
\newcommand{\hatcurSMEiloggeccen}[1]{\ifnum#1=39 %
\hatcurSMEiloggxxxxeccenA
\else
\ifnum#1=40 %
\hatcurSMEiloggxxxxeccenB
\else
\ifnum#1=41 %
\hatcurSMEiloggxxxxeccenC
\else
??????\fi
\fi
\fi
}
\newcommand{\hatcurSMEiteffeccen}[1]{\ifnum#1=39 %
\hatcurSMEiteffxxxxeccenA
\else
\ifnum#1=40 %
\hatcurSMEiteffxxxxeccenB
\else
\ifnum#1=41 %
\hatcurSMEiteffxxxxeccenC
\else
??????\fi
\fi
\fi
}
\newcommand{\hatcurSMEivmaceccen}[1]{\ifnum#1=39 %
\hatcurSMEivmacxxxxeccenA
\else
\ifnum#1=40 %
\hatcurSMEivmacxxxxeccenB
\else
\ifnum#1=41 %
\hatcurSMEivmacxxxxeccenC
\else
??????\fi
\fi
\fi
}
\newcommand{\hatcurSMEivmiceccen}[1]{\ifnum#1=39 %
\hatcurSMEivmicxxxxeccenA
\else
\ifnum#1=40 %
\hatcurSMEivmicxxxxeccenB
\else
\ifnum#1=41 %
\hatcurSMEivmicxxxxeccenC
\else
??????\fi
\fi
\fi
}
\newcommand{\hatcurSMEivsineccen}[1]{\ifnum#1=39 %
\hatcurSMEivsinxxxxeccenA
\else
\ifnum#1=40 %
\hatcurSMEivsinxxxxeccenB
\else
\ifnum#1=41 %
\hatcurSMEivsinxxxxeccenC
\else
??????\fi
\fi
\fi
}
\newcommand{\hatcurSMEizfeheccen}[1]{\ifnum#1=39 %
\hatcurSMEizfehxxxxeccenA
\else
\ifnum#1=40 %
\hatcurSMEizfehxxxxeccenB
\else
\ifnum#1=41 %
\hatcurSMEizfehxxxxeccenC
\else
??????\fi
\fi
\fi
}
\newcommand{\hatcurSMEizfehshorteccen}[1]{\ifnum#1=39 %
\hatcurSMEizfehshortxxxxeccenA
\else
\ifnum#1=40 %
\hatcurSMEizfehshortxxxxeccenB
\else
\ifnum#1=41 %
\hatcurSMEizfehshortxxxxeccenC
\else
??????\fi
\fi
\fi
}
\newcommand{\hatcurTRESgammaeccen}[1]{\ifnum#1=39 %
\hatcurTRESgammaxxxxeccenA
\else
\ifnum#1=40 %
\hatcurTRESgammaxxxxeccenB
\else
\ifnum#1=41 %
\hatcurTRESgammaxxxxeccenC
\else
??????\fi
\fi
\fi
}
\newcommand{\hatcurTRESloggeccen}[1]{\ifnum#1=39 %
\hatcurTRESloggxxxxeccenA
\else
\ifnum#1=40 %
\hatcurTRESloggxxxxeccenB
\else
\ifnum#1=41 %
\hatcurTRESloggxxxxeccenC
\else
??????\fi
\fi
\fi
}
\newcommand{\hatcurTRESnumspececcen}[1]{\ifnum#1=39 %
\hatcurTRESnumspecxxxxeccenA
\else
\ifnum#1=40 %
\hatcurTRESnumspecxxxxeccenB
\else
\ifnum#1=41 %
\hatcurTRESnumspecxxxxeccenC
\else
??????\fi
\fi
\fi
}
\newcommand{\hatcurTRESrvrmseccen}[1]{\ifnum#1=39 %
\hatcurTRESrvrmsxxxxeccenA
\else
\ifnum#1=40 %
\hatcurTRESrvrmsxxxxeccenB
\else
\ifnum#1=41 %
\hatcurTRESrvrmsxxxxeccenC
\else
??????\fi
\fi
\fi
}
\newcommand{\hatcurTRESspaneccen}[1]{\ifnum#1=39 %
\hatcurTRESspanxxxxeccenA
\else
\ifnum#1=40 %
\hatcurTRESspanxxxxeccenB
\else
\ifnum#1=41 %
\hatcurTRESspanxxxxeccenC
\else
??????\fi
\fi
\fi
}
\newcommand{\hatcurTRESteffeccen}[1]{\ifnum#1=39 %
\hatcurTRESteffxxxxeccenA
\else
\ifnum#1=40 %
\hatcurTRESteffxxxxeccenB
\else
\ifnum#1=41 %
\hatcurTRESteffxxxxeccenC
\else
??????\fi
\fi
\fi
}
\newcommand{\hatcurTRESvsinieccen}[1]{\ifnum#1=39 %
\hatcurTRESvsinixxxxeccenA
\else
\ifnum#1=40 %
\hatcurTRESvsinixxxxeccenB
\else
\ifnum#1=41 %
\hatcurTRESvsinixxxxeccenC
\else
??????\fi
\fi
\fi
}
\newcommand{\hatcurTRESzfeheccen}[1]{\ifnum#1=39 %
\hatcurTRESzfehxxxxeccenA
\else
\ifnum#1=40 %
\hatcurTRESzfehxxxxeccenB
\else
\ifnum#1=41 %
\hatcurTRESzfehxxxxeccenC
\else
??????\fi
\fi
\fi
}
\newcommand{\hatcurXdisteccen}[1]{\ifnum#1=39 %
\hatcurXdistxxxxeccenA
\else
\ifnum#1=40 %
\hatcurXdistxxxxeccenB
\else
\ifnum#1=41 %
\hatcurXdistxxxxeccenC
\else
??????\fi
\fi
\fi
}
\newcommand{\hatcurXdistredeccen}[1]{\ifnum#1=39 %
\hatcurXdistredxxxxeccenA
\else
\ifnum#1=40 %
\hatcurXdistredxxxxeccenB
\else
\ifnum#1=41 %
\hatcurXdistredxxxxeccenC
\else
??????\fi
\fi
\fi
}
\newcommand{\hatcurXAveccen}[1]{\ifnum#1=39 %
\hatcurXAvxxxxeccenA
\else
\ifnum#1=40 %
\hatcurXAvxxxxeccenB
\else
\ifnum#1=41 %
\hatcurXAvxxxxeccenC
\else
??????\fi
\fi
\fi
}
\newcommand{\hatcurXsecdureccen}[1]{\ifnum#1=39 %
\hatcurXsecdurxxxxeccenA
\else
\ifnum#1=40 %
\hatcurXsecdurxxxxeccenB
\else
\ifnum#1=41 %
\hatcurXsecdurxxxxeccenC
\else
??????\fi
\fi
\fi
}
\newcommand{\hatcurXsecingdureccen}[1]{\ifnum#1=39 %
\hatcurXsecingdurxxxxeccenA
\else
\ifnum#1=40 %
\hatcurXsecingdurxxxxeccenB
\else
\ifnum#1=41 %
\hatcurXsecingdurxxxxeccenC
\else
??????\fi
\fi
\fi
}
\newcommand{\hatcurXsecondaryeccen}[1]{\ifnum#1=39 %
\hatcurXsecondaryxxxxeccenA
\else
\ifnum#1=40 %
\hatcurXsecondaryxxxxeccenB
\else
\ifnum#1=41 %
\hatcurXsecondaryxxxxeccenC
\else
??????\fi
\fi
\fi
}
\newcommand{\hatcurXsecphaseeccen}[1]{\ifnum#1=39 %
\hatcurXsecphasexxxxeccenA
\else
\ifnum#1=40 %
\hatcurXsecphasexxxxeccenB
\else
\ifnum#1=41 %
\hatcurXsecphasexxxxeccenC
\else
??????\fi
\fi
\fi
}
\newcommand{\hatcurxxxxA}{HAT-P-39}
\newcommand{\hatcurbxxxxA}{HAT-P-39b}
\newcommand{\hatcurcxxxxA}{HAT-P-39c}

\newcommand{\hatcurplanetnumxxxxA}{39}
\newcommand{\hatcurplanetnumxxxxeccenA}{39}

\newcommand{\hatcurRVgammaabsxxxxA}{\hatcurDSgamma{\hatcurplanetnumxxxxA}}                           

\newcommand{\hatcurRVgammarelxxxxA}{\hatcurRVgamma{\hatcurplanetnumxxxxA}}                           

\newcommand{\hatcurCCtassvixxxxA}{\ensuremath{0.58\pm0.15}}                  
\newcommand{\hatcurCCtassvixxxxeccenA}{\ensuremath{0.58\pm0.15}}                  

\newcommand{\hatcurSMEversionxxxxA}{ii}                                       
\newcommand{\hatcurSMEversionxxxxeccenA}{ii}                                       

\newcommand{\hatcurisoshortxxxxA}{YY}
\newcommand{\hatcurisofullxxxxA}{Yonsei-Yale (YY)}
\newcommand{\hatcurisocitexxxxA}{yi:2001}

\newcommand{\hatcurlumindxxxxA}{\arstar}

\newcommand{\hatcurjhkfilsetxxxxA}{ESO}

\newcommand{\hatcurSMEteffxxxxA}{\ifthenelse{\equal{\hatcurSMEversionxxxxA}{i}}{\hatcurSMEiteff{\hatcurplanetnumxxxxA}}{\hatcurSMEiiteff{\hatcurplanetnumxxxxA}}}
\newcommand{\hatcurSMEzfehxxxxA}{\ifthenelse{\equal{\hatcurSMEversionxxxxA}{i}}{\hatcurSMEizfeh{\hatcurplanetnumxxxxA}}{\hatcurSMEiizfeh{\hatcurplanetnumxxxxA}}}
\newcommand{\hatcurSMEzfehshortxxxxA}{\ifthenelse{\equal{\hatcurSMEversionxxxxA}{i}}{\hatcurSMEizfehshort{\hatcurplanetnumxxxxA}}{\hatcurSMEiizfehshort{\hatcurplanetnumxxxxA}}}
\newcommand{\hatcurSMEloggxxxxA}{\ifthenelse{\equal{\hatcurSMEversionxxxxA}{i}}{\hatcurSMEilogg{\hatcurplanetnumxxxxA}}{\hatcurSMEiilogg{\hatcurplanetnumxxxxA}}}
\newcommand{\hatcurSMEvsinxxxxA}{\ifthenelse{\equal{\hatcurSMEversionxxxxA}{i}}{\hatcurSMEivsin{\hatcurplanetnumxxxxA}}{\hatcurSMEiivsin{\hatcurplanetnumxxxxA}}}
\newcommand{\hatcurSMEvmacxxxxA}{\ifthenelse{\equal{\hatcurSMEversionxxxxA}{i}}{\hatcurSMEivmac{\hatcurplanetnumxxxxA}}{\hatcurSMEiivmac{\hatcurplanetnumxxxxA}}}
\newcommand{\hatcurSMEvmicxxxxA}{\ifthenelse{\equal{\hatcurSMEversionxxxxA}{i}}{\hatcurSMEivmic{\hatcurplanetnumxxxxA}}{\hatcurSMEiivmic{\hatcurplanetnumxxxxA}}}

\newcommand{\hatcurSMEteffxxxxeccenA}{\ifthenelse{\equal{\hatcurSMEversionxxxxeccenA}{i}}{\hatcurSMEiteff{\hatcurplanetnumxxxxeccenA}}{\hatcurSMEiiteff{\hatcurplanetnumxxxxeccenA}}}
\newcommand{\hatcurSMEzfehxxxxeccenA}{\ifthenelse{\equal{\hatcurSMEversionxxxxeccenA}{i}}{\hatcurSMEizfeh{\hatcurplanetnumxxxxeccenA}}{\hatcurSMEiizfeh{\hatcurplanetnumxxxxeccenA}}}
\newcommand{\hatcurSMEzfehshortxxxxeccenA}{\ifthenelse{\equal{\hatcurSMEversionxxxxeccenA}{i}}{\hatcurSMEizfehshort{\hatcurplanetnumxxxxeccenA}}{\hatcurSMEiizfehshort{\hatcurplanetnumxxxxeccenA}}}
\newcommand{\hatcurSMEloggxxxxeccenA}{\ifthenelse{\equal{\hatcurSMEversionxxxxeccenA}{i}}{\hatcurSMEilogg{\hatcurplanetnumxxxxeccenA}}{\hatcurSMEiilogg{\hatcurplanetnumxxxxeccenA}}}
\newcommand{\hatcurSMEvsinxxxxeccenA}{\ifthenelse{\equal{\hatcurSMEversionxxxxeccenA}{i}}{\hatcurSMEivsin{\hatcurplanetnumxxxxeccenA}}{\hatcurSMEiivsin{\hatcurplanetnumxxxxeccenA}}}
\newcommand{\hatcurSMEvmacxxxxeccenA}{\ifthenelse{\equal{\hatcurSMEversionxxxxeccenA}{i}}{\hatcurSMEivmac{\hatcurplanetnumxxxxeccenA}}{\hatcurSMEiivmac{\hatcurplanetnumxxxxeccenA}}}
\newcommand{\hatcurSMEvmicxxxxeccenA}{\ifthenelse{\equal{\hatcurSMEversionxxxxeccenA}{i}}{\hatcurSMEivmic{\hatcurplanetnumxxxxeccenA}}{\hatcurSMEiivmic{\hatcurplanetnumxxxxeccenA}}}

\newcommand{\hatcurxxxxB}{HAT-P-40}
\newcommand{\hatcurbxxxxB}{HAT-P-40b}
\newcommand{\hatcurcxxxxB}{HAT-P-40c}

\newcommand{\hatcurplanetnumxxxxB}{40}
\newcommand{\hatcurplanetnumxxxxeccenB}{40}

\newcommand{\hatcurRVgammaabsxxxxB}{\hatcurDSgamma{\hatcurplanetnumxxxxB}}                           

\newcommand{\hatcurRVgammarelxxxxB}{\hatcurRVgamma{\hatcurplanetnumxxxxB}}                           

\newcommand{\hatcurCCtassvixxxxB}{\ensuremath{0.77\pm0.12}}                  
\newcommand{\hatcurCCtassvixxxxeccenB}{\ensuremath{0.77\pm0.12}}                  

\newcommand{\hatcurSMEversionxxxxB}{ii}                                       
\newcommand{\hatcurSMEversionxxxxeccenB}{ii}                                       

\newcommand{\hatcurisoshortxxxxB}{YY}
\newcommand{\hatcurisofullxxxxB}{Yonsei-Yale (YY)}
\newcommand{\hatcurisocitexxxxB}{yi:2001}

\newcommand{\hatcurlumindxxxxB}{\arstar}

\newcommand{\hatcurjhkfilsetxxxxB}{ESO}

\newcommand{\hatcurSMEteffxxxxB}{\ifthenelse{\equal{\hatcurSMEversionxxxxB}{i}}{\hatcurSMEiteff{\hatcurplanetnumxxxxB}}{\hatcurSMEiiteff{\hatcurplanetnumxxxxB}}}
\newcommand{\hatcurSMEzfehxxxxB}{\ifthenelse{\equal{\hatcurSMEversionxxxxB}{i}}{\hatcurSMEizfeh{\hatcurplanetnumxxxxB}}{\hatcurSMEiizfeh{\hatcurplanetnumxxxxB}}}
\newcommand{\hatcurSMEzfehshortxxxxB}{\ifthenelse{\equal{\hatcurSMEversionxxxxB}{i}}{\hatcurSMEizfehshort{\hatcurplanetnumxxxxB}}{\hatcurSMEiizfehshort{\hatcurplanetnumxxxxB}}}
\newcommand{\hatcurSMEloggxxxxB}{\ifthenelse{\equal{\hatcurSMEversionxxxxB}{i}}{\hatcurSMEilogg{\hatcurplanetnumxxxxB}}{\hatcurSMEiilogg{\hatcurplanetnumxxxxB}}}
\newcommand{\hatcurSMEvsinxxxxB}{\ifthenelse{\equal{\hatcurSMEversionxxxxB}{i}}{\hatcurSMEivsin{\hatcurplanetnumxxxxB}}{\hatcurSMEiivsin{\hatcurplanetnumxxxxB}}}
\newcommand{\hatcurSMEvmacxxxxB}{\ifthenelse{\equal{\hatcurSMEversionxxxxB}{i}}{\hatcurSMEivmac{\hatcurplanetnumxxxxB}}{\hatcurSMEiivmac{\hatcurplanetnumxxxxB}}}
\newcommand{\hatcurSMEvmicxxxxB}{\ifthenelse{\equal{\hatcurSMEversionxxxxB}{i}}{\hatcurSMEivmic{\hatcurplanetnumxxxxB}}{\hatcurSMEiivmic{\hatcurplanetnumxxxxB}}}

\newcommand{\hatcurSMEteffxxxxeccenB}{\ifthenelse{\equal{\hatcurSMEversionxxxxeccenB}{i}}{\hatcurSMEiteff{\hatcurplanetnumxxxxeccenB}}{\hatcurSMEiiteff{\hatcurplanetnumxxxxeccenB}}}
\newcommand{\hatcurSMEzfehxxxxeccenB}{\ifthenelse{\equal{\hatcurSMEversionxxxxeccenB}{i}}{\hatcurSMEizfeh{\hatcurplanetnumxxxxeccenB}}{\hatcurSMEiizfeh{\hatcurplanetnumxxxxeccenB}}}
\newcommand{\hatcurSMEzfehshortxxxxeccenB}{\ifthenelse{\equal{\hatcurSMEversionxxxxeccenB}{i}}{\hatcurSMEizfehshort{\hatcurplanetnumxxxxeccenB}}{\hatcurSMEiizfehshort{\hatcurplanetnumxxxxeccenB}}}
\newcommand{\hatcurSMEloggxxxxeccenB}{\ifthenelse{\equal{\hatcurSMEversionxxxxeccenB}{i}}{\hatcurSMEilogg{\hatcurplanetnumxxxxeccenB}}{\hatcurSMEiilogg{\hatcurplanetnumxxxxeccenB}}}
\newcommand{\hatcurSMEvsinxxxxeccenB}{\ifthenelse{\equal{\hatcurSMEversionxxxxeccenB}{i}}{\hatcurSMEivsin{\hatcurplanetnumxxxxeccenB}}{\hatcurSMEiivsin{\hatcurplanetnumxxxxeccenB}}}
\newcommand{\hatcurSMEvmacxxxxeccenB}{\ifthenelse{\equal{\hatcurSMEversionxxxxeccenB}{i}}{\hatcurSMEivmac{\hatcurplanetnumxxxxeccenB}}{\hatcurSMEiivmac{\hatcurplanetnumxxxxeccenB}}}
\newcommand{\hatcurSMEvmicxxxxeccenB}{\ifthenelse{\equal{\hatcurSMEversionxxxxeccenB}{i}}{\hatcurSMEivmic{\hatcurplanetnumxxxxeccenB}}{\hatcurSMEiivmic{\hatcurplanetnumxxxxeccenB}}}

\newcommand{\hatcurxxxxC}{HAT-P-41}
\newcommand{\hatcurbxxxxC}{HAT-P-41b}
\newcommand{\hatcurcxxxxC}{HAT-P-41c}

\newcommand{\hatcurplanetnumxxxxC}{41}
\newcommand{\hatcurplanetnumxxxxeccenC}{41}

\newcommand{\hatcurRVgammaabsxxxxC}{\hatcurDSgamma{\hatcurplanetnumxxxxC}}                           

\newcommand{\hatcurRVgammarelxxxxC}{\hatcurRVgamma{\hatcurplanetnumxxxxC}}                           

\newcommand{\hatcurCCtassvixxxxC}{\ensuremath{0.63\pm0.10}}                  

\newcommand{\hatcurCCtassvixxxxeccenC}{\ensuremath{0.63\pm0.10}}                  

\newcommand{\hatcurSMEversionxxxxC}{ii}                                       

\newcommand{\hatcurSMEversionxxxxeccenC}{ii}                                       

\newcommand{\hatcurisoshortxxxxC}{YY}
\newcommand{\hatcurisofullxxxxC}{Yonsei-Yale (YY)}
\newcommand{\hatcurisocitexxxxC}{yi:2001}

\newcommand{\hatcurlumindxxxxC}{\arstar}

\newcommand{\hatcurjhkfilsetxxxxC}{ESO}

\newcommand{\hatcurSMEteffxxxxC}{\ifthenelse{\equal{\hatcurSMEversionxxxxC}{i}}{\hatcurSMEiteff{\hatcurplanetnumxxxxC}}{\hatcurSMEiiteff{\hatcurplanetnumxxxxC}}}
\newcommand{\hatcurSMEzfehxxxxC}{\ifthenelse{\equal{\hatcurSMEversionxxxxC}{i}}{\hatcurSMEizfeh{\hatcurplanetnumxxxxC}}{\hatcurSMEiizfeh{\hatcurplanetnumxxxxC}}}
\newcommand{\hatcurSMEzfehshortxxxxC}{\ifthenelse{\equal{\hatcurSMEversionxxxxC}{i}}{\hatcurSMEizfehshort{\hatcurplanetnumxxxxC}}{\hatcurSMEiizfehshort{\hatcurplanetnumxxxxC}}}
\newcommand{\hatcurSMEloggxxxxC}{\ifthenelse{\equal{\hatcurSMEversionxxxxC}{i}}{\hatcurSMEilogg{\hatcurplanetnumxxxxC}}{\hatcurSMEiilogg{\hatcurplanetnumxxxxC}}}
\newcommand{\hatcurSMEvsinxxxxC}{\ifthenelse{\equal{\hatcurSMEversionxxxxC}{i}}{\hatcurSMEivsin{\hatcurplanetnumxxxxC}}{\hatcurSMEiivsin{\hatcurplanetnumxxxxC}}}
\newcommand{\hatcurSMEvmacxxxxC}{\ifthenelse{\equal{\hatcurSMEversionxxxxC}{i}}{\hatcurSMEivmac{\hatcurplanetnumxxxxC}}{\hatcurSMEiivmac{\hatcurplanetnumxxxxC}}}
\newcommand{\hatcurSMEvmicxxxxC}{\ifthenelse{\equal{\hatcurSMEversionxxxxC}{i}}{\hatcurSMEivmic{\hatcurplanetnumxxxxC}}{\hatcurSMEiivmic{\hatcurplanetnumxxxxC}}}

\newcommand{\hatcurSMEteffxxxxeccenC}{\ifthenelse{\equal{\hatcurSMEversionxxxxeccenC}{i}}{\hatcurSMEiteff{\hatcurplanetnumxxxxeccenC}}{\hatcurSMEiiteff{\hatcurplanetnumxxxxeccenC}}}
\newcommand{\hatcurSMEzfehxxxxeccenC}{\ifthenelse{\equal{\hatcurSMEversionxxxxeccenC}{i}}{\hatcurSMEizfeh{\hatcurplanetnumxxxxeccenC}}{\hatcurSMEiizfeh{\hatcurplanetnumxxxxeccenC}}}
\newcommand{\hatcurSMEzfehshortxxxxeccenC}{\ifthenelse{\equal{\hatcurSMEversionxxxxeccenC}{i}}{\hatcurSMEizfehshort{\hatcurplanetnumxxxxeccenC}}{\hatcurSMEiizfehshort{\hatcurplanetnumxxxxeccenC}}}
\newcommand{\hatcurSMEloggxxxxeccenC}{\ifthenelse{\equal{\hatcurSMEversionxxxxeccenC}{i}}{\hatcurSMEilogg{\hatcurplanetnumxxxxeccenC}}{\hatcurSMEiilogg{\hatcurplanetnumxxxxeccenC}}}
\newcommand{\hatcurSMEvsinxxxxeccenC}{\ifthenelse{\equal{\hatcurSMEversionxxxxeccenC}{i}}{\hatcurSMEivsin{\hatcurplanetnumxxxxeccenC}}{\hatcurSMEiivsin{\hatcurplanetnumxxxxeccenC}}}
\newcommand{\hatcurSMEvmacxxxxeccenC}{\ifthenelse{\equal{\hatcurSMEversionxxxxeccenC}{i}}{\hatcurSMEivmac{\hatcurplanetnumxxxxeccenC}}{\hatcurSMEiivmac{\hatcurplanetnumxxxxeccenC}}}
\newcommand{\hatcurSMEvmicxxxxeccenC}{\ifthenelse{\equal{\hatcurSMEversionxxxxeccenC}{i}}{\hatcurSMEivmic{\hatcurplanetnumxxxxeccenC}}{\hatcurSMEiivmic{\hatcurplanetnumxxxxeccenC}}}
\newcommand{\hatcur}[1]{\ifnum#1=39 %
\hatcurxxxxA
\else
\ifnum#1=40 %
\hatcurxxxxB
\else
\ifnum#1=41 %
\hatcurxxxxC
\else
??????\fi
\fi
\fi
}
\newcommand{\hatcurb}[1]{\ifnum#1=39 %
\hatcurbxxxxA
\else
\ifnum#1=40 %
\hatcurbxxxxB
\else
\ifnum#1=41 %
\hatcurbxxxxC
\else
??????\fi
\fi
\fi
}
\newcommand{\hatcurc}[1]{\ifnum#1=39 %
\hatcurcxxxxA
\else
\ifnum#1=40 %
\hatcurcxxxxB
\else
\ifnum#1=41 %
\hatcurcxxxxC
\else
??????\fi
\fi
\fi
}
\newcommand{\hatcurCCtassvi}[1]{\ifnum#1=39 %
\hatcurCCtassvixxxxA
\else
\ifnum#1=40 %
\hatcurCCtassvixxxxB
\else
\ifnum#1=41 %
\hatcurCCtassvixxxxC
\else
??????\fi
\fi
\fi
}
\newcommand{\hatcurCCtassvieccen}[1]{\ifnum#1=39 %
\hatcurCCtassvixxxxeccenA
\else
\ifnum#1=40 %
\hatcurCCtassvixxxxeccenB
\else
\ifnum#1=41 %
\hatcurCCtassvixxxxeccenC
\else
??????\fi
\fi
\fi
}
\newcommand{\hatcurisocite}[1]{\ifnum#1=39 %
\hatcurisocitexxxxA
\else
\ifnum#1=40 %
\hatcurisocitexxxxB
\else
\ifnum#1=41 %
\hatcurisocitexxxxC
\else
??????\fi
\fi
\fi
}
\newcommand{\hatcurisofull}[1]{\ifnum#1=39 %
\hatcurisofullxxxxA
\else
\ifnum#1=40 %
\hatcurisofullxxxxB
\else
\ifnum#1=41 %
\hatcurisofullxxxxC
\else
??????\fi
\fi
\fi
}
\newcommand{\hatcurisoshort}[1]{\ifnum#1=39 %
\hatcurisoshortxxxxA
\else
\ifnum#1=40 %
\hatcurisoshortxxxxB
\else
\ifnum#1=41 %
\hatcurisoshortxxxxC
\else
??????\fi
\fi
\fi
}
\newcommand{\hatcurjhkfilset}[1]{\ifnum#1=39 %
\hatcurjhkfilsetxxxxA
\else
\ifnum#1=40 %
\hatcurjhkfilsetxxxxB
\else
\ifnum#1=41 %
\hatcurjhkfilsetxxxxC
\else
??????\fi
\fi
\fi
}
\newcommand{\hatcurlumind}[1]{\ifnum#1=39 %
\hatcurlumindxxxxA
\else
\ifnum#1=40 %
\hatcurlumindxxxxB
\else
\ifnum#1=41 %
\hatcurlumindxxxxC
\else
??????\fi
\fi
\fi
}
\newcommand{\hatcurplanetnum}[1]{\ifnum#1=39 %
\hatcurplanetnumxxxxA
\else
\ifnum#1=40 %
\hatcurplanetnumxxxxB
\else
\ifnum#1=41 %
\hatcurplanetnumxxxxC
\else
??????\fi
\fi
\fi
}
\newcommand{\hatcurRVgammaabs}[1]{\ifnum#1=39 %
\hatcurRVgammaabsxxxxA
\else
\ifnum#1=40 %
\hatcurRVgammaabsxxxxB
\else
\ifnum#1=41 %
\hatcurRVgammaabsxxxxC
\else
??????\fi
\fi
\fi
}
\newcommand{\hatcurRVgammarel}[1]{\ifnum#1=39 %
\hatcurRVgammarelxxxxA
\else
\ifnum#1=40 %
\hatcurRVgammarelxxxxB
\else
\ifnum#1=41 %
\hatcurRVgammarelxxxxC
\else
??????\fi
\fi
\fi
}
\newcommand{\hatcurSMElogg}[1]{\ifnum#1=39 %
\hatcurSMEloggxxxxA
\else
\ifnum#1=40 %
\hatcurSMEloggxxxxB
\else
\ifnum#1=41 %
\hatcurSMEloggxxxxC
\else
??????\fi
\fi
\fi
}
\newcommand{\hatcurSMEloggeccen}[1]{\ifnum#1=39 %
\hatcurSMEloggxxxxeccenA
\else
\ifnum#1=40 %
\hatcurSMEloggxxxxeccenB
\else
\ifnum#1=41 %
\hatcurSMEloggxxxxeccenC
\else
??????\fi
\fi
\fi
}
\newcommand{\hatcurSMEteff}[1]{\ifnum#1=39 %
\hatcurSMEteffxxxxA
\else
\ifnum#1=40 %
\hatcurSMEteffxxxxB
\else
\ifnum#1=41 %
\hatcurSMEteffxxxxC
\else
??????\fi
\fi
\fi
}
\newcommand{\hatcurSMEteffeccen}[1]{\ifnum#1=39 %
\hatcurSMEteffxxxxeccenA
\else
\ifnum#1=40 %
\hatcurSMEteffxxxxeccenB
\else
\ifnum#1=41 %
\hatcurSMEteffxxxxeccenC
\else
??????\fi
\fi
\fi
}
\newcommand{\hatcurSMEversion}[1]{\ifnum#1=39 %
\hatcurSMEversionxxxxA
\else
\ifnum#1=40 %
\hatcurSMEversionxxxxB
\else
\ifnum#1=41 %
\hatcurSMEversionxxxxC
\else
??????\fi
\fi
\fi
}
\newcommand{\hatcurSMEversioneccen}[1]{\ifnum#1=39 %
\hatcurSMEversionxxxxeccenA
\else
\ifnum#1=40 %
\hatcurSMEversionxxxxeccenB
\else
\ifnum#1=41 %
\hatcurSMEversionxxxxeccenC
\else
??????\fi
\fi
\fi
}
\newcommand{\hatcurSMEvmac}[1]{\ifnum#1=39 %
\hatcurSMEvmacxxxxA
\else
\ifnum#1=40 %
\hatcurSMEvmacxxxxB
\else
\ifnum#1=41 %
\hatcurSMEvmacxxxxC
\else
??????\fi
\fi
\fi
}
\newcommand{\hatcurSMEvmaceccen}[1]{\ifnum#1=39 %
\hatcurSMEvmacxxxxeccenA
\else
\ifnum#1=40 %
\hatcurSMEvmacxxxxeccenB
\else
\ifnum#1=41 %
\hatcurSMEvmacxxxxeccenC
\else
??????\fi
\fi
\fi
}
\newcommand{\hatcurSMEvmic}[1]{\ifnum#1=39 %
\hatcurSMEvmicxxxxA
\else
\ifnum#1=40 %
\hatcurSMEvmicxxxxB
\else
\ifnum#1=41 %
\hatcurSMEvmicxxxxC
\else
??????\fi
\fi
\fi
}
\newcommand{\hatcurSMEvmiceccen}[1]{\ifnum#1=39 %
\hatcurSMEvmicxxxxeccenA
\else
\ifnum#1=40 %
\hatcurSMEvmicxxxxeccenB
\else
\ifnum#1=41 %
\hatcurSMEvmicxxxxeccenC
\else
??????\fi
\fi
\fi
}
\newcommand{\hatcurSMEvsin}[1]{\ifnum#1=39 %
\hatcurSMEvsinxxxxA
\else
\ifnum#1=40 %
\hatcurSMEvsinxxxxB
\else
\ifnum#1=41 %
\hatcurSMEvsinxxxxC
\else
??????\fi
\fi
\fi
}
\newcommand{\hatcurSMEvsineccen}[1]{\ifnum#1=39 %
\hatcurSMEvsinxxxxeccenA
\else
\ifnum#1=40 %
\hatcurSMEvsinxxxxeccenB
\else
\ifnum#1=41 %
\hatcurSMEvsinxxxxeccenC
\else
??????\fi
\fi
\fi
}
\newcommand{\hatcurSMEzfeh}[1]{\ifnum#1=39 %
\hatcurSMEzfehxxxxA
\else
\ifnum#1=40 %
\hatcurSMEzfehxxxxB
\else
\ifnum#1=41 %
\hatcurSMEzfehxxxxC
\else
??????\fi
\fi
\fi
}
\newcommand{\hatcurSMEzfeheccen}[1]{\ifnum#1=39 %
\hatcurSMEzfehxxxxeccenA
\else
\ifnum#1=40 %
\hatcurSMEzfehxxxxeccenB
\else
\ifnum#1=41 %
\hatcurSMEzfehxxxxeccenC
\else
??????\fi
\fi
\fi
}
\newcommand{\hatcurSMEzfehshort}[1]{\ifnum#1=39 %
\hatcurSMEzfehshortxxxxA
\else
\ifnum#1=40 %
\hatcurSMEzfehshortxxxxB
\else
\ifnum#1=41 %
\hatcurSMEzfehshortxxxxC
\else
??????\fi
\fi
\fi
}
\newcommand{\hatcurSMEzfehshorteccen}[1]{\ifnum#1=39 %
\hatcurSMEzfehshortxxxxeccenA
\else
\ifnum#1=40 %
\hatcurSMEzfehshortxxxxeccenB
\else
\ifnum#1=41 %
\hatcurSMEzfehshortxxxxeccenC
\else
??????\fi
\fi
\fi
}

\newcounter{planetcounter}

\newboolean{emulateapj}
\setboolean{emulateapj}{true}

\newboolean{rvtablelong}
\setboolean{rvtablelong}{true}

\newboolean{astroph}
\setboolean{astroph}{true}

\shortauthors{Hartman et al.}
\shorttitle{
\setcounter{planetcounter}{1}
\loopand\hatcur{39}\lowercase{b}\loopcommanospace
\setcounter{planetcounter}{2}
\loopand\hatcur{40}\lowercase{b}\loopcommanospace
\setcounter{planetcounter}{3}
\loopand\hatcur{41}\lowercase{b}\loopcommanospace
}
\ifthenelse{\boolean{emulateapj}}{
    \newcommand{\titledag}{$\dagger$}
}{
    \newcommand{\titledag}{\dagger}
}

\begin{document}

\title{
\hatcur{39}\lowercase{b}--\hatcur{41}\lowercase{b}: 
Three Highly Inflated Transiting Hot Jupiters \altaffilmark{\titledag}
}

\author{
    J.~D.~Hartman\altaffilmark{1},
    G.~\'A.~Bakos\altaffilmark{1,2},
    B.~B\'eky\altaffilmark{3},
    G.~Torres\altaffilmark{3},
    D.~W.~Latham\altaffilmark{3},
    Z.~Csubry\altaffilmark{1},
    K.~Penev\altaffilmark{1},
    A.~Shporer\altaffilmark{4,5},
    B.~J.~Fulton\altaffilmark{4},
    L.~A.~Buchhave\altaffilmark{6},
    J.~A.~Johnson\altaffilmark{7},
    A.~W.~Howard\altaffilmark{8},
    G.~W.~Marcy\altaffilmark{8},
    D.~A.~Fischer\altaffilmark{9},
    G.~Kov\'acs\altaffilmark{10,11},
    R.~W.~Noyes\altaffilmark{3},
    G.~A.~Esquerdo\altaffilmark{3},
    M.~Everett\altaffilmark{3},
    T.~Szklen\'ar\altaffilmark{3},
    S.~N.~Quinn\altaffilmark{3,12},
    A.~Bieryla\altaffilmark{3},
    R.~P.~Knox\altaffilmark{13},
    P.~Hinz\altaffilmark{13},
    D.~D.~Sasselov\altaffilmark{3},
    G.~F\H{u}r\'esz\altaffilmark{3},
    R.~P.~Stefanik\altaffilmark{3},
    J.~L\'az\'ar\altaffilmark{14},
    I.~Papp\altaffilmark{14},
    P.~S\'ari\altaffilmark{14},
}
\altaffiltext{1}{Department of Astrophysical Sciences, Princeton
  University, Princeton, NJ 08544; email: gbakos@astro.princeton.edu}

\altaffiltext{2}{Sloan Fellow}

\altaffiltext{3}{Harvard-Smithsonian Center for Astrophysics,
    Cambridge, MA}

\altaffiltext{4}{LCOGT, 6740 Cortona Drive, Santa Barbara, CA}

\altaffiltext{5}{Department of Physics, Broida Hall, UC Santa Barbara, CA}

\altaffiltext{6}{Niels Bohr Institute, University of Copenhagen, DK-2100, Denmark, and Centre for Star and Planet Formation, Natural History Museum of Denmark, DK-1350 Copenhagen}

\altaffiltext{7}{California Institute of Technology, Department of Astrophysics, MC 249-17, Pasadena, CA}

\altaffiltext{8}{Department of Astronomy, University of California,
    Berkeley, CA}

\altaffiltext{9}{Astronomy Department, Yale University, New Haven, CT}

\altaffiltext{10}{Konkoly Observatory, Budapest, Hungary}

\altaffiltext{11}{Department of Physics and Astrophysics, University of North Dakota, Grand Forks, ND}

\altaffiltext{12}{Department of Physics and Astronomy, Georgia State University, Atlanta, GA}

\altaffiltext{13}{Steward Observatory, University of Arizona, Tucson, AZ}

\altaffiltext{14}{Hungarian Astronomical Association, Budapest, 
    Hungary}

\altaffiltext{$\dagger$}{
    Based in part on observations obtained at the W.~M.~Keck
    Observatory, which is operated by the University of California and
    the California Institute of Technology. Keck time has been granted
    by NOAO (A201Hr, A289Hr, A284Hr), NASA (N049Hr, N018Hr, N167Hr,
    N029Hr, N108Hr, N154Hr), and the NOAO Gemini/Keck time-exchange
    program (G329Hr). Based in part on observations made with the
    Nordic Optical Telescope, operated on the island of La Palma
    jointly by Denmark, Finland, Iceland, Norway, and Sweden, in the
    Spanish Observatorio del Roque de los Muchachos of the Instituto
    de Astrofisica de Canarias. Based in part on observations obtained
    with facilities of the Las Cumbres Observatory Global
    Telescope. Observations reported here were obtained at the MMT
    Observatory, a joint facility of the Smithsonian Institution and
    the University of Arizona.
}


\begin{abstract}

\setcounter{footnote}{10}
We report the discovery of three new transiting
extrasolar planets orbiting moderately bright ($V = 11.1$ to $12.4$) F
stars. The planets have periods of $P = \hatcurLCPshort{41}$\,d to
$\hatcurLCPshort{40}$\,d, masses of $\hatcurPPmshort{39}$\,\mjup\ to
$\hatcurPPmshort{41}$\,\mjup, and radii of
$\hatcurPPrshort{39}$\,\rjup\ to $\hatcurPPrshort{40}$\,\rjup.  They
orbit stars with masses between $\hatcurISOmshort{39}$\,\msun\ and
$\hatcurISOmshort{40}$\,\msun. 
The three planets are members of an emerging population of highly
inflated Jupiters with $0.4\,\mjup <
M < 1.5\,\mjup$ and $R > 1.5\,\rjup$.
\setcounter{footnote}{0}
\end{abstract}



\section{Introduction}
\label{sec:introduction}

Transiting exoplanets (TEPs) are key objects for the study of planets
outside the Solar System.  The geometry of these planetary systems
enables measurements of several important physical parameters, such as
planetary masses and radii, or the sky-projected angle between the
orbital axis of a planet and the spin axis of its host star
\citep[e.g.][]{queloz:2000}. The vast majority of well-characterized
TEPs (i.e.~TEPs with measured masses and radii) have been discovered
by dedicated photometric surveys, including the Wide Angle Search for
Planets \citep[WASP;][]{pollacco:2006}, the Hungarian-made Automated
Telescope Network \citep[HATNet;][]{bakos:2004} and its southern
extension \citep[HATSouth;][]{bakos:2012}, Kepler
\citep{borucki:2010}, CoRoT \citep{barge:2008}, OGLE
\citep{udalski:2002}, TrES \citep{alonso:2004}, XO
\citep{mccullough:2006}, the Qatar Exoplanet Survey
\citep[QES;][]{alsubai:2011}, the Kilodegree Extremely Little
Telescope survey \citep[KELT;][]{siverd:2012}, and MEarth
\citep{charbonneau:2009}.

Significant among these are the ground-based, wide-field surveys using
small aperture telescopes, including WASP, HATNet, HATSouth, TrES, XO,
QES, and KELT. While these surveys are heavily biased towards
discovering large planets on short-period orbits compared to the
Kepler and CoRoT space-based surveys, the planets discovered by
ground-based surveys tend to orbit stars that are brighter than those
discovered by the space-based surveys, making such planets more
amenable to detailed characterization and follow-up studies (this is
true as well for the few, but valuable, transiting planets discovered
by RV searches, which are found around even brighter stars than those
discovered by photometric surveys). Additionally the extreme
environments in which these planets are discovered, while perhaps not
representative of most planetary systems, create a natural experiment
for testing theories of planet structure and formation. For example, a
number of gas-giant planets have been discovered with radii that are
substantially larger than theoretically expected
\citep[e.g.][]{mandushev:2007,colliercameron:2007,snellen:2009,hebb:2009,latham:2010,fortney:2011,anderson:2010,anderson:2011,enoch:2011,hartman:2011,
  smalley:2012}. These have been used to empirically determine the
factors affecting the radii of planets \citep[e.g.][]{enoch:2012}
which in turn informs theoretical work on the subject.

In this paper we present the discovery and characterization of three
new transiting planets around the relatively bright stars
\hatcurCCgsc{39}, \hatcurCCgsc{40}, and \hatcurCCgsc{41}, by the
HATNet survey. As members of the growing sample of highly inflated
planets, these objects will provide valuable leverage for understanding the
physics that determines the structure of planets.

In \refsecl{obs} we summarize the detection of the photometric transit
signal and the subsequent spectroscopic and photometric observations
of each star to confirm the planets. In \refsecl{analysis} we analyze
the data to rule out false positive scenarios, and to determine the
stellar and planetary parameters. Our findings are briefly discussed in
\refsecl{discussion}.

\section{Observations}
\label{sec:obs}

The observational procedure employed by HATNet to discover TEPs has
been described in detail in several previous discovery papers
\citep[e.g.][]{bakos:2010,latham:2009}. In the following subsections
we highlight specific details of the procedure that are relevant to
the discoveries of \hatcurb{39} through \hatcurb{41}.

\subsection{Photometric detection}
\label{sec:detection}

\begin{figure}[]
\ifthenelse{\boolean{emulateapj}}{
\epsscale{1.0}
}{
\epsscale{0.5}
}
\plotone{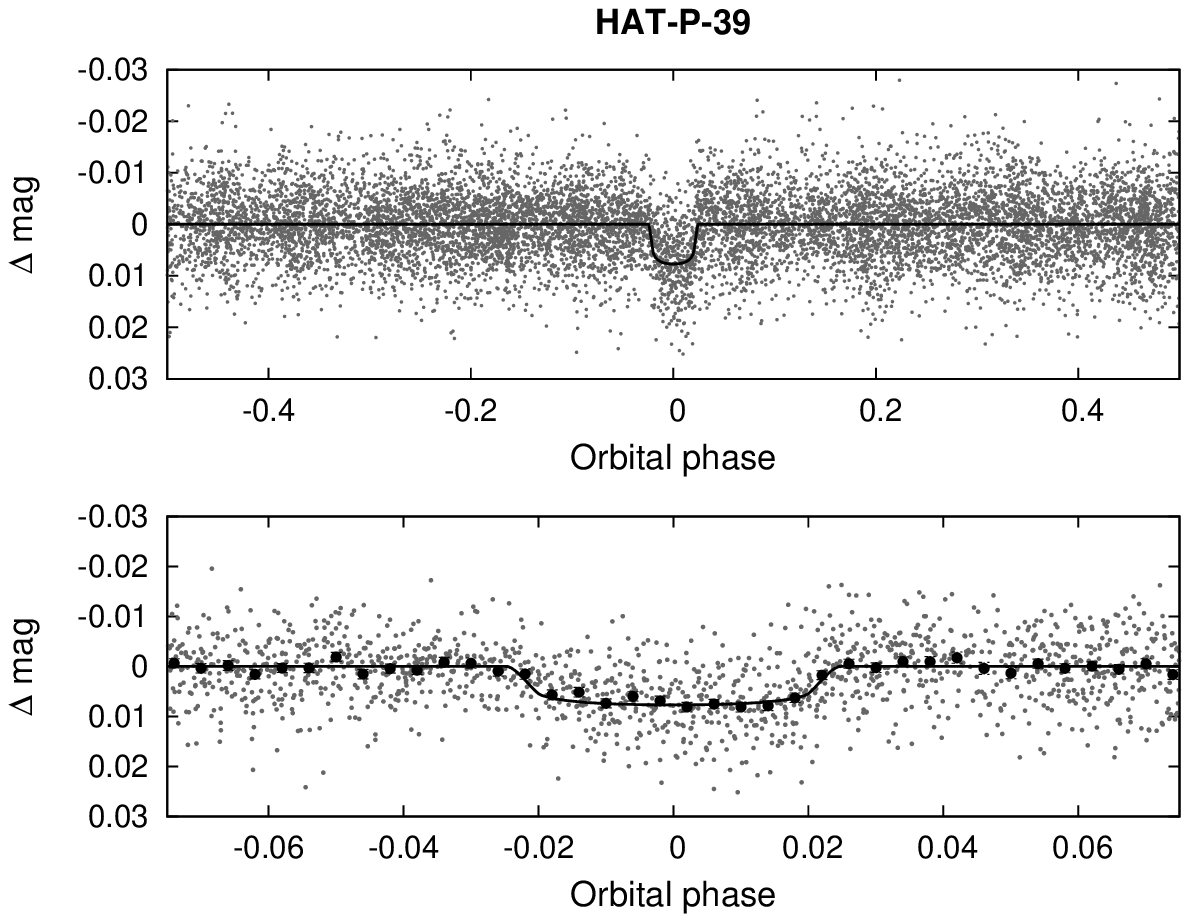}
\plotone{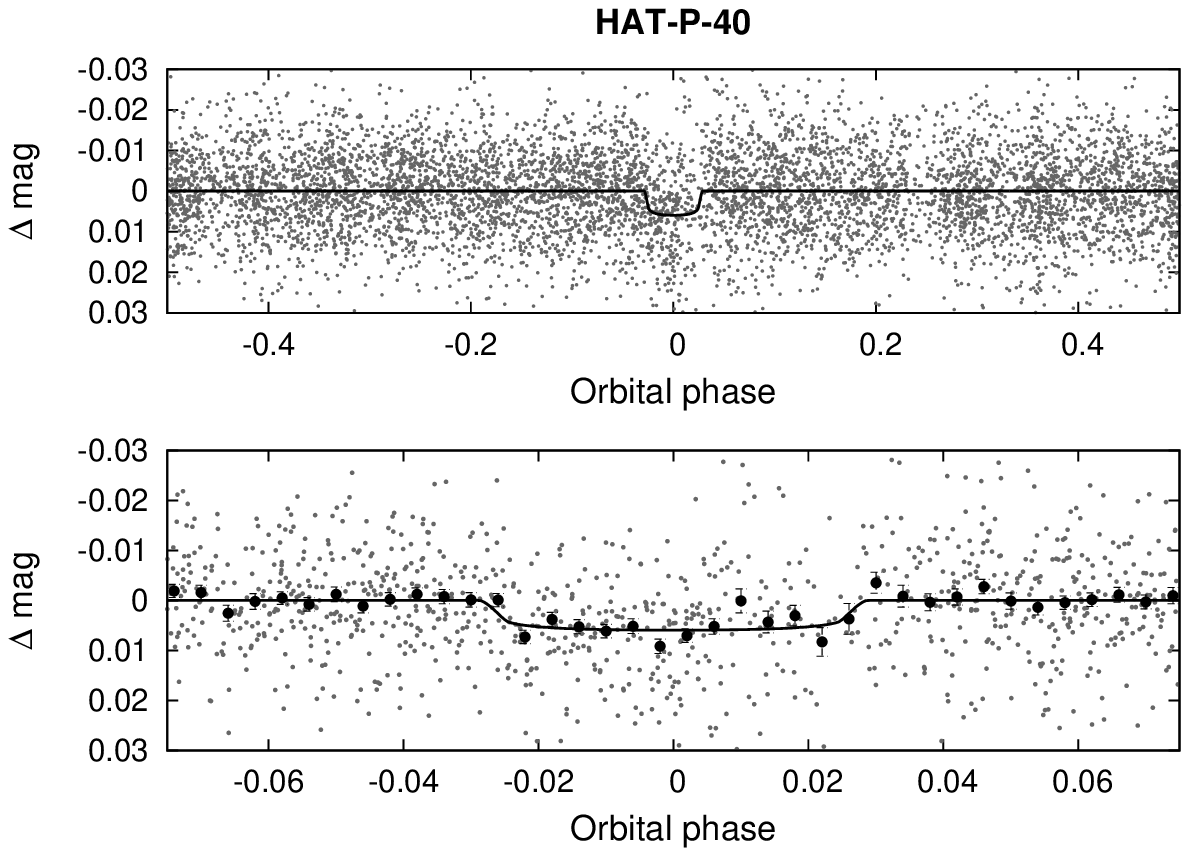}
\plotone{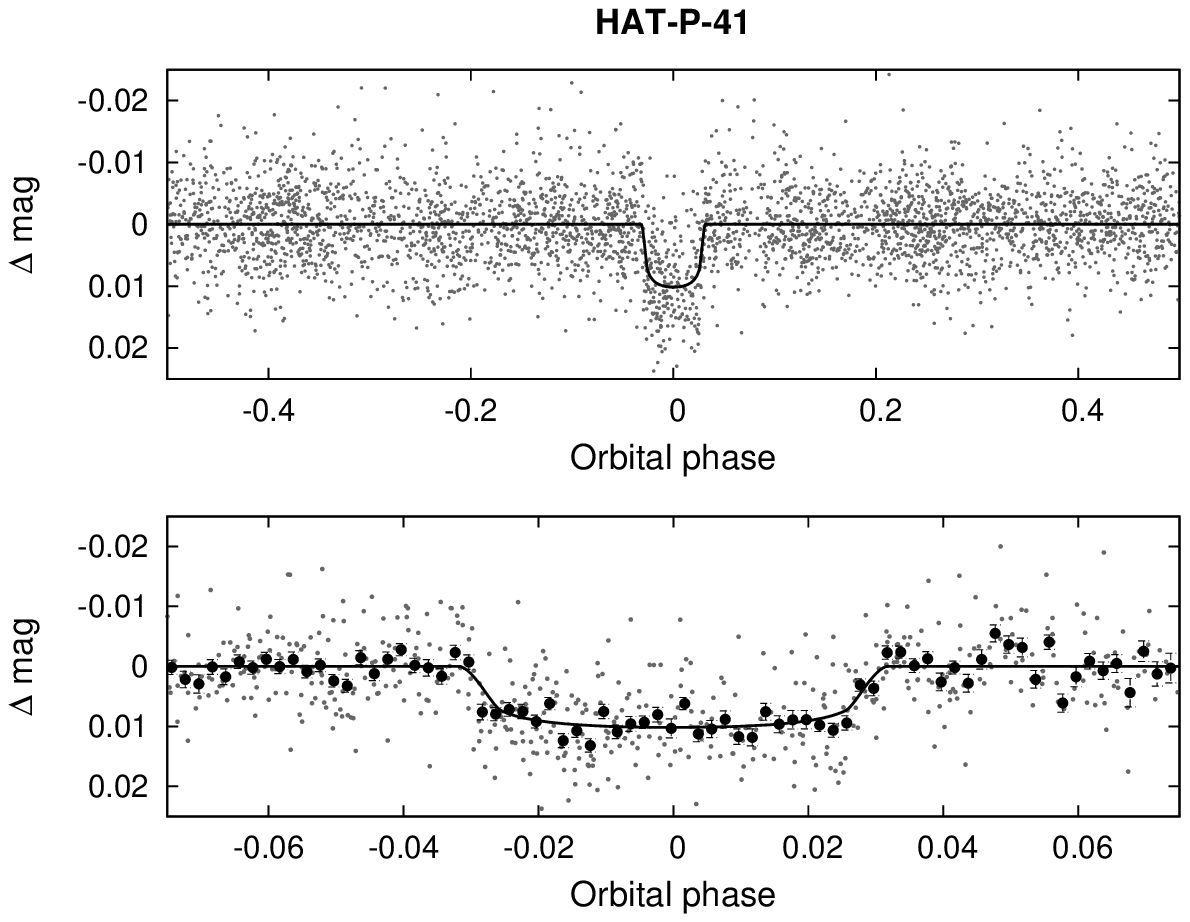}
\caption[]{
    HATNet \lcs{} of \hatcur{39} (top), \hatcur{40} (middle) and
    \hatcur{41} (bottom) phase folded with the transit period. In each
    case we show two panels: the top shows the unbinned light curve,
    while the bottom shows the region zoomed-in on the transit, with
    dark filled circles for the light curve binned in phase with a
    binsize of 0.002. The solid line shows the model fit to the light
    curve.
\label{fig:hatnet}}
\end{figure}

\reftabl{photobs} summarizes the photometric observations of each new
planetary system, including the discovery observations made with the
HATNet system. The HATNet images were processed and reduced to
trend-filtered light curves following the procedure described by
\cite{bakos:2010}.  The \lcs{} were searched for periodic box-shaped
signals using the Box Least-Squares \citep[BLS; see][]{kovacs:2002}
method. We detected significant signals in the \lcs\ of the stars
summarized below (see \reffigl{hatnet}):

\begin{itemize}
\item {\em \hatcur{39}} -- \hatcurCCgsc{39} (also known as
  \hatcurCCtwomass{39}; $\alpha = \hatcurCCra{39}$, $\delta =
  \hatcurCCdec{39}$; J2000; $V=\hatcurCCtassmv{39}$,
  \citealp{droege:2006}). A signal was detected for this star with an
  apparent depth of $\sim$\hatcurLCdip{39}\,mmag, and a period of
  $P=\hatcurLCPshort{39}$\,days.
\item {\em \hatcur{40}} -- \hatcurCCgsc{40} (also known as
  \hatcurCCtwomass{40}; $\alpha = \hatcurCCra{40}$, $\delta =
  \hatcurCCdec{40}$; J2000; $V=\hatcurCCtassmv{40}$,
  \citealp{droege:2006}). A signal was detected for this star with an
  apparent depth of $\sim$\hatcurLCdip{40}\,mmag, and a period of
  $P=\hatcurLCPshort{40}$\,days.
\item {\em \hatcur{41}} -- \hatcurCCgsc{41} (also known as
  \hatcurCCtwomass{41}; $\alpha = \hatcurCCra{41}$, $\delta =
  \hatcurCCdec{41}$; J2000; $V=\hatcurCCtassmv{41}$,
  \citealp{droege:2006}). A signal was detected for this star with an
  apparent depth of $\sim$\hatcurLCdip{41}\,mmag, and a period of
  $P=\hatcurLCPshort{41}$\,days.
\end{itemize}

%
%

\ifthenelse{\boolean{emulateapj}}{
    \begin{deluxetable*}{llrrr}
}{
    \begin{deluxetable}{llrrr}
}
\tablewidth{0pc}
\tabletypesize{\scriptsize}
\tablecaption{
    Summary of photometric observations
    \label{tab:photobs}
}
\tablehead{
    \colhead{~~~~~~~~Instrument/Field~~~~~~~~}  &
    \colhead{Date(s)} &
    \colhead{Number of Images} &
    \colhead{Mode Cadence (min)} &
    \colhead{Filter}
}
\startdata
\sidehead{\textbf{\hatcur{39}}}
~~~~HAT-7/G267 & 2007 Dec--2008 May & 800 & $5.5$ & \band{R} \\
~~~~HAT-8/G267 & 2007 Oct--2008 May & 1850 & $5.5$ & \band{R} \\
~~~~HAT-6/G315 & 2007 Dec--2008 May & 4700 & $5.5$ & \band{R} \\
~~~~HAT-9/G315 & 2007 Oct--2008 May & 2800 & $5.5$ & \band{R} \\
~~~~KeplerCam & 2009 Mar 30 & 257 & $0.89$ & Sloan~\band{i}
\\
~~~~KeplerCam & 2009 Apr 06 & 204 & $0.89$ & Sloan~\band{i}
\\
~~~~KeplerCam & 2009 Dec 17 & 373 & $0.89$ & Sloan~\band{i}
\\
~~~~KeplerCam & 2011 Feb 12 & 240 & $1.0$ & Sloan~\band{i}
\\
\sidehead{\textbf{\hatcur{40}}}
~~~~HAT-5/G159 & 2004 Sep--2006 Jan & 4403 & $5.5$ & \band{I} \\
~~~~HAT-8/G159 & 2005 Nov--2006 Jan & 1591 & $5.5$ & \band{I} \\
~~~~HAT-9/G159 & 2004 Dec--2005 Jan & 411 &  $5.5$ & \band{I} \\
~~~~HAT-11/G159 & 2004 Dec--2005 Jan & 331 & $5.5$ & \band{I} \\
~~~~KeplerCam         & 2010 Sep 16        & 371 & $1.40$ & Sloan~\band{i} \\
~~~~FTN               & 2010 Oct 17        & 241 & $0.50$ & Sloan~\band{i} \\
~~~~FTN               & 2011 Aug 16        & 285 & $0.85$ & Sloan~\band{i} \\
~~~~FTN               & 2011 Aug 25        & 325 & $0.94$ & Sloan~\band{i} \\
~~~~FTN               & 2011 Oct 04        & 177 & $0.94$ & Sloan~\band{i} \\
~~~~BOS               & 2011 Oct 13        & 381 & $1.17$ & Sloan~\band{i} \\
\sidehead{\textbf{\hatcur{41}}}
~~~~HAT-6/G388 & 2009 May--2009 Jul & 176 & $5.5$ & Sloan~\band{r} \\
~~~~HAT-8/G388 & 2009 May--2009 Sep & 3380 & $5.5$ & Sloan~\band{r} \\
~~~~KeplerCam         & 2010 May 27        & 226 & $0.98$ & Sloan~\band{i} \\
~~~~KeplerCam         & 2010 Jun 23        & 438 & $0.49$ & Sloan~\band{i} \\
~~~~KeplerCam         & 2011 May 31        & 307 & $0.73$ & Sloan~\band{i} \\
~~~~KeplerCam         & 2011 Jun 19        & 653 & $0.48$ & Sloan~\band{i} \\
~~~~BOS               & 2011 Jul 24        & 238 & $1.17$ & Sloan~\band{i} \\
\enddata
\ifthenelse{\boolean{emulateapj}}{
    \end{deluxetable*}
}{
    \end{deluxetable}
}

\subsection{Reconnaissance Spectroscopy}
\label{sec:recspec}

High-resolution, low-S/N ``reconnaissance'' spectra were obtained for
\hatcur{39}, \hatcur{40}, and \hatcur{41} using the
Harvard-Smithsonian Center for Astrophysics (CfA) Digital Speedometer
\citep[DS;][]{latham:1992} until it was retired in 2009, and
thereafter the Tillinghast Reflector Echelle Spectrograph
\citep[TRES;][]{furesz:2008}, both on the 1.5\,m Tillinghast Reflector
at the Fred Lawrence Whipple Observatory (FLWO) in AZ. The
reconnaissance spectroscopic observations and results for each system
are summarized in \reftabl{reconspecobs}. The DS observations were
reduced and analyzed following the procedure described by
\cite{torres:2002}, while the TRES observations were reduced and
analyzed following the procedure described by \cite{quinn:2012} and
\cite{buchhave:2010}.

Based on the observations summarized in \reftabl{reconspecobs} we find
that all 3 systems have RV root mean square (rms) residuals consistent
with no detectable RV variation within the precision of the
measurements. All spectra were single-lined, i.e., there is no
evidence that any of these targets consist of more than one star. Note
that while there is a close companion to \hatcur{41} (\refsecl{ao}),
it was resolved by the TRES guider and the light from the companion
did not go down the fiber. The gravities for all of the stars indicate
that none of the stars are giants, though \hatcur{40} may be slightly
evolved.

\ifthenelse{\boolean{emulateapj}}{
    \begin{deluxetable*}{llrrrrr}
}{
    \begin{deluxetable}{llrrrrr}
}
\tablewidth{0pc}
\tabletypesize{\scriptsize}
\tablecaption{
    Summary of reconnaissance spectroscopy observations
    \label{tab:reconspecobs}
}
\tablehead{
    \multicolumn{1}{c}{Instrument}          &
    \multicolumn{1}{c}{$HJD - 2400000$}             &
    \multicolumn{1}{c}{$\teffstar$}         &
    \multicolumn{1}{c}{$\loggstar$}         &
    \multicolumn{1}{c}{$\vsini$}            &
    \multicolumn{1}{c}{$\gamma_{\rm RV}$\tablenotemark{a}} \\
    &
    &
    \multicolumn{1}{c}{(K)}                 &
    \multicolumn{1}{c}{(cgs)}               &
    \multicolumn{1}{c}{(\kms)}              &
    \multicolumn{1}{c}{(\kms)}
}
\startdata
\sidehead{\textbf{\hatcur{39}}}
~~~~DS                & 4 obs $54807$--$54931$ & $6250$  & $4.0$ & $16$ & $28.54 \pm 0.58$ (rms) \\
~~~~TRES              & $54934.6546$ & $6500$  & $4.0$ & $16$ & $29.25$ \\
\sidehead{\textbf{\hatcur{40}}}
~~~~TRES              & $55084.8821$ & $6110 \pm 80$ & $4.21 \pm 0.13$ & $8.7 \pm 0.5$ & $-25.97$ \\
~~~~TRES              & $55131.6811$ & $5940 \pm 170$ & $4.04 \pm 0.26$ & $10.8 \pm 1.5$ & $-25.67$ \\
~~~~TRES              & $55138.6609$ & $5910 \pm 170$ & $3.81 \pm 0.27$ & $12.4 \pm 1.6$ & $-26.08$ \\
~~~~TRES              & $55162.6713$ & $6020 \pm 50$ & $3.93 \pm 0.10$ & $8.0 \pm 0.5$ & $-25.65$ \\
~~~~TRES              & $55168.5775$ & $6120 \pm 80$ & $4.15 \pm 0.14$ & $8.5 \pm 0.5$ & $-25.58$ \\
\sidehead{\textbf{\hatcur{41}}}
~~~~TRES              & $55319.9727$ & $6504 \pm 100$ & $4.3 \pm 0.16$ & $23.9 \pm 0.7$ & $32.32$ \\
~~~~TRES              & $55372.9209$ & $5807 \pm 223$ & $3.94 \pm 0.35$ & $32.1 \pm 2.5$ & $29.84$ \\
~~~~TRES              & $55373.9046$ & $6430 \pm 105$ & $4.28 \pm 0.17$ & $27.5 \pm 0.7$ & $30.44$ \\
\enddata 
\tablenotetext{a}{
    The mean heliocentric RV of the target on the IAU system, with a
    systematic uncertainty of approximately $0.1$\,\kms\ mostly
    limited by how well the velocities of the standard stars have been
    established. We give the mean and rms RV for the four DS
    observations of \hatcur{39}, the velocity and classification for
    each TRES observation of \hatcur{39} through \hatcur{41} is listed
    individually.
}
\ifthenelse{\boolean{emulateapj}}{
    \end{deluxetable*}
}{
    \end{deluxetable}
}

\subsection{High resolution, high S/N spectroscopy}
\label{sec:hispec}

\begin{figure} []
\ifthenelse{\boolean{emulateapj}}{
\epsscale{1.0}
}{
\epsscale{0.5}
}
\plotone{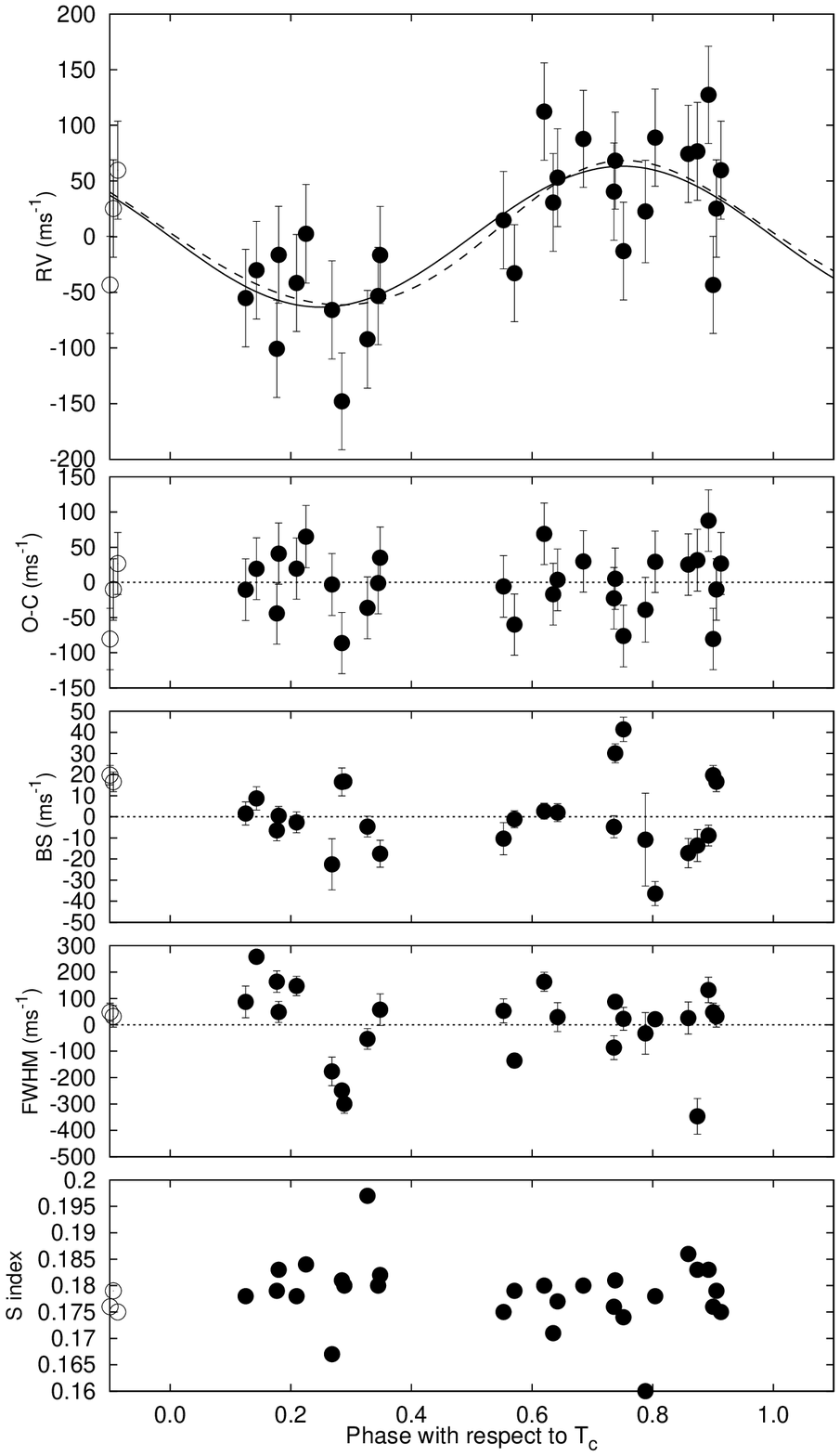}
\ifthenelse{\value{planetcounter}=1}{
\caption{
    {\em Top panel:} Keck/HIRES RV measurements for
    \hbox{\hatcur{39}{}} shown as a function of orbital phase, along
    with our best-fit circular model (solid line; see
    \reftabl{planetparam}), and our best-fit eccentric model (dashed
    line).  Zero phase corresponds to the time of mid-transit.  The
    center-of-mass velocity has been subtracted.
    {\em Second panel:} Velocity $O\!-\!C$ residuals from the best
        fit. The error bars include a component from astrophysical
        jitter (\hatcurRVjitter{39}\,\ms) added in quadrature to
        the formal errors (see \refsecl{globmod}).
    {\em Third panel:} Bisector spans (BS), with the mean value
    subtracted. The measurement from the template spectrum is
    included. The BS uncertainties are internal errors determined for
    each spectrum from the scatter of the individual BS values
    measured on separate orders of the spectrum; they do not include
    the unknown contribution from stellar jitter.
    {\em Fourth panel:} Full width at half maximum (FWHM) of the
    cross-correlation functions computed from the blue regions of the
    Keck/HIRES spectra, with the mean value subtracted.
    {\em Bottom panel:} Chromospheric activity index $S$.
    Note the different vertical scales of the panels. Observations
    shown twice are represented with open symbols.
}}{
\caption{
    Keck/HIRES observations of \hatcur{39}. The panels are as in
    \reffigl{rvbis42}.  The parameters used in the
    best-fit model are given in \reftabl{planetparam}.
}}
\label{fig:rvbis42}
\end{figure}

We proceeded with the follow-up of each candidate by obtaining
high-resolution, high-S/N spectra to characterize the RV variations,
and to refine the determination of the stellar parameters. The
observations were made with HIRES \citep{vogt:1994} on the Keck-I
telescope in Hawaii, and with FIES on the Nordic Optical Telescope on
the island of La Palma, Spain \citep{djupvik:2010}. We used the
high-resolution fiber (providing spectra with a resolution $R=67,\!000$)
for four of the FIES observations, and the medium-resolution fiber
($R=46,\!000$) for five of the FIES observations. The HIRES observations
were reduced to radial velocities in the barycentric frame following
the procedure described by \cite{butler:1996}, while the FIES
observations were reduced following \cite{buchhave:2010}. The RV
measurements and uncertainties are given in \reftabls{rvs42}{rvs44}
for \hatcur{39} through \hatcur{41}, respectively. The period-folded
data, along with our best fit described below in \refsecl{analysis}
are displayed in \reffigls{rvbis42}{rvbis44}.

\setcounter{planetcounter}{1}
%
\setcounter{planetcounter}{2}
%
\begin{figure} []
\ifthenelse{\boolean{emulateapj}}{
\epsscale{1.0}
}{
\epsscale{0.5}
}
\plotone{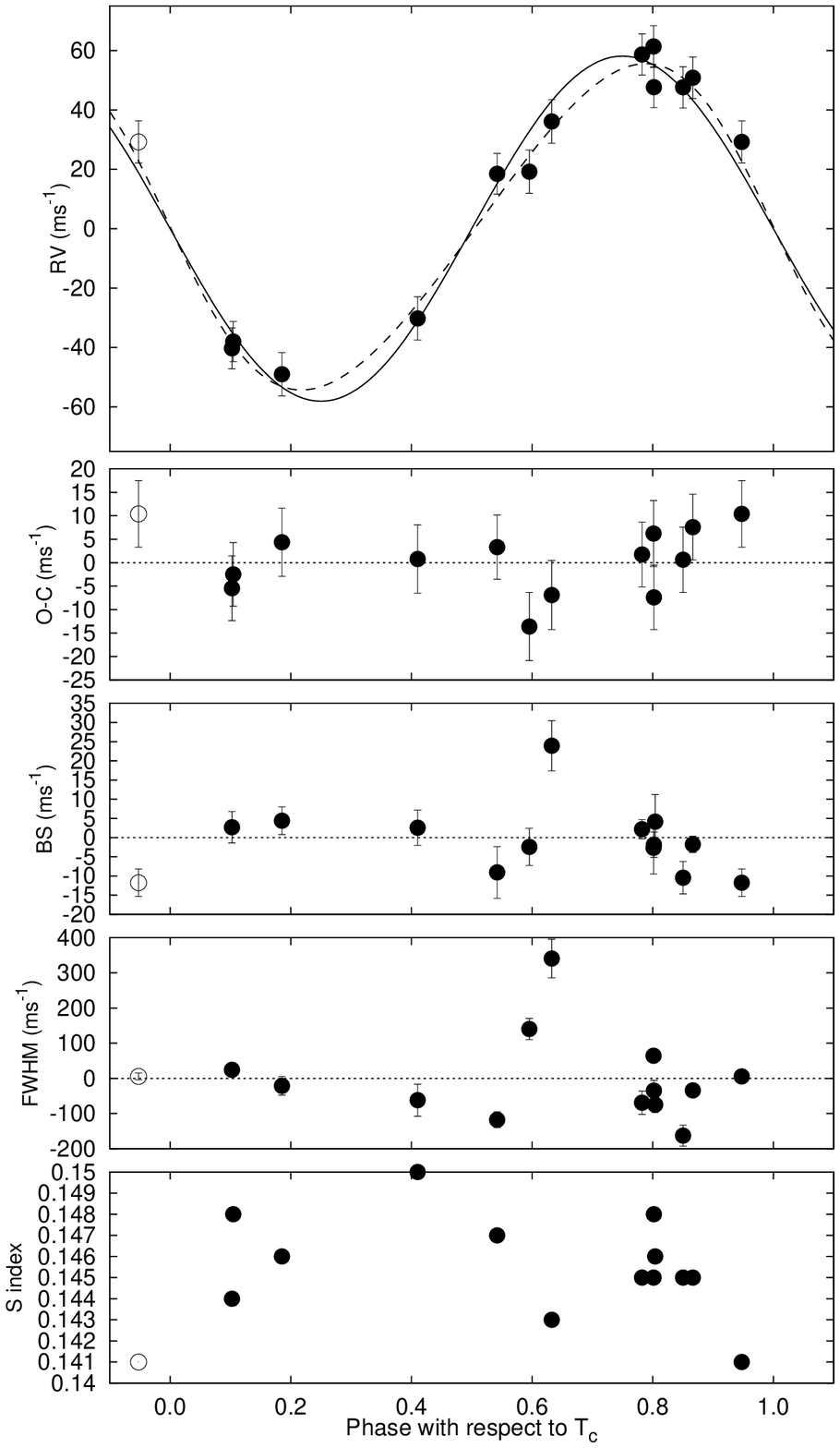}
\ifthenelse{\value{planetcounter}=1}{
\caption{
    {\em Top panel:} Keck/HIRES RV measurements for
    \hbox{\hatcur{40}{}} shown as a function of orbital phase, along
    with our best-fit circular model (solid line; see
    \reftabl{planetparam}), and our best-fit eccentric model (dashed
    line).  Zero phase corresponds to the time of mid-transit.  The
    center-of-mass velocity has been subtracted.
    {\em Second panel:} Velocity $O\!-\!C$ residuals from the best
        fit. The error bars include a component from astrophysical
        jitter (\hatcurRVjitter{40}\,\ms) added in quadrature to
        the formal errors (see \refsecl{globmod}).
    {\em Third panel:} Bisector spans (BS), with the mean value
        subtracted. The measurement from the template spectrum is
        included (see \refsecl{blend}).
    {\em Bottom panel:} Chromospheric activity index $S$
        measured from the Keck spectra.
    Note the different vertical scales of the panels. Observations
    shown twice are represented with open symbols.
}}{
\caption{
    Keck/HIRES observations of \hatcur{40}. The panels are as in
    \reffigl{rvbis42}.  The parameters used in the
    best-fit model are given in \reftabl{planetparam}.
}}
\label{fig:rvbis43}
\end{figure}
\setcounter{planetcounter}{3}
%
\begin{figure} []
\ifthenelse{\boolean{emulateapj}}{
\epsscale{1.0}
}{
\epsscale{0.5}
}
\plotone{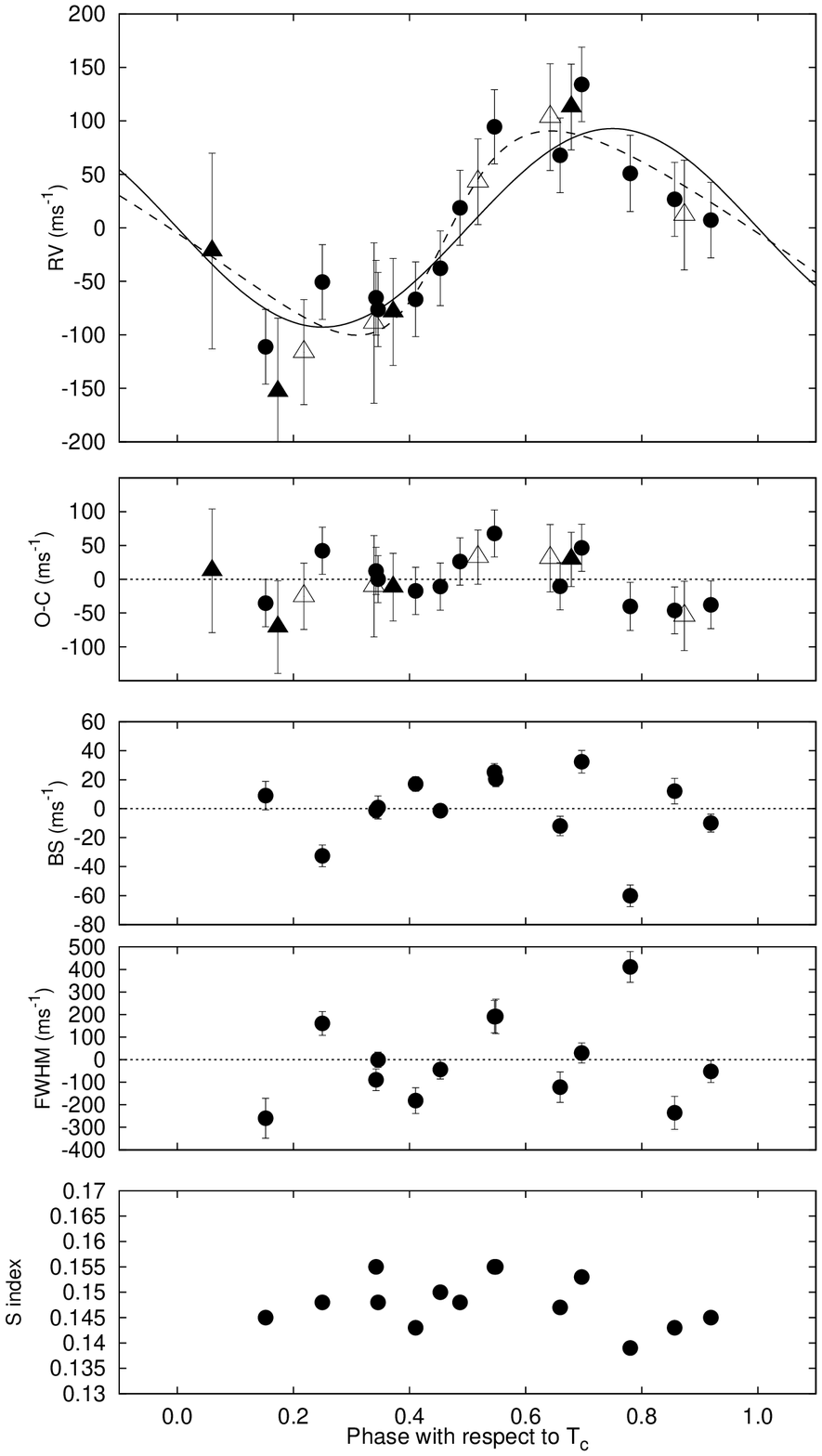}
\ifthenelse{\value{planetcounter}=1}{
\caption{
    {\em Top panel:} Keck/HIRES RV measurements for
        \hbox{\hatcur{41}{}} shown as a function of orbital
        phase, along with our best-fit model (see
        \reftabl{planetparam}).  Zero phase corresponds to the
        time of mid-transit.  The center-of-mass velocity has been
        subtracted.
    {\em Second panel:} Velocity $O\!-\!C$ residuals from the best
        fit. The error bars include a component from astrophysical
        jitter (\hatcurRVjitter{41}\,\ms) added in quadrature to
        the formal errors (see \refsecl{globmod}).
    {\em Third panel:} Bisector spans (BS), with the mean value
        subtracted. The measurement from the template spectrum is
        included (see \refsecl{blend}).
    {\em Bottom panel:} Chromospheric activity index $S$
        measured from the Keck spectra.
    Note the different vertical scales of the panels. Observations
    shown twice are represented with open symbols.
}}{
\caption{
    Keck/HIRES and FIES/NOT observations of \hatcur{41}. The panels
    are as in \reffigl{rvbis42}.  Here we use filled circles to show
    Keck/HIRES observations, filled triangles to show FIES/NOT
    observations with the high resolution fiber, and open triangles to
    show FIES/NOT observations with the low resolution fiber. BS and S
    index measurements are only available for the Keck/HIRES
    observations. The parameters used in the best-fit model are given
    in \reftabl{planetparam}.
}}
\label{fig:rvbis44}
\end{figure}

In each figure we show also the spectral line bisector spans (BSs)
computed from the Keck/HIRES spectra following \cite{torres:2007}, the
full width at half maximum (FWHM) of the Keck/HIRES spectral lines
(computed from the cross-correlation function in a similar manner to
the BSs), and the $S$ activity index calculated following
\citet{isaacson:2010}.

\ifthenelse{\boolean{emulateapj}}{
    \begin{deluxetable*}{lrrrrrr}
}{
    \begin{deluxetable}{lrrrrrr}
}
\tablewidth{0pc}
\tablecaption{
    Relative radial velocities, bisector spans, and activity index
    measurements of \hatcur{39}.
    \label{tab:rvs42}
}
\tablehead{
    \colhead{BJD\tablenotemark{a}} &
    \colhead{RV\tablenotemark{b}} &
    \colhead{\ensuremath{\sigma_{\rm RV}}\tablenotemark{c}} &
    \colhead{BS} &
    \colhead{\ensuremath{\sigma_{\rm BS}}} &
    \colhead{S\tablenotemark{d}} &
    \colhead{Phase}\\
    \colhead{\hbox{(2,454,000$+$)}} &
    \colhead{(\ms)} &
    \colhead{(\ms)} &
    \colhead{(\ms)} &
    \colhead{(\ms)} &
    \colhead{} &
    \colhead{}
}
\startdata
$ 954.81362 $ & $   -51.56 $ & $     7.81 $ & \nodata      & \nodata      & $    0.180 $ & $   0.345 $\\
$ 955.84224 $ & $    32.48 $ & $     8.82 $ & \nodata      & \nodata      & $    0.171 $ & $   0.635 $\\
$ 956.82792 $ & $    61.50 $ & $     9.51 $ & \nodata      & \nodata      & $    0.175 $ & $   0.913 $\\
$ 1107.13759 $ & $   -90.39 $ & $     8.88 $ & $    -4.63 $ & $     4.91 $ & $    0.197 $ & $   0.327 $\\
$ 1109.07573 $ & $    78.38 $ & $     9.05 $ & $   -13.59 $ & $     7.57 $ & $    0.183 $ & $   0.874 $\\
$ 1112.12944 $ & $    42.24 $ & $     7.88 $ & $    -4.74 $ & $     5.19 $ & $    0.176 $ & $   0.736 $\\
$ 1135.07184 $ & $   -39.87 $ & $     6.92 $ & $    -2.64 $ & $     4.96 $ & $    0.178 $ & $   0.210 $\\
$ 1172.95852 $ & $   -41.57 $ & $     7.30 $ & $    19.77 $ & $     4.58 $ & $    0.176 $ & $   0.900 $\\
$ 1187.99450 $ & $   -28.34 $ & $     8.69 $ & $     8.74 $ & $     5.51 $ &      \nodata & $   0.143 $\\
$ 1190.10225 $ & $    70.06 $ & $     7.12 $ & $    30.09 $ & $     4.51 $ & $    0.181 $ & $   0.738 $\\
$ 1192.03996 $ & $  -146.19 $ & $     6.26 $ & $    16.58 $ & $     6.61 $ & $    0.181 $ & $   0.285 $\\
$ 1192.05453 $ & \nodata      & \nodata      & $    16.75 $ & $     3.03 $ & $    0.180 $ & $   0.289 $\\
$ 1193.05376 $ & $   -31.04 $ & $     6.96 $ & $    -1.15 $ & $     4.07 $ & $    0.179 $ & $   0.571 $\\
$ 1193.88084 $ & $    90.74 $ & $     7.48 $ & $   -36.37 $ & $     5.72 $ & $    0.178 $ & $   0.804 $\\
$ 1197.00264 $ & $    89.62 $ & $     7.37 $ & \nodata      & \nodata      & $    0.180 $ & $   0.685 $\\
$ 1198.91700 $ & $     4.34 $ & $    10.58 $ & \nodata      & \nodata      & $    0.184 $ & $   0.225 $\\
$ 1250.89571 $ & $   129.15 $ & $     7.46 $ & $    -8.86 $ & $     4.93 $ & $    0.183 $ & $   0.892 $\\
$ 1251.90337 $ & $   -98.98 $ & $     7.98 $ & $    -6.43 $ & $     4.96 $ & $    0.179 $ & $   0.177 $\\
$ 1289.75983 $ & $    76.11 $ & $     7.50 $ & $   -17.19 $ & $     6.90 $ & $    0.186 $ & $   0.859 $\\
$ 1312.75610 $ & $   -14.73 $ & $     7.40 $ & $   -17.48 $ & $     6.40 $ & $    0.182 $ & $   0.348 $\\
$ 1313.79897 $ & $    54.76 $ & $     8.78 $ & $     2.00 $ & $     4.22 $ & $    0.177 $ & $   0.642 $\\
$ 1466.10653 $ & $   114.14 $ & $     7.74 $ & $     2.68 $ & $     3.75 $ & $    0.180 $ & $   0.620 $\\
$ 1468.09041 $ & $   -14.54 $ & $     6.61 $ & $     0.53 $ & $     4.41 $ & $    0.183 $ & $   0.180 $\\
$ 1470.11601 $ & $   -11.17 $ & $     8.91 $ & $    41.44 $ & $     5.79 $ & $    0.174 $ & $   0.751 $\\
$ 1545.08385 $ & $    27.02 $ & $     8.10 $ & $    16.59 $ & $     4.67 $ & $    0.179 $ & $   0.906 $\\
$ 1545.86194 $ & $   -53.41 $ & $     8.31 $ & $     1.59 $ & $     5.47 $ & $    0.178 $ & $   0.125 $\\
$ 1611.99999 $ & $    24.36 $ & $    16.44 $ & $   -10.80 $ & $    22.01 $ & $    0.160 $ & $   0.788 $\\
$ 1698.75509 $ & $   -64.08 $ & $     9.64 $ & $   -22.52 $ & $    12.11 $ & $    0.167 $ & $   0.268 $\\
$ 1699.76307 $ & $    16.59 $ & $     8.07 $ & $   -10.35 $ & $     7.63 $ & $    0.175 $ & $   0.553 $\\
\enddata
\tablenotetext{a}{
    Barycentric Julian Date calculated directly from UTC, {\em
      without} correction for leap seconds.
}
\tablenotetext{b}{
    The zero-point of these velocities is arbitrary. An overall offset
    $\gamma_{\rm rel}$ fitted to these velocities in \refsecl{globmod}
    has {\em not} been subtracted.
}
\tablenotetext{c}{
    Internal errors excluding the component of astrophysical jitter
    considered in \refsecl{globmod}.
}
\tablenotetext{d}{
    Chromospheric activity index.
}
\ifthenelse{\boolean{rvtablelong}}{
    \tablecomments{
        Note that for the iodine-free template exposures we do not
        measure the RV but do measure the BS and S index.  Such
        template exposures can be distinguished by the missing RV
        value.
    }
}{
    \tablecomments{
        Note that for the iodine-free template exposures we do not
        measure the RV but do measure the BS and S index.  Such
        template exposures can be distinguished by the missing RV
        value.  We exclude BS measurements for a handful of
        measurements which were heavily affected by contamination from
        scattered moonlight. This table is presented in its entirety
        in the electronic edition of the Astrophysical Journal.  A
        portion is shown here for guidance regarding its form and
        content.
    }
} 
\ifthenelse{\boolean{emulateapj}}{
    \end{deluxetable*}
}{
    \end{deluxetable}
}
%
\ifthenelse{\boolean{emulateapj}}{
    \begin{deluxetable*}{lrrrrrr}
}{
    \begin{deluxetable}{lrrrrrr}
}
\tablewidth{0pc}
\tablecaption{
    Relative radial velocities, bisector spans, and activity index
    measurements of \hatcur{40}.
    \label{tab:rvs43}
}
\tablehead{
    \colhead{BJD\tablenotemark{a}} &
    \colhead{RV\tablenotemark{b}} &
    \colhead{\ensuremath{\sigma_{\rm RV}}\tablenotemark{c}} &
    \colhead{BS} &
    \colhead{\ensuremath{\sigma_{\rm BS}}} &
    \colhead{S\tablenotemark{d}} &
    \colhead{Phase}\\
    \colhead{\hbox{(2,454,000$+$)}} &
    \colhead{(\ms)} &
    \colhead{(\ms)} &
    \colhead{(\ms)} &
    \colhead{(\ms)} &
    \colhead{} &
    \colhead{}
}
\startdata
$ 1191.81591 $ & $    18.28 $ & $     3.81 $ & $    -2.42 $ & $     4.85 $ &     \nodata      & $   0.595 $\\
$ 1192.73706 $ & $    46.75 $ & $     2.96 $ & $    -1.85 $ & $     3.29 $ & $    0.148 $ & $   0.802 $\\
$ 1192.74599 $ & \nodata      & \nodata      & $     4.16 $ & $     7.06 $ & $    0.146 $ & $   0.804 $\\
$ 1464.91615 $ & $    49.92 $ & $     3.24 $ & $    -1.75 $ & $     2.12 $ & $    0.145 $ & $   0.867 $\\
$ 1465.96736 $ & $   -41.21 $ & $     3.02 $ & $     2.71 $ & $     4.09 $ & $    0.144 $ & $   0.102 $\\
$ 1467.92695 $ & $    17.59 $ & $     2.93 $ & $    -9.07 $ & $     6.74 $ & $    0.147 $ & $   0.542 $\\
$ 1468.99826 $ & $    57.77 $ & $     3.09 $ & $     2.18 $ & $     2.51 $ & $    0.145 $ & $   0.782 $\\
$ 1469.73453 $ & $    28.31 $ & $     3.45 $ & $   -11.76 $ & $     3.57 $ & $    0.141 $ & $   0.948 $\\
$ 1486.91164 $ & $    60.46 $ & $     3.33 $ & $    -2.67 $ & $     6.82 $ & $    0.145 $ & $   0.801 $\\
$ 1521.81745 $ & $    35.22 $ & $     3.96 $ & $    23.93 $ & $     6.53 $ & $    0.143 $ & $   0.633 $\\
$ 1698.11353 $ & $   -49.94 $ & $     3.80 $ & $     4.39 $ & $     3.60 $ & $    0.146 $ & $   0.185 $\\
$ 1699.11678 $ & $   -31.14 $ & $     3.84 $ & $     2.58 $ & $     4.61 $ & $    0.150 $ & $   0.410 $\\
$ 1701.07726 $ & $    46.71 $ & $     3.15 $ & $   -10.44 $ & $     4.24 $ & $    0.145 $ & $   0.850 $\\
$ 1853.75742 $ & $   -38.93 $ & $     2.73 $ & \nodata      & \nodata      & $    0.148 $ & $   0.105 $\\
\enddata
\tablenotetext{a}{
    Barycentric Julian Date calculated directly from UTC, {\em
      without} correction for leap seconds.
}
\tablenotetext{b}{
    The zero-point of these velocities is arbitrary. An overall offset
    $\gamma_{\rm rel}$ fitted to these velocities in \refsecl{globmod}
    has {\em not} been subtracted.
}
\tablenotetext{c}{
    Internal errors excluding the component of astrophysical jitter
    considered in \refsecl{globmod}.
}
\tablenotetext{d}{
    Chromospheric activity index.
}
\ifthenelse{\boolean{rvtablelong}}{
    \tablecomments{
        Note that for the iodine-free template exposures we do not
        measure the RV but do measure the BS and S index.  Such
        template exposures can be distinguished by the missing RV
        value.
    }
}{
    \tablecomments{
        Note that for the iodine-free template exposures we do not
        measure the RV but do measure the BS and S index.  Such
        template exposures can be distinguished by the missing RV
        value.  This table is presented in its entirety in the
        electronic edition of the Astrophysical Journal.  A portion is
        shown here for guidance regarding its form and content.
    }
} 
\ifthenelse{\boolean{emulateapj}}{
    \end{deluxetable*}
}{
    \end{deluxetable}
}
%
\ifthenelse{\boolean{emulateapj}}{
    \begin{deluxetable*}{lrrrrrrr}
}{
    \begin{deluxetable}{lrrrrrrr}
}
\tablewidth{0pc}
\tablecaption{
    Relative radial velocities, bisector spans, and activity index
    measurements of \hatcur{41}.
    \label{tab:rvs44}
}
\tablehead{
    \colhead{BJD\tablenotemark{a}} &
    \colhead{RV\tablenotemark{b}} &
    \colhead{\ensuremath{\sigma_{\rm RV}}\tablenotemark{c}} &
    \colhead{BS} &
    \colhead{\ensuremath{\sigma_{\rm BS}}} &
    \colhead{S\tablenotemark{d}} &
    \colhead{Phase} &
    \colhead{Instrument\tablenotemark{e}}\\
    \colhead{\hbox{(2,454,000$+$)}} &
    \colhead{(\ms)} &
    \colhead{(\ms)} &
    \colhead{(\ms)} &
    \colhead{(\ms)} &
    \colhead{} &
    \colhead{} &
    \colhead{}
}
\startdata
$ 1375.97050 $ & $    92.63 $ & $     9.38 $ & $    25.18 $ & $     6.00 $ & $    0.155 $ & $   0.546 $ & Keck \\
$ 1375.97664 $ & \nodata      & \nodata      & $    20.47 $ & $     5.29 $ & $    0.155 $ & $   0.549 $ & Keck \\
$ 1378.12394 $ & $   -78.22 $ & $     9.40 $ & $     0.84 $ & $     7.92 $ & $    0.148 $ & $   0.346 $ & Keck \\
$ 1378.58766 $ & $    52.31 $ & $    40.10 $ & \nodata      & \nodata      & \nodata      & $   0.518 $ & FIESm \\
$ 1379.06917 $ & $   132.24 $ & $     9.90 $ & $    32.40 $ & $     7.80 $ & $    0.153 $ & $   0.697 $ & Keck \\
$ 1379.54578 $ & $    21.11 $ & $    51.30 $ & \nodata      & \nodata      & \nodata      & $   0.874 $ & FIESm \\
$ 1380.47420 $ & $  -106.99 $ & $    49.00 $ & \nodata      & \nodata      & \nodata      & $   0.218 $ & FIESm \\
$ 1381.10708 $ & $   -39.66 $ & $    10.23 $ & $    -1.39 $ & $     4.57 $ & $    0.150 $ & $   0.453 $ & Keck \\
$ 1381.61687 $ & $   112.71 $ & $    49.90 $ & \nodata      & \nodata      & \nodata      & $   0.642 $ & FIESm \\
$ 1383.49357 $ & $   -79.79 $ & $    75.00 $ & \nodata      & \nodata      & \nodata      & $   0.339 $ & FIESm \\
$ 1400.84669 $ & $    49.04 $ & $    12.32 $ & $   -60.12 $ & $     7.44 $ & $    0.139 $ & $   0.780 $ & Keck \\
$ 1404.80595 $ & $   -52.53 $ & $    10.18 $ & $   -32.58 $ & $     7.48 $ & $    0.148 $ & $   0.250 $ & Keck \\
$ 1427.51377 $ & $   112.18 $ & $    40.10 $ & \nodata      & \nodata      & \nodata      & $   0.679 $ & FIESh \\
$ 1428.54090 $ & $   -22.42 $ & $    91.50 $ & \nodata      & \nodata      & \nodata      & $   0.060 $ & FIESh \\
$ 1429.38087 $ & $   -79.42 $ & $    50.00 $ & \nodata      & \nodata      & \nodata      & $   0.372 $ & FIESh \\
$ 1431.54013 $ & $  -153.82 $ & $    68.60 $ & \nodata      & \nodata      & \nodata      & $   0.173 $ & FIESh \\
$ 1465.87835 $ & $     5.42 $ & $    11.42 $ & $    -9.93 $ & $     6.17 $ & $    0.145 $ & $   0.919 $ & Keck \\
$ 1467.87264 $ & $    65.94 $ & $    10.34 $ & $   -11.95 $ & $     6.68 $ & $    0.147 $ & $   0.659 $ & Keck \\
$ 1469.89585 $ & $   -68.65 $ & $    10.18 $ & $    17.06 $ & $     4.94 $ & $    0.143 $ & $   0.410 $ & Keck \\
$ 1490.75265 $ & $  -113.04 $ & $    10.38 $ & $     9.02 $ & $     9.83 $ & $    0.145 $ & $   0.152 $ & Keck \\
$ 1500.73246 $ & $    24.76 $ & $     9.14 $ & $    12.08 $ & $     8.78 $ & $    0.143 $ & $   0.857 $ & Keck \\
$ 1704.09441 $ & $   -67.20 $ & $    10.27 $ & $    -1.09 $ & $     5.26 $ & $    0.155 $ & $   0.342 $ & Keck \\
$ 1814.94017 $ & $    16.90 $ & $    10.72 $ & \nodata      & \nodata      & $    0.148 $ & $   0.487 $ & Keck \\
\enddata
\tablenotetext{a}{
    Barycentric Julian Date calculated directly from UTC, {\em
      without} correction for leap seconds.
}
\tablenotetext{b}{
    The zero-point of these velocities is arbitrary. An overall offset
    $\gamma_{\rm rel}$ fitted to these velocities in \refsecl{globmod}
    has {\em not} been subtracted.
}
\tablenotetext{c}{
    Internal errors excluding the component of astrophysical jitter
    considered in \refsecl{globmod}.
}
\tablenotetext{d}{
    Chromospheric activity index.
}
\tablenotetext{e}{
    We indicate separately observations obtained with FIES using the
    medium-resolution fiber, and observations obtained with FIES using
    the high-resolution fiber.
}
\ifthenelse{\boolean{rvtablelong}}{
    \tablecomments{
        Note that for the iodine-free template exposures we do not
        measure the RV but do measure the BS and S index.  Such
        template exposures can be distinguished by the missing RV
        value.
    }
}{
    \tablecomments{
        Note that for the iodine-free template exposures we do not
        measure the RV but do measure the BS and S index.  Such
        template exposures can be distinguished by the missing RV
        value.  This table is presented in its entirety in the
        electronic edition of the Astrophysical Journal.  A portion is
        shown here for guidance regarding its form and content.
    }
} 
\ifthenelse{\boolean{emulateapj}}{
    \end{deluxetable*}
}{
    \end{deluxetable}
}

\subsection{Photometric follow-up observations}
\label{sec:phot}

%
\setcounter{planetcounter}{1}
%
\begin{figure}[]
\ifthenelse{\boolean{emulateapj}}{
\epsscale{1.0}
}{
\epsscale{0.5}
}
\plotone{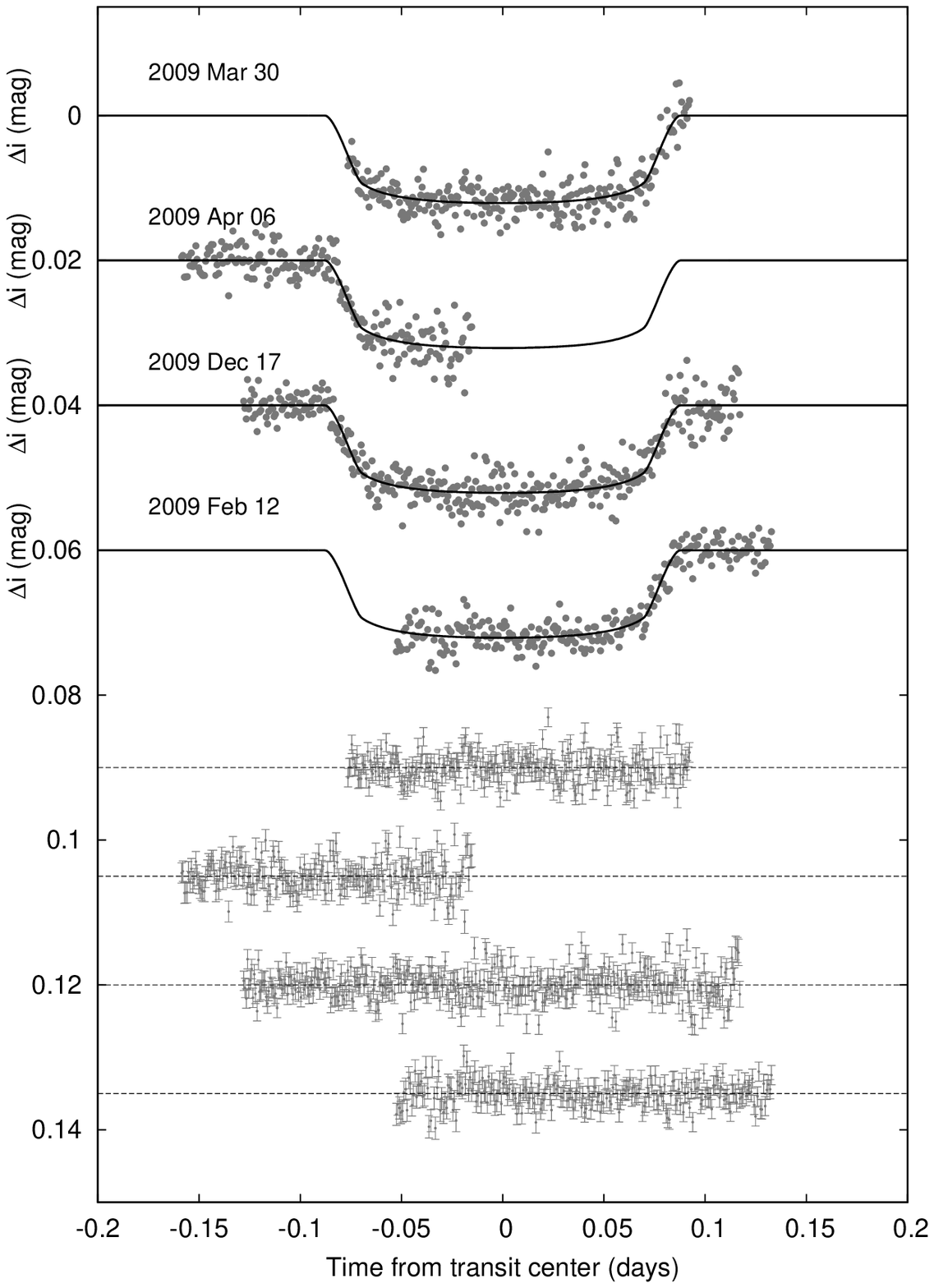}
\ifthenelse{\value{planetcounter}=1}{
\caption{
    Unbinned transit \lcs{} for \hatcur{39}, acquired with
    KeplerCam at the \flwof{} telescope.  The light curves have been
    EPD- and TFA-processed, as described in \refsec{globmod}.
    The dates of the events are indicated.  Curves after the first are
    displaced vertically for clarity.  Our best fit from the global
    modeling described in \refsecl{globmod} is shown by the solid
    lines.  Residuals from the fits are displayed at the bottom, in the
    same order as the top curves.  The error bars represent the photon
    and background shot noise, plus the readout noise.
}}{
\caption{
    Similar to \reffigl{lc42}; here we show the follow-up
    \lcs{} for \hatcur{39}.
}}
\label{fig:lc42}
\end{figure}
\setcounter{planetcounter}{2}
%
\begin{figure}[]
\ifthenelse{\boolean{emulateapj}}{
\epsscale{1.0}
}{
\epsscale{0.5}
}
\plotone{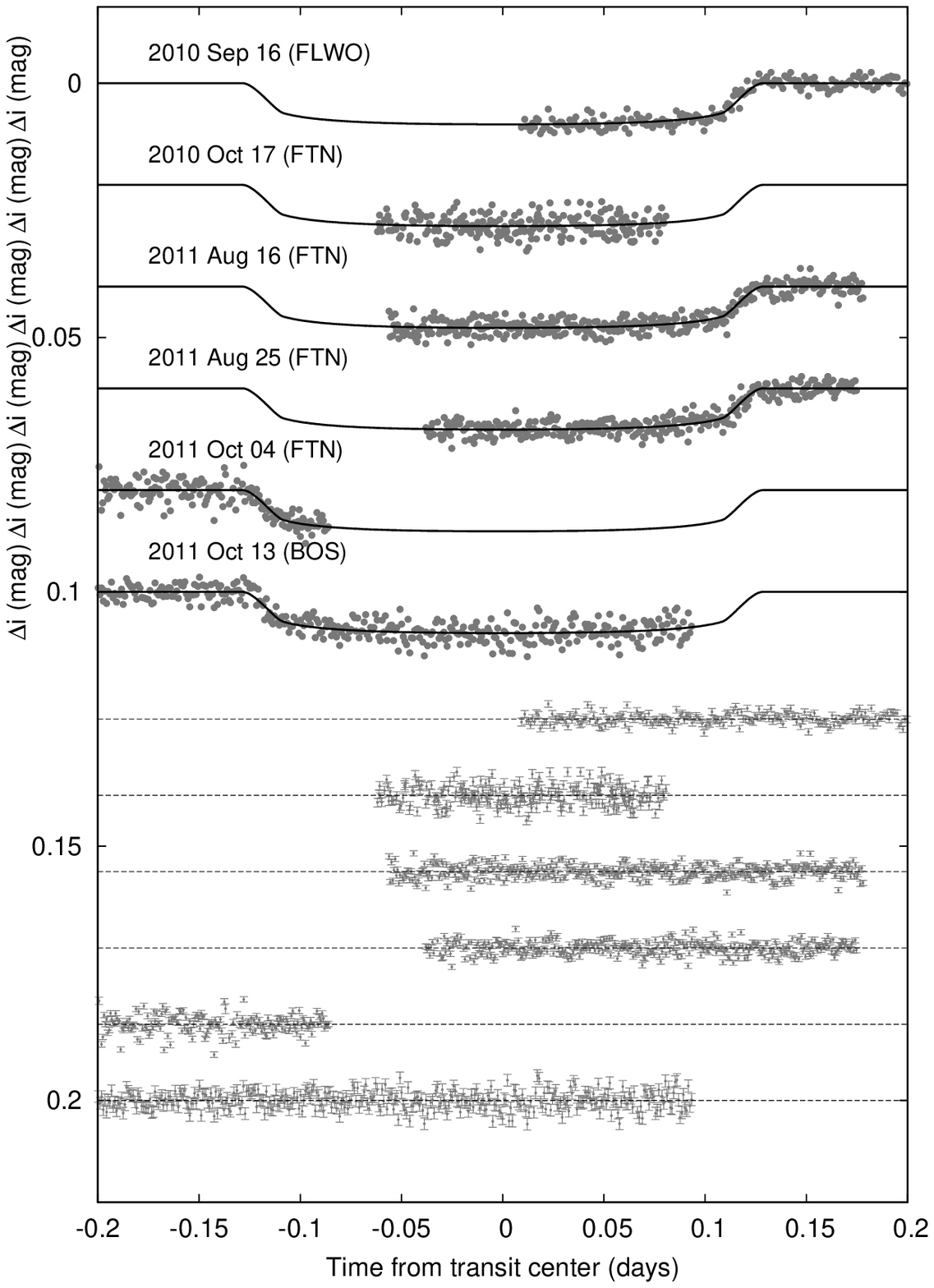}
\ifthenelse{\value{planetcounter}=1}{
\caption{
    Unbinned transit \lcs{} for \hatcur{40}, acquired with
    KeplerCam at the \flwof{} telescope.  The light curves have been
    EPD and TFA processed, as described in \refsec{globmod}.
    The dates of the events are indicated.  Curves after the first are
    displaced vertically for clarity.  Our best fit from the global
    modeling described in \refsecl{globmod} is shown by the solid
    lines.  Residuals from the fits are displayed at the bottom, in the
    same order as the top curves.  The error bars represent the photon
    and background shot noise, plus the readout noise.
}}{
\caption{
    Similar to \reffigl{lc42}; here we show the follow-up \lcs{} for
    \hatcur{40}. The facility used for each each light curve is
    indicated next to the date of the event.
}}
\label{fig:lc43}
\end{figure}
\setcounter{planetcounter}{3}
%
\begin{figure}[]
\ifthenelse{\boolean{emulateapj}}{
\epsscale{1.0}
}{
\epsscale{0.5}
}
\plotone{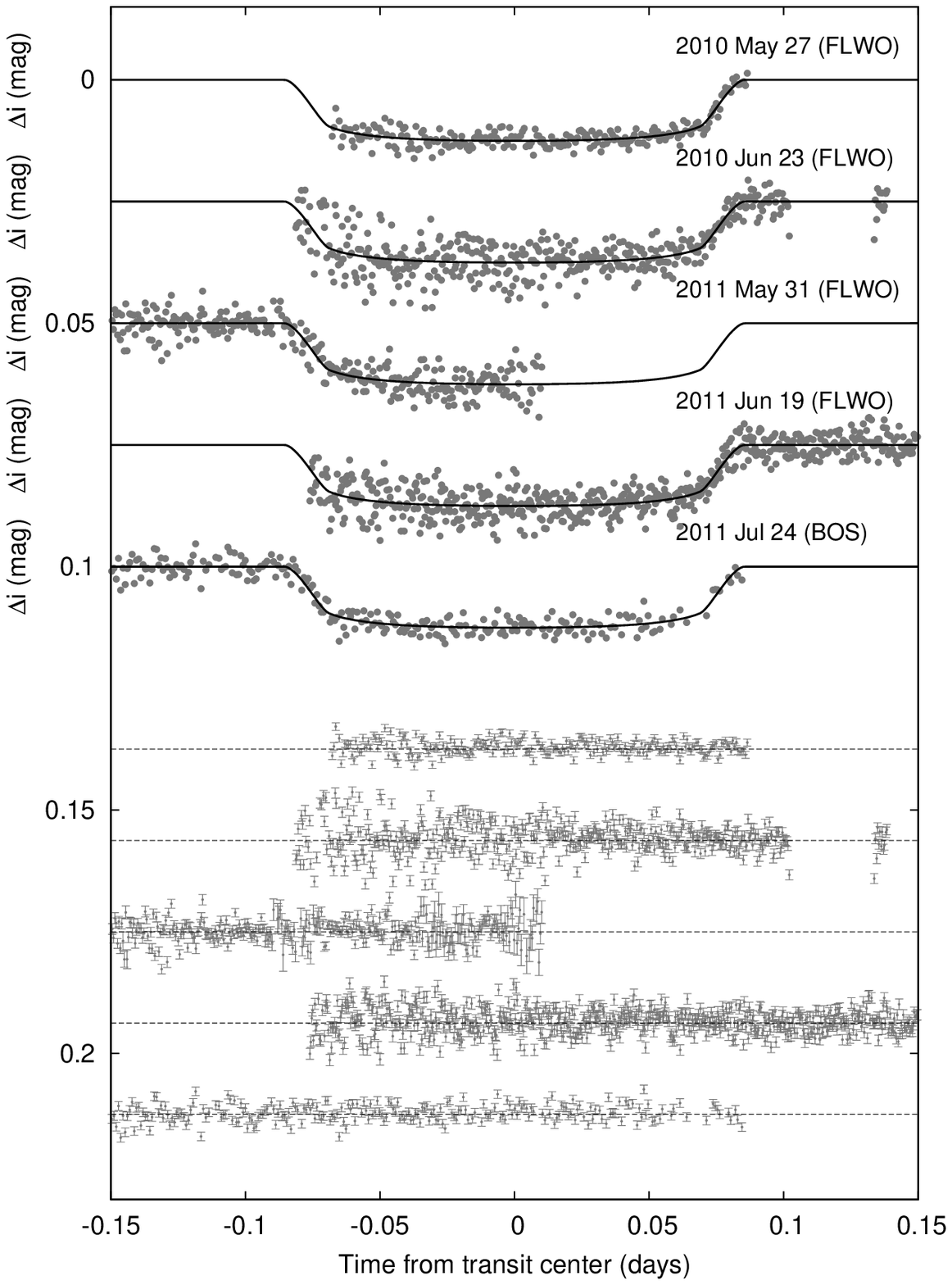}
\ifthenelse{\value{planetcounter}=1}{
\caption{
    Unbinned transit \lcs{} for \hatcur{41}, acquired with
    KeplerCam at the \flwof{} telescope.  The light curves have been
    EPD and TFA processed, as described in \refsec{globmod}.
    The dates of the events are indicated.  Curves after the first are
    displaced vertically for clarity.  Our best fit from the global
    modeling described in \refsecl{globmod} is shown by the solid
    lines.  Residuals from the fits are displayed at the bottom, in the
    same order as the top curves.  The error bars represent the photon
    and background shot noise, plus the readout noise.
}}{
\caption{
    Similar to \reffigl{lc42}; here we show the follow-up
    \lcs{} for \hatcur{41}. The facility used for each each light curve is
    indicated next to the date of the event.
}}
\label{fig:lc44}
\end{figure}

We conducted additional photometric observations of the three stars
with the KeplerCam CCD camera on the \flwof{} telescope, the Spectral
CCD on the 2.0\,m Faulkes Telescope North (FTN) at Haleakala
Observatory in Hawaii, and the CCD imager on the Byrne Observatory at
Sedgwick (BOS) 0.8\,m telescope, at Sedgwick Reserve in the Santa
Ynez Valley, CA.  Both FTN and BOS are operated by the Las Cumbres
Observatory Global Telescope (LCOGT\footnote{http://lcogt.net}; Brown
et al.~2012, in preparation). The observations for each target are
summarized in \reftabl{photobs}.

The reduction of the KeplerCam images to light curves was performed as
described by \citet{bakos:2010}.  The FTN and BOS images were reduced
in a similar manner.  We performed EPD and TFA to remove trends
simultaneously with the light curve modeling (for more details, see
\citet{bakos:2010}).  The final time series, together with our
best-fit transit \lc{} model, are shown in the top portion of
\reffigls{lc42}{lc44}, while the individual measurements are reported
in \reftabls{phfu42}{phfu44}.

\begin{deluxetable}{lrrrr}
\tablewidth{0pc}
\tablecaption{
    High-precision differential photometry of
    \hatcur{39}\label{tab:phfu42}.
}
\tablehead{
    \colhead{BJD\tablenotemark{a}} & 
    \colhead{Mag\tablenotemark{b}} & 
    \colhead{\ensuremath{\sigma_{\rm Mag}}} &
    \colhead{Mag(orig)\tablenotemark{c}} & 
    \colhead{Filter} \\
    \colhead{\hbox{~~~~(2,400,000$+$)~~~~}} & 
    \colhead{} & 
    \colhead{} &
    \colhead{} & 
    \colhead{}
}
\startdata
$ 54921.62065 $ & $   0.00694 $ & $   0.00127 $ & $  11.10550 $ & $ i$\\
$ 54921.62127 $ & $   0.00597 $ & $   0.00127 $ & $  11.10530 $ & $ i$\\
$ 54921.62187 $ & $   0.00631 $ & $   0.00126 $ & $  11.10640 $ & $ i$\\
$ 54921.62248 $ & $   0.00358 $ & $   0.00127 $ & $  11.10490 $ & $ i$\\
$ 54921.62311 $ & $   0.00600 $ & $   0.00127 $ & $  11.10610 $ & $ i$\\
$ 54921.62390 $ & $   0.00762 $ & $   0.00127 $ & $  11.10910 $ & $ i$\\
$ 54921.62452 $ & $   0.00787 $ & $   0.00126 $ & $  11.10670 $ & $ i$\\
$ 54921.62512 $ & $   0.00663 $ & $   0.00126 $ & $  11.10860 $ & $ i$\\
$ 54921.62574 $ & $   0.01035 $ & $   0.00126 $ & $  11.10940 $ & $ i$\\
$ 54921.62637 $ & $   0.01050 $ & $   0.00126 $ & $  11.11030 $ & $ i$\\
\enddata
\tablenotetext{a}{
    Barycentric Julian Date calculated directly from UTC, {\em
      without} correction for leap seconds.
}
\tablenotetext{b}{
    The out-of-transit level has been subtracted. These magnitudes have
    been subjected to the EPD and TFA procedures, carried out
    simultaneously with the transit fit.
}
\tablenotetext{c}{
    Raw magnitude values without application of the EPD and TFA
    procedures.
}
\tablecomments{
    This table is available in a machine-readable form in the online
    journal.  A portion is shown here for guidance regarding its form
    and content.
}
\end{deluxetable}
%
\begin{deluxetable}{lrrrr}
\tablewidth{0pc}
\tablecaption{
    High-precision differential photometry of
    \hatcur{40}\label{tab:phfu43}.
}
\tablehead{
    \colhead{BJD\tablenotemark{a}} & 
    \colhead{Mag\tablenotemark{b}} & 
    \colhead{\ensuremath{\sigma_{\rm Mag}}} &
    \colhead{Mag(orig)\tablenotemark{c}} & 
    \colhead{Filter} \\
    \colhead{\hbox{~~~~(2,400,000$+$)~~~~}} & 
    \colhead{} & 
    \colhead{} &
    \colhead{} & 
    \colhead{}
}
\startdata
$ 55456.60581 $ & $   0.00871 $ & $   0.00080 $ & $  10.22060 $ & $ i$\\
$ 55456.60739 $ & $   0.00713 $ & $   0.00067 $ & $  10.21900 $ & $ i$\\
$ 55456.60830 $ & $   0.00854 $ & $   0.00066 $ & $  10.22070 $ & $ i$\\
$ 55456.60930 $ & $   0.00851 $ & $   0.00066 $ & $  10.22100 $ & $ i$\\
$ 55456.61028 $ & $   0.00935 $ & $   0.00065 $ & $  10.22120 $ & $ i$\\
$ 55456.61124 $ & $   0.00789 $ & $   0.00065 $ & $  10.21980 $ & $ i$\\
$ 55456.61224 $ & $   0.00868 $ & $   0.00065 $ & $  10.22070 $ & $ i$\\
$ 55456.61319 $ & $   0.00695 $ & $   0.00065 $ & $  10.21870 $ & $ i$\\
$ 55456.61416 $ & $   0.00727 $ & $   0.00065 $ & $  10.21880 $ & $ i$\\
$ 55456.61513 $ & $   0.00996 $ & $   0.00064 $ & $  10.22200 $ & $ i$\\
\enddata
\tablenotetext{a}{
    Barycentric Julian Date calculated directly from UTC, {\em
      without} correction for leap seconds.
}
\tablenotetext{b}{
    The out-of-transit level has been subtracted. These magnitudes have
    been subjected to the EPD and TFA procedures, carried out
    simultaneously with the transit fit.
}
\tablenotetext{c}{
    Raw magnitude values without application of the EPD and TFA
    procedures.
}
\tablecomments{
    This table is available in a machine-readable form in the online
    journal.  A portion is shown here for guidance regarding its form
    and content.
}
\end{deluxetable}
%
\begin{deluxetable}{lrrrr}
\tablewidth{0pc}
\tablecaption{
    High-precision differential photometry of
    \hatcur{41}\label{tab:phfu44}.
}
\tablehead{
    \colhead{BJD\tablenotemark{a}} & 
    \colhead{Mag\tablenotemark{b}} & 
    \colhead{\ensuremath{\sigma_{\rm Mag}}} &
    \colhead{Mag(orig)\tablenotemark{c}} & 
    \colhead{Filter} \\
    \colhead{\hbox{~~~~(2,400,000$+$)~~~~}} & 
    \colhead{} & 
    \colhead{} &
    \colhead{} & 
    \colhead{}
}
\startdata
$ 55344.79616 $ & $   0.01161 $ & $   0.00077 $ & $  10.01060 $ & $ i$\\
$ 55344.79679 $ & $   0.00980 $ & $   0.00078 $ & $  10.00870 $ & $ i$\\
$ 55344.79744 $ & $   0.00586 $ & $   0.00078 $ & $  10.00480 $ & $ i$\\
$ 55344.79878 $ & $   0.00847 $ & $   0.00078 $ & $  10.00830 $ & $ i$\\
$ 55344.79946 $ & $   0.01379 $ & $   0.00078 $ & $  10.01310 $ & $ i$\\
$ 55344.80015 $ & $   0.01168 $ & $   0.00078 $ & $  10.01140 $ & $ i$\\
$ 55344.80083 $ & $   0.00987 $ & $   0.00077 $ & $  10.00820 $ & $ i$\\
$ 55344.80151 $ & $   0.01021 $ & $   0.00077 $ & $  10.00830 $ & $ i$\\
$ 55344.80219 $ & $   0.01031 $ & $   0.00076 $ & $  10.00870 $ & $ i$\\
$ 55344.80288 $ & $   0.01025 $ & $   0.00077 $ & $  10.00800 $ & $ i$\\
\enddata
\tablenotetext{a}{
    Barycentric Julian Date calculated directly from UTC, {\em
      without} correction for leap seconds.
}
\tablenotetext{b}{
    The out-of-transit level has been subtracted. These magnitudes have
    been subjected to the EPD and TFA procedures, carried out
    simultaneously with the transit fit.
}
\tablenotetext{c}{
    Raw magnitude values without application of the EPD and TFA
    procedures.
}
\tablecomments{
    This table is available in a machine-readable form in the online
    journal.  A portion is shown here for guidance regarding its form
    and content.
}
\end{deluxetable}

\subsection{Adaptive Optics Imaging}
\label{sec:ao}
\begin{figure}[]
\ifthenelse{\boolean{emulateapj}}{
\epsscale{1.0}
}{
\epsscale{0.5}
}
\plotone{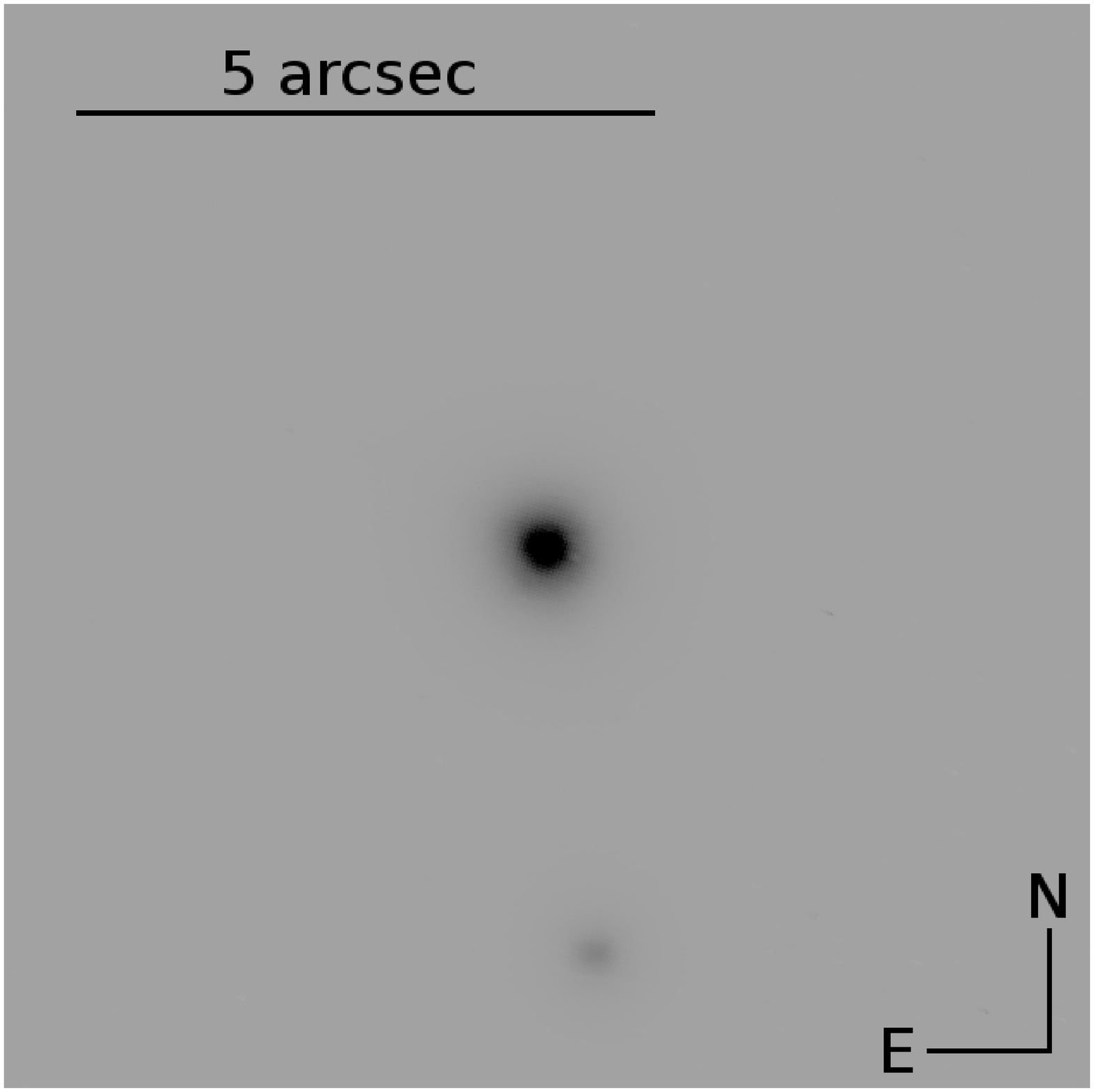}
\caption[]{
    $H$-band AO image of \hatcur{41} obtained with MMT/Clio2 showing the
    $\sim 3\farcs 5$ neighbor to the south.
\label{fig:mmtclio}}
\end{figure}

We obtained high-resolution imaging of \hatcur{41} on the night of 21
June 2011 using the Clio2 near-IR imager on the MMT 6.5\,m telescope
in AZ. Observations were obtained with the adaptive optics (AO) system
in $H$-band and in $L^{\prime}$-band. \reffigl{mmtclio} shows the
resulting $H$-band image of \hatcur{41} which easily resolves the $3
\farcs 56 \pm 0 \farcs 02$ neighbor.

Based on these observations we measure the $H$- and $L^{\prime}$-band
magnitudes of the neighbor relative to \hatcur{41} to be $\Delta H =
2.46 \pm 0.06$\,mag and $\Delta L^{\prime} = 2.6 \pm
2.0$\,mag. Assuming the star is a physical companion to \hatcur{41},
these magnitude differences are consistent with the neighbor being a
$\sim 0.7$\,\msun\ K-dwarf, or roughly $\sim 3.5$\,mag fainter than
\hatcur{41} in \band{i}. 

The KeplerCam observations of \hatcur{41}
described in \refsecl{phot} show \hatcur{41} to have an elongation
in the wing of its PSF due to the companion, but the seeing is not
good enough to resolve the stars given the magnitude difference. We
use the {\sc daophot} and {\sc allstar} programs \citep{stetson:1987,
  stetson:1992} to obtain PSF-fitting photometry for the two stars,
and find that the neighbor is fainter in \band{i} by $\Delta i \approx
3.3$\,mag, with an uncertainty of at least $0.1$\,mag. We conclude
that the broad-band photometry is consistent with \hatcur{41} having a
K-dwarf binary companion. At the distance of \hatcur{41}, the $3\farcs 5$
angular separation corresponds to a projected physical separation of
$\sim 1000$\,AU between the two stars.

\section{Analysis}
\label{sec:analysis}

\subsection{Properties of the parent star}
\label{sec:stelparam}

We measured the stellar atmospheric parameters for each star using the
Keck/HIRES Iodine-free template spectra, together with the Spectroscopy
Made Easy \citep[SME;][]{valenti:1996} package, and the
\cite{valenti:2005} atomic line database.  For each star, we obtained
the following {\em initial} values and uncertainties:
\begin{itemize}
\item {\em \hatcur{39}} --
effective temperature $\teffstar=\hatcurSMEiteff{39}$\,K, 
metallicity $\feh=\hatcurSMEizfeh{39}$\,dex,
stellar surface gravity $\loggstar=\hatcurSMEilogg{39}$\,(cgs), and
projected rotational velocity $\vsini=\hatcurSMEivsin{39}$\,\kms.
\item {\em \hatcur{40}} --
effective temperature $\teffstar=\hatcurSMEiteff{40}$\,K, 
metallicity $\feh=\hatcurSMEizfeh{40}$\,dex,
stellar surface gravity $\loggstar=\hatcurSMEilogg{40}$\,(cgs), and
projected rotational velocity $\vsini=\hatcurSMEivsin{40}$\,\kms.
\item {\em \hatcur{41}} --
effective temperature $\teffstar=\hatcurSMEiteff{41}$\,K, 
metallicity $\feh=\hatcurSMEizfeh{41}$\,dex,
stellar surface gravity $\loggstar=\hatcurSMEilogg{41}$\,(cgs), and
projected rotational velocity $\vsini=\hatcurSMEivsin{41}$\,\kms.
\end{itemize}


As described in our previous papers \citep[e.g.][]{bakos:2010}, these
initial values were used to determine the quadratic limb-darkening
coefficients for each star from the \cite{claret:2004} tables. We then
used the mean stellar density, determined from the normalized
semimajor axis \arstar, together with the effective temperature and
metallicity to determine an initial estimate of the mass and radius of
each star from the \hatcurisofull{39} isochrones
\citep{\hatcurisocite{39}}. This provided a refined estimate of the
stellar surface gravity, which we fixed in a second iteration of SME
for each star. For each system a third iteration did not change
\loggstar\ appreciably, so we adopted the values from the second
iteration as the final spectroscopic parameters for each star.  These
parameters are listed in \reftabl{stellar}. In this same table we also
list the available broad-band photometric magnitudes from the
literature, and physical parameters, such as the stellar masses and
radii, which are determined from the spectroscopic parameters together
with the model isochrones. As discussed in \refsecl{globmod} we adopt
the parameters assuming a circular orbit for each planet. Some of the
parameters, especially the derived stellar masses and radii, depend on
the eccentricity; \reftabl{stellareccen} lists the values for these
parameters when the eccentricity is allowed to vary. 

\begin{figure}[]
\ifthenelse{\boolean{emulateapj}}{
\epsscale{1.0}
}{
\epsscale{0.5}
}
\plotone{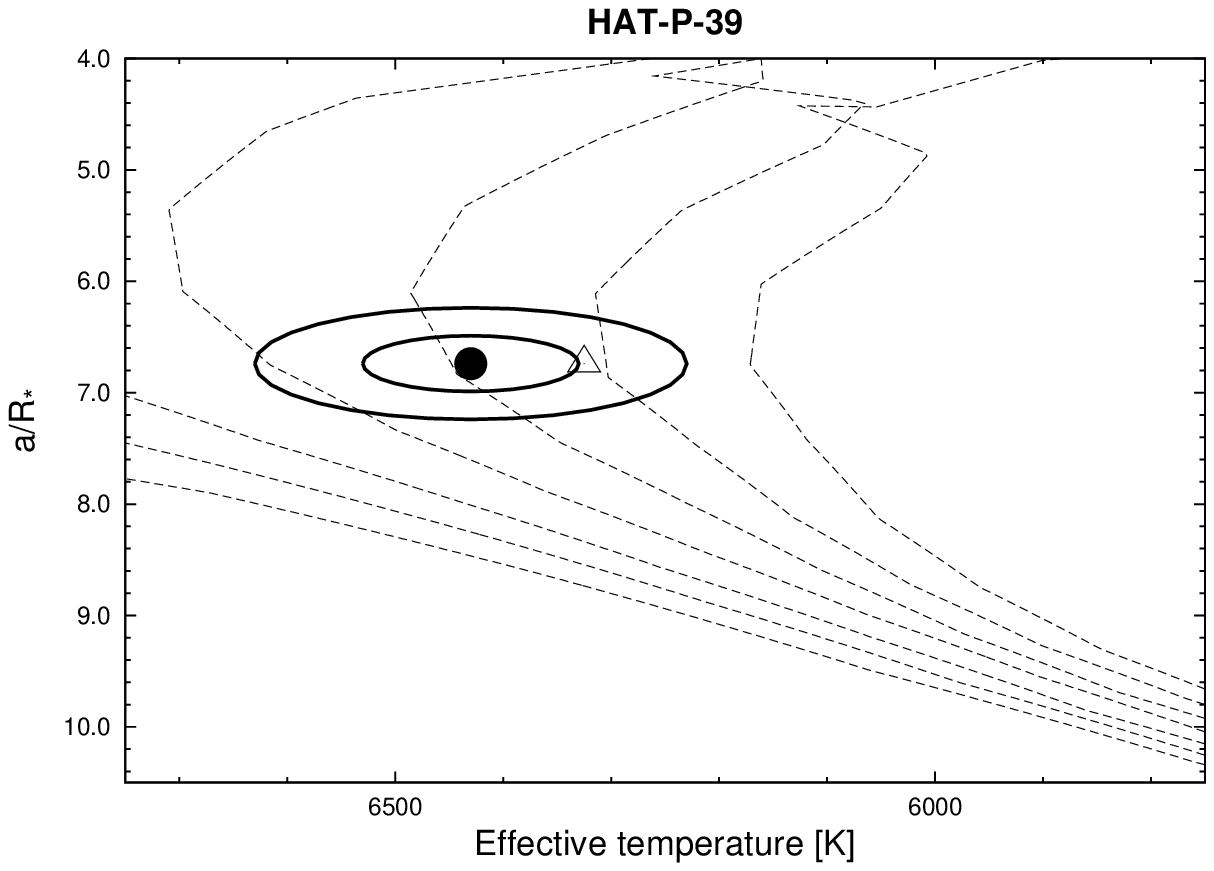}
\plotone{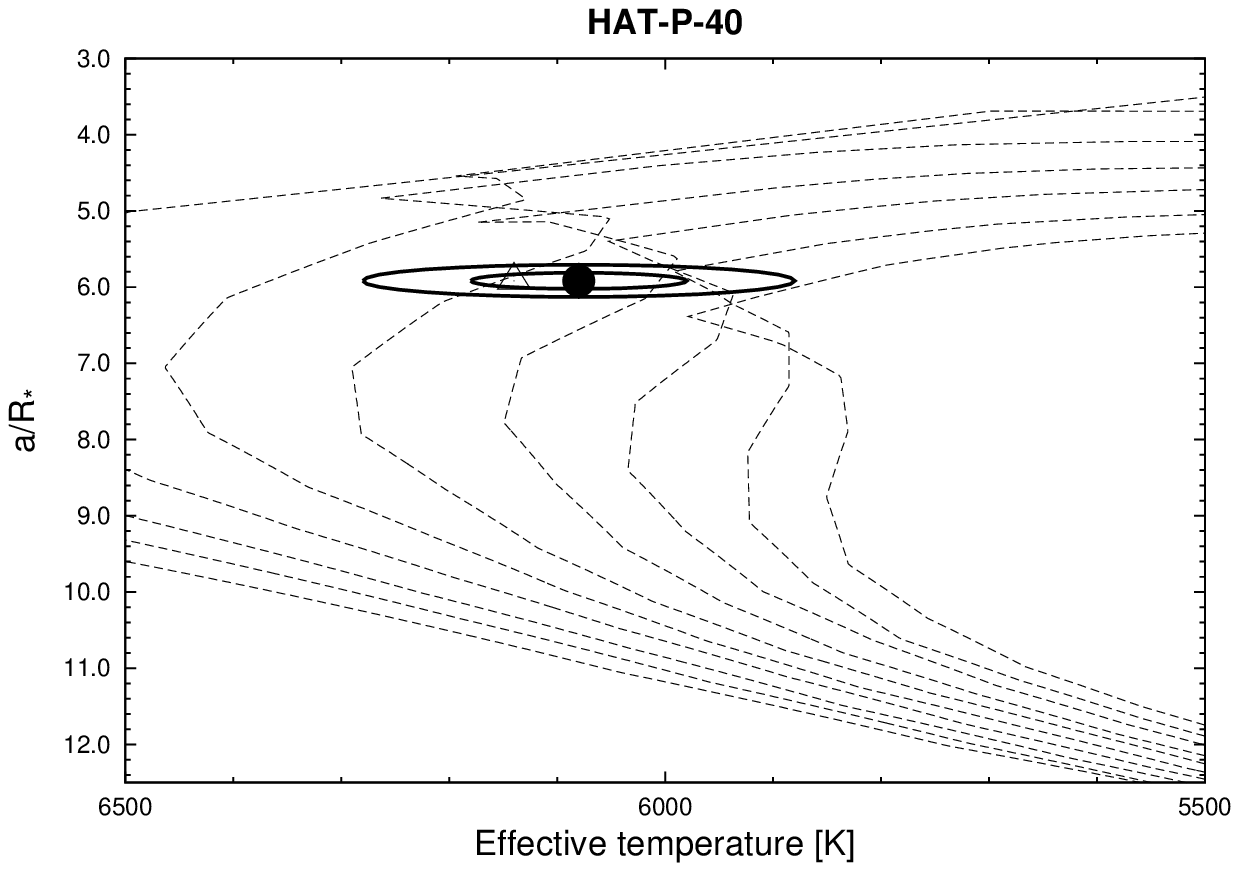}
\plotone{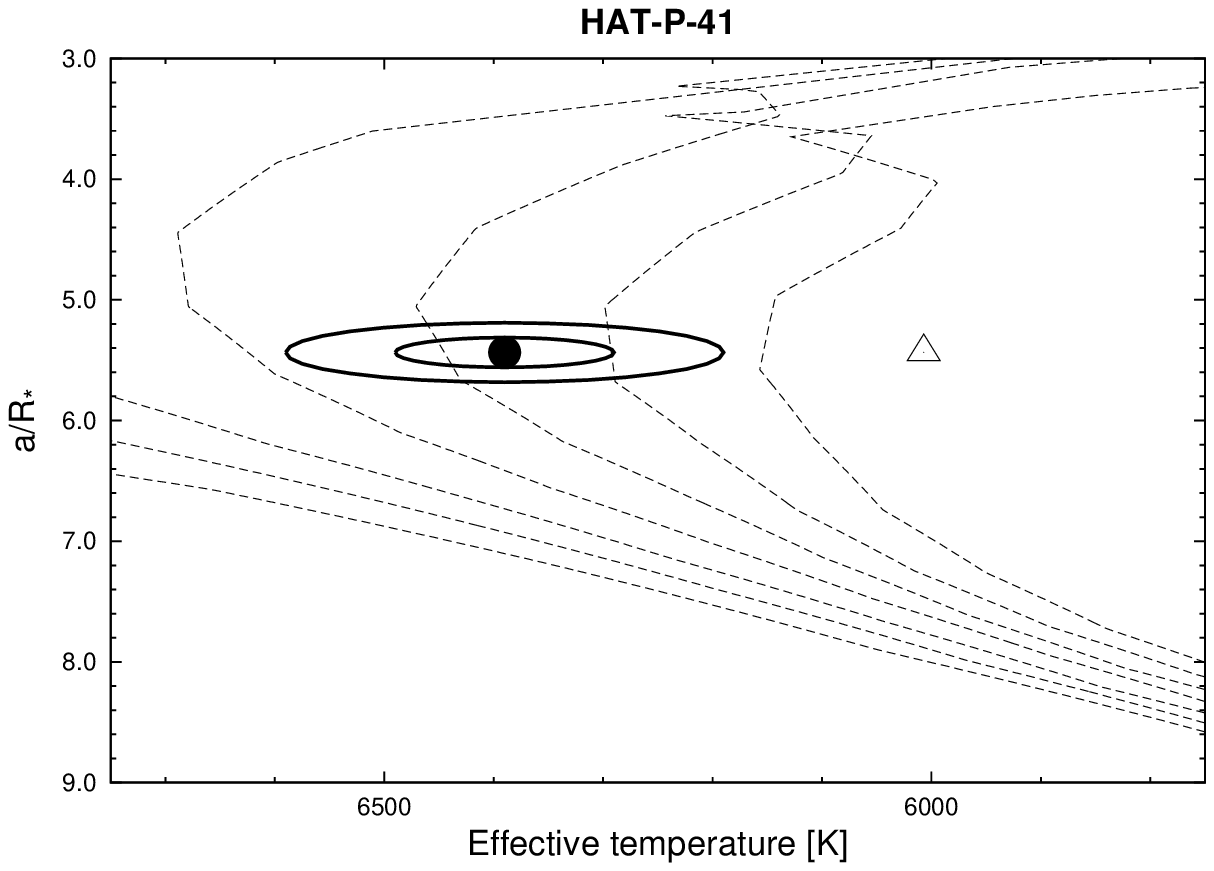}
\caption[]{
    Model isochrones from \cite{\hatcurisocite{39}} for the
    metallicities of \hatcur{39} (top), \hatcur{40} (center) and
    \hatcur{41} (bottom). For \hatcur{39} and \hatcur{41} the
    isochrones are shown for ages of 0.2\,Gyr, 0.6\,Gyr, and 1.0\,Gyr
    to 3.0\,Gyr in steps of 0.5\,Gyr (left to right), while for
    \hatcur{40} the isochrones are shown for ages of 0.2\,Gyr,
    0.6\,Gyr and 1.0\,Gyr to 4.5\,Gyr in steps of 0.5\,Gyr (left to
    right). The adopted values of $\teffstar$ and \arstar\ are shown
    together with their 1$\sigma$ and 2$\sigma$ confidence
    ellipsoids. In each plot the initial values of
    \teffstar\ and \arstar\ from the first SME and \lc\ analyses are
    represented with a triangle.
\label{fig:iso}}
\end{figure}

The inferred location of each star in a diagram of \arstar\ versus
\teffstar, analogous to the classical H-R diagram, is shown in
\reffigl{iso}.  In each case the stellar properties and their
1$\sigma$ and 2$\sigma$ confidence ellipsoids are displayed against
the backdrop of model isochrones for a range of ages, and the
appropriate stellar metallicity.  For comparison, the locations
implied by the initial SME results are also shown (in each case with a
triangle).

We determine the distance and extinction to each star by comparing the
$J$, $H$ and $K_{S}$ magnitudes from the 2MASS Catalogue
\citep{skrutskie:2006}, and the $V$ and $I_{C}$ magnitudes from the
TASS Mark IV Catalogue \citep{droege:2006}, to the expected magnitudes
from the stellar models. We use the transformations by
\citet{carpenter:2001} to convert the 2MASS magnitudes to the
photometric system of the models (ESO), and use the
\citet{cardelli:1989} extinction law, assuming a total-to-selective
extinction ratio of $R_{V} = 3.1$, to relate the extinction in each
band-pass to the $V$-band extinction $A_{V}$. The resulting $A_{V}$
and distance measurements are given in \reftabl{stellar}. We find
total \band{V} extinctions of $A_{V} = \hatcurXAv{39}$,
$\hatcurXAv{40}$, and $\hatcurXAv{41}$\,mag for \hatcur{39} through
\hatcur{41}, respectively. For comparison, the total line of sight
extinctions in each direction, estimated from the \cite{schlegel:1998}
dust maps, are: $0.146$\,mag, $0.724$\,mag, and
$0.513$\,mag. Following \cite{bonifacio:2000}, we estimate the
expected distance-corrected extinction to each source to be
$0.113$\,mag, $0.306$\,mag, and $0.157$\,mag, respectively. For
\hatcur{39} and \hatcur{40} the measured and expected values are
consistent. For \hatcur{41} the broad-band photometry appears to point
to a slightly redder star than expected based on the spectroscopic
temperature and expected extinction. As noted in \refsecl{ao},
\hatcur{41} has a close companion which is unresolved in the 2MASS or
TASS catalogs, this companion is the probable cause of the discrepancy
between the expected and observed magnitudes of \hatcur{41}.

\ifthenelse{\boolean{emulateapj}}{
  \begin{deluxetable*}{lcccr}
}{
  \begin{deluxetable}{lcccr}
}
\tablewidth{0pc}
\tabletypesize{\scriptsize}
\tablecaption{
    Adopted stellar parameters for \hatcur{39}--\hatcur{41} assuming circular orbits
    \label{tab:stellar}
}
\tablehead{
    \multicolumn{1}{c}{} &
    \multicolumn{1}{c}{{\bf HAT-P-39}} &
    \multicolumn{1}{c}{{\bf HAT-P-40}} &
    \multicolumn{1}{c}{{\bf HAT-P-41}} &
    \multicolumn{1}{c}{} \\
    \multicolumn{1}{c}{~~~~~~~~Parameter~~~~~~~~} &
    \multicolumn{1}{c}{Value}                     &
    \multicolumn{1}{c}{Value}                     &
    \multicolumn{1}{c}{Value}                     &
    \multicolumn{1}{c}{Source}                    
}
\startdata
\noalign{\vskip -3pt}
\sidehead{Spectroscopic properties}
~~~~$\teffstar$ (K)\dotfill         &  \hatcurSMEteff{39}   &  \hatcurSMEteff{40}   &  \hatcurSMEteff{41}   & SME\tablenotemark{a}\\
~~~~$\feh$\dotfill                  &  \hatcurSMEzfeh{39}   &  \hatcurSMEzfeh{40}   &  \hatcurSMEzfeh{41}   & SME                 \\
~~~~$\vsini$ (\kms)\dotfill         &  \hatcurSMEvsin{39}   &  \hatcurSMEvsin{40}   &  \hatcurSMEvsin{41}   & SME                 \\
~~~~$\vmac$ (\kms)\dotfill          &  \hatcurSMEvmac{39}   &  \hatcurSMEvmac{40}   &  \hatcurSMEvmac{41}   & SME                 \\
~~~~$\vmic$ (\kms)\dotfill          &  \hatcurSMEvmic{39}   &  \hatcurSMEvmic{40}   &  \hatcurSMEvmic{41}   & SME                 \\
~~~~$\gamma_{\rm RV}$ (\kms)\dotfill&  \hatcurDSgamma{39}   &  \hatcurTRESgamma{40}   &  \hatcurTRESgamma{41}   & DS or TRES                  \\
\sidehead{Photometric properties}
~~~~$V$ (mag)\dotfill               &  \hatcurCCtassmv{39}  &  \hatcurCCtassmv{40}  &  \hatcurCCtassmv{41}  & TASS                \\
~~~~$\vic$ (mag)\dotfill            &  \hatcurCCtassvi{39}  &  \hatcurCCtassvi{40}  &  \hatcurCCtassvi{41}  & TASS                \\
~~~~$J$ (mag)\dotfill               &  \hatcurCCtwomassJmag{39} &  \hatcurCCtwomassJmag{40} &  \hatcurCCtwomassJmag{41} & 2MASS           \\
~~~~$H$ (mag)\dotfill               &  \hatcurCCtwomassHmag{39} &  \hatcurCCtwomassHmag{40} &  \hatcurCCtwomassHmag{41} & 2MASS           \\
~~~~$K_s$ (mag)\dotfill             &  \hatcurCCtwomassKmag{39} &  \hatcurCCtwomassKmag{40} &  \hatcurCCtwomassKmag{41} & 2MASS           \\
\sidehead{Derived properties}
~~~~$\mstar$ ($\msun$)\dotfill      &  \hatcurISOmlong{39}   &  \hatcurISOmlong{40}   &  \hatcurISOmlong{41}   & \hatcurisoshort{39}+\hatcurlumind{39}+SME\tablenotemark{b}\\
~~~~$\rstar$ ($\rsun$)\dotfill      &  \hatcurISOrlong{39}   &  \hatcurISOrlong{40}   &  \hatcurISOrlong{41}   & \hatcurisoshort{39}+\hatcurlumind{39}+SME         \\
~~~~$\loggstar$ (cgs)\dotfill       &  \hatcurISOlogg{39}    &  \hatcurISOlogg{40}    &  \hatcurISOlogg{41}    & \hatcurisoshort{39}+\hatcurlumind{39}+SME         \\
~~~~$\lstar$ ($\lsun$)\dotfill      &  \hatcurISOlum{39}     &  \hatcurISOlum{40}     &  \hatcurISOlum{41}     & \hatcurisoshort{39}+\hatcurlumind{39}+SME         \\
~~~~$M_V$ (mag)\dotfill             &  \hatcurISOmv{39}      &  \hatcurISOmv{40}      &  \hatcurISOmv{41}      & \hatcurisoshort{39}+\hatcurlumind{39}+SME         \\
~~~~$M_K$ (mag,\hatcurjhkfilset{39})\dotfill &  \hatcurISOMK{39} &  \hatcurISOMK{40} &  \hatcurISOMK{41} & \hatcurisoshort{39}+\hatcurlumind{39}+SME         \\
~~~~Age (Gyr)\dotfill               &  \hatcurISOage{39}     &  \hatcurISOage{40}     &  \hatcurISOage{41}     & \hatcurisoshort{39}+\hatcurlumind{39}+SME         \\
~~~~$A_{V}$ (mag)\tablenotemark{c}\dotfill           &  \hatcurXAv{39}\phn  &  \hatcurXAv{40}    &  \hatcurXAv{41}    & \hatcurisoshort{39}+\hatcurlumind{39}+SME\\
~~~~Distance (pc)\dotfill           &  \hatcurXdistred{39}\phn  &  \hatcurXdistred{40}\phn  &  \hatcurXdistred{41}\phn  & \hatcurisoshort{39}+\hatcurlumind{39}+SME\\
~~~~$\log R^{\prime}_{\rm HK}$\tablenotemark{d}\dotfill & $-4.85 \pm 0.07$ & $-5.12 \pm 0.16$ & $-5.04 \pm 0.04$ & Keck/HIRES\\ [-1.5ex]
\enddata
\tablenotetext{a}{
    SME = ``Spectroscopy Made Easy'' package for the analysis of
    high-resolution spectra \citep{valenti:1996}.  These parameters
    rely primarily on SME, but have a small dependence also on the
    iterative analysis incorporating the isochrone search and global
    modeling of the data, as described in the text.
}
\tablenotetext{b}{
    \hatcurisoshort{39}+\hatcurlumind{39}+SME = Based on the \hatcurisoshort{39}\
    isochrones \citep{\hatcurisocite{39}}, \hatcurlumind{39}\ as a luminosity
    indicator, and the SME results.
}
\tablenotetext{c}{
  \band{V} extinction determined by comparing the measured 2MASS and
  TASS photometry for the star to the expected magnitudes from the
  \hatcurisoshort{39}+\hatcurlumind{39}+SME model for the star. We use
  the \citet{cardelli:1989} extinction law.  
}
\tablenotetext{d}{
  Chromospheric activity index defined in \citet{noyes:1984}
  determined from the Keck/HIRES spectra following
  \cite{isaacson:2010}. In each case we give the average value and the
  standard deviation from the individual spectra.
}
\ifthenelse{\boolean{emulateapj}}{
  \end{deluxetable*}
}{
  \end{deluxetable}
}

\ifthenelse{\boolean{emulateapj}}{
  \begin{deluxetable*}{lcccr}
}{
  \begin{deluxetable}{lcccr}
}
\tablewidth{0pc}
\tabletypesize{\scriptsize}
\tablecaption{
    Derived stellar parameters for \hatcur{39}--\hatcur{41} allowing eccentric orbits\tablenotemark{a}
    \label{tab:stellareccen}
}
\tablehead{
    \multicolumn{1}{c}{} &
    \multicolumn{1}{c}{{\bf HAT-P-39}} &
    \multicolumn{1}{c}{{\bf HAT-P-40}} &
    \multicolumn{1}{c}{{\bf HAT-P-41}} &
    \multicolumn{1}{c}{} \\
    \multicolumn{1}{c}{~~~~~~~~Parameter~~~~~~~~} &
    \multicolumn{1}{c}{Value}                     &
    \multicolumn{1}{c}{Value}                     &
    \multicolumn{1}{c}{Value}                     &
    \multicolumn{1}{c}{Source}                    
}
\startdata
\noalign{\vskip -3pt}
~~~~$\mstar$ ($\msun$)\dotfill      &  \hatcurISOmlongeccen{39}   &  \hatcurISOmlongeccen{40}   &  \hatcurISOmlongeccen{41}   & \hatcurisoshort{39}+\hatcurlumind{39}+SME \\
~~~~$\rstar$ ($\rsun$)\dotfill      &  \hatcurISOrlongeccen{39}   &  \hatcurISOrlongeccen{40}   &  \hatcurISOrlongeccen{41}   & \hatcurisoshort{39}+\hatcurlumind{39}+SME         \\
~~~~$\loggstar$ (cgs)\dotfill       &  \hatcurISOloggeccen{39}    &  \hatcurISOloggeccen{40}    &  \hatcurISOloggeccen{41}    & \hatcurisoshort{39}+\hatcurlumind{39}+SME         \\
~~~~$\lstar$ ($\lsun$)\dotfill      &  \hatcurISOlumeccen{39}     &  \hatcurISOlumeccen{40}     &  \hatcurISOlumeccen{41}     & \hatcurisoshort{39}+\hatcurlumind{39}+SME         \\
~~~~$M_V$ (mag)\dotfill             &  \hatcurISOmveccen{39}      &  \hatcurISOmveccen{40}      &  \hatcurISOmveccen{41}      & \hatcurisoshort{39}+\hatcurlumind{39}+SME         \\
~~~~$M_K$ (mag,\hatcurjhkfilset{39})\dotfill &  \hatcurISOMKeccen{39} &  \hatcurISOMKeccen{40} &  \hatcurISOMKeccen{41} & \hatcurisoshort{39}+\hatcurlumind{39}+SME         \\
~~~~Age (Gyr)\dotfill               &  \hatcurISOageeccen{39}     &  \hatcurISOageeccen{40}     &  \hatcurISOageeccen{41}     & \hatcurisoshort{39}+\hatcurlumind{39}+SME         \\
~~~~$A_{V}$ (mag)\dotfill           &  \hatcurXAveccen{39}\phn  &  \hatcurXAveccen{40}    &  \hatcurXAveccen{41}    & \hatcurisoshort{39}+\hatcurlumind{39}+SME\\
~~~~Distance (pc)\dotfill           &  \hatcurXdistredeccen{39}\phn  &  \hatcurXdistredeccen{40}\phn  &  \hatcurXdistredeccen{41}\phn  & \hatcurisoshort{39}+\hatcurlumind{39}+SME\\
\enddata
\tablenotetext{a}{
    Quantities and abbreviations are as in \reftabl{stellar}, which gives our adopted values, determined assuming circular orbits. We do not list parameters that are independent of the eccentricity.
}
\ifthenelse{\boolean{emulateapj}}{
  \end{deluxetable*}
}{
  \end{deluxetable}
}

\subsection{Excluding blend scenarios}
\label{sec:blend}

The analyses of our reconnaissance spectroscopic observations
discussed in \refsecl{recspec} rule out the most obvious astrophysical
false positive scenarios for \hatcur{39} through
\hatcur{41}. Additionally the spectral-line bisector span (BS)
analyses which we conducted (\reffigls{rvbis42}{rvbis44}) provide
constraints on more subtle blend scenarios similar to that presented
in \cite{torres:2004}. However, because \hatcur{39} and \hatcur{41}
have high RV jitter, and consequently high BS scatter ($\sim
80$\,\ms\ and $\sim 30$\,\ms, respectively), relative to their RV
semiamplitudes ($\hatcurRVK{39}$\,\ms\ and $\hatcurRVK{41}$\,\ms,
respectively), the BS test provides a less stringent constraint on
possible blend scenarios than it does in a case such as \hatcur{40}
(BS scatter of $\sim 10$\,\ms\ and RV semiamplitude of
$\hatcurRVK{40}$\,\ms). To provide additional support for the
planetary interpretation of the observations of each system we conduct
detailed blend analyses of the light curves (including both the HATNet
discovery light curves, and all available follow-up light curves) and
absolute photometry following \cite{hartman:2011}. 

\begin{figure}[]
\ifthenelse{\boolean{emulateapj}}{
\epsscale{1.0}
}{
\epsscale{0.5}
}
\plotone{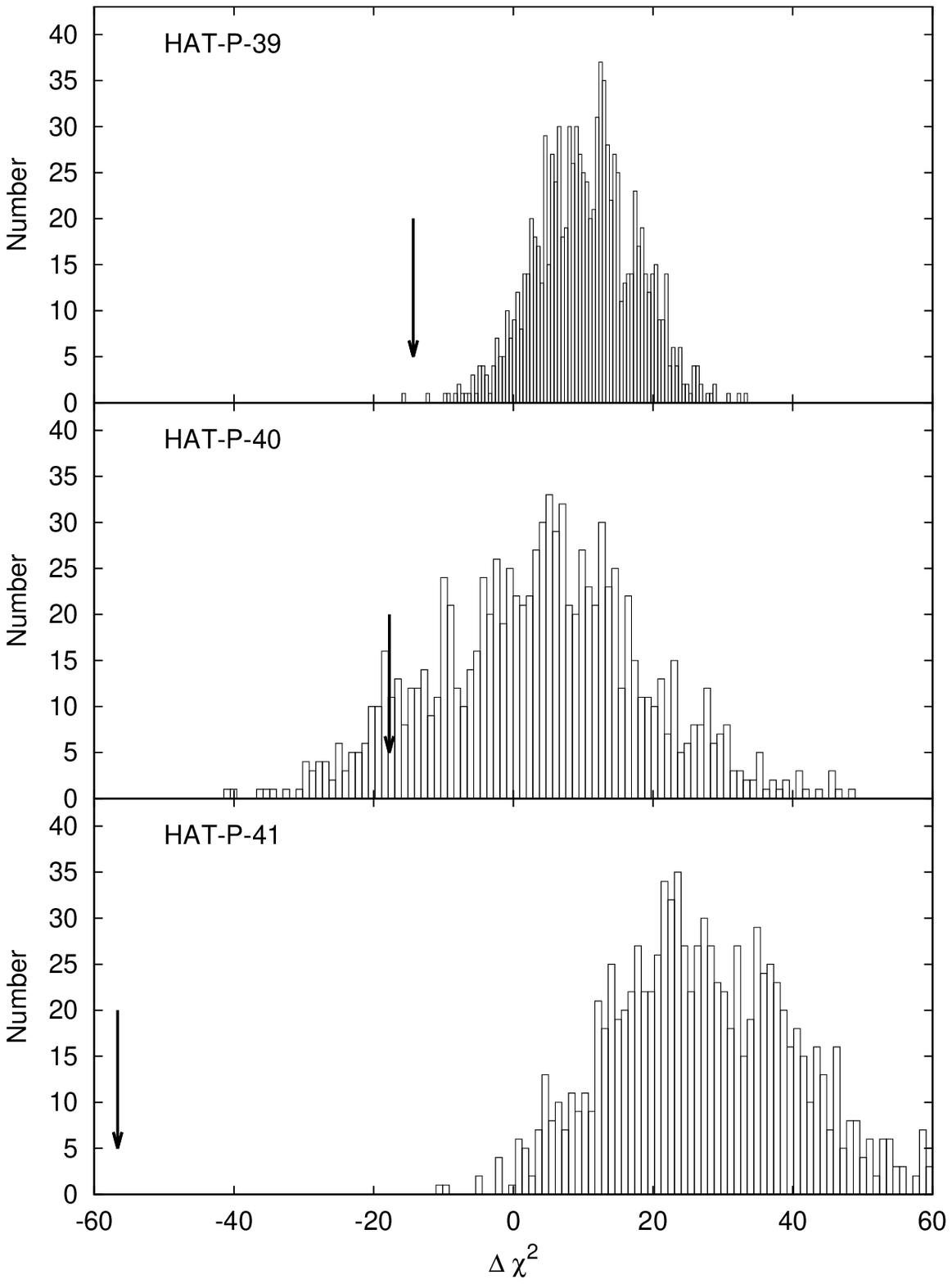}
\caption[]{
The histogram of $\Delta \chi^{2}$ values between the best-fit
transiting planet model and the best-fit blend model for simulated
data sets having the same noise properties as the observed residuals
from the best-fit blend model (see \citealp{hartman:2011} for a more
detailed discussion). The histogram is shown separately for each
system. In each panel the arrow shows the $\chi^{2}$ difference
between the models fitted to the actual observations. For \hatcur{41}
the best-fit blend scenario is rejected based on the photometry
with greater than $5\sigma$ confidence. For \hatcur{39} and
\hatcur{40}, blend scenarios which cannot be rejected based solely on
the photometry are rejected based on spectroscopic information (limits
on variations in RV, and in the BS and FWHM of the spectral line
profiles).
\label{fig:blendchi2}}
\end{figure}

In \reffigl{blendchi2} we show, for each system, the histogram of
$\Delta \chi^{2}$ values between the best-fit transiting planet model
and the best-fit blend model for simulated data sets having the same
noise properties as the observed residuals from the best-fit blend
model (see \citealp{hartman:2011} for a more detailed discussion). In
this same figure we also show the $\Delta \chi^{2}$ difference between
the best-fit models applied to the observations.  We find that for
\hatcur{39} and \hatcur{41} we can reject blend scenarios involving
combinations of three stars with greater than $3\sigma$ and $5\sigma$
confidence, respectively, based solely on the photometry. For
\hatcur{39} the detailed shape of the transit, as determined from the
follow-up light curves, contributes most of the $\chi^{2}$ difference
between the models, while for \hatcur{41} it is the lack of
out-of-transit variations, as determined from the HATNet light curve,
that contributes most of the $\chi^{2}$ difference. For \hatcur{40} we
are unable to rule out blend scenarios based solely on the photometry,
however in this case the lack of BS or FWHM variations rules out such
blends. To quantify this, we simulate the cross-correlation function
(CCF) of blended systems which could plausibly fit the photometric
data (configurations which cannot be rejected with greater than
$5\sigma$ confidence) and find that in all cases either the RV or FWHM
of the blended configuration varies by several \kms, or the BS varies
by greater than $\sim 100$\,\ms, greatly exceeding the observational
limits on any such variations. Similarly for \hatcur{39} we find that
stellar blend configurations which cannot be rejected with greater
than $5\sigma$ confidence predict greater than $500$\,\ms\ variations
in the RV or BS of the Keck spectra, which are well above the
observational constraints.

As discussed in \refsecl{ao}, \hatcur{41} has a close neighbor which
we estimate to be $\sim 3.5$\,mag fainter in \band{i}. While such a
neighbor could, in principle, be eclipsed by an object that would
produce a $\sim 1$\% dip in the blended light curve, as we have shown
here, the detailed shape of the light curve cannot be produced using
physically possible combinations of stars (i.e.~stars with parameters
determined from stellar evolution models). We note that the KeplerCam
observations show no evidence for variations in the flux centroid that
correlate with the photometry, providing further evidence that the
observed variation is not due to a deep eclipse in the poorly resolved
neighbor.

While we can rule out, for each of the systems, blend scenarios
involving only stellar-mass components, we cannot rule out scenarios
involving binary star systems with one component hosting a transiting
planet. Indeed, for \hatcur{41} we find that including a $\sim
0.7$\,\msun\ star in the system provides a slightly better fit to the
photometric observations. In this case we actually know that there is
a faint companion (though it is unresolved in our light curves or the
available absolute broad-band photometry measurements), so it is
reassuring to find that the blend analysis of the photometric data
also points to the existence of this companion. For \hatcur{39} we can
rule out binary companion stars with $M < 1.24\,\msun$ while for
\hatcur{40} and \hatcur{41} we cannot rule out binary companions of
any mass up to that of the mass inferred for the brighter star in the
system. While massive binary companions in general should be easier to
detect in the spectrum, if the two stars have very similar average
velocities, the resulting variations in BS, FWHM, and RV measurements
can be less than the constraints set by the observations.

We conclude that each system presented here contains a transiting
planet, however we cannot definitively claim that these are all single
stars. There is no evidence that either \hatcur{39} or \hatcur{40} is
a binary system, so we treat each of these as single stars in the
analysis that follows. For \hatcur{41} there is a resolved neighbor
which we account for in our analysis of the system.

\subsection{Global modeling of the data}
\label{sec:globmod}

We modelled the HATNet photometry, the follow-up photometry, and the
high-precision RV measurements using the procedure described by
\citet{bakos:2010}. Following the discussion by \citet{eastman:2012},
we made two important changes to our analysis procedure compared to
what was done in \citet{bakos:2010}. As noted in section 3.5.3 of
\citet{eastman:2012}, there is a common mistake in the implementation
of the Metropolis-Hastings (M-H) algorithm for conducting a Markov
Chain Monte Carlo (MCMC) analysis whereby the Markov chain is not
increased when a proposed transition is rejected. We discovered that
the implementation we have been using for the analysis of HATNet
planets has made this mistake, and have corrected it for the analysis
of the planets presented in this paper. We found that this bug tends
to inflate the errorbars on the determined parameters by a factor of a
few parts in a hundred. Errors given in previous discovery papers may
thus be slightly overestimated. The second significant change that we
have made is to use $\sqrt{e} \cos \omega$ and $\sqrt{e} \sin \omega$
as jump parameters, rather than $e \cos \omega$ and $e \sin
\omega$. Previously we had been assuming uniform priors on the latter
jump parameters, which amounts to assuming a linear prior on the
eccentricity $e$, creating a bias towards measuring nonzero
eccentricities. As other authors have noted, using $\sqrt{e} \cos
\omega$ and $\sqrt{e} \sin \omega$ leads to a uniform prior on $e$. We
found that this change had a much more significant impact on our
determined parameters than did correcting the bug in our
implementation of M-H. For example, for \hatcurb{39} the eccentricity
that we find is $\hatcurRVecceneccen{39}$ compared with $0.161 \pm 0.094$
when using our old jump parameters.

We also made a few minor changes to the analysis for the particular
planets presented in this paper. For the analysis of \hatcur{41} we
allowed independent RV zero-points for the Keck/HIRES RVs, the
high-resolution NOT/FIES RVs, and the medium-resolution NOT/FIES
RVs. To account for the contribution from the neighbor to the
photometry, we also fixed the third light in the \band{i} to $4\%$,
based on our PSF-fitting analysis of the KeplerCam observations of
this system.

For each planet we performed the fit both allowing eccentricity to
vary and fixing it to zero. The resulting parameters for each system
are listed in \reftabl{planetparam}, assuming circular orbits, and in
\reftabl{planetparameccen} allowing eccentric orbits. We use a
\citet{lucy:1971} test to determine the significance of the measured
eccentricities of each system. We find that the observations of all
three systems are consistent with the planets being on circular orbits
(the circular orbit hypothesis is rejected with 45\% confidence, 86\%
confidence, and 91\% confidence, for \hatcurb{39}, \hatcurb{40} and
\hatcurb{41}, respectively; for reference the confidence level would
need to be greater than 99.7\% for the detection of a nonzero
eccentricity to be significant at the $3\sigma$ level, which is
generally taken as a minimum level of significance).  In the past we
have generally presented parameters for systems allowing the
eccentricity to vary, even in cases where the observations are
consistent with a circular orbit, on the grounds that this provides a
more conservative estimate of the errors. However, as discussed
recently by \citet{anderson:2012}, the best-fit parameters which
result from allowing the eccentricity to vary are often biased
relative to the circular orbit values, and in most cases further
follow-up observations, such as occultation observations, reveal the
planets to be on circular orbits after all. We therefore suggest
adopting the circular orbit parameters as the most probable values for
each planet. These are given in \reftabl{planetparam}. For reference
\reftabl{planetparameccen} lists the resulting parameters when the
eccentricities are varied.


%
%
\ifthenelse{\boolean{emulateapj}}{
  \begin{deluxetable*}{lccc}
}{
  \begin{deluxetable}{lccc}
}
\tabletypesize{\scriptsize}
\tablecaption{Adopted orbital and planetary parameters for \hatcurb{39}--\hatcurb{41} assuming circular orbits\label{tab:planetparam}}
\tablehead{
    \multicolumn{1}{c}{} &
    \multicolumn{1}{c}{{\bf HAT-P-39b}} &
    \multicolumn{1}{c}{{\bf HAT-P-40b}} &
    \multicolumn{1}{c}{{\bf HAT-P-41b}} \\
    \multicolumn{1}{c}{~~~~~~~~Parameter~~~~~~~~} &
    \multicolumn{1}{c}{Value}                     &
    \multicolumn{1}{c}{Value}                     &
    \multicolumn{1}{c}{Value}                     
}
\startdata
\noalign{\vskip -3pt}
\sidehead{\Lc{} parameters}
~~~$P$ (days)             \dotfill    & $\hatcurLCP{39}$              & $\hatcurLCP{40}$              & $\hatcurLCP{41}$              \\
~~~$T_c$ (${\rm BJD}$)    
      \tablenotemark{a}   \dotfill    & $\hatcurLCT{39}$              & $\hatcurLCT{40}$              & $\hatcurLCT{41}$              \\
~~~$T_{14}$ (days)
      \tablenotemark{a}   \dotfill    & $\hatcurLCdur{39}$            & $\hatcurLCdur{40}$            & $\hatcurLCdur{41}$            \\
~~~$T_{12} = T_{34}$ (days)
      \tablenotemark{a}   \dotfill    & $\hatcurLCingdur{39}$         & $\hatcurLCingdur{40}$         & $\hatcurLCingdur{41}$         \\
~~~$\arstar$              \dotfill    & $\hatcurPPar{39}$             & $\hatcurPPar{40}$             & $\hatcurPPar{41}$             \\
~~~$\zrstar$\tablenotemark{b}              \dotfill    & $\hatcurLCzeta{39}$\phn       & $\hatcurLCzeta{40}$\phn       & $\hatcurLCzeta{41}$\phn       \\
~~~$\rpl/\rstar$          \dotfill    & $\hatcurLCrprstar{39}$        & $\hatcurLCrprstar{40}$        & $\hatcurLCrprstar{41}$        \\
~~~$b^2$                  \dotfill    & $\hatcurLCbsq{39}$            & $\hatcurLCbsq{40}$            & $\hatcurLCbsq{41}$            \\
~~~$b \equiv a \cos i/\rstar$
                          \dotfill    & $\hatcurLCimp{39}$            & $\hatcurLCimp{40}$            & $\hatcurLCimp{41}$            \\
~~~$i$ (deg)              \dotfill    & $\hatcurPPi{39}$\phn          & $\hatcurPPi{40}$\phn          & $\hatcurPPi{41}$\phn          \\

\sidehead{Limb-darkening coefficients \tablenotemark{c}}
~~~$c_1,i$ (linear term)  \dotfill    & $\hatcurLBii{39}$             & $\hatcurLBii{40}$             & $\hatcurLBii{41}$             \\
~~~$c_2,i$ (quadratic term) \dotfill  & $\hatcurLBiii{39}$            & $\hatcurLBiii{40}$            & $\hatcurLBiii{41}$            \\
~~~$c_1,r$               \dotfill    & $\cdots$             & $\cdots$             & $\hatcurLBir{41}$             \\
~~~$c_2,r$               \dotfill    & $\cdots$            & $\cdots$            & $\hatcurLBiir{41}$            \\
~~~$c_1,I$               \dotfill    & $\cdots$             & $\hatcurLBiI{40}$             & $\cdots$             \\
~~~$c_2,I$               \dotfill    & $\cdots$            & $\hatcurLBiiI{40}$            & $\cdots$            \\
~~~$c_1,R$               \dotfill    & $\hatcurLBiR{39}$             & $\cdots$             & $\cdots$             \\
~~~$c_2,R$               \dotfill    & $\hatcurLBiiR{39}$            & $\cdots$            & $\cdots$            \\

\sidehead{RV parameters}
~~~$K$ (\ms)              \dotfill    & $\hatcurRVK{39}$\phn\phn      & $\hatcurRVK{40}$\phn\phn      & $\hatcurRVK{41}$\phn\phn      \\
~~~$e$                    \dotfill    & $0$ (fixed)          & $0$ (fixed)          & $0$ (fixed)          \\
~~~RV jitter (\ms)\tablenotemark{d}        \dotfill    & \hatcurRVjitter{39}           & \hatcurRVjitter{40}           & \hatcurRVjitter{41}           \\

\sidehead{Planetary parameters}
~~~$\mpl$ ($\mjup$)       \dotfill    & $\hatcurPPmlong{39}$          & $\hatcurPPmlong{40}$          & $\hatcurPPmlong{41}$          \\
~~~$\rpl$ ($\rjup$)       \dotfill    & $\hatcurPPrlong{39}$          & $\hatcurPPrlong{40}$          & $\hatcurPPrlong{41}$          \\
~~~$C(\mpl,\rpl)$
    \tablenotemark{e}     \dotfill    & $\hatcurPPmrcorr{39}$         & $\hatcurPPmrcorr{40}$         & $\hatcurPPmrcorr{41}$         \\
~~~$\rhopl$ (\gcmc)       \dotfill    & $\hatcurPPrho{39}$            & $\hatcurPPrho{40}$            & $\hatcurPPrho{41}$            \\
~~~$\log g_p$ (cgs)       \dotfill    & $\hatcurPPlogg{39}$           & $\hatcurPPlogg{40}$           & $\hatcurPPlogg{41}$           \\
~~~$a$ (AU)               \dotfill    & $\hatcurPParel{39}$           & $\hatcurPParel{40}$           & $\hatcurPParel{41}$           \\
~~~$T_{\rm eq}$ (K)\tablenotemark{f}        \dotfill   & $\hatcurPPteff{39}$           & $\hatcurPPteff{40}$           & $\hatcurPPteff{41}$           \\
~~~$\Theta$\tablenotemark{g} \dotfill & $\hatcurPPtheta{39}$          & $\hatcurPPtheta{40}$          & $\hatcurPPtheta{41}$          \\
~~~$\langle F \rangle$ ($10^{9}$\ergscmsq) \tablenotemark{h}
                          \dotfill    & $\hatcurPPfluxavg{39}$        & $\hatcurPPfluxavg{40}$        & $\hatcurPPfluxavg{41}$        \\ [-1.5ex]
\enddata
\tablenotetext{a}{
    Reported times are in Barycentric Julian Date calculated directly
    from UTC, {\em without} correction for leap seconds.
    \ensuremath{T_c}: Reference epoch of mid transit that
    minimizes the correlation with the orbital period.
    \ensuremath{T_{14}}: total transit duration, time
    between first to last contact;
    \ensuremath{T_{12}=T_{34}}: ingress/egress time, time between first
    and second, or third and fourth contact.
}
\tablenotetext{b}{
    Reciprocal of the half duration of the transit used as a jump
    parameter in our MCMC analysis in place of $\arstar$. It is
    related to $\arstar$ by the expression $\zrstar = \arstar
    (2\pi(1+e\sin \omega))/(P \sqrt{1 - b^{2}}\sqrt{1-e^{2}})$
    \citep{bakos:2010}.
}
\tablenotetext{c}{
    Values for a quadratic law, adopted from the tabulations by
    \cite{claret:2004} according to the spectroscopic (SME) parameters
    listed in \reftabl{stellar}.
}
\tablenotetext{d}{
    Error term, either astrophysical or instrumental in origin, added
    in quadrature to the formal Keck/HIRES RV errors such that
    $\chi^{2}$ per degree of freedom is unity. For \hatcur{41} we did
    not add a jitter term to the FIES/NOT RV errors because the formal
    errors for these observations exceeded the scatter in the RV
    residuals.
}
\tablenotetext{e}{
    Correlation coefficient between the planetary mass \mpl\ and radius
    \rpl.
}
\tablenotetext{f}{
    Planet equilibrium temperature averaged over the orbit, calculated
    assuming a Bond albedo of zero, and that flux is reradiated from
    the full planet surface.
}
\tablenotetext{g}{
    The Safronov number is given by $\Theta = \frac{1}{2}(V_{\rm
    esc}/V_{\rm orb})^2 = (a/\rpl)(\mpl / \mstar )$
    \citep[see][]{hansen:2007}.
}
\tablenotetext{h}{
    Incoming flux per unit surface area, averaged over the orbit.
}
\ifthenelse{\boolean{emulateapj}}{
  \end{deluxetable*}
}{
  \end{deluxetable}
}
%

\ifthenelse{\boolean{emulateapj}}{
  \begin{deluxetable*}{lccc}
}{
  \begin{deluxetable}{lccc}
}
\tabletypesize{\scriptsize}
\tablecaption{Orbital and planetary parameters for \hatcurb{39}--\hatcurb{41} allowing eccentric orbits\tablenotemark{a}\label{tab:planetparameccen}}
\tablehead{
    \multicolumn{1}{c}{} &
    \multicolumn{1}{c}{{\bf HAT-P-39b}} &
    \multicolumn{1}{c}{{\bf HAT-P-40b}} &
    \multicolumn{1}{c}{{\bf HAT-P-41b}} \\
    \multicolumn{1}{c}{~~~~~~~~Parameter~~~~~~~~} &
    \multicolumn{1}{c}{Value}                     &
    \multicolumn{1}{c}{Value}                     &
    \multicolumn{1}{c}{Value}                     
}
\startdata
\noalign{\vskip -3pt}
\sidehead{\Lc{} parameters}
~~~$\arstar$              \dotfill    & $\hatcurPPareccen{39}$             & $\hatcurPPareccen{40}$             & $\hatcurPPareccen{41}$             \\
~~~$\zrstar$              \dotfill    & $\hatcurLCzetaeccen{39}$\phn       & $\hatcurLCzetaeccen{40}$\phn       & $\hatcurLCzetaeccen{41}$\phn       \\
~~~$i$ (deg)              \dotfill    & $\hatcurPPieccen{39}$\phn          & $\hatcurPPieccen{40}$\phn          & $\hatcurPPieccen{41}$\phn          \\

\sidehead{RV parameters}
~~~$K$ (\ms)              \dotfill    & $\hatcurRVKeccen{39}$\phn\phn      & $\hatcurRVKeccen{40}$\phn\phn      & $\hatcurRVKeccen{41}$\phn\phn      \\
~~~$\sqrt{e} \cos \omega$ 
                          \dotfill    & $\hatcurRVrkeccen{39}$\phs          & $\hatcurRVrkeccen{40}$\phs          & $\hatcurRVrkeccen{41}$\phs          \\
~~~$\sqrt{e} \sin \omega$
                          \dotfill    & $\hatcurRVrheccen{39}$              & $\hatcurRVrheccen{40}$              & $\hatcurRVrheccen{41}$              \\
~~~$e \cos \omega$ 
                          \dotfill    & $\hatcurRVkeccen{39}$\phs          & $\hatcurRVkeccen{40}$\phs          & $\hatcurRVkeccen{41}$\phs          \\
~~~$e \sin \omega$
                          \dotfill    & $\hatcurRVheccen{39}$              & $\hatcurRVheccen{40}$              & $\hatcurRVheccen{41}$              \\
~~~$e$                    \dotfill    & $\hatcurRVecceneccen{39}$          & $\hatcurRVecceneccen{40}$          & $\hatcurRVecceneccen{41}$          \\
~~~$\omega$ (deg)         \dotfill    & $\hatcurRVomegaeccen{39}$\phn      & $\hatcurRVomegaeccen{40}$\phn      & $\hatcurRVomegaeccen{41}$\phn      \\
~~~RV jitter (\ms)        \dotfill    & \hatcurRVjittereccen{39}           & \hatcurRVjittereccen{40}           & \hatcurRVjittereccen{41}           \\

\sidehead{Secondary eclipse parameters}
~~~$T_s$ (BJD)            \dotfill    & $\hatcurXsecondaryeccen{39}$       & $\hatcurXsecondaryeccen{40}$       & $\hatcurXsecondaryeccen{41}$       \\
~~~$T_{s,14}$              \dotfill   & $\hatcurXsecdureccen{39}$          & $\hatcurXsecdureccen{40}$          & $\hatcurXsecdureccen{41}$          \\
~~~$T_{s,12}$              \dotfill   & $\hatcurXsecingdureccen{39}$       & $\hatcurXsecingdureccen{40}$       & $\hatcurXsecingdureccen{41}$       \\

\sidehead{Planetary parameters}
~~~$\mpl$ ($\mjup$)       \dotfill    & $\hatcurPPmlongeccen{39}$          & $\hatcurPPmlongeccen{40}$          & $\hatcurPPmlongeccen{41}$          \\
~~~$\rpl$ ($\rjup$)       \dotfill    & $\hatcurPPrlongeccen{39}$          & $\hatcurPPrlongeccen{40}$          & $\hatcurPPrlongeccen{41}$          \\
~~~$C(\mpl,\rpl)$
         \dotfill    & $\hatcurPPmrcorreccen{39}$         & $\hatcurPPmrcorreccen{40}$         & $\hatcurPPmrcorreccen{41}$         \\
~~~$\rhopl$ (\gcmc)       \dotfill    & $\hatcurPPrhoeccen{39}$            & $\hatcurPPrhoeccen{40}$            & $\hatcurPPrhoeccen{41}$            \\
~~~$\log g_p$ (cgs)       \dotfill    & $\hatcurPPloggeccen{39}$           & $\hatcurPPloggeccen{40}$           & $\hatcurPPloggeccen{41}$           \\
~~~$a$ (AU)               \dotfill    & $\hatcurPPareleccen{39}$           & $\hatcurPPareleccen{40}$           & $\hatcurPPareleccen{41}$           \\
~~~$T_{\rm eq}$ (K)        \dotfill   & $\hatcurPPteffeccen{39}$           & $\hatcurPPteffeccen{40}$           & $\hatcurPPteffeccen{41}$           \\
~~~$\Theta$ \dotfill & $\hatcurPPthetaeccen{39}$          & $\hatcurPPthetaeccen{40}$          & $\hatcurPPthetaeccen{41}$          \\
~~~$\langle F \rangle$ ($10^{9}$\ergscmsq)
                          \dotfill    & $\hatcurPPfluxavgeccen{39}$        & $\hatcurPPfluxavgeccen{40}$        & $\hatcurPPfluxavgeccen{41}$        \\ [-1.5ex]
\enddata
\tablenotetext{a}{
    Quantities and definitions are as in \reftabl{planetparam}, which gives our adopted values, determined assuming circular orbits. Here we do not list parameters that are effectively independent of the eccentricity.
}
\ifthenelse{\boolean{emulateapj}}{
  \end{deluxetable*}
}{
  \end{deluxetable}
}
%



\section{Discussion}
\label{sec:discussion}

\begin{figure*}[]
\ifthenelse{\boolean{emulateapj}}{
\epsscale{1.0}
}{
\epsscale{1.0}
}
\includegraphics[width=7.0in]{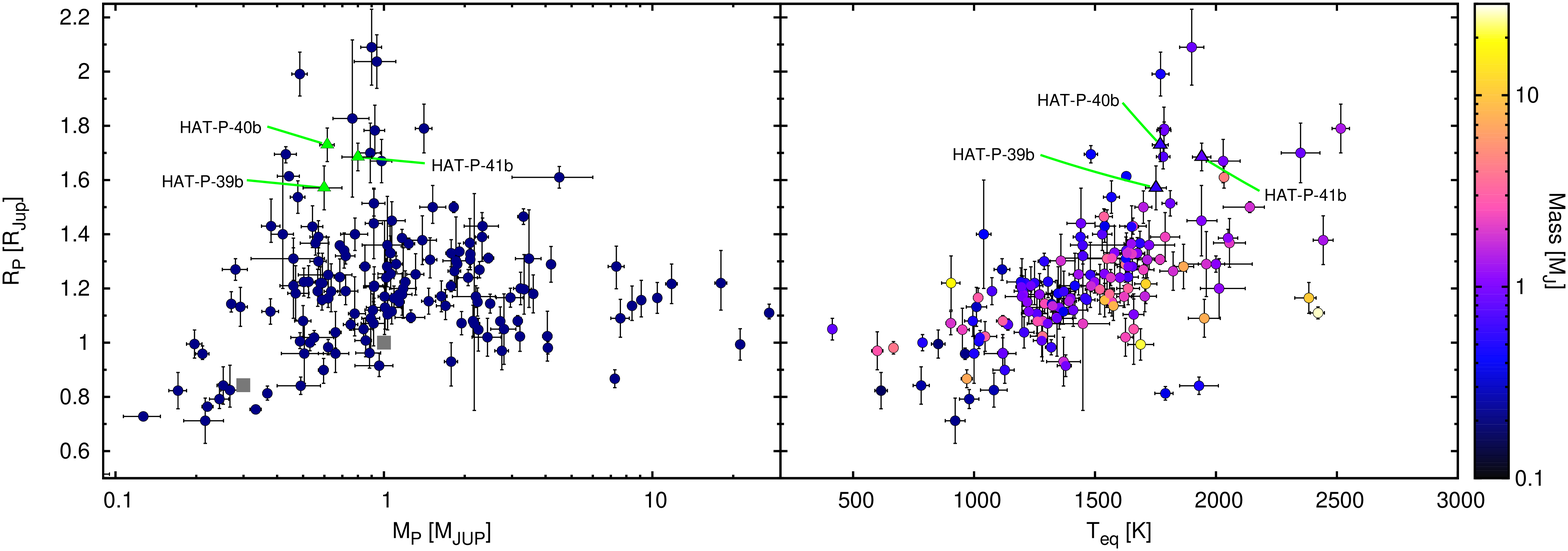}
\caption{
    (Left): Mass--radius diagram of TEPs. \hatcurb{39} through
  \hatcurb{41} are indicated. Jupiter and Saturn are marked with filled squares.
  (Right): Equilibrium temperature versus radius, the mass of each
  planet is indicated by the color of its symbol.
\label{fig:exomr}}
\end{figure*}

We have presented the discovery of three new transiting planets which
we show on mass--radius and equilibrium temperature--radius diagrams
in \reffigl{exomr}. As seen in the mass--radius diagram planets
generally have radii with $0.6\,\rjup < R < 1.5\,\rjup$ over a broad
mass-range spanning over two orders of magnitude, except for in the
range $0.4\,\mjup < M < 1.5\,\mjup$ where planets are found with radii
as large as $\sim 2.0\,\rjup$ (if WASP-12b is excluded, then the mass
range is $0.4\,\mjup < M < 1.0\,\mjup$). Applying the
Kolmogorov-Smirnov test \citep[e.g.][]{press:1992}, we find that there
is only a $0.7\%$ chance that the masses of planets with $M >
0.4\,\mjup$ and $R > 1.5\,\rjup$ are drawn from the same distribution
as the masses of planets with $M > 0.4\,\mjup$ and $R <
1.5\,\rjup$. The three planets presented here fall in the population
of large radius ($R > 1.5\,\rjup$), sub-Jupiter-mass planets which we
refer to as highly inflated planets.

As has been repeatedly noted \citep[the earliest reference
  being][]{guillot:2005} the radii of close-in gas-giant planets are
strongly correlated with the degree of irradiation (variously traced
by the planet equilibrium temperature estimated by adopting a constant
albedo, typically zero, for all planets, and making an assumption
about the heat redistribution, or traced by the bolometric surface
flux). As is evident in \reffigl{exomr}, the degree to which planets
are inflated depends on their masses, with lower mass planets
showing a stronger correlation between temperature and radius. This
has also been previously noted \citep[e.g.][]{enoch:2012} and has
been taken as evidence for some theoretical models of the inflation
process \citep[e.g.][]{batygin:2011,laughlin:2011}. The planets
presented here generally follow the established empirical trends,
though they are somewhat more inflated than other planets with
comparable equilibrium temperatures and semimajor axes. The empirical
relation given by \citet{enoch:2012}, which gives a prediction for the
radius as a function of $T_{\rm eq}$ and $a$ for planets with
$0.5\,\mjup < \mpl < 2.0\,\mjup$, yields radii of $1.52\,\rjup$,
$1.63\,\rjup$, and $1.58\,\rjup$ for \hatcurb{39} through
\hatcurb{41}, respectively. The formula given by \citet{beky:2011},
which uses $T_{\rm eq}$ and [Fe/H] as indepedent variables and was
derived for planets with $0.3\,\mjup < \mpl < 0.8\,\mjup$ predicts
radii of $1.31\,\rjup$, $1.30\,\rjup$ and $1.37\,\rjup$. In all cases
the predicted radii are smaller than the measured values
(\hatcurPPrshort{39}\,\rjup, \hatcurPPrshort{40}\,\rjup, and
\hatcurPPrshort{41}\,\rjup), though the \citet{enoch:2012} relation,
which was determined including more recent discoveries, gives values
that are closer to the observations.

We find that \hatcur{41} has a neighbor which has near-IR photometry
consistent with it being a $0.7\,\msun$ star at the same distance as
\hatcur{41}. \hatcur{41} is thus one of number of hot Jupiter host
stars with close visually-resolved neighbors (e.g. HD~189733,
\citealp{bakos:2006}; HAT-P-1, \citealp{bakos:2007}; and many others),
though it is not known what fraction of these are physical
companions. Such companions may be responsible for driving planets
into close-in orbits via the Kozai Mechanism \citep{fabrycky:2007}, a
hypothesis which has gained traction recently with the discovery that
many close-in planets are on high obliquity orbits
\citep[e.g.][]{triaud:2010,albrecht:2012}, but a complete survey to
determine the frequency of companions to hot Jupiter host stars in a
statistically robust way has not yet been published.

Finally we note that all three of these planets are good targets for
measuring the Rossiter-McLaughlin effect
\citep[R-M;][]{rossiter:1924,mclaughlin:1924} as they orbit bright
stars with relatively high projected rotation velocities. For
\hatcurb{39} and \hatcurb{41} the expected R-M semiamplitude is over
$100\,\ms$. For \hatcurb{40} the signal amplitude is lower, but this
is compensated by the long duration of the transit. The three stars
are also positioned closely below (\hatcur{40}) and above (\hatcur{39}
and \hatcur{41}) the 6250\,K effective temperature threshold found by
\citet{winn:2010} to separate planets on orbits that are well aligned
with the spin axes of their host stars from planets that are on high
obliquity orbits \citep[see also][]{albrecht:2012}.

\acknowledgements 

\paragraph{Acknowledgements}
HATNet operations have been funded by NASA grants NNG04GN74G,
NNX08AF23G and SAO IR\&D grants. J.H.~acknowledges partial support
from NSF grant AST-1108686. G.\'A.B., Z.C. and K.P. acknowledge
partial support from NASA grant NNX09AB29G. GT acknowledges partial
support from NASA grant NNX09AF59G. We acknowledge partial support
also from the Kepler Mission under NASA Cooperative Agreement
NCC2-1390 (D.W.L., PI). G.K.~thanks the Hungarian Scientific Research
Foundation (OTKA) for support through grant K-81373. This research has
made use of Keck telescope time granted through NOAO (A201Hr, A289Hr,
A284Hr), NASA (N049Hr, N018Hr, N167Hr, N029Hr, N108Hr, N154Hr), and
the NOAO Gemini/Keck time-exchange program (G329Hr). This paper
presents observations made with the Nordic Optical Telescope, operated
on the island of La Palma jointly by Denmark, Finland, Iceland,
Norway, and Sweden, in the Spanish Observatorio del Roque de los
Muchachos of the Instituto de Astrofisica de Canarias. This paper uses
observations obtained with facilities of the Las Cumbres Observatory
Global Telescope.  The Byrne Observatory at Sedgwick (BOS) is operated
by the Las Cumbres Observatory Global Telescope Network and is located
at the Sedgwick Reserve, a part of the University of California
Natural Reserve System. Data presented in this paper are based on
observations obtained at the HAT station at the Submillimeter Array of
SAO, and the HAT station at the Fred Lawrence Whipple Observatory of
SAO. The authors wish to recognize and acknowledge the very
significant cultural role and reverence that the summit of Mauna Kea
has always had within the indigenous Hawaiian community. We are most
fortunate to have the opportunity to conduct observations from this
mountain.



\clearpage
\begin{thebibliography}{}

\bibitem[Albrecht et al.(2012)]{albrecht:2012}
Albrecht, S., Winn, J.~N., Johnson, J.~A., et al.\ 2012, submitted to
\apj, arXiv:1206.6105

\bibitem[Alonso et al.(2004)]{alonso:2004}
Alonso, R., et al.\ 2004, \apj, 613, L153

\bibitem[Alsubai et al.(2011)]{alsubai:2011}
Alsubai, K.~A., Parley, N.~R., Bramich, D.~M., et al.\ 2011, \mnras, 417, 709

\bibitem[Anderson et al.(2010)]{anderson:2010}
Anderson, D.~R., et al.\ 2010, \apj, 709, 159

\bibitem[Anderson et al.(2011)]{anderson:2011}
Anderson, D.~R., Collier Cameron, A., Hellier, C., et al.\ 2011, \aap, 531, 60

\bibitem[Anderson et al.(2012)]{anderson:2012}
Anderson, D.~R., Collier Cameron, A., Gillon, M., et al.\ 2012, \mnras, 422, 1988

\bibitem[Bakos et al.(2004)]{bakos:2004}
 Bakos, G.~\'A., Noyes, R.~W., Kov\'acs, G., Stanek, K.~Z.,
 Sasselov, D.~D., \& Domsa, I.~2004, \pasp, 116, 266

\bibitem[Bakos et al.(2006)]{bakos:2006}
Bakos, G.~\'A., P\'al, A., Latham, D.~W., et al.\ 2006, \apj, 641, L57

\bibitem[Bakos et al.(2007)]{bakos:2007}
Bakos, G.~\'A., Noyes, R.~W., Kov\'acs, G., et al.\ 2007, \apj, 656, 552

\bibitem[Bakos et al.(2010)]{bakos:2010} Bakos, G.~{\'A}., et al.~2010,
\apj, 710, 1724

\bibitem[Bakos et al.(2012)]{bakos:2012} Bakos, G.~{\'A}., Csubry, Z., Penev, K., et al.\ 2012, submitted to \pasp, arXiv:1206.1391

\bibitem[Barge et al.(2008)]{barge:2008} Barge, P., et al.\ 2008, \aap, 482, L17

\bibitem[Batygin et al.(2011)]{batygin:2011}
Batygin, K., Stevenson, D.~J., \& Bodenheimer, P.~H.\ 2011, \apj, 738, 1

\bibitem[B\'eky et al.(2011)]{beky:2011}
B\'eky, B., Bakos, G.~\'A., Hartman, J., et al.\ 2011, \apj, 734, 109

\bibitem[Bonifacio et al.(2000)]{bonifacio:2000}
Bonifacio, P., Monai, S., \& Beers, T.~C.\ 2000, \aj, 120, 2065

\bibitem[Borucki et al.(2010)]{borucki:2010}
Borucki, W.~J., et al.\ 2010, \apj, 713, L126

\bibitem[Buchhave et al.(2010)]{buchhave:2010}
Buchhave, L.~A., Bakos, G.~\'A., Hartman, J.~D., et al.\ 2010, \apj, 720, 1118

\bibitem[Butler et al.(1996)]{butler:1996} 
Butler, R.~P.~et al.~1996, \pasp, 108, 500

\bibitem[Cardelli et al.(1989)]{cardelli:1989} Cardelli, J.~A.,
  Clayton, G.~C., \& Mathis, J.~S.\ 1989, \apj, 345, 245

\bibitem[Carpenter(2001)]{carpenter:2001} Carpenter, J.~M.~2001, \aj, 121, 2851 

\bibitem[Charbonneau et al.(2009)]{charbonneau:2009} Charbonneau, D., et al.\ 2009, \nat, 462, 891

\bibitem[Claret(2004)]{claret:2004}
 Claret, A.~2004, \aap, 428, 1001

\bibitem[Collier Cameron et al.(2007)]{colliercameron:2007} Collier Cameron, A., et al.\ 2007, \mnras, 375, 951

\bibitem[Djupvik \& Andersen(2010)]{djupvik:2010} Djupvik, A.~A., \&
  Andersen, J.\ 2010, in ``Highlights of Spanish Astrophysics V''
  eds. J.~M.~Diego, L.~J.~Goicoechea, J.~I.~Gonz\'alez-Serrano, \&
  J.~Gorgas (Springer: Berlin), p.\ 211

\bibitem[Droege et al.(2006)]{droege:2006}
Droege, T.~F., Richmond, M.~W., \& Sallman, M.~2006, \pasp, 118, 1666

\bibitem[Eastman et al.(2012)]{eastman:2012}
Eastman, J., Gaudi, B.~S., \& Agol, E.\ 2012, submitted to \pasp, arXiv:1206.5798

\bibitem[Enoch et al.(2011)]{enoch:2011}
Enoch, B., Anderson, D.~R., Barros, S.~C.~C., et al.\ 2011, \aj, 142, 86

\bibitem[Enoch et al.(2012)]{enoch:2012}
Enoch, B., Collier Cameron, A., \& Horne, K.\ 2012, \aap, 540, 99

\bibitem[Fabrycky \& Tremaine(2007)]{fabrycky:2007}
Fabrycky, D., \& Tremaine, S.\ 2007, \apj, 669, 1298

\bibitem[Fortney et al.(2011)]{fortney:2011} Fortney, J.~J., Demory, B.-O., D\'esert, J.-M., et al.\ 2011, \apjs, 197, 9

\bibitem[F\H{u}r\'{e}sz(2008)]{furesz:2008} F\H{u}r\'esz, G.\ 2008, Ph.D. thesis, University of Szeged, Hungary

\bibitem[Guillot(2005)]{guillot:2005} Guillot, T.\ 2005, Annual Review of Earth and Planetary Sciences, 33, 493

\bibitem[Hansen \& Barman(2007)]{hansen:2007} Hansen, B.~M.~S., \& Barman, T.~2007, \apj, 671, 861 

\bibitem[Hartman et al.(2011)]{hartman:2011} Hartman, J.~D., Bakos,
  G.~\'A., Torres, G., et al.\ 2011, \apj, 742, 59

\bibitem[Hebb et al.(2009)]{hebb:2009} Hebb, L., et al.\ 2009, \apj, 693, 1920

\bibitem[Isaacson \& Fischer(2010)]{isaacson:2010}
Isaacson, H., \& Fischer, D.\ 2010, \apj, 725, 875

\bibitem[Kov\'acs et al.(2002)]{kovacs:2002}
Kov\'acs, G., Zucker, S., \& Mazeh, T.~2002, \aap, 391, 369

\bibitem[Latham(1992)]{latham:1992}
 Latham, D.~W.~1992, in IAU Coll.~135, Complementary Approaches to
 Double and Multiple Star Research, ASP Conf.~Ser.~32, 
 eds.~H.~A.~McAlister \& W.~I.~Hartkopf (San Francisco: ASP), 110

\bibitem[Latham et al.(2009)]{latham:2009} Latham, D.~W., et al.~2009, 
\apj, 704, 1107

\bibitem[Latham et al.(2010)]{latham:2010} Latham, D.~W., Borucki, W.~J., Koch, D.~G., et al.\ 2010, \apj, 713, L140

\bibitem[Laughlin et al.(2011)]{laughlin:2011}
Laughlin, G., Crismani, M., \& Adams, F.~C.\ 2011, \apj, 729, L7

\bibitem[Lucy \& Sweeney(1971)]{lucy:1971}
Lucy, L.~B., \& Sweeney, M.~A.\ 1971, \aj, 76, 544

\bibitem[Mandushev et al.(2007)]{mandushev:2007}
Mandushev, G., et al.\ 2007, \apj, 667, L195

\bibitem[McCullough et al.(2006)]{mccullough:2006}
McCullough, P.~R., et al.\ 2006, \apj, 648, 1228

\bibitem[McLaughlin(1924)]{mclaughlin:1924}
McLaughlin, D.~B.\ 1924, \apj, 60, 22

\bibitem[Noyes et al.(1984)]{noyes:1984} Noyes, R.~W., Hartmann, L.~W.,
Baliunas, S.~L., Duncan, D.~K., \& Vaughan, A.~H.~1984, \apj, 279, 763

\bibitem[Pollacco et al.(2006)]{pollacco:2006} Pollacco, D.~L., Skillen, I., Collier Cameron, A., et al.\ 2006, \pasp, 118, 1407

\bibitem[Press et al.(1992)]{press:1992} Press, W.~H., Teukolsky,
  S.~A., Vetterling, W.~T., \& Flannery, B.~P.\ 1992, Numerical
  Recipes in C. The art of scientific computing, 2nd edition. New
  York: Cambridge University Press. p 623.

\bibitem[Queloz et al.(2000)]{queloz:2000} Queloz, D., Eggenberger, A., Mayor, M., Perrier, C., Beuzit, J.~L., Naef, D., Sivan, J.~P., \& Udry, S.\ 2000, \aap, 359, L13

\bibitem[Quinn et al.(2012)]{quinn:2012} Quinn, S.~N., Bakos, G.~\'A., Hartman, J.~D., et al.\ 2012, \apj, 745, 80

\bibitem[Rossiter(1924)]{rossiter:1924} Rossiter, R.~A.\ 1924, \apj, 60, 15

\bibitem[Schlegel et al.(1998)]{schlegel:1998}
Schlegel, D.~J., Finkbeiner, D.~P., \& Davis, M.\ 1998, \apj, 500, 525

\bibitem[Siverd et al.(2012)]{siverd:2012}
Siverd, R.~J., Beatty, T.~G., Pepper, J., et al.\ 2012, arXiv:1206.1635

\bibitem[Skrutskie et al.(2006)]{skrutskie:2006} Skrutskie, M.~F., et 
al.~2006, \aj, 131, 1163

\bibitem[Snellen et al.(2009)]{snellen:2009} Snellen, I.~A.~G., et al.\ 2009, \aap, 497, 545

\bibitem[Smalley et al.(2012)]{smalley:2012} Smalley, B., Anderson, D.~R., Collier Cameron, A., et al.\ 2012, arXiv:1206.1177

\bibitem[Stetson(1987)]{stetson:1987}
 Stetson, P.~B.\ 1987, \pasp, 99, 191

\bibitem[Stetson(1992)]{stetson:1992}
 Stetson, P.~B.\ 1992, JRASC, 86, 71

\bibitem[Torres et al.(2002)]{torres:2002}
Torres, G., Neuh\"auser, R., \& Guenther, E.~W.\ 2002, \aj, 123, 1701

\bibitem[Torres et al.(2004)]{torres:2004}
 Torres, G., Konacki, M., Sasselov, D.~D., \& Jha, S.\ 2004, \apj, 614, 979

\bibitem[Torres et al.(2007)]{torres:2007}
 Torres, G.~et al.~2007, \apjl, 666, 121

\bibitem[Triaud et al.(2010)]{triaud:2010}
Triaud, A.~H.~M.~J., Collier Cameron, A., Queloz, D., et al.\ 2010, \aap, 524, 25

\bibitem[Udalski et al.(2002)]{udalski:2002}
 Udalski, A., Paczynski, B., Zebrun, K., et al.\ 2002, AcA, 52, 1

\bibitem[Valenti \& Fischer(2005)]{valenti:2005}
 Valenti, J.~A., \& Fischer, D.~A. 2005, \apjs, 159, 141

\bibitem[Valenti \& Piskunov(1996)]{valenti:1996}
 Valenti, J.~A., \& Piskunov, N.~1996, \aaps, 118, 595

\bibitem[Vogt et al.(1994)]{vogt:1994}
 Vogt, S.~S.~et al.~1994, Proc.~SPIE, 2198, 362

\bibitem[Winn et al.(2010)]{winn:2010}
Winn, J.~N., Fabrycky, D., Albrecht, S., \& Johnson, J.~A.\ 2010, \apj, 718, L145

\bibitem[Yi et al.(2001)]{yi:2001}
 Yi, S.~K.~et al.~2001, \apjs, 136, 417

\end{thebibliography}
\end{document}